\documentclass{scthesis}
\usepackage{psfig}
\usepackage[dvips]{epsfig}
\usepackage{axodraw}
\usepackage{pifont}
\usepackage{mathptmx}
\usepackage{graphicx}
\usepackage{psfrag}
\DeclareGraphicsExtensions{{}}

\newcommand{\be}{\begin{eqnarray}}
\newcommand{\ee}{\end{eqnarray}}
\newcommand{\bx}{{\bf x}}
\newcommand{\by}{{\bf y}}
\newcommand{\bz}{{\bf z}}
\newcommand{\br}{{\hat {\bf r}}}
\newcommand{\bs}{{\hat {\bf s}}}
\newcommand{\bt}{{\hat {\bf t}}}
\begin{document}

\submissionmonth{}
\submissionyear{2003}
\author{\bf {DIPANKAR CHAKRAVORTY} \\ SAHA INSTITUTE OF NUCLEAR PHYSICS \\ KOLKATA}
\title{NON-PERTURBATIVE STUDIES IN QUANTUM CHROMODYNAMICS}
\maketitle

%
%
%
\centerline{ \bf  ACKNOWLEDGEMENTS}
\vskip .5in
{\it It gives me great pleasure to acknowledge my sincere gratitude to
my supervisor Prof. A. Harindranath for
providing me all the inspiration, encouragement and guidance 
to carry out my research work and for his  prompt and sincere
 help whenever I needed it most.
 I thank him for his constant endeavor to show me 
what is happening and what can happen in the present and future ``light cone''.

I would also like to thank my collaborators Prof. Asit K. De and Prof. James P. Vary
for their active participation  and illuminating discussions that I had with them. 
 I am thankful to
Rajen Kundu and Asmita Mukherjee for their ungrudging cooperation at 
different levels of
my research program. 

It is a mammoth task to individually mention  the names of my numerous 
friends whose friendship I cherish.
I would like to thank  them all, 
 both in and outside of  Saha Institute, for their 
cooperation and help 
whenever I approached them, whether it be  academic or nonacademic. 
Last, but not the least, I would  like to thank the members of
my extended family, especially my parents who stood by me all along 
 my voyage to  this point.
}
\vskip 2.5cm
\noindent Bidhan Nagar, \hspace{8.6cm} Dipankar Chakrabarti\\
\noindent November, 2003 \hspace{8.2cm} Theory Group, SINP


%


\tableofcontents
\listoffigures
\listoftables

%
%

\chapter{Introduction} \label{intro}
\section{Motivation}
Four types of force govern the universe, namely, Strong, Electromagnetic, 
 Weak and Gravitation. 
One of the most challenging and 
least understood problems in modern particle physics 
is  the dynamics of the fundamental particles responsible for the strong
interactions. Quantum Chromodynamics (QCD), the SU(3) color gauge theory 
provides the theory of strong interactions. But even after more than 
thirty years of the formulation of QCD, it is not 
well understood and not yet solved.  The fundamental particles 
of this theory are called {\it quarks}. They can appear in six flavors 
{\it up, down, strange, charm, bottom and top} and in three colors, say, 
{\it red, green and blue}.  The quanta of the non-Abelian gauge field in QCD, 
called {\it gluon}, mediates color interactions between quarks.  
 Gluons are 
flavor blind but carry  color charges  and hence interact among themselves.
This can be contrasted with the Abelian gauge theory, Quantum Electrodynamics
(QED), which describes the theory of electromagnetic interactions. Photons
which mediate electromagnetic interactions between charged particles in QED
do not have any electric charge and hence cannot interact among
themselves.  
This non-Abelian property of the gauge fields in QCD makes the game 
completely different from QED. Again as the interactions are "strong" in QCD
(the relative strengths of the four types of interaction at hadronic scale
 are  $1$, $10^{-2}$, $10^{-7}$ and $10^{-39}$ ), 
the perturbative treatment which is very 
successful to solve QED,  is not applicable to solve the theory for 
strong interaction. 
The most important and challenging part of QCD is to understand and solve 
the low energy or large distance physics where it shows "confinement" and is
 completely nonperturbative in nature as discussed below.   

In nature, quarks can only be found in  colorless
bound states, mesons and baryons which are combinedly called hadrons. 
The color symmetry imposed on QCD says that
only colorless states can be physically realized and hence a free quark which
carries color charge cannot be observed. According to this postulate 
all hadrons are required to be in the singlet of color SU(3). It is just a 
kinematical constraint to eliminate non-singlet (colored) states. However, it is 
believed that the quark confinement is a dynamical consequence of QCD and thus it 
requires more attention and investigations.     
Owing to the non-Abelian nature of gluons QCD possesses an interesting
property called "asymptotic freedom" \cite{0asym} according to which   
the interactions between quarks
become weak at very high energy (or at short distance). 
Thus, QCD enjoys asymptotic freedom at short distances while shows
confinement at large distances (of the order of hadronic length scale).
One can address the
scattering experiments at very high energy (such as Deeply Inelastic 
Scattering (DIS), Deeply
Virtual Compton Scattering (DVCS), etc) through perturbative QCD.
But, investigation of low energy (or  long distance) physics 
(such as hadronic  bound state
problem, nucleon-nucleon interactions), where the coupling strength become large,   
is not possible by perturbative theory and one needs a non-perturbative tool to
compute the QCD bound state spectrum. It is also desirable to have
informations about the hadronic wavefunctions which are essential to 
 calculate the QCD observables such as form factors,
 structure functions, distribution functions, decay rates etc.  
We do not know how to exactly solve QCD. So, various approximations go into 
QCD calculations. But, no approximation is valid for all length scales of QCD.
 For example,
 perturbation theory which is applicable in high energy 
domain is not at all applicable for low energy calculations and  low 
energy models assume dominance of few particle states.
Again, the issues of 
confinement  and hadronic bound states are much more difficult to be addressed with 
proper confidence due to the lack of adequate nonperturbative methods. 
 To have control over
 realistic QCD calculations it is desirable to assess different nonperturbative 
methods in terms of their strengths and weaknesses. In this thesis, we will
 mainly concentrate  on this aspect for different nonperturbative methods.

Till date, the most practiced 
non-perturbative  method is the lattice gauge theory \cite{0LGT}. But
lattice gauge theory has its own difficulties and limitations. In this method,
without fixing any particular gauge,
one calculates different n-point Green functions by path integral formalism in
Euclidean space, but can not have any firsthand knowledge about the
bound state wavefunctions.     
To explore the structure of hadrons ( baryons or mesons) in terms of their 
constituent degrees of freedom, 
a straightforward way is to solve the Hamiltonian
eigenvalue equation 
\be
H \mid\Psi\rangle = E \mid \Psi\rangle \label{HE}
\ee
expanding the 
eigenstate into multiparticle Fock states. The conventional Fock state expansion 
becomes  intractable in equal-time framework (by equal-time framework we refer 
to the {\it instant form} of Hamiltonian dynamics with quantization surface given 
by $x^0=0$ \cite{0dirac})  due to the complicated vacuum
of the relativistic quantum field theory.    Again   
in the equal-time framework the  square root operator in 
$ E= \sqrt{{\vec P}^2 + M^2}$ brings in severe mathematical difficulties. 
Even if one can solve these problems once, the eigensolutions are found 
in the rest frame, finding the solutions in a moving frame is highly nontrivial.
This is because  boosts are interaction dependent (dynamical) operators 
and boosting the system is as complicated as solving the theory. 

An elegant way to avoid all the difficulties mentioned above is to choose the
light-front (also called light-cone) framework \cite{0dirac} where the quantization 
surface is chosen to
be the tangential plane to the light-cone. The light-cone or light-front coordinates
 are defined as
$$ x^\pm = x^0 \pm x^3, ~ x^\perp = \{x^1, x^2\} $$ 
and the quantization surface is now chosen to be light-front time $x^+ = 0$ instead 
of $x^0 = 0$ in equal-time. The operator conjugate to $x^+$ is the light-front 
Hamiltonian $P^-$ and  $P^+$ which is conjugate to $x^-$ is the light-front 
longitudinal momentum. 
(The detail of light-front coordinate system will be discussed in the next chapter.)
 The eigenvalue $E$ in Eq. (\ref{HE})
does not involve any square root operator and takes
the form $ E = \frac {{P^\perp}^2 + M^2}{P^+}$, the vacuum structure is
relatively simple  and the boosts are interaction independent (kinematical) 
operators. Thus, unlike the usual equal-time Hamiltonian formalism,  
it is possible to
have a frame independent (boost invariant) description of the bound 
state wavefunctions
in the light-front Hamiltonian approach. Since the vacuum state of the 
free Hamiltonian
is also an eigenstate of the full QCD Hamiltonian, the Fock space expansion on this
vacuum  provides a complete relativistic many-body basis for diagonalization of the
full QCD Hamiltonian. Again, as the many-body states are also high energy states
(see the discussion on light-front in Chapter \ref{lfcoords}), one can hope to have few
body description of  the bound states and can reconcile relativistic field theory
like QCD in one hand with constituent quark model (CQM) in the other hand. [The CQM
was motivated by the hadron spectroscopy. 
According to CQM, mesons are made of a quark and an antiquark and baryons  are  
made of three quarks or three antiquarks bounded by  some empirical 
(phenomenological) potential.]

Two light-front non-perturbative methods that one can take up to address the
 QCD bound
 state problem are
Similarity Renormalization Group (SRG) approach \cite{0srg1,0srg2} to construct an 
effective theory with few constituents and the light-front transverse lattice (LFTL)
 approach \cite{0lftl1} which is a 
clever combination of lattice gauge theory and light-front Hamiltonian formalism. 
We will devote this thesis to investigate these two non-perturbative methods in 
the context of meson bound state problem.

In (3+1) dimensions, 
Euclidean lattice gauge theory predicts linear confinement for quarks while the 
lowest order calculation in SRG scheme produces logarithmic confinement which again
 violates rotational symmetry. Automatically, questions come. {\it Is the confinement
 produced  by SRG an artifact of the scheme?
What type of confinement does it produce in (2+1) 
dimensions?  Does that also violate rotational symmetry? }
In light-front field theory ultraviolet (UV) divergence issue is complicated due to
different power counting rule on the light-front.    In (2+1) dimensions this issue 
becomes simplified due to the absence of ultraviolet divergences except in mass
corrections. Since in (2+1) dimensions QCD is superrenormalizable, the coupling 
does not run 
and one can keep the coupling arbitrarily small and study the
structure of the bound states in a weakly coupled theory.
In (2+1) dimensions one component of the gauge field remains dynamical
and  one
can systematically study the effects of dynamical gluons without additional 
complications of (3+1) dimensions. One can also hope to    
enlarge the Fock space sector and investigate
their effect on restoring Lorentz invariance. It is expected that such
investigations are more viable in (2+1) dimensions compared to (3+1) dimensions
due to less severe demand on computational resources.  A major part of this work is 
devoted to understand these issues in (2+1) dimensional QCD using SRG scheme.  SRG is
a modification over Bloch effective theory \cite{0bleff}. A study of meson bound 
state problem using
Bloch effective Hamiltonian can serve as benchmark to assess the strengths and 
weaknesses of SRG approach.

As we have already 
mentioned, another nonperturbative approach in light-front framework  is 
the light-front transverse lattice formalism. In this approach, one keeps the 
light-front time ($x^+$) and longitudinal direction  ($x^-$)  continuous while the 
transverse plane ($x^\perp =(x^1,x^2)$) is discretized on a square lattice. In the 
light-front field theories, UV divergences come only from small transverse 
separations.
 Lattice provides the gauge invariant UV cutoff on the transverse space. 
Thus, in the 
light-cone gauge $A^+=0$, one can still preserve $x^-$ independent residual gauge
 invariance on the transverse plane.  
In this approach one avails the advantages of  gauge invariant UV cutoff coming 
from the lattice on the transverse plane and the  beautiful features of light-front 
framework. But, it is well known  that formulation of fermions on a lattice is 
complicated due 
to the notorious problem of species doubling. In the usual (Euclidean or Minkowski) 
lattice gauge theory extra species of fermions are generated from the corners of the 
Brillouin zone. {\it Is there any fermion doubling on light-front transverse lattice?
 If yes, do they also come from the corners of the Brillouin zone as usual lattice?
 What are the 
possible ways to remove the doublers on the transverse lattice? }
In usual  lattice gauge theory there are rigorous theorems and anomaly arguments 
regarding fermion doubling.  In standard
lattice gauge theory, some chiral symmetry needs to be broken in the
kinetic part of the action to avoid the doublers. In the light-front chirality is 
the same as helicity even for a massive fermion. The constraint equation for fermion 
in the light-front field theory violates the usual chiral symmetry.
 Here one should ask the question,
{\it is it still possible to relate light-front chirality and fermion doubling on the
 transverse lattice?  Since the notion of light-front chirality is different from 
usual chirality, is there any  way of formulating fermions on a light-front 
transverse lattice without generating extra species of fermions?}  There are many 
such questions one has to answer in order to use this method  as a practical tool for 
QCD calculations.  Our work shows that one can exploit the 
constraint
 equation for the fermionic field to formulate fermions in two different ways 
on the transverse lattice. In one formulation where we use forward and backward 
lattice derivatives without spoiling the Hermiticity of the Hamiltonian  doublers 
do not appear.  In the other way of formulating fermions people use symmetric lattice 
derivatives and encounters  doublers. We have investigated both ways of 
formulating fermions on the transverse lattice and tried to understand the origin of 
doublers, possible ways to remove them and  the symmetry relevant for doubling.  
Only when one understands the properties of fermions on the transverse lattice, 
 realistic QCD calculations become viable.  

Since we have two 
 two possible ways of formulating 
fermions on a
light-front  transverse lattice (a) with forward and backward lattice
derivatives and (b) with symmetric lattice derivative, we need a comparative study of 
these two approaches 
in order to decide which one is best suited for QCD calculations. We take meson bound
 state problem in (3+1) dimensions for this purpose. 
Again we focus on the comparative analysis of the 
strengths and weaknesses of the different fermion formulations to deal with different
 QCD interactions rather than fitting data.

One important issue in nonperturbative analysis is the numerical procedure. Since 
analytic solutions  of the nonpertubative bound state equations are not possible, 
one needs to solve them numerically. In Hamiltonian formalism, we need to 
diagonalize the Hamiltonian numerically in a suitable basis. There are 
several numerical procedures 
to diagonalize a matrix. When one deals with a theory like QCD, several 
complicated interactions come with different ultraviolet and infrared singularities.
It is very important to know the efficiency of the numerical procedures in handling 
singular interactions. In the meson bound state problem using SRG approach we use 
Gauss Quadrature  method to convert the integral equations into a matrix eigenvalue 
problem and discuss the efficiency of this method in handling different 
infrared singular
 terms.
 Due to longitudinal dynamics, in the bound state equation for transverse lattice
 Hamiltonian, linear light-front infrared divergences and 
logarithmic infrared divergences in self energy diagrams arise and one
 need to
 add counterterms to cancel them on the computer. We also discuss the efficiency of 
counterterms to cancel the divergences on a computer. 

\section{Organization of the thesis}
The main objective of our work is to assess different nonperturbative approaches in the 
light-front formalism. It is very important to know the strengths and weaknesses of 
different approaches when one wants to do any nonperturbative QCD calculation. 
We investigate the 
meson bound state problem in light-front QCD using different nonperturbative methods.
 Whenever it is possible, we make comparative studies of different
approaches. 
 We should again emphasize 
that our main aim is not to  fit data but to assess the different nonperturbative 
approaches in terms of their strengths and weaknesses to have control over 
the QCD calculations.

In Chapter \ref{lfcoords}, we discuss the basic features of light-front
field theory which will heavily be  used in this thesis.
Then, we embark on detailed investigations of the meson bound state problem
in light-front QCD.
We can broadly categorize the thesis into two parts, namely, (1) SRG approach and 
(2) transverse lattice formalism.  

In the SRG scheme,  starting from a bare cutoff Hamiltonian and by some similarity 
transformations one arrives 
at a low energy  effective Hamiltonian which is band diagonal in Fock basis.  
The effective Hamiltonian is then diagonalized nonperturbatively. 
We have already mentioned that SRG is a 
modification over Bloch effective theory and a study of Bloch theory can serve as a benchmark
to assess the SRG approach.
 In Chapter \ref{chapbloch}, we investigate the meson bound state problem with Bloch
 effective Hamiltonian in (2+1) dimensions. We start with a brief discussion on the basic 
ideas about the effective field theory (EFT). We observe that Bloch theory is 
infected with infrared
 divergences  in (2+1) dimensions and as one cranks up the strength of the coupling 
constant 
eigenvalues diverge bringing instability in the system. Thus we define a reduced model 
which is free from the divergences but still has confinement in the lowest 
nontrivial order of the expansion in coupling constant. 
This allows us the opportunity to study
 the manifestation   and possible violation of rotational
 symmetry in the context of light-front field theory.

After having the knowledge of difficulties and shortcomings of Bloch effective theory in 
the context of (2+1) dimensional meson bound state problem, we investigate the 
same meson bound state problem using similarity renormalization approach in Chapter 
\ref{chapsrg}. After a brief review of renormalization approach of constructing a 
low energy effective theory, we move on to the detailed analysis of SRG scheme. We 
immediately see the improvements due to SRG over Bloch effective theory. The bound state 
integral equation is converted into a matrix diagonalization problem by Gauss Quadrature
 method. The Gauss Quadrature method is  quite efficient in handling divergences. 
We find that SRG produces linear confinement in transverse direction but only square
 root potential in the longitudinal direction. Thus, it severely violates the rotational 
symmetry.  Our results show that higher order calculations are essential to investigate 
whether rotational invariance can be restored or not.
By tuning the similarity cutoff and the quark mass
we also study the interplay between SRG generated confining interaction and the 
Coulomb interaction which in (2+1) dimensions gives rotationally invariant logarithmic 
confinement. 
 
In Chapter \ref{tlchap1}, we introduce the other promising nonperturbative approach in 
the light-front framework, namely, light-front transverse lattice (LFTL). We have already
 emphasized  the problem of fermion doubling in usual lattice gauge theory and the 
relevant questions on the transverse lattice  one should ask.  In this chapter, we propose a 
new   method of putting fermions on a light-front transverse lattice which is free from 
fermion doubling. In this approach we use forward and backward lattice derivative in such a 
way that the Hermiticity of the action is not spoiled. In that case an irrelevant helicity 
flip interaction survives in the free field limit. We also discuss the  violation 
of rotational symmetry on the transverse lattice.
Then we discuss the transverse lattice
 Hamiltonian with symmetric lattice derivative where one encounters doublers.  Our results 
show that the origin of doublers on the transverse lattice is not the same as usual lattice 
gauge theory. Here, the doublers appear due to decoupling of even and odd sub-lattices. 
The removal of doublers  by two different ways, namely staggered fermion formulation and 
Wilson fermion formulation are discussed. We identify a even-odd helicity flip 
symmetry on the transverse lattice relevant for fermion doubling.

Once we understand the properties of fermions, in Chapter \ref{tlchap2}
we investigate the meson bound state problem 
in (3+1)  dimensions with  fermions formulated with forward and backward
 lattice derivatives and fermions formulated with symmetric lattice derivative.  
In the case of symmetric lattice derivative, we add  a Wilson term to remove the doublers.
In our investigation we use one link approximation, i.e., quark and antiquark at most 
can sit one lattice spacing apart on the transverse plane.  We compare and contrast the two 
ways of fermion formulation in the context of meson bound state problem.
 The major difference between the two approaches is that, 
with forward and backward lattice derivatives,  hopping of quark (or antiquark) 
in the transverse plane with  helicity flip interferes with helicity non flip hopping, while 
there is no such interference with symmetric lattice derivatives. The consequence of this 
interference in the spectrum is also studied.

Summary and conclusions are given in Chapter 7.
To ease the reading, we provide references at the end of each chapter. 
Several appendices are provided to clarify the notations and formalisms used 
and also to elucidate different intermediate steps.


\chapter{Some Basic Features of Light-Front Field Theory}\label{lfcoords}
In this Chapter, we  introduce the basic features of light-front
field theory, in the context of light-front QCD (LFQCD).  Since in the next few
chapters we will heavily use the light-front coordinates and the features of 
field theory in this formalism, it is instructive to discuss  
its advantages and the
problems that one must understand for a successful 
practical application of the theory. As, in Chapter \ref{intro},
we have already emphasized the advantages of 
using Hamiltonian formalism to investigate the hadronic bound state problem and as we 
will perform all our investigations in the light-front Hamiltonian approach,
here we shall only be concerned with the Hamiltonian formulation of
LFQCD.
 
In 1949, Dirac \cite{0adirac} showed that there are   three 
independent parametrizations of the space and time that can not be mapped
 on each other by a Lorentz transformations and discussed 
three forms of Hamiltonian dynamics. 
In the equal-time Hamiltonian formulation of field theory, quantization
conditions in the form of commutator (or anticommutaor) of dynamical fields 
and their conjugate momenta
are specified on the space-like hypersurface $x^0=0$ and the 
Hamiltonian generates the time-evolution of the system (Dirac called it   
{\it instant form} as the kinematical part of the Lorentz group leaves the instant invariant). 
In the {\it front form},
the quantization conditions are specified on a light-like
hypersurface $x^+=x^0+x^3=0$ (called a light-front) and  the
light-front Hamiltonian generates the evolution for a new time ($x^+$). This
formulation  is known as the light-front Hamiltonian field theory. 
Another form that Dirac mentioned is the {\it point form} of Hamiltonian dynamics where 
the quantization hypersurface is given by the hyperboloid
$x^\mu x_\mu = \kappa^2$ with  $ x^0 > 0$ and $\kappa^2 >0$,  and
the Lorentz group leaves a point invariant. However, later on 
 two more possibilities of parametrization of the space and time were found
\cite{0aLeuSt}. The quantization 
hypersurfaces for these two parametrizations are given by $x_0^2-x_1^2-x_2^2 = \kappa^2 >0$ 
with $ x_0 > 0$ and  $x_0^2-x_3^2 = \kappa^2 >0$ with $ x_0 > 0$. Among all  the parametrizations,
the front form has the largest stability group, the subgroup of the Poincare group that maps the
quantization hypersurface onto itself.

There is no well defined guideline to decide which parametrization one should use.
High energy experiments (e.g., deep inelastic scattering) probe the hadrons near 
the light-cone. It motivates people to use light-front parametrization of space and time 
to explore the QCD observables. One may hope that for highly relativistic systems
in which cases the world-line lies very close to the light-cone, physics will be more
transparent and  it will be relatively easy to extract them 
if one uses light-front field theory.

In the context of current algebra, Fubini and Furlan \cite{0aFF} introduced another notion
of Lorentz frame known as {\it Infinite-Momentum Frame (IMF)} as a limit of a 
reference frame 
moving with almost the speed of light. Weinberg \cite{0awein}  using old-fashioned 
perturbation theory for scalar meson showed that  vacuum structures  
become simplified in the infinite-momentum limit.  Later, Susskind  \cite{0asusk}
 established that 
although the Lorentz transformation required to arrive at IMF is evidently
singular ($\gamma=1/\sqrt{1-\frac{v^2}{c^2}}\rightarrow \infty ~\,{\rm as}
\,~v\rightarrow c$), the
singularity cancels in the calculation of physical objects (like Poincare
generators) and results in
an effective coordinate change given by
\begin{equation}
x^{\pm}=x^0 \pm x^3,\quad\quad {\bf x}^\perp=\{x^1,x^2\},
\label{lfcord}
\end{equation}
same as the light-front coordinate we defined in Chapter \ref{intro}. 
Thus, one can see the fact
that what one obtains after going through singular limiting procedure in IMF
is built in quite naturally in the light front field theory. That is why, light-front
 field theories are also sometimes referred as field theories in the infinite-momentum frame.
 But, we should reemphasize that the formulation here is as
prescribed by Dirac and has no connection with any singular limiting
procedure. For a review and exhaustive list of references on light-front field theories
see Ref. \cite{0alfb}.

\subsection{ LF dispersion relation}
The  
inner product between two four-vectors is defined on the light front as 
\be
x\cdot y = {1\over 2}x^+y^- + {1\over 2}x^-y^+ - x^\perp\cdot y^\perp.
\ee
In analogy with the light-front space-time variables,  the light-front four momenta are 
defined as
\be
k^{\pm}=k^0 \pm k^3,\quad\quad {\bf k}^\perp=\{k^1,k^2\},
\ee
 where $k^-$ being
conjugate to $x^+$ is the light front energy and $k^+$ which is conjugate to
$x^-$ is the light-front longitudinal
momentum. With the above definitions, the  dispersion relation, i.e., the
relation between light-front energy $k^-$ and the spatial components of 
momenta $(k^+,{\bf
k}^\perp)$, for an on mass-shell  particle of mass $m$, is given by, 
\be
 k^- =
{{ k_\perp}^2 + m^2\over k^+}\ .
\label{ekd}
\ee
One of the remarkable features of this relativistic dispersion relation 
is that there is no square root involved 
in contrast to  
the relativistic equal-time  dispersion relation $E=\sqrt{\vec{k}^2+m^2}$. 
This  provide great 
simplification when one tries to solve eigenvalue equation which we have
 already emphasized in chapter \ref{intro}.
 Secondly, the numerator in Eq. (\ref{ekd})
being always positive implies that the particles with positive light-front 
energy ($k^-$) always carry positive longitudinal momentum ($k^+ \ge 0$). 
As usual, the 
particles with negative $k^-$ which must have negative $k^+$
are mapped to antiparticles with positive $k^-$ and $k^+$. As a consequence,
we always have $k^+\ge0$ for real particles. 
Thirdly, $k^-$ becomes large for the
large value of $k^\perp$ as well as very small values  of $k^+$.
This makes light front renormalization aspects very different 
from the usual one.
Lastly, the dependence on the transverse momenta $k_\perp$ is just 
 like a nonrelativistic dispersion relation. 
We shall see later in this Chapter the crucial implications  
of  these novel features of the light-front dispersion relation in the light-front 
field theory. 

\subsection{  The light-front vacuum}
The above dispersion relation has profound consequence in the vacuum structure of 
light-front field theory.
 Vacuum state is
always an eigenstate of the longitudinal momentum  $\hat{P^+}\mid0\rangle =0$.
 The positivity condition of $k^+$ ($k^+\ge0$) implies that the 
vacuum $\mid 0\rangle$ is
either a no particle state or, at most can have particles with longitudinal
momenta exactly equal to zero.
Now, if we consider a cut-off theory where
longitudinal momentum is restricted to be $k^+\ge \epsilon$, the vacuum state  
$\mid 0\rangle$ becomes completely devoid of any particle
and therefore, an eigenstate of the full interacting Hamiltonian with zero
eigenvalue.
Thus, the light-front vacuum  becomes  {\it trivial}. It should be contrasted with
equal-time case  where the vacuum has highly complicated structure.
 In equal-time case, vacuum can contain  infinite number of particles moving 
with positive and negative momenta adding up to zero.
 Another aspect of the 
cutoff $k^+ \ge \epsilon $ is that  
 it automatically puts a restriction on the number of constituent particles
 a state with finite $P^+$ can have.  A composite state  with total longitudinal 
momentum $P^+$ 
 now can have at most $P^+/\epsilon$ constituents. This again simplifies the Fock
space expansion for the hadronic bound states.     
Again, as small $k^+$ means high energy (large $k^-$), one can hope to have a 
few body description for the low lying hadron states and reconcile QCD with CQM which
is beyond hope in equal-time formalism.

 On the other hand, a complicated vacuum structure  is
supposed to be responsible for spontaneous chiral symmetry breaking or
confinement in QCD. 
It seems that with the trivial vacuum structure in
light-front theory after removing the zero modes ($k^+ = 0$), we may lose these
important aspects in our theory. 
It should be emphasized
that we have not simply removed the zero modes from our theory. The
longitudinal momentum cut-off ($\epsilon$) 
should be removed from the theory at the end of
any calculation by adding necessary counter terms in the effective
Hamiltonian to render the observables
independent of $\epsilon$. Thus, we expect to get back all the effects of
zero mode as an effective interaction in the  Hamiltonian through
renormalization. 

\subsection{ Poincare generators in light-front}
In equal-time theory, out of the  ten Poincare generators (Hamiltonian ($P^0$), {\it 
three}  linear
momenta ($\vec P$), {\it three} angular momenta ($\vec J$) and {\it three} 
boosts 
($\vec K$)) {\it six} are kinematical $ \{{\vec P}, {\vec J}\}$, i.e., they do not 
depend on the dynamics (interactions) and other {\it four} are dynamical 
 $\{ P^0,  {\vec K}\}$.

In light-front, Poincare generators can be constructed in the same way as 
in equal-time case.
 Starting from Lagrangian density we construct the
energy momentum stress tensor $T^{\mu\nu}$ and from the stress tensor we construct the 
four-momentum $P^\mu$ and the generalized angular momentum $M^{\mu\nu}$ defined 
in the following way. 
\be
P^\mu = {1\over 2}\int dx^-d^2x^\perp T^{+\mu}\, ,\\
M^{\mu\nu} = {1\over 2}\int dx^-d^2x^\perp \big[ x^\nu T^{+\mu}-x^\mu
T^{+\nu}\big]\, .
\ee
 In light-front dynamics $P^-$ is the Hamiltonian and $P^+$ and
$P^i$ with $(i=1,2)$ are the longitudinal and transverse momenta. 
$M^{+-}=2K^3$ and $M^{+i}=E^i$ are the
boost operators and $M^{12}=J^3$ and $M^{-i}=F^i$ are generators for rotations. 
In light-front theory, boost operators ($K^3$ and $E^i$) are kinematical. Longitudinal
boost is like a scale transformation and the transverse boosts behave like
Gallilean boosts in the nonrelativistic theory.
To elucidate it further let us consider the boost along  the 3-axis ($K^3$)
 as an example. In equal-time, $K^3$ transforms the time ($x^0$) and the 3-axis ($x^3$) 
but leaves the transverse space invariant.  
\be
{\tilde x}^0= \gamma(x^0-\beta x^3),~~~
{\tilde x}^3= \gamma(x^3-\beta x^0),~~~
{\tilde x}^{1,2}= x^{1,2},
\ee
where $\beta={v\over c}$ and $\gamma={1\over \sqrt {1-\beta^2}}$.
From the above equations we see that $K^3$ changes  the quantization
surface $x^0=0$  and hence, $K^3$ is a dynamical generator 
in equal-time theory.  
Introducing the parameter $\phi$ such that $\gamma=\cosh\phi$ and 
$\beta\gamma=\sinh\phi$, we see that, in the light-front
\be
{\tilde x}^{+}= {\tilde x}^0+{\tilde x}^3=
e^{-\phi}~x^+,~~~{\tilde x}^{-}={\tilde x}^0-{\tilde x}^3 = e^{\phi}~x^-,~~~
{\tilde x}^{1,2}= x^{1,2}\, .
\ee
It clearly shows that $K^3$, which is  known as generator of 
longitudinal boost in light-front, behaves like 
a scale transformation. In particular, it keeps the
quantization surface $x^+=0$ invariant. Therefore, it is a kinematical
generator in light-front theory. 
On the other hand, two
rotations about transverse axes ($F^1$ and $F^2$)
 which are kinematical in equal-time case
become dynamical in light-front theory.  Thus, in the light-front theory,
we have {\it seven} kinemetcal ( 3 boosts, 3 translations and rotation about 3-axis),
and {\it three} dynamical (Hamiltonian and two rotations about the transverse axes)
 generators.   

Notice that the boost
generators form a closed algebra among themselves which is similar  to the generators of 
non-relativistic dynamics:
\be
\big[ E^1, E^2\big]=0,~~\big[ K^3, E^i\big]= iE^i,~~
\ee
and 
\be
\big[J^3, E^i\big]=i\epsilon^{ij}E^j,
\ee
where $\epsilon^{12} = -\epsilon^{21} =1$ and $\epsilon^{11} = \epsilon^{22} =0$.
Also, $F^1,~F^2$ and $J^3$  form a closed algebra.
\be
\big [ F^1, F^2 \big ] =0,~~~\big [J^3, F^i \big ] = i\epsilon^{ij}F^j.
\ee
For more details of the Poincare algebra
in light-front see Ref. \cite{0alfb}.

Since the  kinematical subgroup of the Poincare group enlarges and contains seven
generators in light-front theory, it is expected that defining   a 
system will be easier in the light-front theory as we  can fix more variables of
 the system irrespective of its dynamics. Moreover, since different set of
generators are kinematical in light-front compared to the equal-time theory,
it is worth pursuing this theory, for certain things difficult to study in
equal-time may just become simpler here. One such example is the feasibility
of representing the QCD-bound states in terms of just a few 
boost invariant multi-particle wave-functions in the Fock-space expansion, 
which we discuss next. 

\subsection{ Basic strategy for bound state problem} 
The starting point is the Hamiltonian eigenvalue equation
\be
  P^- \mid \Psi \rangle = {{P_\bot}^2 + {\cal M}^2\over P^+} \mid \Psi \rangle
\ee
where $ {\cal M}^2$ is the invariant mass-squared of the state $\mid \Psi\rangle$.
As it is mentioned already, trivial structure of light-front vacuum makes it 
feasible to study the hadronic bound states in Fock language.
Since the Fock-states form a complete basis, any state vector, in
principle, can be
expanded in terms of that basis introducing corresponding amplitude for each
Fock-basis. 
The bound state of a hadron on light-front can be simply 
expanded in terms of the Fock states as
\be
        \mid \Psi \rangle = \sum_{n,\lambda_i} \int' dx_i d^2\kappa_{\bot i} 
		 ~\mid  n, x_i, x_iP_{\bot}+ 
		\kappa_{\bot i}, \lambda_i \rangle ~\Phi_n 
		(x_i,\kappa_{\bot i}, \lambda_i) \, , \label{2lfwf} 
\ee
where $n$ represents $n$ constituents contained in the Fock state 
 $\mid n, x_i, x_i P_{\bot} + \kappa_{\bot i}, \lambda_i \rangle$, 
$\lambda_i$ is the helicity of the $i$-th constituent,
 $x_i$ is the fraction of the total longitudinal momentum 
carried by the $i$-th constituent, and $\kappa_{\bot i}$ is its relative 
transverse   momentum with respect to the center of mass frame. 
\be
   x_i = { p_i^+ \over P^+}~~, ~~~ \kappa_{i\bot} = p_{i\bot} - x_i P_{\bot}\, , 
\ee
with $p_i^+, \,p_{i\bot}$ being the longitudinal and transverse momenta
of the $i$-th constituent
 
\be  \label{2lfspc}
        \sum_i x_i = 1, ~~ {\rm and} ~~~ \sum_i \kappa_{\bot i} = 0
\ee 
 and $\int'$ denotes 
the integral over the space.
 $\Phi_n (x_i,\kappa_{\bot i},\lambda_i)$ 
is the amplitude of the Fock state $\mid n, x_i, x_iP_{\bot}+ \kappa_{\bot 
i}, \lambda_i \rangle $, i.e., the {\it multi-parton wave function},
which is boost invariant and satisfies the normalization condition: 
\be
        \sum_{n,\lambda_i} \int' dx_i d^2\kappa_{\bot i} 
		|\Phi_n (x_i,\kappa_{\bot i},\lambda_i)|^2 = 1.
\ee
For example, if we consider the meson bound state problem, then 
 after expanding the eigenstate in the basis of Fock states, the light-front bound state 
equation can be written as,
\begin{equation}
        \Big({\cal M}^2 - \sum_{i=1}^n { \kappa_{i\bot}^2 + m_i^2 \over x_i} 
		\Big) \left[\begin{array}{c} \Phi_{q{\bar q}} \\
                \Phi_{q{\bar q}g} \\ \vdots \end{array} \right]
                  = \left[ \begin{array}{ccc} \langle q{\bar q}
                | H_{int} | q{\bar q} \rangle & \langle q{\bar q} | H_{int}
                | q {\bar q}g \rangle & \cdots \\ \langle q{\bar q} g
                | H_{int} | q {\bar q} \rangle & \cdots & ~~  \\ \vdots &
                \ddots & ~~ \end{array} \right] \left[\begin{array}{c}
                \Phi_{q{\bar q}} \\ \Phi_{q{\bar q}g} \\ \vdots \end{array}
                \right] . \label{bdseq}
\end{equation}
Here $H_{int}$ is the interaction part of the light-front QCD 
Hamiltonian.

The expansion has  infinite number of terms and it is  
impossible to solve the bound state equation, Eq. (\ref{bdseq}), which is an 
infinite dimensional coupled
equation. To make any practical
calculation viable using Fock-expansion, one needs to truncate the
expansion at a suitable maximum particle number 
with the hope that a first few terms in the
expansion may give useful information. We know two important informations 
from the light-front dispersion
relation, one is that   longitudinal momentum $p^+$ is always positive and
secondly, states with  small $p^+$ are high energy states (large $p^-$). Since 
 $p^+$ is  always positive, constituents of a many particles state with a fixed 
longitudinal momentum $P^+$  carry  only small amounts of  $p_i^+$ and hence the state 
is of high energy. Since high energy states are weakly coupled one can hope that the
dynamics of the bound states is dominated by few particle states and multiparticle states
can be considered in a bound state perturbation theory in a consistent manner.  

 The situation should again be contrasted with equal-time approach.
As each Fock-state is obtained by operating
various creation operator(s) on the vacuum of the theory, if  the vacuum
already has a complicated structure (as is the case in equal-time theory),
which may contain arbitrary number of particles and thereby,  the vacuum itself 
needs a 
Fock-expansion. This, in effect, render the Fock-expansion in
equal-time theory meaningless for any practical application. This is not the
case in light-front theory due to the simplicity of the vacuum.

\subsection{ Renormalization aspects}
In light-front field theory in the Hamiltonian framework, the renormalization 
is a more 
complicated issue mainly due to the noncovariant structure of the theory
and is quite different compared to the usual covariant one. 
This is due to the fact that the {\it power 
counting in light front
is very different}. For a detailed discussion on light-front power counting, 
see the Ref. \cite{0aWilson}. Here we notice the fact that only transverse
directions $x^\perp$ carry the mass dimension, while the longitudinal
direction $x^-$ has no mass dimension. Thus, one has to treat transverse and
longitudinal directions  separately in determining the superficial degree of 
divergence of a divergent integral by power counting, in contrast to the covariant case
where all the space-time directions are treated democratically. This is
also evident in the single particle dispersion relation
$k^-={(k^\perp)^2+m^2\over k^+}$, which shows that there are two sources 
of divergences: $k^+\rightarrow 0+$ and $k^\perp\rightarrow\infty$. The
divergence coming from $k^+\rightarrow 0+$ is referred as infrared (IR) 
divergence, whereas $k^\perp\rightarrow\infty$ is known as the 
ultraviolate divergence (UV) in light-front theory. 
 
For the above reason, dimensional regularization, which is so elegant and
commonly used in covariant perturbation theory, is of very little importance in
light-front theory. Only in the transverse direction, one may use
dimensional regularization.   
Since we know that the lattice gauge theory provides gauge invariant UV cutoff,
another way to regulate the UV divergences is to discretize the transverse plane on
a square lattice \cite{0abrp}.
IR divergences
are also regularized by putting a small longitudinal momentum cut-off, 
which is equivalent to
using principal value prescription for the integration over longitudinal
momenta. Also the fact that the light-front theory being gauge fixed and
noncovariant, leads to new type of divergences like quadratic divergences 
(if we are using cut-off instead of transverse dimensional regularization) 
in mass renormalization or mixed divergences involving both IR and UV ones. To
remove these divergences one has to add counter terms to the canonical
Hamiltonian, which are often nonlocal and help to restore the invariance of
the theory that might be broken in the process of manipulation. For detailed
discussion on this subject,  see the
Refs. \cite{0aWilson, 0azhari2, 0azhari3, 0aPerry}. 
Another method specially designed to address the bound state problem in light-front, 
is that of
{\it similarity renormalization} introduced by Glazek and Wilson \cite{0aglaz} and Wegner
\cite{0aweg}, where
first an effective Hamiltonian is obtained perturbatively, by performing a
similarity transformation on the bare UV cutoff Hamiltonian.  Similarity 
renormalization  approach will be discussed in detail in  Chapter \ref{chapsrg} and 
in Appendix \ref{APsrg}. Notations and conventions are given in Appendix A.



\chapter{Bloch Effective Hamiltonian and Bound State Problem
 in $(2+1)$-Dimensional Light-Front QCD  }\label{chapbloch}

\section{Introduction }

One of the very well known techniques to extract  relativistic bound state solutions 
is the Bethe-Salpeter formalism \cite{1B_S}. 
Though Bethe-Salpeter equation is formally an exact equation for bound state
 problem,
provides a  covariant formalism 
 and successful in  quantitative understanding of bound states in different
models \cite{1naka}  
and positronium
 bound states in QED \cite{1murota},  the calculations in this approach are 
very complicated and 
almost out of control beyond the ladder approximation (for a review in the 
context of QCD see \cite{1LSG}).

The straightforward way to extract the relativistic and nonperturbative wavefunctions is
 the Hamiltonian approach where one solves the eigenvalue equation 
$ H |\Psi\rangle = E   |\Psi\rangle $.  But the 
  straightforward diagonalization of the Hamiltonian has two major problems
namely {\it (1) it involves infinitely many energy scales and (2) the rapid growth of 
the dimension of the Hamiltonian matrix with particle number}.  
In the spirit to diagonalize the Hamiltonian in a single step,
 one may 
implement Discretized Light Cone Quantization (DLCQ)  \cite{1Brodsky:1998de}.
 DLCQ has been quite successful in two dimensional models, but for QCD this
approach may be quite ambitious. 
Typical Hamiltonians of interest
couple low energy scales with high energy scales which results in
ultraviolet divergences. Furthermore, Hamiltonian couples every particle
number sector allowed by symmetries and at strong coupling, brute force 
particle truncation can fail miserably. 
 An alternative 
approach will be to use an  effective Hamiltonian that operates in a few 
particle basis. 

Effective field theory (EFT) \cite{1books_WG} relies on the assumption
that physics at a low
energy scale is insensitive to the microscopic details of the underlying
physics  at a high energy scale.  
  EFT provides a powerful framework 
to study  low-energy phenomena where one can replace the microscopic degrees
 of freedom and their interactions by effective macroscopic degrees of freedom and 
their effective interactions.
The basic procedure of  EFT is to separate out the 
important field components
 and redefine the theory with those fields within a  certain range of energy 
and momentum, so that,
acting on a limited Hilbert space it produces the same result as the original 
theory. In other words, to construct a low energy effective theory one needs to
 ``integrate out'' the high energy degrees of freedom (degrees of freedom above 
the scale considered) from the theory. The effective interactions are renormalized
 accordingly to incorporate the
effects of the  degrees of freedom above the cutoff. 
 An effective theory  describes the  main features  of the original theory
 below the 
scale one considers in a simpler way.   Effective field theory has wide
 applications in different wings of 
physics such as condense matter physics, nuclear physics, high energy physics, 
etc. 

\begin{figure}[h]
\begin{center}
\begin{picture}(360,120)(0,20)
\ArrowLine(30,60)(120,60)
\ArrowLine(30,100)(120,100)
\GlueArc(75,100)(20,0,180){4}{4}
\ArrowLine(150,60)(240,60)
\ArrowLine(150,100)(240,100)
\GlueArc(195,100)(20,0,180){4}{4}
\Gluon(170,60)(205,100){4}{4}
\ArrowLine(270,60)(360,60)
\ArrowLine(270,100)(360,100)
\GlueArc(315,100)(20,0,180){4}{4}
\Gluon(345,60)(305,100){4}{4}
\DashLine(80,52)(80,130){5}
\DashLine(185,52)(185,130){5}
\DashLine(320,52)(320,130){5}
\Text(75,40)[]{(a)}
\Text(195,40)[]{(b)}
\Text(320,40)[]{(c)}
\end{picture}
\caption[ $x^+$-ordered Hamiltonian diagrams for self energy and vertex corrections.]{ $x^+$-ordered 
Hamiltonian diagrams for (a) self energy,
 (b) and (c) vertex correction. Diagrams (b) and (c) are not allowed by 
Tamm-Dancoff
truncation as they involve two gluons in the intermediate state (shown by 
dashed lines).
\label{td}}
\end{center} 
\end{figure}
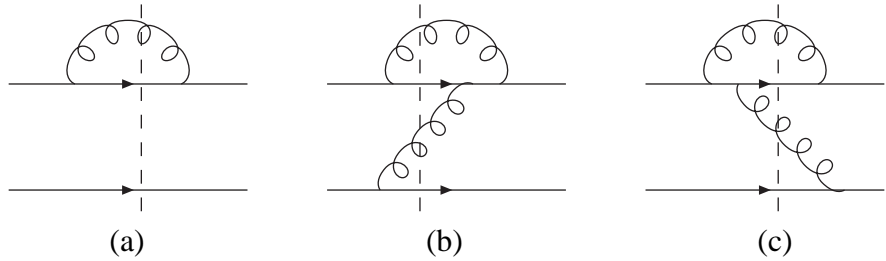

The strength or validity of an effective theory
depends   on how accurately the effective interactions mimic the  effects of the 
degrees of freedom thrown out from the theory.
There are many approaches  to  construct the effective Hamiltonian which acts 
on few particle states.  It is also well known that there are some or the
other drawbacks in  all 
 effective Hamiltonians.
As the light-front framework is very
 much suitable for Hamiltonian formalism and due to triviality of the vacuum 
one can expand the bound states in Fock basis states, several attempts have 
been made for nonperturbative diagonalization of the light-front Hamiltonian
for relativistic bound states (for a review see, Ref. \cite{1Brodsky:1998de}).
One of the first attempt was to implement the Tamm-Dancoff 
truncation \cite{1Perry:1990mz} or Bloch-Horowitz effective 
Hamiltonian \cite{1bheft}.
Though  Tamm-Dancoff  was successful in tackling
(1+1) dimensional gauge theories, its deficiencies become apparent when
attempts were made in (3+1) dimensions. First and foremost is the lack of
confinement in the case of QCD in the first non-trivial order. 
Second is the appearance of the bound state eigenvalue in the
energy denominators. This has two undesirable consequences. Firstly, a
light-front singularity of the type ${ 1 \over k^+}$, where $k^+$ is the
light front longitudinal momentum of the exchanged gluon, remains in the
bound state equation, which would have canceled if free energies appeared
in the energy denominators. Secondly, for example,  consider the meson
 bound state problem and truncate the Fock space with 
$q{\bar q}$ and $q{\bar q}g$ states.
 From the fermion self energy
contribution (Fig.\ref{td}(a)), in addition to the mass divergence
 another ultraviolet
divergence appears (for an example in the context of (3+1) dimensional Yukawa 
model see Ref. \cite{1Glazek:1993bs}) 
which contributes to the renormalization of the coupling. 
This contribution 
is also infrared divergent and can be identified as arising from fermion wave
function renormalization ($Z_2$). It is the Fock space truncation that has
 produced this unphysical divergence which would otherwise have been
 canceled by
 vertex renormalization ($Z_1$) (Fig.\ref{td}(b) and (c)) in a strict order by
 order perturbative calculation.  Thus,
 it severely violates the gauge invariance (gauge invariance
 demands $Z_1 = Z_2$) and one has to abandon the Tamm-Dancoff formalism
in more than $(1+1)$ dimensions. 
Another approach mentioned above is the Bloch-Horowitz effective Hamiltonian
 approach. But one of the major drawbacks of 
Bloch-Horowitz formula for effective Hamiltonian is that the transformation rule 
does not preserve the ortho-normalization condition of the wavefunctions which is 
very much important for observable calculations.
 Bloch effective  Hamiltonian \cite{1Bloch}(also reinvented  by Wilson
\cite{1Wilson:1970tp} in the context of renormalization group) is a modification of
 Bloch-Horowitz Hamiltonian 
and some of the deficiencies of the Bloch-Horowitz formalism are
absent in the Bloch effective Hamiltonian ( for a review see \cite{1Perry:1994mn}).

Bloch Hamiltonian has 
 two desired properties, 
namely, the effective Hamiltonian is {\it (1) Hermitian and (2) involves only 
unperturbed energies in the energy denominator}. The basic formalism of 
calculating Bloch effective Hamiltonian is 
discussed in  Appendix \ref{APbloch}.
Use of Bloch effective Hamiltonian eliminates two major problems of the
Tamm-Dancoff approach to gauge theories mentioned above.
 However, Bloch effective Hamiltonian involves  the undesirable  
 vanishing energy denominators.
To the best 
of our knowledge, Bloch effective Hamiltonian was never assessed in terms of
its strengths and weaknesses  in the study of
bound state problems in field theory. We first explore the 
Bloch effective Hamiltonian  in the context of meson bound states 
in $(2+1)$ dimensional light-front QCD.

A major feature of gauge theories on the light-front is severe light-front
infrared divergence of the type ${ 1 \over (k^+)^2}$ where $k^+$ is the
exchanged gluon longitudinal momentum which appears in instantaneous
four-fermion, two-fermion two-gluon, and four-gluon interactions. In
old-fashioned perturbation theory these divergences are canceled by
transverse gluon interactions. In similarity perturbation theory
\cite{1Similar} which is considered in the next Chapter, the
cancellation is only partial and singular interactions survive. Before
embarking on a detailed study of effective Hamiltonian in the similarity
renormalization approach which is a modification of the Bloch effective 
Hamiltonian, it is quite instructive to study the Bloch effective
Hamiltonian itself. 
The result of such a study serves as benchmark against which one can
evaluate the merits of similarity renormalization scheme.
This  also provides us quantitative measures on the
strengths and weaknesses of numerical procedures in handling singular
interactions (in the context of light-front field theory) 
on the computer. It is crucial to have such quantitative
measures in order to study the effects of similarity cutoff factors on the
nature of the spectrum.  

Just as the Tamm-Dancoff or the Bloch-Horowitz formalism, Bloch effective
Hamiltonian of QCD in the first non-trivial order also does not exhibit
confinement in (3+1) dimensions. Since one of our major concerns is the study
of spectra for confining interactions, we go to (2+1) dimensions. In this case, 
in the limit of heavy
fermion mass, a logarithmic confining potential emerges. 
There are several other reasons also to study light-front QCD in (2+1) 
dimensions. They
arise from both theoretical and computational issues which we discuss next.

First of all, issues related to ultraviolet divergence become more
complicated in the light-front approach since power counting is
different \cite{1Wilson:1994fk}  on the 
light front. We get products of ultraviolet
and infrared divergent  factors which complicate the renormalization
problem. Going to two space one time dimensions greatly simplifies this
issue due to the absence of ultraviolet divergences except in mass
corrections. An extra complication is that Fock space truncation introduces
extra ultraviolet divergences which complicate the situation in
non-perturbative bound state computations \cite{1Glazek:1993bs}. Such special
divergences do not occur in (2+1) dimensions. A third complication one
faces in (3+1) dimensions is that on enlarging the Fock space in a bound state
calculation, one soon faces the running of the coupling constant. At low
energy scales, the effective coupling grows 
resulting in a strongly coupled theory \cite{1Glazek:1998gt} making the weak
coupling approach with a perturbatively determined Hamiltonian unsuitable or 
making it mandatory to invent mechanisms like non-zero gluon mass to stop the
drastic growth \cite{1Wilson:1994fk}. In (2+1) dimensional QCD we do not face this
problem since the coupling constant is dimensionful in this
superrenormalizable field theory and does not run due to ultraviolet
divergence. We {\it can keep} the coupling arbitrarily small and study the
structure of the bound states in a weakly coupled theory. 

Secondly, in (1+1) dimensions, in the gauge 
$A^+=0$, dynamical gluons are absent and
their effect is felt only through instantaneous interactions between
fermions. Further, recall that in light front theory, vacuum is trivial. As
a result, the Fock space structure of the bound states are remarkably
simple. For example, the ground state meson is just a $q {\bar q}$ pair both
at weak and strong couplings. In contrast, in (2+1) dimensions, one component
of the gauge field remains dynamical and one
can systematically study the effects of dynamical gluons. Also note that (2+1)
dimensions are the lowest dimensions where glueball states are possible and
offer an opportunity to study their structure in the Fock space language
without additional complications of (3+1) dimensions.


A third reason deals with aspects of rotational symmetry. 
(2+1) dimensions offer the first opportunity to investigate violations of
Lorentz invariance introduced by various cutoffs (momenta and/or
particle number) in the context of bound state calculations. This is to be
contrasted with (1+1) dimensions where the sole Lorentz generator, namely
boost, is kinematical in light-front field theory. Since in (2+1) dimensions
we have a superrenormalizable field theory, violations introduced by
transverse momentum cutoffs are minimal. Thus in contrast to (3+1) dimensions, 
one can
study the violations caused by truncation of particle number alone and
longitudinal momentum cutoffs. It is
also conceivable that one can enlarge the Fock space sector and investigate
their effect on restoring Lorentz invariance. It is expected that such
investigations are more viable in (2+1) dimensions compared to (3+1) dimensions
due to less severe demand on computational resources.   



A fourth reason concerns similarity renormalization approach.
In (3+1) dimensions it has been shown that  similarity 
renormalization group
approach \cite{1Similar} to effective Hamiltonian in QCD leads to 
logarithmic confining interaction \cite{1Perry:1994kp}.  It is of interest 
to investigate corresponding effective
Hamiltonian in  (2+1) dimensions especially since the canonical Hamiltonian
already leads to logarithmic confinement in the nonrelativistic limit in
this case. It is also known that in (3+1) dimensions the confining part of the
effective Hamiltonian violates rotational symmetry. Does the violation of
rotational symmetry occur also in (2+1) dimensions? If so, how does it
manifest itself?   

In \cite{1DH1} we initiated a systematic study of light-front QCD in
(2+1) dimensions to investigate the various issues discussed above.

\section{Canonical Hamiltonian}


In this section we present the canonical light front Hamiltonian of (2+1) 
dimensional QCD. The Lagrangian density is given by
\be
{\cal L} =\Big [ - { 1 \over 4} (F_{\lambda \sigma a})^2 +
{\overline \psi} ( \gamma^\lambda (i\partial_\lambda + gA_\lambda) - m) \psi \Big ]
\ee
with 
\be
F^{\mu \lambda a} = \partial^\mu A^{\lambda a} - \partial^\lambda A^{\mu
a} + g f^{abc} A^{\mu b} A^{\lambda c}.
\ee
We have the equations of motion,
\be
\Big [ i \gamma^\mu \partial_\mu + g \gamma^\mu A_\mu - m \Big ] \psi &=&0,
\\
\partial_\mu F^{\mu \nu a} + g f^{abc} A_{\mu b} F^{\mu \nu}_c + g {\bar
\psi} \gamma^\nu T^a \psi & =&0.
\ee
Because we are in (2+1) dimensions, we immediately face an ambiguity  since
there are no $\gamma$ matrices in (2+1) dimensions. In the literature both
two component \cite{1bitar} and four component representation 
\cite{1Burkardt:1991xf} have been in use. 
For simplicity, we pick the two component representation. Explicitly,
\be
\gamma^0 =\sigma_2 = \pmatrix{ 0 & -i \cr
                               i & 0 }, ~~\gamma^1 = i \sigma_3 = 
\pmatrix{i & 0 \cr
         0 & -i},~~  \gamma^2 = i \sigma_1 = \pmatrix{0 & i \cr
                                                    i & 0 }.
\ee
\be
\gamma^{\pm} = \gamma^0 \pm \gamma^2,~~ \gamma^+=\pmatrix{ 0 & 0 \cr
                                                          2i & 0}, ~~
\gamma^- = \pmatrix{0 & -2i \cr
                    0 & 0}.
\ee
\be
\Lambda^\pm = { 1 \over 4} \gamma^\mp \gamma^\pm, ~~ \Lambda^+ = \pmatrix{1
& 0 \cr
0 & 0}, ~~ \Lambda^- = \pmatrix{0 & 0 \cr
                                0 & 1}.
\ee
Fermion field operator $ \psi^\pm = \Lambda^\pm \psi$.  We have
\be
\psi^+ = \pmatrix{ \eta \cr
                    0}, ~~ \psi^- = \pmatrix{0 \cr
                                             \xi}
\ee
where $ \xi$ and $\eta$ are one component fields. 
We choose the light front gauge $A^{+a}=0$.
From the equation of motion, we get the equation of constraint for fermion
\be
i \partial^+ \psi^- = \Big [ \alpha^1 (i \partial^1 + g A^1)+ \gamma^0 m \Big
] \psi^+.
\ee
Thus the fermion constrained field 
\be
\xi = { 1 \over \partial^+} \Big [ - (i \partial^1 + g A^1) + i m \Big ]
\eta.
\ee
 The equation 
of constraint for the gauge fields is
\be
- { 1 \over 2} (\partial^+)^2 A^{-a} = - \partial^1 \partial^+ A^{1a} 
- g f^{abc} A^{1b} \partial^+ A^{1c} - 2 g \eta^\dagger T^a \eta.
\ee
Using these equations of constraint, we eliminate $\psi^-$ and $A^-$ in favor
of dynamical field $\psi^+$ and $A^1$ and arrive at 
the canonical Hamiltonian given by
\be
H = H_0 + H_{int} = \int dx^- dx^1 ({\cal H}_0 + {\cal H}_{int}).
\ee
The free Hamiltonian density is given by
\be
{\cal H}_0 = \eta^\dagger { -(\partial^1)^2 + m^2 \over i \partial^+} \eta
+ { 1 \over 2} \partial^1 A^{1a} \partial^1 A^{1a}.
\ee
The interaction Hamiltonian density is given by
\be
{\cal H}_{int} = {\cal H}_1 + {\cal H}_2
\ee
with
\be
{\cal H}_1 & =& g \eta^\dagger A^1{\partial^1 \over \partial^+} \eta + g
\eta^\dagger {\partial^1 \over \partial^+}(A^1 \eta) \nonumber \\
&~&~- g m \eta^\dagger A^1 { 1 \over \partial^+}\eta
+ gm \eta^\dagger {1 \over \partial^+}(A^1 \eta) \nonumber \\
&~&~ -2g {1 \over \partial^+}(\partial^1 A^{1a}) \eta^\dagger T^a \eta 
+ g f^{abc} \partial^1 A^{1a} { 1 \over \partial^+}(A^{1b} \partial^+
A^{1c})
\ee 
and
\be
{\cal H}_2 &=& -2 g^2 \eta^\dagger T^a \eta 
\left({1 \over \partial^+}\right )^2 \eta^\dagger
T^a \eta + g^2 \eta^\dagger A^1 { 1 \over \partial^+} (A^1 \eta) \nonumber \\
&~& + 2 g^2 f^{abc} { 1 \over \partial^+} (\eta^\dagger T^a \eta) { 1 \over
\partial^+} (A^{1b} \partial^+ A^{1c}) \nonumber \\
&~& + {1 \over 2} g^2 f^{abc} f^{ade} { 1 \over  
\partial^+} (A^{1b} \partial^+ A^{1c}) 
{ 1 \over  
\partial^+} (A^{1d} \partial^+ A^{1e}) . 
\ee
The one component fermion field is given by
\be
\eta(x^+=0, x^-, x^1) = \int {dk^+ dk^1 \over 2 (2 \pi)^2 \sqrt{k^+}}
\Big [  b(k)e^{-ik \cdot x} + d^\dagger(k) e^{ik \cdot x} \Big]. \label{ffe}
\ee
The Fock operators obey the anti commutation relation
\be \{b(k), b^\dagger (q) \} = 2 (2 \pi)^2 k^+ \delta^2(k-q), ~~  
\{d(k), d^\dagger (q) \} = 2 (2 \pi)^2 k^+ \delta^2(k-q),
\ee
other anti commutators being zero.
 Note that in two component representation, 
light front fermions do not carry helicity in (2+1) dimensions.

In free field theory, the equation of motion of the dynamical field
$A^1$ is the same as that of a free 
massless scalar field \cite{1Binegar:1982gv} and hence we
can write
\be
A^{1}(x^+=0, x^-, x^1) = \int {dk^+ dk^1 \over 2 (2 \pi)^2 k^+} \Big [
a(k) e^{-ik\cdot x} + a^\dagger(k) e^{ - i k \cdot x} \Big ]. \label{bfe}
\ee
The Fock operators obey the commutation relation
\be
[a(k), a^\dagger(q)] = 2 (2 \pi)^2 k^+ \delta^2(k-q),
\ee
other commutators being zero.

We substitute the Fock expansions, Eqs. (\ref{ffe}) and (\ref{bfe}) into the
Hamiltonian and treat all the terms to be normal ordered. Thus we arrive at
the canonical Hamiltonian in the Fock basis.

\section{Bloch effective Hamiltonian in the meson  
sector and the bound state equation} 

In this section we evaluate the Block effective Hamiltonian to the lowest
non-trivial order for a meson state and derive  the effective bound state
equation. 
We define the $P$ space to be $q {\bar q}$ sector of the Fock space and $Q$
space to be the rest of the space (see Appendix \ref{APbloch} for details).
In the lowest non-trivial order, the Bloch effective Hamiltonian is given by
\be
\langle a \mid H_{eff} \mid b \rangle& = &\langle a \mid (H_0+H_{int}) \mid b
\rangle \nonumber \\ 
& ~&~~+{ 1 \over 2}
\sum_k \langle a \mid H_{int} \mid k \rangle \langle k \mid H_{int} \mid b 
\rangle \Big
[{ 1 \over \epsilon_a - \epsilon_k} + { 1 \over \epsilon_b - \epsilon_k}
\Big ].
\ee
The states $ \mid a \rangle$ and $\mid b \rangle$ are, explicitly,
\be
\mid a \rangle &=& b^\dagger(p_1, \alpha) d^\dagger (p_2, \alpha) \mid 0
\rangle , \nonumber \\
\mid b \rangle &=& b^\dagger(p_3, \beta) d^\dagger (p_4, \beta) \mid 0
\rangle,
\ee
where $p_1$, $p_2$ denote momenta and $\alpha$, $ \beta$ denote color which
is summed over. Explicitly, $p_1 = (p_1^+, p_1^1)$ etc.,   where $p_1^+$ is
the longitudinal component and $p_1^1$ is the transverse component. For simplicity of
notation, we will denote the transverse component of momenta without the
superscript 1.  
  
The free part of the Hamiltonian leads to the matrix element
\be \langle a \mid H \mid b \rangle = \left [ {m^2 + p_1^2 \over p_1^+ }
+ {m^2 + p_2^2 \over p_2^+} \right ] 2 (2 \pi)^2 p_1^+ \delta^2(p_1-p_3)
2 (2 \pi)^2 p_2^+ \delta^2(p_2- p_4) \delta_{\alpha \beta} .
\ee
From the four fermion interaction, we get the contribution
\be -4 g^2  (T^a T^a)_{\alpha \alpha} { 1 \over (p_1^+ -
p_3^+)^2} 2 (2 \pi)^2 \sqrt{p_1^+ p_2^+ p_3^+ p_4^+} \delta^2(p_1+p_2- p_3 -
p_4)~ \delta_{\alpha \beta}.
\ee 

Next we evaluate the contribution from the second order term. 
The intermediate state $ \mid k \rangle$ is any state in higher Fock
space, {\rm e.g.},
$q {\bar q} g,~q {\bar q} g g, ~q {\bar q} q {\bar q},$ and so on. 
In the lowest nontrivial order we take $ \mid k \rangle$
 a quark, anti-quark, gluon ($q {\bar q} g$) state. 
This intermediate state gives rise to both self energy and gluon exchange
contributions.  

The self energy contributions are
\be
&& g^2 ~C_f ~\delta_{\alpha \beta} ~p_1^+ 2 (2 \pi)^2 \delta^2(p_1 - p_3)
~p_2^+ 2 (2 \pi)^2 \delta^2(p_2 - p_4) \nonumber \\
&& \int { dk_1^+ dk_1 \over 2 (2 \pi)^2 (p_1^+ - k_1^+)} 
\left \{ -2 {(p_1 - k_1) \over (p_1^+ - k_1^+)} + {k_1 \over k_1^+} + 
{p_1 \over p_1^+} -i { m \over k_1^+} + i {m \over p_1^+} \right \} { 1 \over
{\cal E}_1} \nonumber \\
&&~~~~~~~~~~  \left \{ -2 {(p_1 - k_1) \over (p_1^+ - k_1^+)} 
+ {k_1 \over k_1^+} + 
{p_1 \over p_1^+} +i { m \over k_1^+} - i {m \over p_1^+} \right \} \nonumber
\\
&& +g^2 ~C_f ~\delta_{\alpha \beta} ~p_1^+ 2 (2 \pi)^2 \delta^2(p_1 - p_3)
~p_2^+ 2 (2 \pi)^2 \delta^2(p_2 - p_4) \nonumber \\
&& \int { dk_2^+ dk_2 \over 2 (2 \pi)^2 (p_2^+ - k_2^+)} 
\left \{ -2 {(p_2 - k_2) \over (p_2^+ - k_2^+)} + {k_2 \over k_2^+} + 
{p_2 \over p_2^+} -i { m \over k_2^+} + i {m \over p_2^+} \right \} { 1 \over
{\cal E}_2} \nonumber \\
&&~~~~~~~~~~  \left \{ -2 {(p_2 - k_2) \over (p_2^+ - k_2^+)} 
+ {k_2 \over k_2^+} + 
{p_2 \over p_2^+} +i { m \over k_2^+} - i {m \over p_2^+} \right \} ,
\ee
with
\be
{\cal E}_1 &=& {p_1^2 + m^2 \over p_1^+} - {m^2 + k_1^2 \over k_1^+} - {(p_1 - k_1)^2
\over (p_1^+ - k_1^+)}, \nonumber \\
{\cal E}_2 &=& {p_2^2 + m^2 \over p_2^+} - {m^2 + k_2^2 \over k_2^+} - {(p_2 - k_2)^2
\over (p_2^+ - k_2^+)}, 
\ee 
 and the color factor $C_f = (T^a T^a)={N^2-1 \over 2N}$ for $N$ number of colors.
The gluon exchange contributions are
\be
&& -g^2 ~C_f~ 2 (2 \pi)^2 \delta^2 (p_1 +p_2 - p_3 -
p_4) \sqrt{p_1^+ p_2^+ p_3^+ p_4^+} \nonumber \\
&&~~ \left \{  -2 {(p_1 - p_3) \over (p_1^+ - p_3^+)} + {p_3 \over p_3^+} + 
{p_1 \over p_1^+} -i { m \over p_3^+} + i {m \over p_1^+} \right \} 
\left \{  -2 {(p_1 - p_3) \over (p_1^+ - p_3^+)} + {p_2 \over p_2^+} + 
{p_4 \over p_4^+} +i { m \over p_2^+} - i {m \over p_4^+} \right \}  \nonumber
\\
&&~~{1 \over 2}{\theta(p_1^+ - p_3^+) \over (p_1^+ - p_3^+)} \left \{
{1 \over {m^2 + p_4^2 \over p_4^+} - {(p_1 - p_3)^2 \over (p_1^+ - p_3^+)} -
{m^2 + p_2^2 \over p_2^+}} +  
 {1 \over {m^2 + p_1^2 \over p_1^+} - {(p_1 - p_3)^2 \over (p_1^+ - p_3^+)} -
{m^2 + p_3^2 \over p_3^+}} \right \}\nonumber \\
&& -g^2 ~C_f~ 2 (2 \pi)^2 \delta^2 (p_1 +p_2 - p_3 -
p_4) \sqrt{p_1^+ p_2^+ p_3^+ p_4^+} \nonumber \\
&&~~ \left \{  -2 {(p_3 - p_1) \over (p_3^+ - p_1^+)} + {p_3 \over p_3^+} + 
{p_1 \over p_1^+} -i { m \over p_3^+} + i {m \over p_1^+} \right \} 
\left \{  -2 {(p_3 - p_1) \over (p_3^+ - p_1^+)} + {p_2 \over p_2^+} + 
{p_4 \over p_4^+} +i { m \over p_2^+} - i {m \over p_4^+} \right \}  \nonumber
\\
&&~~{1 \over 2}{\theta(p_3^+ - p_1^+) \over (p_3^+ - p_1^+)} \left \{
{1 \over {m^2 + p_2^2 \over p_2^+} - {(p_3 - p_1)^2 \over (p_3^+ - p_1^+)} -
{m^2 + p_4^2 \over p_4^+}} +  
 {1 \over {m^2 + p_3^2 \over p_3^+} - {(p_3 - p_1)^2 \over (p_3^+ - p_1^+)} -
{m^2 + p_1^2 \over p_1^+}} \right \}.
\ee

After the construction of $H_{eff}$ in the two particle space, we proceed as
follows. Consider the bound state equation
\be H_{eff} \mid \Psi \rangle = {M^2 +P^2 \over P^+} \mid \Psi \rangle
\ee
where $P^+$, $P$, and $M$ are the longitudinal momentum, the 
transverse momentum
and the invariant mass of the state respectively. The two particle ($q \bar q$) bound 
state $\mid \Psi \rangle$
is given by
\newpage          
\be 
\mid \Psi \rangle && = \sum_\beta ~\int {dp_3^+ dp_3 \over \sqrt{2 (2
\pi)^2 p_3^+}}~ \int {dp_4^+ dp_4 \over \sqrt{2 (2
\pi)^2 p_4^+}}~ \phi_2(P; p_3, p_4)~ b^\dagger(p_3, \beta) d^\dagger(p_4,
\beta) \mid 0 \rangle \nonumber \\
&& ~~~~~~~~~~~~~~~~ \sqrt{2 (2 \pi)^2 P^+} \delta^2(P-p_3 -p_4)
\ee
with the normalization 
\be
\langle \Psi(Q) \mid \Psi(P) \rangle =2(2\pi)^2 P^+ \delta^2(P-Q)
\ee
provided
\be
\int\int dp_1^+dp_1^1 \mid \phi_2(P; p_1,P-p_1)\mid^2 = 1. 
\ee
 We symbolically represent the above state as
\be
\mid \Psi \rangle = \sum_j \phi_{2j} \mid j \rangle.
\ee
Taking projection with the state
$ \langle i  \mid  = \langle 0 \mid d(p_2, \alpha) b(p_1, \alpha) $, 
we get the effective bound state equation,
\be {M^2 + P^2 \over P^+} \phi_{2i} = H_{0i} \phi_{2i} + \sum_j \langle i
\mid  H_{Ieff} \mid j \rangle ~ \phi_{2j}.
\ee
Introduce the internal momentum variables $ (x,k)$ and  $(y,q)$ via 
$p_1^+ = xP^+$, $p_1= xP+ k$, 
$ p_2^+ = (1-x)P^+$, $ p_2 = (1-x)P-k$, $p_3^+ = yP^+$, $p_3=yP+q$, $p_4^+ = 
(1-y)P^+$, $ p_4 = (1-y)P -q$ and the amplitude $\phi_2(P; p_1,p_2) ={1
\over \sqrt{P^+}} \psi_2(x, k) $.

The fermion momentum fractions $x$ and $y$ range from 0 to 1. To handle end
point singularities, we introduce the cutoff $ \epsilon \le x,y \le 1 $. This
does not prevent the gluon longitudinal momentum fraction $(x-y)$ from
becoming zero and we introduce the regulator $\delta$ such that $ \mid x-y
\mid \ge \delta $. To regulate ultraviolet divergences, we introduce the
cutoff $\Lambda$  on the relative transverse momenta $k$ and $q$. We remind
the reader that in the superrenormalizable field theory under study, only
ultraviolet divergence is in the fermion self energy contribution which we
remove by a counterterm before discretization.

The bound state equation is
\be
\Big [ M^2 - {m^2 + k^2 \over x (1-x)} \Big ] \psi_2(x,k) &=& { S} \
\psi_2(x,k) \ 
- 4 {g^2 \over 2 (2 \pi)^2}C_f \int dy dq ~\psi_2(y,q) ~{ 1 \over (x-y)^2}
\nonumber \\
&~& - {g^2 \over 2 (2 \pi)^2}C_f \int dy dq ~\psi_2(y,q) ~{ 1 \over 2}  
{ {\cal V} \over {\cal E}}. \label{ebe1}
\nonumber \\
\ee
The self energy contribution
\be
{ S} &=&- {g^2 \over 2 (2 \pi)^2 } C_f \Big [\int_0^x dy \int  dq
~ xy~ { \Big [ \Big ({q \over y} +{ k \over x} - {2 (k-q) \over (x-y)}
\Big )^2 + {m^2 (x-y)^2 \over x^2 y^2} \Big ] \over
(ky-qx)^2 + m^2 (x-y)^2} \nonumber \\
&~&+ \int_x^1 dy \int  dq
~ (1-x)(1-y) ~ { \Big [ \Big ({q \over 1-y} +{ k \over 1-x} + 
{2 (q-k
) \over (y-x)}
\Big )^2 + {m^2 (y-x)^2 \over (1-x)^2 (1-y)^2} \Big ] \over
[k(1-y)-q(1-x)]^2 + m^2 (x-y)^2} \Big ].
\ee 
The boson exchange contribution 
\be
{{\cal V} \over {\cal E}} &=& {\theta (x-y) \over (x-y)} \left [ 
{1 \over {m^2 + q^2 \over y}+ {(k-q)^2 \over (x-y)} - {m^2 + k^2 \over x}} 
+ {1 \over {m^2 + k^2 \over 1-x} + {(k-q)^2 \over x-y} - {m^2 +q^2 \over 1-y}}
\right ] \nonumber \\
&& ~~\times \Big [ K(k,x,q,y) ~ + ~i V_I  \Big ]
\nonumber \\
&~& + {\theta (y-x) \over (y-x)} \left [ {1 \over
{m^2 + k^2 \over x}+ {(q-k)^2 \over (y-x)} - {q^2 +m^2 \over y}}+  
{1 \over {m^2 + q^2 \over 1-y} + {(q-k)^2 \over y-x} - {m^2 +k^2 \over 1-x}}
\right ] \nonumber \\
&& ~~\times \Big [ K(q,y,k,x)~ + ~ i V_I  \Big ] ,
\ee
where
\be
K(k,x,q,y) =  \Big ( {q \over y} + {k \over x} - 
 {2(k-q) \over (x-y)} \Big )
\Big ( { q \over 1-y} + { k \over 1-x} + {2 (k-q) \over (x-y)} \Big )
- { m^2 (x-y)^2 \over x y (1-x) (1-y)},
\ee
\be
V_I = - { m \over x y (1-x) (1-y)} [ q (2-y-3x) + k(3y+x-2)].
\ee 
 

\section{Divergence Structure}

Now, let us  analyze  the divergence
structure of the effective bound state equation. We encounter both infrared
and ultraviolet divergences. 

\subsection{Ultraviolet Divergences}

First consider ultraviolet divergences. In the super renormalizable field
theory under consideration, with the terms appearing in the canonical
Hamiltonian as normal ordered, 
ultraviolet divergence is encountered only in the self energy contributions.
To isolate the ultraviolet divergence, we rewrite 
the self energy integrals as
\be
S & =& - {g^2 \over 2 (2 \pi)^2 } C_f \int_0^x dy \int_{-\Lambda}^{+\Lambda}  dq
   \left [ {(x+y)^2 \over x y (x-y)^2}- { 4 m^2 \over
(ky-qx)^2 + m^2 (x-y)^2} \right ] \nonumber \\
&~&~- {g^2 \over 2 (2 \pi)^2 } C_f \int_x^1 dy \int_{- \Lambda}^{+\Lambda}  dq
\nonumber \\
&&~~~~~~~~~  \left [ {(2 -x -y)^2 \over (y-x)^2 (1-x)(1-y)} -   
{4 m^2  \over
[k(1-y)-q(1-x)]^2 + m^2 (x-y)^2} \right ] .
\ee 
The first term inside the square brackets in the above equation is
ultraviolet divergent, which we cancel by adding an ultraviolet counterterm
given by
\be CT =  + {g^2 \over 2 (2 \pi)^2 } C_f  \int_{-\Lambda}^{+\Lambda} 
 dq \left [\int_0^x dy
    {(x+y)^2 \over x y (x-y)^2} + \int_x^1 dy{(2 -x -y)^2 \over (y-x)^2
(1-x)(1-y)} \right ] . \label{ct} 
\ee
After the addition of this counterterm, the bound state equation
is ultraviolet finite.


\subsection{Infrared Divergences}

The infrared divergences (IR) that appear in the bound state equation are of two
types: (1) light front infrared divergences that arise from the gluon
longitudinal momentum fraction $x_g=0$, (2) true infrared divergences that
arise from gluon transverse momentum $k_g=0$ and gluon longitudinal momentum
fraction $x_g=0$. The IR divergences of type (1) are generated  due to
 elimination of the constrained degrees of freedom.


\subsubsection{Cancellation of Light-front Infrared Divergences 
in the Effective Bound State Equation}

First consider light front infrared divergences.
The effective bound state equation, Eq. ({\ref{ebe1}), explicitly has a linear
light front infrared divergent term ${1 \over (x-y)^2}$ coming from
instantaneous gluon exchange. The most divergent
part of the numerator of the transverse gluon exchange term in this equation is
$-4{(k-q)^2 \over (x-y)^2}$. After combining the terms, the linear infrared
divergent term is completely canceled and the resultant effective bound
state equation takes the form
\be
\Big [ M^2 - {m^2 + k^2 \over x (1-x)} \Big ] \psi_2(x,k) &=&  S_1 \
\psi_2(x,k) 
- {g^2 \over 2 (2 \pi)^2}C_f 
\int dy dq ~\psi_2(y,q) \nonumber \\
&~&~~~~~~~~ \times {1 \over 2}\left [ {{\tilde V}_1 \over E_1}
+ {{\tilde V}_2 \over E_2}+ i V_I 
 \left ({1 \over E_1} + { 1 \over E_2} \right ) \right ].  
 \label{ebe2}
\ee 
The self energy contribution, made ultraviolet finite by the addition of the
counterterm is
\be
S_1 & =& + {g^2 \over 2 (2 \pi)^2 } C_f \int_0^x dy \int_{-\Lambda}^{+\Lambda}
  dq
   { 4 m^2 \over
(ky-qx)^2 + m^2 (x-y)^2}  \nonumber \\
&~&~ +{g^2 \over 2 (2 \pi)^2 } C_f \int_x^1 dy \int_{- \Lambda}^{+\Lambda}  dq
{4 m^2  \over
[k(1-y)-q(1-x)]^2 + m^2 (x-y)^2} . \label{se1}
\ee
The energy denominator factors are 
\be
{1 \over E_1}   = {x y \over [ky-qx]^2 + m^2 (x-y)^2}, ~~
{ 1 \over E_2} = 
{(1-x)(1-y) \over [k(1-y) - q(1-x)]^2 + m^2 (x-y)^2}.
\ee
The vertex terms are
\be
{\tilde V}_1 = \theta(x-y)  {\tilde U}(k,x,q,y) ~+~ \theta(y-x) {\tilde U}(q,y,k,x) ,
\ee
\be
{\tilde V}_2  = \theta(x-y) {\tilde U}(k,1-x,q,1-y)  
~+~ \theta(y-x) {\tilde U}(q,1-y,k,1-x) ,
\ee
with
\be
{\tilde U} (k,x,q,y) & =&
4 { m^2 \over x y} - { m^2 (x-y)^2 \over x y (1-x) (1-y)} \nonumber \\ 
&~&~+ {q^2 \over y (1-y)} + {k^2 \over x (1-x)} 
- 2 {k^2 \over (x -y)}{1 \over x (1-x)} + 2 {q^2 \over (x-y)}{1 \over y (1-y)}
\nonumber \\  
&~&~~~+ { kq \over x (1-y)} + {kq \over y (1-x)} + 2 {kq \over (x-y)} \Big [ 
{1 -2 y \over y (1-y)} - {1 - 2 x \over x(1-x)} \Big ].
\ee
In addition to the ${1 \over x_g^2}$ singularity which is canceled, 
transverse gluon exchange 
contributions also contain ${ 1 \over x_g}$ singularity which is removed by 
the principal value prescription. 
Cancellation of this singularity is an appealing feature of the Bloch 
effective Hamiltonian in contrast to the Tamm-Dancoff effective Hamiltonian 
where the singularity cancellation does not occur because of the presence of 
invariant mass in the energy denominator \cite{1Krautgartner:1992xz}. 

\subsubsection{$``$True" infrared divergences}


Next we consider true infrared divergences. Consider the self energy
integrals. The energy denominators in these expressions vanish when $k=q$
and $x=y$ which correspond to vanishing gluon momentum. By carrying out the
integrals explicitly, in the limit $ \Lambda \rightarrow \infty $ we get,
\be
{S}_1 ~= ~{mg^2 \over 2 \pi}~C_f~ \Big [ ~{ 1 \over x}~ 
{\rm ln} \ {x \over \delta}~ +~ {
1 \over 1-x} ~{\rm ln} \ {1-x \over \delta}~ \Big ].
\ee 
Thus the singular part of self energy is 
\be
{S}_{1~ singular} ~= ~- { m g^2 \over 2 \pi}~C_f ~ { 1 \over x (1-x)}~{\rm ln}\
\delta.
\ee 
The infrared divergent contribution from self energy gives a positive
contribution to the fermion mass.
It is important to note that the vanishing of energy denominator is possible
also in (3+1) dimensions, but in that case we do not encounter any divergence.
It is the peculiarity of (2+1) dimensions that the vanishing energy denominators
cause a severe infrared divergence problem.   

The same vanishing energy denominators occur also in the one gluon exchange
contributions. Let us now consider various terms in the numerator
separately. The terms proportional to $4m^2$ arose from the denominator of
the transverse gluon exchange. A straightforward calculation shows that this
term leads to both finite and infrared divergent contributions. The infrared
divergent contribution is given by
\be
{ m g^2 \over 2 \pi}~C_f ~ { 1 \over x (1-x)}~{\rm ln}\ \delta
\ee 
which exactly cancels the infrared divergent contribution from self energy.
The finite part, in the nonrelativistic limit, can be shown to give rise to
the logarithmically confining potential. Next we have to consider the
remaining terms in the numerator. Rest of the terms proportional to $m^2$
are multiplied by $(x-y)^2$ so that they do not lead to an infrared
divergence problem. The numerator of the imaginary part vanishes at $k=q$,
and $x=y$ and hence is also infrared finite. It is easy to verify that the
rest of the (transverse momentum dependent) terms in the numerator 
does not vanish when 
the denominator vanishes and hence the resulting bound state equation is
inflicted with infrared divergences arising from the vanishing energy
denominator. This problem was first noted in the context of QED in (2+1)
dimensions by Tam, Hamer, and Yung \cite{1Tam:1995qk} but was not
investigated by these authors. 
We remind the reader that this is a peculiarity of (2+1)
dimensions which provides us a unique opportunity to 
explore the consequences of the vanishing energy denominator problem. 


\section{Numerical study of the bound state equation}

Once we derive the effective Hamiltonian, we need to diagonalize it 
nonperturbatively. For that,
we convert the integral equation into a matrix equation with the use of
Gaussian Quadrature. (For details of the numerical procedure see Appendix
\ref{APnumpro}.)  The color factor $C_f$ is set to 1 for all the numerical
 calculations presented.
As mentioned before, an important feature of gauge theories on the
light-front is the presence of linear infrared divergences. They appear in
the canonical Hamiltonian in instantaneous four fermion interaction term.
\begin{figure}[h]
\centering
\includegraphics[height=10cm]{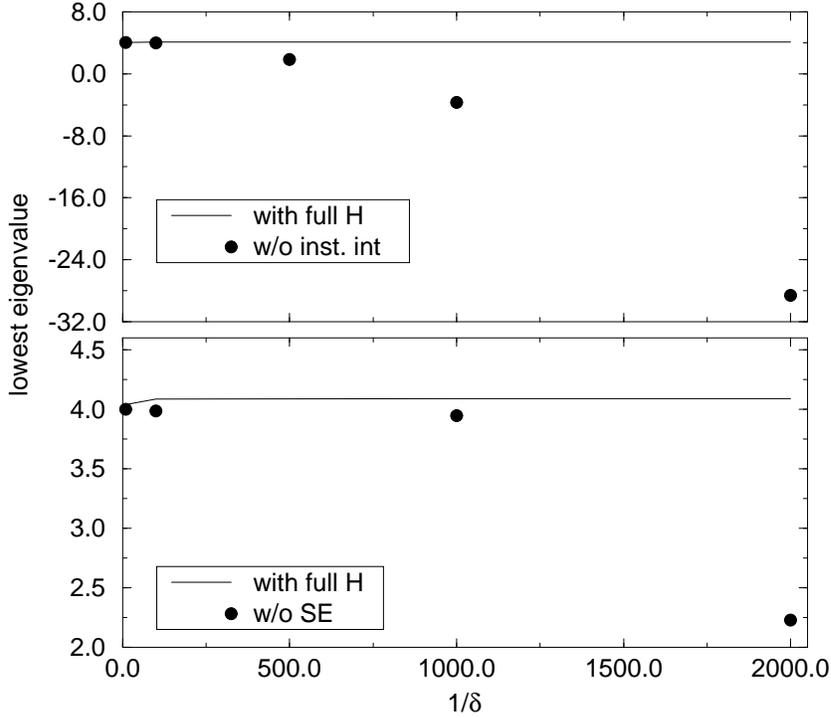}
\caption[Cancellation of infrared divergences in Bloch effective Hamiltonian]
{Cancellation of infrared divergence. Full line
denotes the full Hamiltonian. (a) shows the cancellation 
of light-front infrared divergence by switching on and off the
instantaneous interaction. Filled circles - without instantaneous
interaction. (b) shows the cancellation of logarithmic
infrared divergence by switching on and off the self energy term. Filled
circles - without self energy. The parameters are $g=0.2$, $\epsilon = 0.00001$,
$m=1$, $\kappa=20$, $n_1=40$, $n_2=50$.}  
\label{fig1} 
\end{figure}
When the $q {\bar q}g$ states are integrated out {\em completely} in
perturbation theory, they also appear in the effective four fermion
interaction and cancel against each other. Non-cancellation of this
divergence is a major feature of similarity renormalization approach. 
We first address the issue of how linear divergences manifest in the
non-uniform grid of the Gaussian Quadrature and how well it can handle 
linear light front infrared divergence. We have studied numerically discretized
versions of Eq. (\ref{ebe1}) where the divergences are present separately in the
discretized version together with the counterterm given in Eq. (\ref{ct}). 
For $g=0.2$, we have calculated the
eigenvalues with and without the instantaneous interaction. The result
presented in Fig. \ref{fig1}(a) for the lowest eigenvalue shows 
that  the Gaussian Quadrature can handle
the cancellation very efficiently.

After the cancellation of linear light-front infrared divergence, a
logarithmic infrared divergence which arises from the vanishing energy
denominator survives in the bound state equation. Here we have to
distinguish two types of terms. First type, where the coefficient of the 
logarithmic infrared divergence is independent of the fermion transverse
momentum and the second type where the coefficient is dependent. Self energy
and Coulomb interaction are of the first type. In the weak coupling limit,
since the wavefunction is dominated by very low transverse momentum, we
anticipate that contributions of the second type will be dynamically
suppressed even though both are multiplied by the same coupling constant.
This is especially true of any discrete grid which automatically imposes a
lower limit on the smallest longitudinal momentum fraction allowed.
Thus at weak coupling, even if there are uncanceled infrared divergences
(divergences of the second type),
they may not be significant numerically whereas divergences of the first
type are significant. By
switching the self energy contribution off and on, we have studied this
interplay. The lowest eigenvalue with and without self energy contribution
is plotted in Fig. \ref{fig1}(b). This shows the cancellation of the 
dominant logarithmic infrared divergence. Since there are still
uncanceled infrared divergences in the bound state equation (with
coefficient proportional to fermion transverse momenta) these figures further
illustrate the fact that such divergences are {\em not} numerically significant at
weak coupling.      
\begin{table}[h]
\begin{center}
 \begin{tabular}{||c|c|c|c|c|c|c||}
\hline \hline
  $g$  & $\delta$   &  \multicolumn{5}{c||}  {eigenvalues ($M^2$)} \\
\hline    
     & 0.01     & 4.0870 & 4.0972 & 4.0972 & 4.0973 & 4.0973
\\
\cline{2-7}
     & 0.005    & 4.0901 & 4.1066 & 4.1099 & 4.1100 & 4.1112
\\
\cline{2-7}
 0.2  & 0.001   & 4.0913 & 4.1113 & 4.1122 & 4.1181 & 4.1209
\\
\cline{2-7}
     & 0.0001   & 4.0913 & 4.1113 & 4.1122 & 4.1181 &
 4.1209 \\
\cline{2-7}
     & 0.00001 & 4.0913 & 4.1113 & 4.1122 & 
4.1181 & 4.1209 \\
\hline
     & 0.01    & 4.5735 & 4.7337 & 4.7667 & 4.7832 & 4.8277
\\
\cline{2-7}
     & 0.005   & 1.9094 & 1.9415 & 3.1393 & 3.1399 & 4.5697
\\
\cline{2-7}
 0.6 & 0.001   & -187230.4 & -187225.4 & -186664.9 & -186664.8
& -31506.9 \\
\cline{2-7}
     & 0.0001   & -187230.4 & -187225.4 & -186664.9 &
 -186664.8 & -31506.9 \\
\hline
\hline
\end{tabular}
\end{center}
\caption[ Variation with $\delta$ of the  full Bloch effective Hamiltonian.]
{ Variation with $\delta$ of the  full Hamiltonian. The parameters
are 
$n_1$=40, $n_2$=50, $\epsilon$=0.00001, $ \kappa$=20.0 in $k={1 \over \kappa}
tan({u \pi\over 2})$
\label{table1}}
\end{table}

As the strength of the interaction grows, wavefunction develops medium to
large transverse momentum components and the infrared catastrophe triggered
by the vanishing energy denominator becomes manifest numerically. This is
illustrated in Table \ref{table1} where we present the variation with $\delta$ of the
first five eigenvalues for two different choices of the coupling $g$. The
table clearly shows that on a discrete grid, the uncanceled infrared
divergences due to the vanishing energy denominator problem are not
numerically significant at weak coupling but their effect is readily felt at
a stronger coupling.


\section{Reduced Model}\label{reduce}

In this section we consider a model Hamiltonian free from infrared divergences
constructed by dropping the transverse momentum dependent terms from the
numerator of the effective Hamiltonian. For convenience, we further drop the
terms proportional to $(x-y)^2$ and the imaginary part. This defines our
reduced model which is also ultraviolet finite. The equation governing the
model is given by
\be
\left [ M^2 - {m^2 + k^2 \over x (1-x)} \right ] \psi_2(x,k) = {S}_1 \
 \psi_2(x,k) \ + {\cal B} .  
 \label{rbe2}
\ee 
The self energy contribution ${ S}_1$ is the same as given in Eq. (\ref{se1}).
The boson exchange contribution ${\cal B}$ is given by
\be
{\cal B}& =& - {g^2 \over 4 (2 \pi)^2 } C_f \int_0^1 
dy \int_{-\Lambda}^{+\Lambda}
  dq
   { 4 m^2 \over
(ky-qx)^2 + m^2 (x-y)^2} \  \psi_2(y,q) \nonumber \\
&~&- {g^2 \over 4 (2 \pi)^2 } C_f \int_0^1 
dy \int_{- \Lambda}^{+\Lambda}  dq
{4 m^2  \over
[k(1-y)-q(1-x)]^2 + m^2 (x-y)^2} \ \psi_2(y,q). \label{rme}
\ee
Note  that  in the above approximations we dropped only the term sick with 
vanishing energy denominator and not so important imaginary terms and
Eq. (\ref{rbe2})  still represents a {\em relativistic} bound state equation. 
Though  the rotational symmetry is not manifest in this equation, 
 Eq. (\ref{rbe2}) in the 
nonrelativistic limit reduces to a  Schr\"{o}dinger
equation with explicit rotational symmetry (see Appendix \ref{APnrbe}).
This model provides us an opportunity to study
the simplest manifestation and possible violation of rotational symmetry in
the context of light-front field theory.

\subsection{Numerical study of the reduced model }
Again we discretize the Eq. (\ref{rbe2}) by Gaussian Quadrature. The convergence
of the eigenvalues as a function of the number of grid points is presented
in Table \ref{table2}. In this table we also present the (in)dependence of eigenvalues on
the momentum cutoff.
\begin{table}[h]
\begin{center}
\begin{tabular}{||c|c|c|c|c|c|c||}
\hline \hline
$ n_1$ & $ n_2$ & 
\multicolumn{5}{c||}   {eigenvalues (lowest five) ($ \kappa $ =10.0)} \\
\hline 
 ~20~ &  ~20~ & ~4.08926~ & ~4.10605~ &  ~4.10768~ & ~4.11061~ &
 ~4.11085~ \\
  ~30~ &  ~30~ & 4.09045 & 4.10909 & 4.11038 & 4.11516 &
 4.11699 \\
  40 &  30 & 4.09045 & 4.10913 & 4.11035 & 4.11524 &
 4.11697 \\
  40  &  40  & 4.09102 & 4.11052 & 4.11154 & 4.11711 &
 4.11951 \\
  40  &  50  & 4.09136 & 4.11133 & 4.11222 & 4.11811 &
 4.12096 \\
  50 &  50 & 4.09136 & 4.11135 & 4.11219 & 4.11816 & 
4.12095 \\
  50  & 60 & 4.09158 & 4.11188 & 4.11263 & 4.12189 &
 4.12290 \\
  46  &  60 & 4.09158 & 4.11187 & 4.11264 & 4.11877 &
 4.12189 \\
  46 &  66 & 4.09168 & 4.11212 & 4.11284 & 4.11905 &
 4.12231 \\
  46 &  74 & 4.09179 & 4.11237 & 4.11305 & 4.11934 &
 4.12276 \\
\hline
$n_1$ & $ n_2$ &   \multicolumn{5}{c||}{eigenvalues (lowest five)    
      ($ \kappa $=20.0)} \\
\hline
  46 &  74 & 4.09179 & 4.11240 & 4.11301 & 4.11940 &
 4.12273 \\
\hline
\hline
\end{tabular}
\end{center}
\caption[Convergence of eigenvalue with $n_1$ and $n_2$ (reduced model).]
{ Convergence of eigenvalue with $n_1$ and $n_2$ (reduced model).
 The parameters
are 
 $m$=1.0, $g$=0.2, $\epsilon=0.00001$.
\label{table2}}
\end{table}

(2+1) dimensions provide an opportunity to study the manifestation and 
violation of rotational symmetry in light front field theory in a simpler
setting compared to (3+1) dimensions. The absence of spin further facilitates
this study. Rotational symmetry in this case simply implies degeneracy with
respect to the sign of the azimuthal quantum number $l$ (see Appendix \ref{APnrbe}). 
Thus we expect all
$ l \neq 0$ states to be two fold degenerate.
\begin{table}[h]
\begin{center}
\begin{tabular}{||c|c|c|c|c||}
\hline \hline
 $g$  &              \multicolumn{4}{c||}    {eigenvalues} \\
\hline
    & This   & 4.0918 & (4.1124, 4.1130) & 4.1194 \\ 
0.2    & work   & (4.1227, 4.1235) & (4.1268, 4.1273) &
  (4.1298, 4.1303) \\
\cline{2-5}
    &Koures   & 4.0925~~$(l=0$)& 4.1144~~$(l=1)$&  4.1214~~$(l=0)$ \\
    &(Ref. \cite{1Koures:1995qp}) &  4.1260~~($l=2$) &  4.1303~~$(l=1)$ & 4.1340
~~$(l=3)$ \\
\hline
    & This   & 4.5856 & (4.7741, 4.7821) & 4.8390 \\
0.6    & work  & (4.8767, 4.8816)& (4.9094, 4.9184) &
 (4.9458, 4.9481) \\
\cline{2-5}
    &Koures  & 4.5806 ~~ $(l=0)$ & 4.7777~~$(l=1)$ & 4.8409~~$(l=0)$ \\ 
    &(Ref. \cite{1Koures:1995qp})   & 4.8827~~ $(l=2)$ & 4.9205~~$(l=1)$ &
 4.9545~~ $(l=3)$ \\
\hline
\hline
\end{tabular}
\end{center}
\caption[Comparison of reduced model results with other work.]
{Reduced model. 
The parameters are $n_1$=46, $n_2$=74,  $\epsilon=0.00001,$  $m$=1.0. 
$ k=tan(q \pi/2)/\kappa,~~ \kappa=20.0$. Eigenvalues within () are 
$\pm l$ degenerate (broken) states.
\label{table3}}
\begin{center}
\begin{tabular}{||c|cc|cccccc||}
\hline \hline
      &                   $ n_1$&  $n_2$ & \multicolumn{6}{c||}
      {eigenvalues} \\
\hline
$I$ &   40&  50
&    18.217 &$\pmatrix {30.702 \cr 33.499}$   & 35.206 &
                        $\pmatrix { 39.955 \cr  41.159}$&
 $\pmatrix{41.332 \cr 43.271}$  &  $\pmatrix{44.134 \cr 45.272}$\\
\cline{2-9}                     
            &               46 & 70 &   18.276 & $\pmatrix{30.774 \cr
33.616}$ &
 35.318 &$\pmatrix{ 40.106\cr 41.331} $&$\pmatrix{ 41.483\cr 43.477}$&
                   $\pmatrix{ 44.375\cr 45.503)}$\\
\hline
$II $      &   40 & 50 &    18.980 &
  $\pmatrix{ 31.507\cr  34.219}$& 35.826 &  $\pmatrix{40.406\cr 41.888}$ & 
$\pmatrix{41.921 \cr 43.788}$ &
$\pmatrix{ 44.345\cr 45.163}$ \\
\cline{2-9}
 &                  46 &  70 & 
  19.008 &$\pmatrix{31.542\cr 34.319}$ & 35.935 &
$\pmatrix{40.626\cr 42.031}$
& $\pmatrix{42.088\cr 44.010}$ &  $\pmatrix{44.647\cr 45.780}$ \\
\hline
\hline
\end{tabular}
\end{center}
\caption[First few eigenvalues of reduced model with large coupling.] 
{First few eigenvalues in the reduced model. The parameters are 
$g$=5.0, $m$=1.0, $\epsilon$=0.00001. $(I)$ for the parametrization 
$k=u\Lambda m/((1-u^2)\Lambda+m)$, $\Lambda=40.0.$ $(II)$ for the
parametrization $k=tan(u\pi/2)/\kappa$, $\kappa=10.0.$  
 Eigenvalues within ( ) are $\pm l$ 
degenerate (broken) states.
\label{table4}}
\end{table}
By a suitable change of variables, one can easily show that our reduced
model, in the nonrelativistic limit reduces to Schr\"{o}dinger equation in two
space dimensions with a logarithmic confining potential. 
In the weak coupling limit, since $C_f$ is set to 1, we can compare our results
of the reduced model (where we do not make any nonrelativistic
approximation)
with the spectra obtained in nonrelativistic $QED_{2+1}$. Tam {\it
et al.} \cite{1Tam:1995qk} solved the radial Schr\"{o}dinger equation in momentum 
space for $l=0$ states and Koures \cite{1Koures:1995qp} solved the coordinate 
space radial Schr\"{o}dinger equation
for general $l$. Since we are solving the light front bound state equation,
rotational symmetry is not at all manifest. However, at weak coupling we
expect that the
spectra exhibit rotational symmetry to a very good approximation. 
Our numerical results are compared
with those of Koures in Table \ref{table3} for two 
values of the coupling. At $g=0.2$ we
find reasonable agreement with the degeneracy in the spectrum. Even at $g=0.6$
the violation of rotational symmetry is very small. Splitting of levels
which are supposed to be degenerate become more visible at very strong
coupling as can be seen from Table \ref{table4} for $g=5.0$. 

Along with the eigenvalues, the diagonalization process also yields
wavefunctions. We have plotted the wavefunctions corresponding to the first
four eigenvalues in Fig. \ref{fig2} as a function of $x$ and $k$.
\begin{figure}[h]
\centering
\fbox{\includegraphics[height=12.0cm]{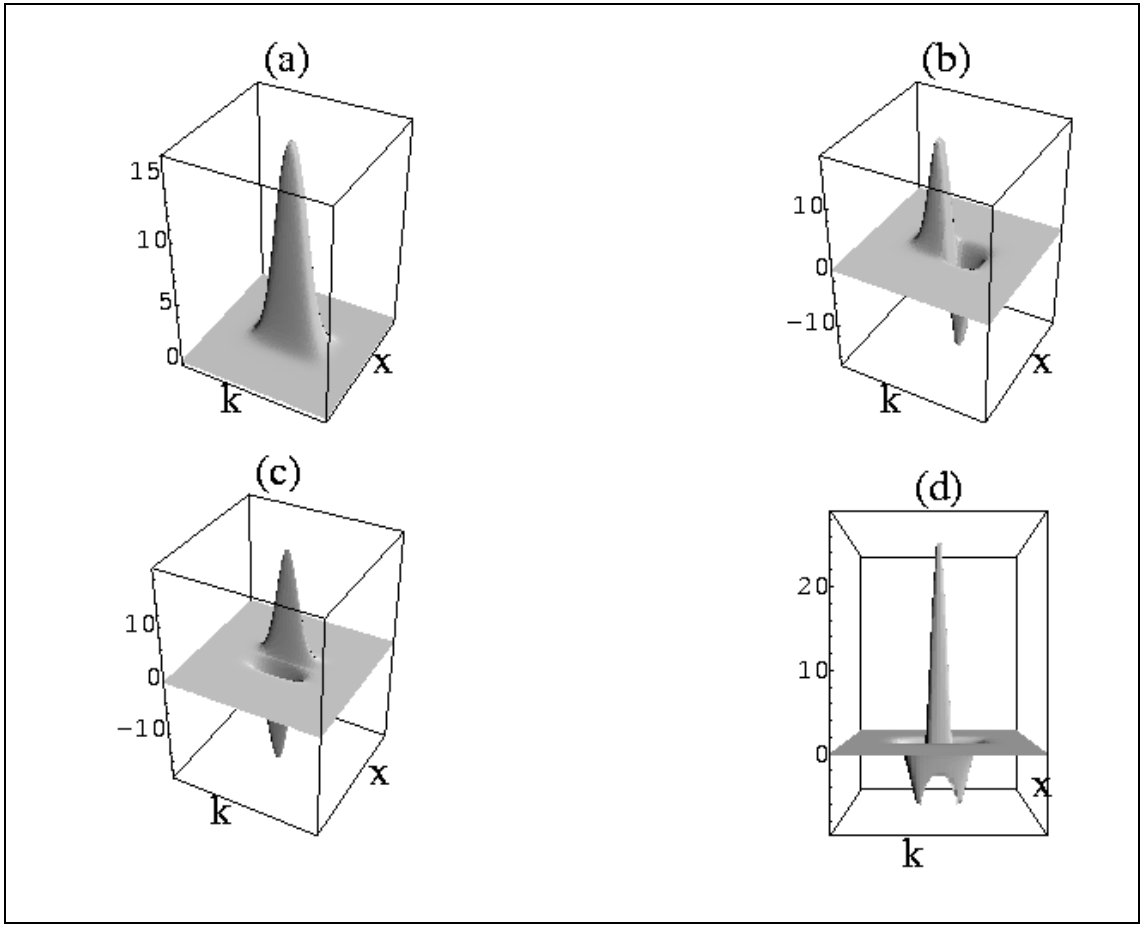}}
\caption[The wavefunctions corresponding to the lowest four
eigenvalues of the reduced model as a function of $x$ and $k$]
{The wavefunctions corresponding to the lowest four
eigenvalues of the reduced model as a function of $x$ and $k$. 
The parameters are 
$g=.2$, $\epsilon=0.00001$,
$m=1$, $\kappa=10$, $n_1=46$, $n_2=74$. 
 (a) Lowest state, (b) first excited state, (c) second
excited state, (d) third excited state. The first and second excited states
should be degenerate in the absence of violation of rotational symmetry.}
\label{fig2}
\end{figure}
 All wavefunctions
are normalized to be $ \int_0^1 dx \int dk ~ \psi^2(x,k)=1$. 
The lowest state is
nodeless and corresponds to $l=0$. The next two states correspond to $l=1$
and have one node. It is interesting to note the way the node appears in the
wavefunctions which correspond to degenerate levels. Since the rotational
symmetry cannot be  manifest in the variables $x$ and $k$, how can the
wavefunctions still indicate this? From Fig. \ref{fig2},  it is clear that 
the way this problem is 
resolved is by
one wavefunction having a node in $k$ and the other wavefunction having a
node in $x$. Thus even if we did not know about the underlying symmetry from
other means, the light-front wavefunctions have a subtle way of indicating
the symmetry.       
  
\section{Summary}
  The numerical solutions of the bound state equations are performed by using
 the Gaussian Quadrature (GQ) which is  a
straightforward procedure to solve the integral equation by converting it 
into a matrix equation.  The investigation shows the efficiency of the GQ method
in handling linear and logarithmic light-front infrared divergences. 
The manifestation of rotational invariance in light-front framework is  
demonstrated very clearly in the reduced model. But, 
our study of the Bloch effective Hamiltonian  indicates that in the context 
of Fock space based effective Hamiltonian
methods to tackle gauge theories in (2+1) dimensions, approaches like 
similarity renormalization
method are mandatory  due to uncanceled infrared divergences caused by
the vanishing energy denominator problem. It is important to recall that
Bloch effective Hamiltonian is generated by completely integrating out the
intermediate gluons irrespective of whether they are low energy or high
energy which has no clear justification specially in a confining theory.
Once we have obtained quantitative measures of the vanishing energy
denominator problem and the nature of the spectra at weak coupling of the
Bloch effective Hamiltonian, the next step is to study QCD$_{2+1}$ in the
similarity renormalization approach which avoids the vanishing energy
denominator problem. An important issue here is the nature of new effective
interactions generated by the similarity approach. It has been shown that in
(3+1) dimensions, similarity approach generates logarithmic confining
interactions \cite{1Perry:1994kp} which however breaks rotational symmetry.
 It is interesting to
investigate the corresponding situation in (2+1) dimensions.


\chapter{Similarity Renormalization Group Approach to Meson Sector in
$(2+1)$ Dimensions}\label{chapsrg}

\section{A Brief Review of Renormalization Group Approach}
In the previous chapter we have studied the meson bound state problem in (2+1)
 dimensional QCD   with Bloch effective Hamiltonian where we encountered the problem
 of vanishing energy denominators.  We also concluded that Similarity Renormalization
 Group (SRG) approach is mandatory to get rid of the problem. 
In this chapter we 
 discuss the same bound state problem in  SRG scheme. But before that, we 
 recapitulate the basic concept of  Renormalization Group and explain why SRG is
preferred over original Wilsonian Renormalization Group approach.
 
It is well established that the most important tool to construct a low
energy effective field 
theory is the  Renormalization Group (RG).
The concept of Renormalization Group was first introduced by
Stueckelberg and Peterman \cite{2stueck} and Gell-Mann and Low
\cite{2gell} and further developed by Bogoliubov and Shirkov \cite{2bog}.
 RG as a practical tool to construct effective theory was developed by 
K.G. Wilson 
\cite{2wilsonrg,2wk,2wilnobel}. The aim of  RG is to simplify the problem with many 
 energy (length) scales involving many degrees of 
freedom which are coupled through the interactions. In the Wilsonian 
approach the cutoff on energy is lowered and the number of degrees of freedom 
is reduced in
 an iterative way and in each step one has to construct the effective interactions
for the  effective degrees of freedom. The simplification of
Renormalization Group lies 
in the hope that the effective interactions are local interactions  i.e., only  nearby 
degrees of freedom are directly coupled by the interactions which holds true 
for original local theories we normally deal with \cite{2wk}.
Construction of nondiagrammatic RG transformations
enables one to solve them  numerically in a computer and hence the
problems which 
cannot be solved by  Feynman diagrams  can be solved by using RG. 

The starting point of the RG transformations is a bare  Hamiltonian
$H_0$  with cutoff 
$\Lambda_0$.  The transformation $\tau$  converts  $H_0$
 to  $H_1$,   $H_1$ to  $H_2$ etc.  as the cutoff is lowered $\Lambda_0 > \Lambda_1
 > \Lambda_2$ in each step and thins the degrees of freedom.  This transformation 
is to be iterated until one gets the effective 
Hamiltonian at the desired low energy scale $\lambda$.  RG
transformation is the evolution operator of the Hamiltonian as the
cutoff changes.
Here I should mention that the  RG transformation is free of (UV or IR)
divergences, since in 
each step, a momentum integral involves only  a finite range of momentum. But the 
divergences occur as a result of  many iterations of the RG
transformation. 
The logic of  RG transformation is best explained by the 
{\it ``triangle of Renormalization''} \cite{2wilnobel}.
  Suppose one is interested  to solve a theory at the
 energy scale 1 in some suitable unit. Consider, for example, that in each step we 
lower the cutoff by a factor of 1/2, i.e.,
we have a discrete set of
cutoffs, $\Lambda=2^N$ for $N=1,2,3,4,\cdots, \infty$.  For each $N$, we apply the 
RG transformation to produce a sequence of effective Hamiltonians $H_0^N, ~H_1^N,
~H_2^N,\cdots $ until we reach $H_N^N$ with cutoff  $\Lambda=1$. 
 Thus the transformation  produces a triangle.
\begin{tabbing}
~~~~~~~~~~~~~~~~~~\=~~~~~~~~~~~~~ $\Lambda_0 =4$ ---~~~~\=$H_0^2$~~~$\cdots$~~~\=$H_{N-2}^N$\\
\>      \> $\downarrow~\tau$   \>$\downarrow~\tau$\\
\>$\Lambda_0=2$---~~~~$H_0^1$ \> $H_1^2$~~~$\cdots$  \>$H_{N-1}^N$\\
\>~~~~~~~~~~~~~~~~~~~~$\downarrow~\tau$  \> $\downarrow~\tau$  \>$\downarrow~\tau$ \\
$~~~~\Lambda_0=1$---~~~~$H_0^0$ \>~~~~~~~~~~~~~~~~~~~$H_1^1$ \>$H_2^2$~~$\cdots$
\> $H_N^N$
\end{tabbing}
The $N \rightarrow \infty$ limit along any row produces the infinite
cutoff limit or the renormalized Hamiltonian. For example, the  $N
\rightarrow \infty$ limit of $\Lambda_0=1$ row generates the Hamiltonian
renormalized at the scale $\Lambda = 1$.

 The fixed point of the 
transformation is defined by 
\be
\tau(H^{\star}) =H^{\star}.
\ee
The fixed point of a transformation is a property of $\tau$ and does not depend on
 the initial Hamiltonian $H_0$. 

But,  this formalism to construct an effective low
energy Hamiltonian acting on a limited Fock space is again plagued with
the problem of
vanishing energy denominator. If we view the Hamiltonian as a matrix,
the energy cutoff limits the size of the matrix and as one lowers the
cutoff, the matrix size is also reduced as shown in Fig. \ref{RG1}(a)
\cite{2szpigel99}. 
The matrix elements near the diagonal region (gray region in
Fig.\ref{RG1}(a)) involve states with  almost
same energy (nearly degenerate)  and involve  degenerate perturbation
theory for calculation of the effective interactions. In nonperturbative
theory like QCD we  do not even know how to do that.
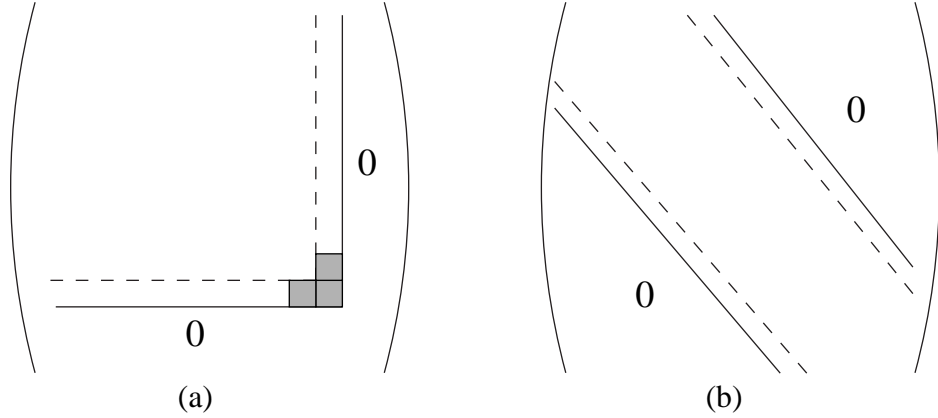
\begin{figure}[h]
\begin{center}
\begin{picture}(360,120)(0,18)
\CArc(300,60)(270,165,195)
\DashLine(45, 25)(145,25){5}
\DashLine(145,25)(145,125){5}
\Line(47,15)(155,15)
\Line(155,15)(155,125)
\GBox(145,15)(155,25){0.7}
\GBox(135,15)(145,25){0.7}
\GBox(145,25)(155,35){0.7}
\CArc(-90,60)(270,345,15)
\CArc(500,60)(270,165,195)
\Line(235,90)(320,-10)
\DashLine(235,100)(330,-10){5}
\DashLine(285,125)(370,20){5}
\Line(295,125)(370,30)
\CArc(110,60)(270,345,15)
\Text(165,70)[]{{\large 0}}
\Text(100,5)[]{{\large 0}}
\Text(270,20)[]{{\large 0}}
\Text(350,90)[]{{\large 0}}
\Text(100,-20)[]{(a)}
\Text(300,-20)[]{(b)}
\end{picture}
\end{center}
\caption[Structure of the effective Hamiltonian matrix as the cutoff is lowered.]{Effective
Hamiltonian 
as the cutoff is lowered (solid to dashed line). (a) energy
cutoff, (b) cutoff on energy difference.
\label{RG1}}   
\end{figure}
To overcome this problem  G{\l}azek and Wilson \cite{2Similar1} and
Wegner \cite{2Similar2} independently developed the similarity
renormalization group (SRG) technique to calculate low energy effective
Hamiltonian. In this approach the cutoff is not on the energy of the
states but  on the energy difference of the states. In place of
removing the states, off-diagonal matrix elements involving large energy
transfer are removed.  
 If the free energy
difference between two states is larger than the cutoff, interactions
between those two states are then removed from the effective
Hamiltonian. The working prescription for SRG is as the following.
Again, as in the standard Wilsonian RG, the  starting point is a finite 
and bare cutoff Hamiltonian
$H_B$ at some ultra-violet cutoff $\Lambda$. Define a similarity transformation that
converts $H_B$ into a band-diagonal $H_{\sigma_1}$  
 as the  energy scale is lowered to $\sigma_1$ and removes the
coupling between the states with energy difference greater than
$\sigma_1$. This process should be repeated until one produces an
effective Hamiltonian $H_{\sigma}$ at a low energy scale $\sigma$. It
can again be viewed as  a
new ``triangle of renormalization'' \cite{2SRGWilson:1994fk} as discussed
above in the context of Wilsonian RG transformations.
The way it works can be compared with the standard numerical algorithm
for matrix diagonalization \cite{2Recipes} where to keep control over 
the complexities, the matrix is first
brought to a tri-diagonal form which is then diagonalized. In the SRG
approach the Hamiltonian is brought perturbatively into a band-diagonal
form (see Fig. \ref{RG1}(b)) which is then diagonalized
nonperturbatively.   Since  it does not
involve any energy jump below the energy scale $\sigma$, the energy denominator
cannot be smaller than $\sigma$  and is free from the problem of
vanishing energy denominator.

%

{\it The diagonalization of the Hamiltonian using SRG is a two step process}.
In the first step, by
removing  any direct interactions between states
with energy difference larger than the cutoff one
arrives at an effective Hamiltonian which is in a band diagonal form. At
this step, one can identify the ultraviolet divergent part of the 
counterterms needed to be added to the
Hamiltonian to remove ultraviolet divergences. It is quite advantageous to
treat the ultraviolet divergences perturbatively especially in gauge
theories since one can avoid pitfalls of other effective Hamiltonian
approaches. For example, it is well known that a simple truncation of the 
Fock space (like a Tamm-Dancoff truncation) leads to uncanceled divergences 
as a result of violations of gauge symmetry.

In the second step, the effective Hamiltonian is diagonalized exactly. It is
important to note that  in the process of removing the interactions
with very large energy exchange we integrate out  small
$x$ gluons, i.e., gluons
having small longitudinal momentum fraction. Since the vacuum is trivial, it
is hoped that, as a result of integrating out small $x$ gluons which are
sensitive to long distance physics on the light front, the effective 
Hamiltonian may contain
interactions responsible for low energy properties of QCD. Indeed
Perry \cite{2SRGPerry} found a logarithmic confining interaction in the $q {\bar q}$
sector in the lowest order effective interaction. 
 
Initial bound state studies in the similarity renormalization approach
worked in either the non-relativistic limit \cite{2Brisudova} or in the heavy
quark
effective theory formalism \cite{2Zhang:1997dd} to investigate heavy-quark
systems. Only in last few years, some works have been done  
\cite{2Allen:2000kx} in the context of glueball spectrum
to address many practical problems, especially the numerical ones that one faces 
in this approach. Since the conceptual and technical problems one encounters
in QCD are numerous, we  initiated a study of bound state problems in
QCD in (2+1) dimensions \cite{2dh1,2DH2}. Our main motivation is not the fitting of
data but a critical evaluation of the strengths and weaknesses of the
various assumptions and approximations made in the similarity approach.

In the previous chapter  we have discussed the meson
sector of (2+1) dimensional 
light front QCD using a Bloch effective Hamiltonian \cite{2BlochH}
in the first non-trivial order. The resulting two 
dimensional integral equation was converted into a matrix equation and solved 
numerically. We have already discussed in detail 
the problem of vanishing energy denominator  which leads to severe 
infrared divergences in (2+1) dimensions and
came to the conclusion that in the context of Fock space based effective Hamiltonian
methods to tackle gauge theories in (2+1) dimensions, approaches like similarity
renormalization method is mandatory  due to uncanceled infrared divergences 
caused by the vanishing energy denominator problem.  
Now, we discuss the similarity renormalization approach 
in the first non-trivial order 
to the same problem. The
detail of similarity renormalization theory for the Effective Hamiltonian
in both G{\l}azek-Wilson and Wegner approaches is discussed in Appendix
\ref{APsrg}.

\section{Effective bound state equation in the $q {\bar q}$ sector in SRG scheme}
In similarity renormalization approach due to G{\l}azek and Wilson, 
to second order, the interacting part of the
effective Hamiltonian at a scale $\sigma$  is given by (see Appendix
\ref{APsrg} for details)
\be
H_{I\sigma ij}^{(2)} = - \sum_k H_{BI ik} H_{BI kj} \left [
{g_{\sigma ijk} \over P_k^- - P_j^-} + { g_{\sigma jik} \over P_k^- - P_i^-}
 \right ], \label{gweh_2}
\ee
where
\begin{eqnarray}
g_{\sigma ijk} &&= f_{\sigma ij} ~ \int_{\sigma}^\infty d\sigma' ~
f_{\sigma' ik}~
{ d \over d \sigma'} f_{\sigma' jk}, \nonumber \\
g_{\sigma jik} &&= f_{\sigma ij}~ \int_{\sigma}^\infty d\sigma'~
f_{\sigma'jk}
{ d \over d \sigma'} f_{\sigma'ik}.\label{gfact}
\end{eqnarray}.
The similarity factor  $f_{\sigma ij}(x)$ is such that 
\begin{eqnarray}
\nonumber
{\rm when} \ {\sigma^2 } >> \Delta M^2_{ij} ,
&&\quad  f(x) = 1\qquad \qquad \qquad
{\it (near\ diagonal\ region)}; \nonumber \\
{\rm when} \  {\sigma^2} << \Delta M^2_{ij},
&& \quad  f(x)
 =0  \qquad \qquad \qquad {\it (far \ off\  diagonal\  region)};
 \nonumber \\
{\rm in \ between}  \qquad \qquad  f(x) ~~~~&& ~~~~~
{\rm drops \ from \ 1 \ to \ 0}
\qquad 
{\it (transition \ region)}  .\label{fx}
\end{eqnarray}
Here $\Delta M^2_{ij} = (M^2_i - M^2_j)$  denotes the difference of invariant
masses of states
$i$ and $j$. 
We restrict ourselves
to the $q {\bar q}$ sector. Then the states involved in the matrix elements
$i$ and $j$  refer to $q {\bar q}$ states and
$k$  refer to $q{\bar q}g$ states.   

Following the steps similar to the ones outlined in Chapter \ref{chapbloch}, we 
arrive at the bound state equation 
\be
\Big [ M^2 - {m^2 + k^2 \over x (1-x)} \Big ] \psi_2(x,k) &=& { S} \
\psi_2(x,k) \ 
- 4 {g^2 \over 2 (2 \pi)^2}C_f \int dy ~ \int dq ~ f_{\sigma ij}~ \psi_2(y,q) ~{ 1 \over (x-y)^2}
\nonumber \\
&~& - {g^2 \over 2 (2 \pi)^2}C_f \int dy ~ \int dq ~\psi_2(y,q)   
{{\cal V} \over {\cal E}}. \label{Sebe1}
\ee
Here $x$ and $y$ are the longitudinal momentum fractions and $k$ and $q$ are
the relative transverse momenta. We introduce the cutoff $\epsilon$ such that 
$ \epsilon \le x,y \le 1 - \epsilon$. We further introduce the regulator $\delta$
such that $ \mid x -y \mid \ge \delta$. Ultraviolet divergences are
regulated by the introduction of the cutoff $\Lambda$ on the relative
transverse momenta $k$ and $q$. 
The self energy contribution
\be
{ S} &=&- {g^2 \over 2 (2 \pi)^2 } C_f \int_0^1 dy \int  dq ~ \theta(x-y)~~
[ 1 - f_{\sigma ik}^2]
~ xy~ { \Big [ \Big ({q \over y} +{ k \over x} - {2 (k-q) \over (x-y)}
\Big )^2 + {m^2 (x-y)^2 \over x^2 y^2} \Big ] \over
(ky-qx)^2 + m^2 (x-y)^2} \nonumber \\
&~&~- {g^2 \over 2 (2 \pi)^2 } C_f \int_0^1 dy \int  dq~\theta(y-x)~
[1 - f_{\sigma ik}^2]
~ (1-x)(1-y) \nonumber \\
&~& ~~~~~~~~~~ \times { \Big [ \Big ({q \over 1-y} +{ k \over 1-x} + 
{2 (q-k
) \over (y-x)}
\Big )^2 + {m^2 (y-x)^2 \over (1-x)^2 (1-y)^2} \Big ] \over
[k(1-y)-q(1-x)]^2 + m^2 (x-y)^2} . \label{sei}
\ee 
The boson exchange contribution 
\be
{{\cal V} \over {\cal E}} &=& {\theta (x-y) \over (x-y)} \left [ 
{g_{\sigma jik} \over {m^2 + q^2 \over y}+ {(k-q)^2 \over (x-y)} - 
{m^2 + k^2 \over x}} 
+ {g_{\sigma ijk} \over {m^2 + k^2 \over 1-x} + {(k-q)^2 \over x-y} - 
{m^2 +q^2 \over 1-y}}
\right ] \nonumber \\
&~& ~~\times \Big [ K(k,x,q,y) ~ + ~i V_I  \Big ]
\nonumber \\
&~& + {\theta (y-x) \over (y-x)} \left [ {g_{\sigma jik} \over
{m^2 + k^2 \over x}+ {(q-k)^2 \over (y-x)} - {q^2 +m^2 \over y}}+  
{g_{\sigma ijk} \over {m^2 + q^2 \over 1-y} + {(q-k)^2 \over y-x} - {m^2 +k^2 \over 1-x}}
\right ] \nonumber \\
&~& ~~\times \Big [ K(q,y,k,x)~ + ~ i V_I  \Big ] ,
\ee
where
\be
K(k,x,q,y) &=&  \left ( {q \over y} + {k \over x} - 
2 {(k-q) \over (x-y)} \right )
\left ( { q \over 1-y} + { k \over 1-x} + {2 (k-q) \over (x-y)} \right )
\nonumber \\
&~&~~~~~~~~~- { m^2 (x-y)^2 \over x y (1-x) (1-y)}, \label{bei}
\ee
\be
V_I = - { m \over x y (1-x) (1-y)} [ q (2-y-3x) + k(3y+x-2)].
\ee 
For all the $f$ and $g$ factors, 
\be
M^2_i = {k^2 + m^2 \over x(1-x)} ~ {\rm and} ~
M^2_j = {q^2 + m^2 \over y (1-y)}.
\ee
\be
{\rm For} ~ x > y,~ M_k^2 = {(k-q)^2 \over
x-y} + {q^2 + m^2 \over y} + {k^2 + m^2 \over 1- x} 
\ee
and 
\be {\rm for}~  y > x,
~ M_k^2 = {(q-k)^2 \over y-x} + {q^2 + m^2 \over 1-y} + {k^2 + m^2 \over
x}. 
\ee
 
Before proceeding further, we perform the ultraviolet renormalization. The
only ultraviolet divergence  arises from the term involving the factor 1
inside the square
bracket in Eq. (\ref{sei}). We isolate the ultraviolet divergent term
which is given by
\be
{S}_{divergent} &=& - {g^2 \over 2 (2 \pi)^2} ~ C_f~ \Bigg [ \int_0^{x -
\delta} dy~
\int_{- \Lambda}^{+ \Lambda} dq ~ {(x+y)^2 \over x y (x-y)^2} \nonumber \\
&~& ~~~~~~~~+~ \int_{x + \delta}^1 dy~
\int_{- \Lambda}^{+ \Lambda} dq ~ {(2-x-y)^2 \over (1-x)(1-y) (x-y)^2}
\Bigg ]
\ee
which is canceled by adding a counterterm.   
\section{Similarity factors}
 Up to this point, we have not chosen any particular form of the
similarity factor $f(x)$. Any function that satisfies the criterion
(\ref{fx}) is a suitable candidate for the similarity factor. Here we
consider three possible choices of similarity factor for comparative
studies in the context of meson bound state problem.
  
\subsection{Parameterization I}
In the SRG study of glueball \cite{2Allen:2000kx}, the following form for the 
similarity factor has been chosen:
\be
f_{\sigma ij} = e^{ - {{(\Delta M^2_{ij})^2} \over \sigma^4}}
\ee
with $ \Delta {M^2_{ij}} = M^2_i - M^2_j$ where $M^2_i$ denotes the invariant
mass of the state $i$, {\it i.e.}, $M^2_i = \Sigma_i{(\kappa_i^\perp)^2 + m_i^2
\over x_i}$. 
Then 
\be
g_{\sigma ijk} &=& f_{\sigma ij} \int_\sigma^\infty d\sigma' ~f_{\sigma' ik}
~ { d \over d \sigma'} f_{\sigma' jk} \nonumber \\
& =& e^{ - {(\Delta M^2_{ij})^2 \over \sigma^4}} ~{(\Delta M^2_{jk})^2 
\over (\Delta
M^2_{ik})^2 + (\Delta M^2_{jk})^2}~ \left [ 1 - 
e^{ - {\Big((\Delta M_{ik}^2)^2 + (\Delta M_{jk}^2)^2\Big)\over \sigma^4}} \right ]~. 
\ee

For the self energy contribution, $i=j$ and we get
\be
g_{\sigma ijk} = g_{\sigma jik} = g_{\sigma iik} ={ 1 \over 2} \left [ 
~1 - e^{ - {2{(\Delta M^2_{ik})^2} \over \sigma^4}} ~ \right ].
\ee
Due to the sharp fall of $f$ with $ \sigma$, the effective Hamiltonian has a
strong dependence on $\sigma$.  Note that this parameterization emerges
naturally in the Wegner formalism (see Appendix \ref{APsrg}).
\subsection{Parameterization II}
St. G{\l}azek has proposed the following form \cite{2Glazek:1998sd} 
for $ f_{\sigma ij}$. 
\be
f_{\sigma ij} = { 1 \over \left [1 + \left ({u_{\sigma ij} (1 - u_0) \over
u_0 ( 1 - u_{\sigma ij})} \right )^{2^{n_g}} \right ]} 
\ee
with
\be
u_{\sigma ij} = 
{ {\Delta M^2_{ij}} \over {\Sigma M^2_{ij}} + \sigma^2}, \label{usg}
\ee
$u_0$ a small parameter, and $n_g$ an integer. The {\em mass sum} $ \Sigma
M^2_{ij} = M^2_i + M^2_j$. 
The derivative
\be
{df_{\sigma ij} \over d \sigma} = 2^{n_g} { 2 \sigma \over {\Sigma M^2_{ij}} +
\sigma^2} \left({ u_{\sigma ij} \over u_0}\right)^{2^{n_g}}
{(1 - u_0)^{2^{n_g}} \over ( 1 - u_{\sigma ij})^{2^{n_g}+1}}
{ 1 \over \left [1 + \left ({u_{\sigma ij} (1 - u_0) \over
u_0 ( 1 - u_{\sigma ij})} \right )^{2^{n_g}} \right ]^2}~.
\ee

Note that for small $u$, both $ 1 - f(u)$ and ${d f \over d\sigma}$
vanish like $ u^{2^{n_g}}$.
\subsection{Parameterization III}

For analytical calculations it is convenient to choose \cite{2SRGPerry} 
a step function cutoff for the similarity factor:
\be
f_{\sigma ij} = \theta(\sigma^2 - {\Delta M^2_{ij}}).
\ee
Then 
\be
g_{\sigma ijk} = \theta(\sigma^2 - {\Delta M^2_{ij}})~ \theta(
{\Delta M^2_{jk}} - \sigma^2)~ \theta({\Delta M^2_{jk}} - {\Delta M^2_{ik}}).
\ee
It is the factor $ \theta({\Delta M^2_{jk}} - \sigma^2)$ in $g_{\sigma ijk}$
that prevents the energy denominator from becoming small.
\section{Analytical calculations with the step function similarity
factor}
\label{stepfunc}
In this section we perform analytical calculations to understand the nature
of the effective interactions generated by the similarity factor. The
differences in the divergences at some places with those discussed in
Chapter \ref{chapbloch} come due to the similarity factors associated
with different terms in SRG scheme. As we have already emphasized that
SRG is a modification over Bloch perturbation theory, we will see here
how the divergences are made softer and their consequences in similarity
group transformed effective bound state equation.   
Since there are no divergences associated with $\epsilon$ and $\Lambda$, 
we suppress their presence in the limits of integration in the 
following equations.

\subsection{Self energy contributions}
Consider the self energy contributions to the bound state equation
Eq. (\ref{sei}). Rewriting the energy denominators to expose the most singular
terms, we have,
\be
{S} &=&- {g^2 \over 2 (2 \pi)^2 } C_f \int_0^1 dy \int  dq ~
{\theta(x-\delta -y)
\over x-y}~~
\{ 1 - f_{\sigma ik}^2 \}
 { \Big [ \Big ({q \over y} +{ k \over x} - {2 (k-q) \over (x-y)}
\Big )^2 + {m^2 (x-y)^2 \over x^2 y^2} \Big ] \over
{(k-q)^2 \over (x-y)} +{q^2 +m ^2 \over y} -{k^2 +m^2 \over x}
} \nonumber \\
&~&~~- {g^2 \over 2 (2 \pi)^2 } C_f \int_0^1 dy \int  dq~{\theta(y-x-\delta)
\over y-x}~
\{1 - f_{\sigma ik}^2\}
~  \nonumber \\
&~& ~~~~~~~~~~ \times { \Big [ \Big ({q \over 1-y} +{ k \over 1-x} + 
{2 (q-k
) \over (y-x)}
\Big )^2 + {m^2 (y-x)^2 \over (1-x)^2 (1-y)^2} \Big ] \over
{(q-k)^2 \over y-x} +{q^2 + m^2 \over 1-y} - {k^2 +m^2 \over 1-x} 
} .
\ee 
The terms associated with 1 in the curly brackets are the same as in
Bloch effective Hamiltonian and lead to ultraviolet
linear divergent terms which we cancel by counterterms.  
They also lead to an infrared
divergent term \cite{2dh1} which remains uncanceled. 
Explicitly this contribution is given by
\be
 4 {g^2 m^2 \over 2 (2 \pi)^2} ~ C_f ~ \int_0^{x- \delta} dy ~ \int dq
{ 1 \over [ky-qx]^2 + m^2 (x-y)^2} \qquad \qquad \nonumber \\
+ 4 {g^2 m^2 \over 2 (2 \pi)^2} ~ C_f ~ \int_{x+ \delta}^1  dy ~ \int dq
{ 1 \over [k(1-y)-q(1-x)]^2 + m^2 (x-y)^2}. \label{selfir}
\ee 
This is simply
indicative of the fact that terms associated with 1 in the curly bracket
still has a vanishing energy denominator problem. We will address the
resolution of this problem shortly.

Let us next consider new infrared divergences that arise as a result of  
the modifications due to similarity factor.
\subsubsection{Leading singular terms}

Keeping only the most infrared singular terms in the numerators (i.e., for $
x > y$, $4 {(k-q)^2 \over (x-y)^2}$ and for $ y > x$,
$4 {(q-k)^2 \over (y-x)^2}$) and 
denominators (i.e., for $
x > y$, $ {(k-q)^2 \over (x-y)}$ and for $ y > x$,
$ {(q-k)^2 \over (y-x)}$), we have,
\be
{S}_1 & =&  {g^2 \over 2 (2 \pi)^2 } C_f \int_0^1 dy \int  dq ~ {\theta(x -
\delta -y)
}~  f_{\sigma ik}^2~
4 {1 \over (x-y)^2}  \nonumber \\
&& ~~~    
~~+ {g^2 \over 2 (2 \pi)^2 } C_f \int_0^1 dy \int  dq~\theta(y-x-\delta)~
f_{\sigma ik}^2~ 4 {1 \over (y-x)^2} .
\ee
The
integral with $\theta$-function similarity factor is given by 
\be
 \int dq \Bigg  [ \int_0^{x - \delta} 
dy  { 1 \over (x-y)^2
} \theta \left(\sigma^2 - {(k-q)^2 \over x-y} \right) + 
\int_{x+ \delta}^1 dy  { 1 \over (y-x)^2
} \theta\left (\sigma^2 - {(k-q)^2 \over y-x} \right) \Bigg  ] .
\ee
We change the transverse momentum variable, $ p=k-q$. For $ x - \delta >y
$, 
we set $ x-y=z$ and for $ y > x + \delta$ we set $ y-x=z$. Then, we have,
\be
4 {g^2 \over 2 (2 \pi)^2 } C_f \Big [ \int_\delta^x {dz \over z^2}
\int dp
~ \theta(\sigma^2 - {p^2 \over z}) &+& 
\int_\delta^{1-x} {dz \over z^2} \int dp 
~\theta(\sigma^2 - {p^2 \over z}) \Big ] = \nonumber \\ 
&~& \!\!\!\!\!\!\!\!\!\!\!\!\!\!\!\!\!\!
{16 g^2 \over 2 (2 \pi)^2}~ C_f ~\sigma ~\left [ 
{ 2 \over \sqrt{\delta}} - {1 \over \sqrt{x}} - { 1 \over \sqrt{1-x}}
\right ] . \label{sss} 
\ee
\subsubsection{Sub-leading singular terms}
Next we study sub-leading singular terms containing ${1 \over x-y}$ 
in self energy generated by the
similarity transformation. They are given by 
\begin{eqnarray}
{S}_2 &=& - 4 {g^2 \over 2 (2 \pi)^2 } ~C_f \Bigg [ ~\int_0^{x - \delta} 
dy ~ \int dq 
~\theta\left (\sigma^2 - {(k-q)^2 \over x-y} \right ) ~{ 1 \over x-y}~ \Big ( {k^2 \over x} - {q^2 \over
y} \Big ) ~  {1 \over (k-q)^2} \nonumber \\
&~& -  ~ \int_{x+ \delta}^1 dy ~ \int dq 
~ \theta\left (\sigma^2 - {(q-k)^2 \over y-x} \right ) ~{ 1 \over y-x}~
\Big ( {k^2 \over 1-x} - {q^2 \over
1-y} \Big ) ~  {1 \over (q-k)^2} \Bigg ] 
\end{eqnarray}
where we have kept only $(k-q)^2$ term in the denominator since the rest
vanish in the limit $ x \rightarrow y$.
As before, for $ x - \delta > y $, we put $ x-y=z$, $ k-q=p$. With the
symmetric integration in $p$, terms linear in $p$ do not contribute. Only
potential source of $\delta$ divergence is the $p^2$ term in the integrand.
Since $p_{max}= \sigma \sqrt{z}$, after $p$ integration ${ 1 \over z}$ is
converted into ${ 1 \over \sqrt{z}}$ which is an integrable singularity. Same
situation occurs for $ y > x$. Thus 
there are no terms divergent in  $\delta$ coming from sub-leading singular
terms.      
\subsection{Gluon exchange contributions}
Let us next consider the effect of similarity factors on
gluon exchange terms. 
\subsubsection{Instantaneous gluon exchange}
From instantaneous interaction we have,
\be
V_{inst}~= ~- 4 ~{g^2 \over 2 (2 \pi)^2}~ C_f~ \int dy~ \int dq ~ 
\psi_2(y,q)~f_{\sigma ij}~ { 1 \over (x-y)^2}. \label{iinit}
\ee
For the sake of clarity, it is convenient to rewrite this as
\begin{eqnarray}
V_{inst}~ &=&~ -4~ { g^2 \over 2 (2 \pi)^2 }~ {1 \over 2} ~C_f ~\int dy ~
\int dq ~f_{\sigma ij} ~ \psi_2(y,q) \nonumber \\
&~& ~~~~\Bigg [ {\theta(x-y - \delta) \over x-y} 
\Bigg \{
{  {(k-q)^2 \over x-y} +({q^2 \over y} -
{k^2 \over x})+ m^2({1 \over y} - { 1 \over x}) 
\over   
{(k-q)^2} +({q^2 \over y} -
{k^2 \over x})(x-y)+ m^2({1 \over y} - { 1 \over x}) (x-y) 
} \nonumber \\
&~&~~~~+~{  {(k-q)^2 \over x-y} -({q^2 \over 1-y} -
{k^2 \over 1-x})- m^2({1 \over 1-y} - { 1 \over 1-x}) 
\over   
{(k-q)^2} -({q^2 \over 1-y} -
{k^2 \over 1-x})(x-y)- m^2({1 \over 1-y} - { 1 \over 1-x}) (x-y) 
}
\Bigg \} \nonumber \\
&~& ~~~~+ ~ {\theta(y-x - \delta) \over y-x} 
\Bigg \{
{  {(q-k)^2 \over y-x} -({q^2 \over y} -
{k^2 \over x})+ m^2({1 \over x} - { 1 \over y}) 
\over   
{(q-k)^2} -({q^2 \over y} -
{k^2 \over x})(y-x)+ m^2({1 \over x} - { 1 \over y}) (y-x) 
} \nonumber \\
&~& ~~~~ + ~ {  {(q-k)^2 \over y-x} +({q^2 \over 1-y} -
{k^2 \over 1-x})+ m^2({1 \over 1-y} - { 1 \over 1-x}) 
\over   
{(q-k)^2} +({q^2 \over 1-y} -
{k^2 \over 1-x})(y-x)+ m^2({1 \over 1-y} - { 1 \over 1-x}) (y-x) 
}
\Bigg \}
\Bigg ].
\ee  
We have to seperately analyze the three types of terms in the numerator.

First consider terms proportional to $m^2$ in the numerator. They are given
by
\be
- 4 ~ {g^2 m^2 \over 2 (2 \pi)^2 } ~ { 1 \over 2} ~C_f ~ \int dy ~\int dq~
f_{\sigma ij} ~\psi_2(y,q) \qquad \qquad \qquad \qquad \qquad \qquad 
\nonumber \\
\times ~\Big [ 
{ 1 \over [ky-qx]^2 + m^2(x-y)^2} + { 1 \over[k(1-y) - q(1-x)]^2 + m^2
(x-y)^2} 
\Big ] \label{coul}
\ee
which leads to the logarithmic confining interaction in the nonrelativistic
limit. Note, however, that Eq. (\ref{coul}) is affected by a logarithmic
infrared divergence arising from the vanishing energy denominator problem.
The logarithmic infrared divergence is canceled by the self energy
contribution, Eq. (\ref{selfir}). Thus it explicitly shows that the
logarithmically confining Coulomb interaction survives similarity
transformation but the associated infrared divergence is canceled by self
energy contribution.

Next we look at the most singular term in the numerator in the limit $ x
\rightarrow y$. In this limit we keep only the leading term in the
denominator and we get
\be
- 4 {g^2 \over 2 (2 \pi)^2} ~ C_f~ \int dy ~ \int dq~ \psi_2(y,q) ~
f_{\sigma ij} \left \{ {\theta (x-y) \over (x-y)^2 } + {\theta (y-x) \over
(y-x)^2} \right \}. \label{mostsi}
\ee
Lastly we look at the rest of the terms in the instantaneous exchange. Since
we are interested only in the singularity structure, we keep only the
leading term in the denominator and we get
\be
- 2 {g^2 \over 2 (2 \pi)^2} ~ C_f ~ \int dy ~ \int dq ~ f_{\sigma ij} ~ { 1
\over (k-q)^2} \qquad \qquad \qquad \qquad \qquad \qquad  
\nonumber \\
\times ~ \Bigg [ {\theta (x-y) \over x -y} \Big [{q^2(1-2y) \over y (1-y)} - {k^2(1-2x)
\over x (1-x)} \Big ] 
+ {\theta (y-x) \over y -x} \Big [ {k^2(1-2x) \over x (1-x)} - {q^2(1-2y)
\over y (1-y)} \Big ]. \label{nextsi}
\ee

\subsubsection{Transverse gluon exchange}
First, consider the most singular terms.

Keeping only the most singular terms, the gluon exchange contribution is
\be
&& -{ g^2 \over 2 (2 \pi)^2} ~C_f~ \int dy ~\int dq ~\psi_2(y,q)~ \times
 ~~~~~~~~~~~~
 \nonumber \\
&& \Bigg \{
 {\theta(x- \delta -y) \over x-y} \Bigg [ { g_{\sigma jik} +
g_{\sigma ijk} \over 
{(k-q)^2 \over (x-y)}} (-4) {(k-q)^2 \over (x-y)^2} \Bigg ] 
+ {\theta(y- x - \delta) \over y-x} \Bigg [ { g_{\sigma jik} +
g_{\sigma ijk} \over 
{(q-k)^2 \over (y-x)}} (-4) {(q-k)^2 \over (y-x)^2} \Bigg ] 
\Bigg \}. \nonumber  \\      
&& =  
- { g^2 \over 2 (2 \pi)^2} ~C_f~ \int dy ~\int dq ~\psi_2(y,q) ~ \times 
\qquad \qquad \nonumber \\ 
&& \Bigg \{
 {\theta(x- \delta -y) \over (x-y)^2} \Bigg [  g_{\sigma jik} +
g_{\sigma ijk}  \Bigg ] 
+ {\theta(y- x - \delta) \over (y-x)^2} \Bigg [  g_{\sigma jik} +
g_{\sigma ijk}  \Bigg ] 
\Bigg \} .
\ee      
Explicitly, for $ x > y$,
\be
g_{\sigma ijk} &=& \theta(\sigma^2 - M^2_{ij})~ \theta(M^2_{jk} - M^2_{ik}) ~
\theta (M^2_{jk} - \sigma^2), \nonumber \\
g_{\sigma jik} &=& \theta(\sigma^2 - M^2_{ij})~ \theta(M^2_{ik} - M^2_{jk}) ~
\theta (M^2_{ik} - \sigma^2).
\ee
We are interested in the situation $x$ near $y$ and $ i $ near $j$. Then
$ \theta(M^2_{jk} - M^2_{ik}) ={ 1 \over 2} = \theta(M^2_{ik} - M^2_{jk})$
and $ \theta(\sigma^2 - M^2_{ij}) =1$.
Then the gluon exchange contribution is
\be
 4 { g^2 \over 2 (2 \pi)^2} &C_f& \int dy~ \int dq ~\psi_2(y,q) 
 \Bigg [ {\theta(x- \delta -y) \over (x-y)^2} \Bigg \{
1 - \theta \left (\sigma^2 - {(k-q)^2 \over x-y}\right ) \Bigg
\}\nonumber \\
&~&~~~+ {\theta(y-x- \delta ) \over (y-x)^2} \Bigg \{
1 - \theta \left (\sigma^2 - {(q-k)^2 \over y-x} \right ) \Bigg \} \Bigg ]
\ee   
where we have used $ \theta(x) = 1 - \theta(-x) $.
Combining with the most singular part of the instantaneous contribution
given in Eq. (\ref{mostsi}) we arrive at
\be
&&~~~~~~ - 4 {g^2 \over 2 (2 \pi)^2}~ C_f ~\int dy~ \int dq ~\psi_2(y,q)
~ \times \nonumber \\
&&\Bigg [ {\theta(x- \delta -y) \over (x-y)^2} \theta \left 
( \sigma^2 - {(k-q)^2
\over x-y} \right ) + {\theta(y- x - \delta) \over (y-x)^2} 
\theta \left (\sigma^2 -
{(q-k)^2 \over y-x} \right ) \Bigg ].
\ee
For convenience we change variables. For $ x > y$, we put $ x-y = {p^+ \over
P^+}$ and $k-q = p^1$ and  for $y > x$, we put $y-x = {p^+ \over P^+}$ and 
$ q-k =
p^1$ where $P^+$ is the total longitudinal momentum. Thus we arrive at
\be
&& - 4 {g^2 \over 2 (2 \pi)^2} ~C_f ~P^+~\Bigg [ 
\int dp \int_{P^+\delta}^{P^+x}dp^+ \psi_2(x-{p^+ \over P^+}, k-p^1) 
{ 1 \over (p^+)^2} \theta \left ( \sigma^2  - {(p^1)^2 P^+ \over p^+}
\right ) \nonumber \\
&& ~~~~~~~~~ +  
\int dp \int_{P^+\delta}^{P^+(1-x)}dp^+ \psi_2(x+{p^+ \over P^+}, k+p^1) 
{ 1 \over (p^+)^2} \theta \left (\sigma^2  - {(p^1)^2 P^+ \over p^+} \right ) \Bigg ].
\ee
Consider the Fourier transform
\be
V(x^- , x^\perp) &=& - 4 {g^2 \over 2 (2 \pi)^2 }~ C_f~ P^+ 
\Bigg [ 
\int_{P^+ \delta}^{P^{+}x}{dp^{+} \over (p^{+})^2} \int_{-
{p^1}_{max}}^{+p^1_{max}} ~dp^1 ~e^{{i \over 2} p^+ x^- - i p^1 x^1}
\theta \left ( \sigma^2 - {(p^{1})^2 P^{+} \over p^+} \right ) \nonumber \\
&~&~~~~~~ \int_{P^{+} \delta}^{P^+(1-x)}{dp^+ \over (p^+)^2} \int_{-
{p^1}_{max}}^{+{p^1}_{max}} ~dp^1 ~e^{{i \over 2} p^+ x^- - i p^1 x^1}
\theta \left (\sigma^2 - {(p^1)^2 P^+ \over p^{+}} \right ) \Bigg ]
\ee
where $ p^1_{max} = \sigma \sqrt{p^+ \over P^+}$.
We are interested in the behavior of $V(x^-, x^1)$ for  large $x^-, x^1$.
For large $x^-$, nonnegligible contribution to the integral comes from the
region $ q^+ < { 1 \over \mid x^- \mid}$. For large $x^1$, we need $p^1_{max} x^1$ to
be small, i.e., $ (p^1_{max})^2 < { 1 \over (x^1)^2}$, i.e., $p^+ < 
{P^+ \over (x^1)^2 \sigma^2}$. Thus we have the requirements,
$ p^+ < { 1 \over \mid x^- \mid}$, $p^+ < 
{P^+ \over (x^1)^2 \sigma^2}$. We make the approximations 
\be
\int_{-p^1_{max}}^{+p^1_{max}} dp^1 e^{- i p^1 x^1} \approx 2 p^1_{max}
\ee
and $e^{ {i \over 2} q^+ x^- } \approx 1 $.

For large $x^-$, we have $p^+ < { 1 \over \mid x^- \mid} < {P^+ \over (x^1)^2
\sigma^2}$, the upper limit of  $p^+$ integral is cut off by ${ 1 \over
\mid x^- \mid}$. Adding the contributions from both the integrals (which are equal),
for large $x^-$, we have
\be
V(x^-, x^1)~ \approx~ 32~ {g^2 \over 2 (2 \pi)^2} ~ C_f~ \sigma ~\Big [ 
\sqrt{P^+ \mid x^- \mid } - {1 \over \sqrt{\delta}} \Big ]. \label{xlp}
\ee
Thus for large $x^-$ the similarity factors have produced a square root
potential but it is also infrared singular.

For large  $x^1$  the upper limit of $p^+$ integral is cut off by ${P^+
\over (x^1)^2 \sigma^2}$ and we get,
\be
V(x^-, x^1) ~ \approx ~32~ {g^2 \over 2 (2 \pi)^2} ~ C_f~ \sigma ~\Big [ 
\mid x^1 \mid \sigma  - { 1 \over \sqrt{\delta}} \Big ]. \label{xtp}
\ee
For large $x^1$, similarity factors have produced a linear confining
potential which is also infrared singular.
We note that the rotational symmetry is violated in the finite part of the
potential. In both cases, however, the infrared singular part is $
-32 {g^2 \over 2 (2 \pi)^2} ~C_f ~\sigma { 1 \over \sqrt{\delta}}$ which is
exactly canceled by the infrared contribution generated by similarity
transformation from self energy, Eq. (\ref{sss}).     

Lastly, we consider the terms that go like ${ 1 \over x-y}$. Keeping only
the leading term in the energy denominator, we have,
\be
-2 {g^2 \over 2(2 \pi)^2} ~C_f~ \int dy ~ \int dq ~ \psi_2(y,q)~
\times \qquad \qquad \qquad \qquad \qquad \qquad  \nonumber \\
\Bigg [
{\theta(x-y) \over x-y}~ {g_{\sigma jik} + g_{\sigma ijk} \over (k-q)^2}
\Big [ {k^2(1-2x) \over x (1-x)} - {q^2 (1-2y) \over y (1-y)} \Big ]
\qquad \qquad \qquad \qquad \nonumber \\  
+{\theta(y-x) \over y-x} ~{g_{\sigma jik} + g_{\sigma ijk} \over (q-k)^2}
\Big [ {q^2(1-2y) \over y (1-y)} - {k^2 (1-2x) \over x (1-x)} \Big ]
\Bigg ]. \label {nexttg}
\ee
With the step function cut off we have
\be
g_{\sigma jik} + g_{\sigma ijk} \approx f_{\sigma ij} \theta (\Delta M^2_{ik}
- \sigma^2).
\ee
Then, combining Eq. (\ref{nextsi}) and Eq. (\ref{nexttg}) for the
sub-leading divergences, we get,
\be
 - 2 {g^2 \over 2(2 \pi)^2} ~ C_f~ \int dy ~ \int dq~ ~ f_{\sigma ij}
~ \psi_2(y,q) ~ \times \qquad \qquad \qquad \qquad \qquad \qquad \qquad \qquad \nonumber \\
\Bigg [
{\theta(x-y) \over x-y}~ {1\over (k-q)^2} 
\theta \left (\sigma^2 - {(k-q)^2 \over x-y} \right ) 
\Big [ {k^2(1-2x) \over x (1-x)} - {q^2 (1-2y) \over y (1-y)} \Big ]
\nonumber \\  
+{\theta(y-x) \over y-x} ~{1 \over (q-k)^2}
\theta \left (\sigma^2 - {(q-k)^2 \over y-x} \right ) 
\Big [ {q^2(1-2y) \over y (1-y)} - {k^2 (1-2x) \over x (1-x)} \Big ]
\Bigg ]. 
\ee
Taking the Fourier transform of this interaction, a straightforward
calculation shows that no $log ~\delta$ divergence arise from this
term.  
\subsubsection{Summary of divergence analysis}
The logarithmic confining Coulomb interaction of (2+1) dimensions is 
unaffected by similarity
transformation and is  still affected by a logarithmic divergence which is
however canceled by a logarithmic divergence from self energy contribution.
Similarity transformation leads to a non-cancellation of the most singular
(${1 \over (x-y)^2}$) term between instantaneous and transverse gluon
interaction terms. This leads to a linear confining interaction for large
transverse seperations and a square root confining interaction for large
longitudinal separations. The confining interactions generated by
similarity transformations violates rotational symmetry in lowest order
of perturbation theory and needs higher
order calculations to see if restoration of the symmety occurs.   
 However, non-cancellation also leads to ${ 1 \over
\sqrt{\delta}}$ divergences where $ \delta $ is the cutoff on $ \mid x-y \mid
$. This divergence is cancelled by new contributions from self energy
generated by similarity transformation. The subleading singular ${ 1 \over x
-y}$ terms do not lead to any divergence in $ \delta$.    
\section{Numerical studies}
The integral equation is converted in to a matrix equation using Gaussian
Quadrature. The matrix is numerically diagonalized using standard LAPACK
routines \cite{2laug}. We follow the same procedure as what we adopted for
study with Bloch effective theory (see Chapter \ref{chapbloch}) and the  
details of numerical procedure are discussed in  Appendix \ref{APnumpro}. 
With the exponential form and the step function form of the similarity
factor, the integral over the scale in the definition of $g_{\sigma}$
factors in Eq. (\ref{gfact}) can be performed analytically as shown in
Sec. \ref{stepfunc}. For parametrization II, we perform the integration
numerically using $n_s$ quadrature points.  

The first question we address is the cancellation of divergences which are
of two types: (1) the $ ln ~ \delta$ divergence in the self energy and
Coulomb interaction which has its source in the vanishing energy denominator
problem that survives the similarity transformation and (2) ${ 1
\over \sqrt{\delta}}$ divergences in the self energy and gluon exchange
generated by the similarity transformation. Here I should remind the
reader once again that the main motivation of our work is not to fit
data but to assess the strengths and weaknesses of SRG scheme  over Bloch
perturbation theory and comparative study of different choices for
similarity factor. 
\begin{table}[h]
\begin{center}
\begin{tabular}{||c|c|c|c|c|c||}
\hline \hline
 $\delta$   &  \multicolumn{5}{c||}  {Parametrization I} \\
\hline                                                           
     0.1 & 4.89535 & 4.90359 & 4.90420 & 4.90420 & 4.90482 \\
     0.01 & 5.62612 & 6.38083 & 6.82963 & 7.20037 & 7.36414 \\
     0.001 & 5.68417 & 6.42147 & 6.90879 & 7.30609 & 7.64650 \\ 
     0.0001 & 5.68432 & 6.42148 & 6.90909 & 7.30611 & 7.64677 \\
     0.00001 & 5.68432 & 6.42148 & 6.90909 & 7.30611 & 7.64677 \\
\hline
   $\delta$   &  \multicolumn{5}{c||}  {Parametrization II} \\
\hline
      0.1 & 4.55364 & 4.55668 & 4.55668 & 4.55668 & 4.55669 \\
      0.01 & 4.86066 & 5.33491 & 5.49693 & 5.59838 & 5.79111 \\
      0.001 & 4.87607 & 5.35671 & 5.59226 & 5.64476 & 5.88613 \\
      0.0001 & 4.87604 & 5.35671 & 5.59236 & 5.64477 & 5.88615 \\
     ~ 0.00001~ & ~4.87604~ & ~5.35671~ & ~5.59236~ & ~5.64477~ & 
    ~5.88615~ \\ 
\hline
 $\delta$   &  \multicolumn{5}{c||}  {Parametrization III} \\
\hline
     0.1 & 5.12410 & 5.13039 & 5.13101 & 5.13101 & 5.13754 \\
     0.01 & 6.02600 & 6.94326 & 7.50445 & 7.98054 & 8.39927 \\
     0.001 & 6.00968 & 6.97160 & 7.55376 & 8.07749 & 8.49199 \\
     0.0001 & 5.96636 & 6.97160 & 7.51814 & 8.07751 & 8.46524 \\
     0.00001 & 5.96636 & 6.97160 & 7.51814 & 8.07751 & 8.46524 \\
\hline
\hline
\end{tabular}
\end{center}
\caption[Variation of first five eigenvalues of the  full Hamiltonian with $\delta$]
 {Variation with $\delta$ of the first five eigenvalues of the 
 full Hamiltonian (excluding the less
significant imaginary term). The parameters
are $m=1.0$,
$g=0.6$, $n_1=58$, $n_2=58$, $\epsilon=0.00001$, $ \Lambda=20.0$, $\sigma=4.0$,
($u_0=0.1,$ $n_g=2$ and $ n_s=500$ (for $\sigma$ integration) in parametrization
II) \label{srgtab1}}
\end{table}
In Table \ref{srgtab1} we present
the $\delta$ independence of the first five eigenvalues of the Hamiltonian
for $ g$=0.6. Results are presented for three
parametrizations of the similarity factor, namely, the exponential form, the
form proposed by St. G{\l}azek and the step function form used in our
analytical studies. It is clear that the Gaussian Quadrature
effectively achieves the cancellation of $ \delta$ divergences. Recall
that in the study of the same problem using Bloch approach in
Chapter \ref{chapbloch},
negative eigenvalues appeared for $g$=0.6 when $\delta$ was sufficiently
small (for example, 0.001) which was caused by the vanishing energy
denominator problem. Our results in the similarity approach for the same
coupling shows that this problem is absent in the latter approach.

Next we study the convergence of eigenvalues with quadrature points. In
Table \ref{srgtab2} we present the results for all three parametrizations of the
similarity factor for the coupling $g$=0.2
 with the transverse space discretized using
$k={1 \over \kappa} tan{u \pi \over 2}$ 
where $u$'s are the quadrature
points (see Appendix \ref{APnumpro}). 
\begin{table}
\begin{center}
\begin{tabular}{||c|c|c|c|c|c|c||}
\hline \hline
$n_1$ &  $n_2$ &   \multicolumn{5}{c||}{Parametrization I     } \\
\hline
  ~10~ & ~10~ & ~4.320~ & ~4.353~ & ~4.357~ & ~4.357~ & 
  ~4.361~ \\
  20 & 20 & 4.375 & 4.442 & 4.484 & 4.484 & 4.485 \\
  20 & 30 & 4.398 & 4.482 & 4.546 & 4.583 & 4.610 \\
  20 & 40 & 4.411 & 4.502 & 4.570 & 4.615 & 4.656 \\
  30 & 40 & 4.412 & 4.503 & 4.570 & 4.615 & 4.655 \\
  40 & 50 & 4.420 & 4.515 & 4.585 & 4.634 & 4.678 \\
  40 & 60 & 4.426 & 4.524 & 4.594 & 4.645 & 4.692 \\
  40 & 80 & 4.434 & 4.535 & 4.607 & 4.661 & 4.709 \\  
\hline
 $n_1$ &  $n_2$ & 
\multicolumn{5}{c||}   {Parametrization II}\\
\hline 
 ~~10~~ & ~~10~~ & ~4.163~ & ~4.194~ & ~4.194~ & ~4.194~ &
 ~4.203~\\
 ~20~ &  ~20~ & ~4.186~ & ~4.244~ &  ~4.276~ & ~4.276~ &
 ~4.277~ \\
  ~20~ &  ~30~ & 4.192 & 4.256 & 4.296 & 4.323 &
 4.329 \\
  20 &  40 & 4.195 & 4.262 & 4.304 & 4.335 &
 4.344 \\
  30  &  40  & 4.195 & 4.262 & 4.304  & 4.335 &
 4.344 \\
  40 & 50 &  4.197 & 4.266 & 4.308 & 4.341 & 4.353 \\
  40 & 60 & 4.199 & 4.268 & 4.311 & 4.345 & 4.359  \\
  40 & 80 & 4.201 & 4.272 & 4.315 & 4.350 & 4.367 \\
\hline
$n_1$ &  $n_2$ &   \multicolumn{5}{c||}{Parametrization III} \\
\hline
 10 & 10 & 4.360 & 4.379 & 4.381 & 4.381 & 4.391 \\
 20 & 20 & 4.469 & 4.533 & 4.572 & 4.572 & 4.572 \\
 20 & 30 & 4.503 & 4.604 & 4.674 & 4.721 &  4.749 \\
 20 & 40 & 4.520 & 4.632 & 4.703 & 4.768 & 4.810 \\
 30 & 40 & 4.528 & 4.636 & 4.714 & 4.768 & 4.811 \\
 40 & 50 & 4.542 & 4.657 & 4.736 & 4.797 & 4.848 \\
 40 & 60 & 4.548 & 4.668 & 4.748 & 4.813 & 4.866 \\
 40 & 80 & 4.556 & 4.683 & 4.764 & 4.833 & 4.889\\
\hline
\hline
\end{tabular}
\end{center}
\caption[Convergence of eigenvalues with $n_1$ and $n_2$ for different
similarity factors.]
 {Convergence of eigenvalues with $n_1$ and $n_2$ for the parametrization
$k={1\over \kappa}tan(q\pi/2)$. The parameters
are 
 $m$=1.0, $g$=0.2, $\epsilon=0.00001$, $\delta=0.00001$, $\kappa=10.0$,
$\sigma=4.0$, ($u_0$=0.1, $n_g$=2 and the number of quadrature points $n_s=500$
for $\sigma$ integration for parametrization II).
  \label{srgtab2} }
\end{table}
The table show that for $m=1$,
convergence is rather slow for all three choices of the similarity factor
compared to the results in Bloch approach.  Among the three choices, 
parameterization II shows better convergence. 

\begin{figure}[h]
\centering
\fbox{\includegraphics[height=10cm]{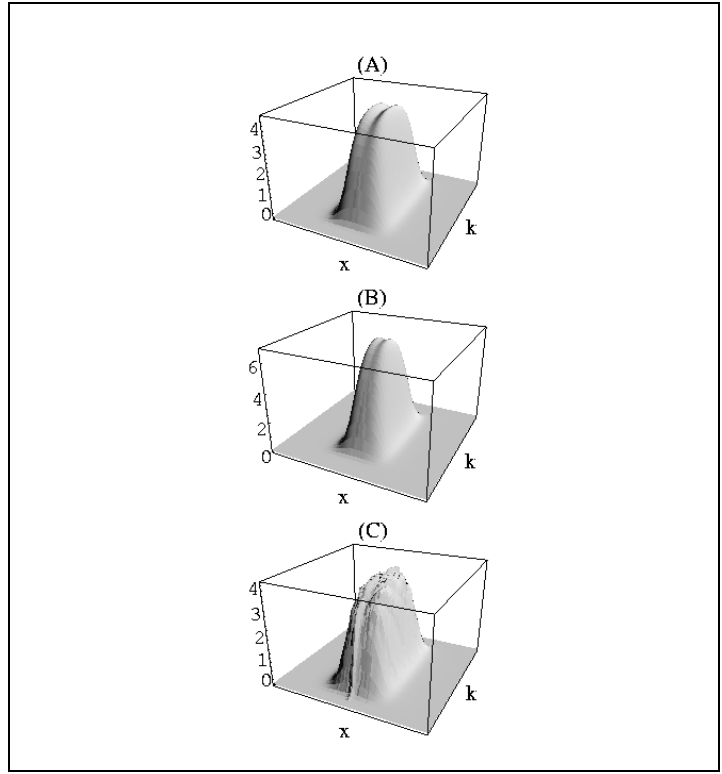}}
\caption[The ground state wavefunction  
for different choices of the similarity factor.]
{ The ground state wavefunction  
for different choices of the similarity factor using the parametrization
$k={1\over \kappa}tan(q\pi/2)$
for transverse momentum grid and for $n_1=40$, $n_2=80$, $m$=1.0,
$g=0.2$, $\sigma=4.0$ and $\epsilon = \delta =10^{-5}$, $\kappa=10.0$ 
as a function of $x$ and $k$. (A) Parametrization I, (B) Parametrization II
with $n_g=2$, $u_0=0.1$, $n_s=500$,
(C) Parametrization III. \label{srgfig1}}
\end{figure}
Let us now discuss the nature of low lying levels and wavefunctions. First we 
show the ground state wavefunctions for all three similarity factors for a 
given choice of parameters in Fig. \ref{srgfig1}.  
As is anticipated, step function choice
produces a non-smooth wavefunction. For parameterizations I and II,
the wavefunctions show some structure near $x=0.5$.
 From our previous 
experience with calculations in the Bloch formalism, we believe that the 
structures indicate poor convergence with the number of grid points.

Now consider the structure of low lying levels. Recall that in the 
Bloch formalism discussed in the previous chapter, the ordering of levels
was $l=0,~1,~0,~ \ldots $ 
corresponding to  logarithmic potential in the nonrelativistic limit (see
Appendix \ref{APnrbe}). 
\begin{figure}[h]
\centering
\fbox{\includegraphics[height=10cm]{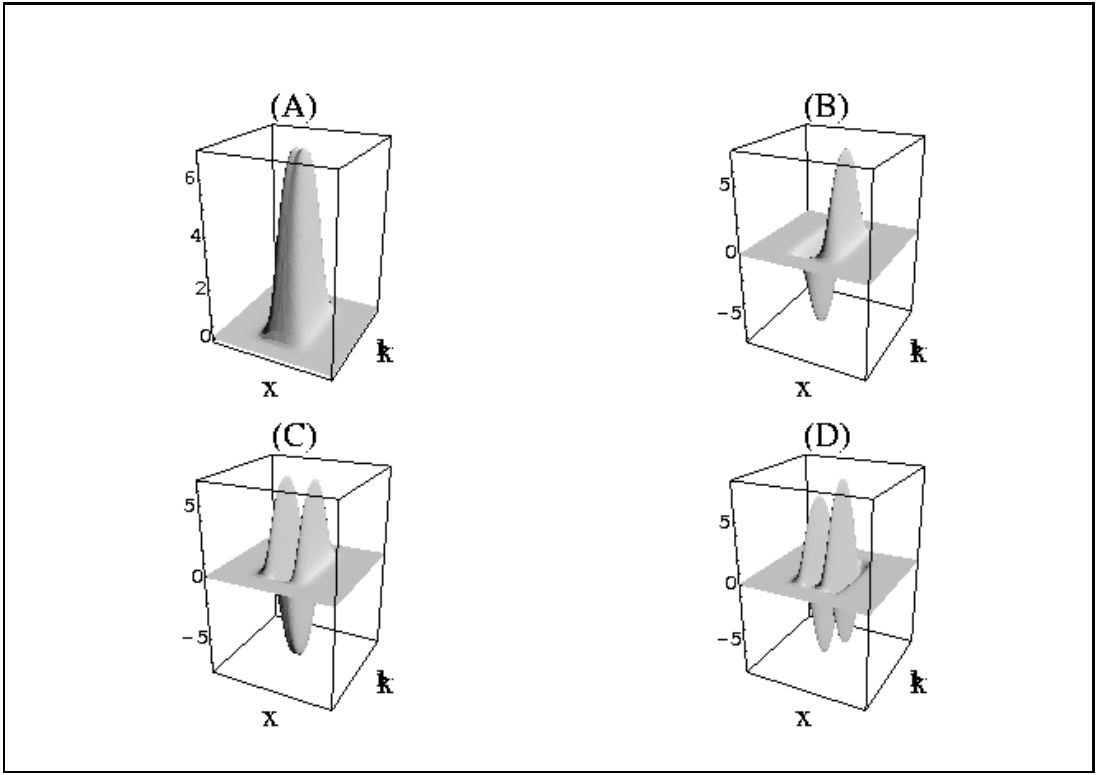}}
\caption[The wavefunctions corresponding to the lowest
four eigenvalues  as a function of $x$ and $k$ with parametrization II. ]
{ The wavefunctions corresponding to the lowest
four eigenvalues  as a function of $x$ and $k$ for parametrization II.
The parameters are as in FIG. \ref{srgfig1}. (A) Lowest state, (B) first excited
state, (C) second excited state, (D) third excited state. \label{srgfig2a}}
\end{figure}
In the presence of effective interactions generated by the similarity 
transformation, obviously the level ordering changes. Now we have 
additional confining interactions which, however, act differently in 
longitudinal and transverse directions. From our analytic calculation in
Sec. \ref{stepfunc} we know that the confining potential generated by
similarity transformation is linear in $x^1$ and square root in
$x^-$. Thus a node in  $x^1$ costs more energy than a node in
$x^-$ and thus the states with nodes in $x^1$ (i.e., in $k$) will be of
higher energy  compared to states with nodes in $x^-$.  This
feature is well manifested in 
Fig. \ref{srgfig2a} where the wavefunctions 
for the first four low lying levels are presented for parametrization II. 

There is extra freedom in parametrization II due to the presence of 
$\Sigma M^2_{ij}$ in the definition of $u_{\sigma ij}$, Eq. (\ref{usg}). For zero 
transverse momentum of constituents, $\Sigma M^2_{ij}$ has the minimum value 
$8m^2$. Thus relative insensitivity of parametrization to $\sigma$ in 
parametrization II for small values of $\sigma$ may be due to this factor. When 
we consider the heavy fermion mass limit, presence of $8m^2$ in 
$u_{\sigma ij}$ enhances the effect of similarity factor. In Fig. \ref{srgfig2b} we 
present the wavefunctions corresponding to first four levels for 
parametrization II with $8m^2$ subtracted from $\Sigma M^2_{ij}$ in 
$u_{\sigma ij}$. From Figs. \ref{srgfig2a} and \ref{srgfig2b}, 
note that the fourth level is different 
for parametrization II with and without $8m^2$ in $u_{\sigma ij}$. 
\begin{figure}[h]
\centering
\fbox{\includegraphics[height=10cm]{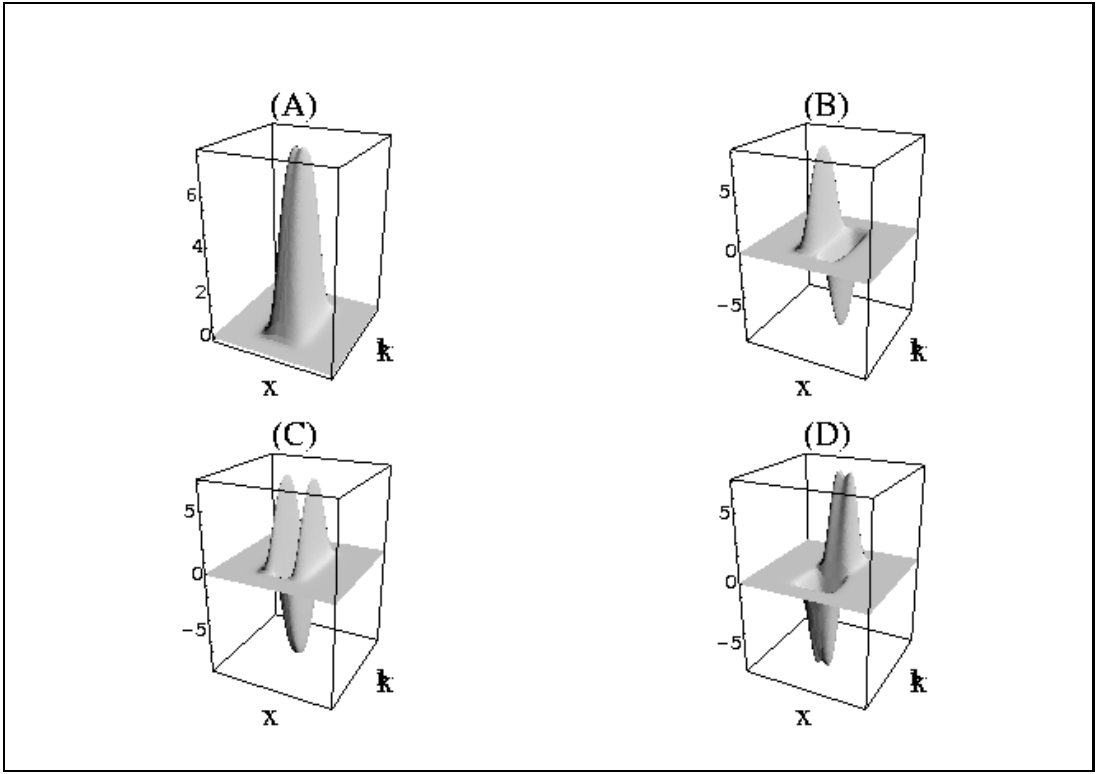}}
\caption
{ Same as in FIG. \ref{srgfig2a} but with $8m^2$ subracted from 
$ \Sigma M^2_{ij}$ . \label{srgfig2b}}
\end{figure}

By suitable choice of parameters we can study the interplay of rotationally 
symmetric logarithmically confining interaction and effective interactions 
generated by similarity transformation. 
Since for a given coupling
constant $g$, strength of the
logarithmic interaction and similarity generated interactions are 
determined by $m$ and $\sigma$ respectively, for $m >> \sigma$ we should 
recover the Bloch spectrum (Fig. \ref{fig2}). Upto what levels the recovery occurs, 
of course depends on the exact value of $m$ and the energy scale
$\sigma$ at which the effective Hamiltonian is constructed.
 As we have  already observed, 
for parametrization II this will happen only if $8m^2$ is subtracted from
$\Sigma M^2_{ij}$. For this case, we present the first four levels for 
$m=10.0$ and $\sigma =4.0$ in Fig. \ref{srgfig3} which clearly shows the level
 spacing corresponding to the Bloch spectrum presented in Fig. \ref{fig2}.
\begin{figure}[h]
\centering
\fbox{\includegraphics[height=10cm]{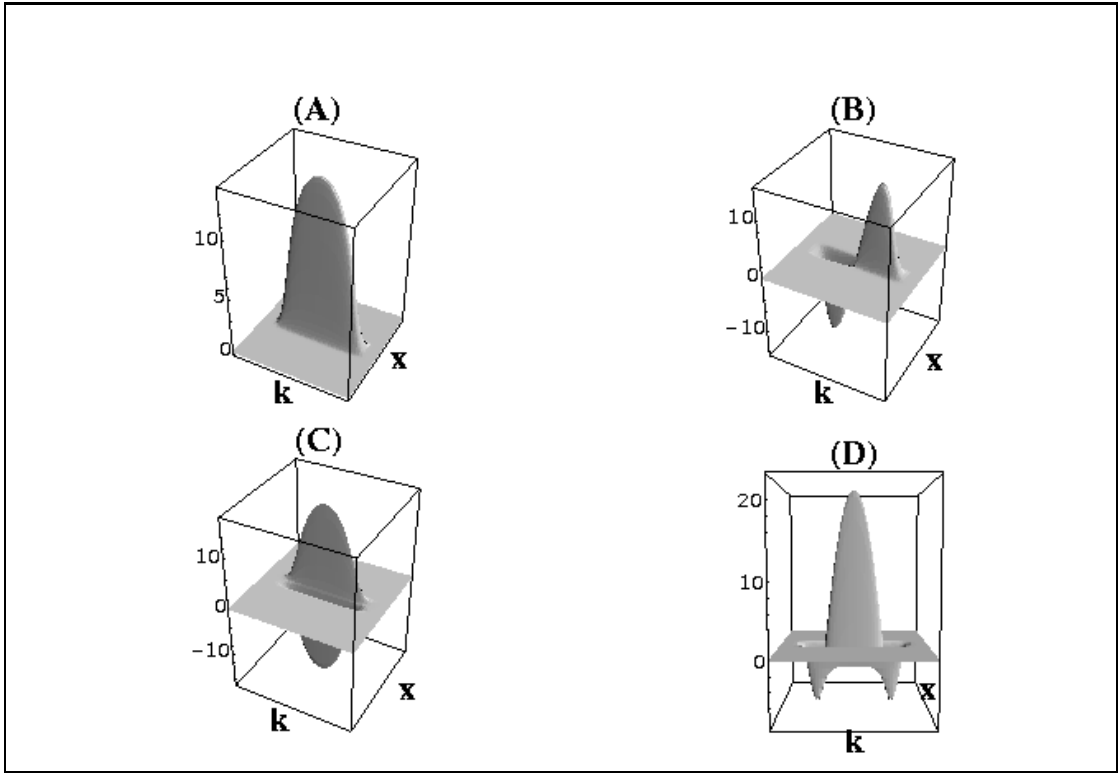}}
\caption[The wavefunctions corresponding to the lowest four
eigenvalues as a function of $x$ and $k$ with parametrization II for large
 fermion mass.] 
{The wavefunctions corresponding to the lowest four
eigenvalues as a function of $x$ and $k$ for large fermion mass 
with parametrization II with $8m^2$ subtracted from $\Sigma M^2_{ij}$.
The parameters are $m=10.0$
$g=0.2$, $\epsilon=\delta=0.00001$,
 $\kappa=10.0$, $n_g=2$, $u_0=0.1$, $n_1=40$, $n_2=80$.  
(A) Lowest state, (B) first excited
state, (C) second excited state, (D) third excited state.\label{srgfig3} }
\end{figure}
                
\begin{table}[h]
\begin{center}
\begin{tabular}{||c|c|c|c|c|c||}
\hline \hline
& \multicolumn{5}{c||}{Eigenvalues ($M^2$)} \\
\hline
 $\sigma$   &  \multicolumn{5}{c||}  {Parametrization I} \\
\hline                                                           
     2.0 & 4.214 & 4.285 & 4.329 & 4.365 & 4.393\\ 
     4.0 & 4.434 & 4.535 & 4.607 & 4.661 & 4.709\\
     6.0 & 4.701 & 4.821 & 4.914 & 4.980 & 5.043\\
\hline
   $\sigma$   &  \multicolumn{5}{c||}  {Parametrization II} \\
\hline
      ~2.0~ & ~4.157~ & ~4.214 ~ & ~4.248~ & ~4.260 ~ & ~4.276~ \\
      4.0 & 4.201 & 4.272 & 4.315 & 4.350 & 4.367 \\
      6.0 &  4.266 & 4.350 & 4.404 & 4.446 & 4.483 \\
\hline
 $\sigma$   &  \multicolumn{5}{c||}  {Parametrization III} \\
\hline
      2.0 & 4.254  & 4.346  & 4.395  & 4.443 & 4.477 \\
      4.0 & 4.556 & 4.683  & 4.764  & 4.833 & 4.889  \\
      6.0 & 4.927 & 5.075 & 5.179  & 5.262 & 5.333 \\

\hline
\hline
\end{tabular}
\end{center}
\caption
[ Variation with $\sigma$ of the  full SRG Hamiltonian for $g =0.2$ and
$k={1\over
\kappa}tan(q\pi/2)$.] 
{ Variation with $\sigma$ of the  full Hamiltonian (excluding the 
 imaginary term). The parameters
are $m=1.0$,
$g=0.2$, $n1=40$, $n2=80$, $\epsilon=0.00001$, $ \kappa=10.0$,
$(k={1\over
\kappa}tan(q\pi/2))$, $\delta=0.00001$,
($u_0=0.1$, $n_g=2$, and $ n_s=500$ for $\sigma$ integration for
parameterization II).\label{srgtab3}}
\end{table}
\begin{table}[h]
\begin{center}
\begin{tabular}{||c|c|c|c|c|c||}
\hline \hline
& \multicolumn{5}{c||}{Eigenvalues ($M^2$)} \\
\hline
 $\sigma$   &  \multicolumn{5}{c||}  {Parametrization I} \\
\hline                                                           
     2.0 & 4.941 & 5.418 & 5.708 & 5.713 & 5.950 \\ 
     4.0 & 5.684 & 6.421 & 6.909 & 7.306 & 7.647 \\
     6.0 & 6.678 & 7.617 & 8.274 & 8.801 & 9.267 \\
\hline
   $\sigma$   &  \multicolumn{5}{c||}  {Parametrization II} \\
\hline
     ~ 2.0~ &~ 4.769 ~& ~5.152~ & ~5.226~ &~ 5.372~ &~ 5.485~ \\
      4.0 & 4.876 & 5.357 & 5.592 & 5.645 & 5.886 \\
      6.0 &  5.071 & 5.655 & 6.016 & 6.159 & 6.316 \\
\hline
 $\sigma$   &  \multicolumn{5}{c||}  {Parametrization III} \\
\hline
      2.0 & 4.888  & 5.589  & 5.882  & 6.120 & 6.263 \\
      4.0 & 5.966 & 6.972  & 7.518  & 8.077 & 8.465  \\
      6.0 & 7.359 & 8.603 & 9.360  & 10.083 & 10.621 \\

\hline
\hline
\end{tabular}
\end{center}
\caption
[ Variation with $\sigma$ of the  full SRG Hamiltonian for $g=0.6$ and
$k={q\Lambda m
\over (1-q^2)\Lambda +m}$.]
{ Variation with $\sigma$ of the  full hamiltonian (excluding the 
 imaginary term). The parameters
are $m=1.0$,
$g=0.6$, $n1=58$, $n2=58$, $\epsilon=0.00001$, $ \Lambda=20.0$, 
$(k={q\Lambda m
\over (1-q^2)\Lambda +m})$, $\delta=0.00001$,
($u_0=0.1$, $n_g=2$,and $ n_s=500$ for $\sigma$ integration for
parameterization II). \label{srgtab4}}
\end{table}
Finally, we discuss the sensitivity of the spectra to the similarity scale
$\sigma$. Ideally the low lying energy levels should be insensitive to 
$\sigma$. However we have calculated the effective Hamiltonian to only order
$g^2$ and we expect significant sensitivity to $ \sigma$. In Tables \ref{srgtab3}
 and \ref{srgtab4} we present
the lowest five eigenvalues for all three parametrizations of the similarity
factor for $g=0.2$ and $g=0.6$ respectively. As expected $\sigma$ dependence
is greater for larger value of $g$.

\begin{table}[h]
\begin{center} 
\begin{tabular}{||c|c|c|c|c|c|c||}
\hline \hline
& & \multicolumn{5}{c||}{Eigenvalues ($M^2$)} \\
\hline
  $g$ & $\sigma$   &  \multicolumn{5}{c||} {Parametrization II} \\
\hline
        & 2.0 & 4.126 & 4.166 & 4.176 & 4.189 & 4.201 \\
  ~0.2~ &~4.0~ & ~4.172~ & ~4.235~ &~ 4.273~ & ~4.298~ & ~4.304~ \\
     & 6.0 &  4.241 & 4.321 & 4.371 & 4.411 & 4.445 \\
\hline
\hline
      & 2.0 & 4.739 & 5.003 & 5.020 & 5.134 & 5.193 \\ 
  0.6  & 4.0 & 4.802 & 5.222 & 5.350 & 5.470 & 5.626 \\
    & 6.0 & 4.993 & 5.541 & 5.876 & 5.940 & 6.156 \\
\hline
\hline
\end{tabular}
\end{center}
\caption[ Variation with $\sigma$ of the  full Hamiltonian (excluding the 
 imaginary term) after subtracting $8m^2$ from $\Sigma M^2_{ij}$ in
the definition of $u_{\sigma ij}$ and $n_g=2$.]
{ Variation with $\sigma$ of the  full Hamiltonian (excluding the 
 imaginary term) after subtracting $8m^2$ from $\Sigma M^2_{ij}$ in
the definition of $u_{\sigma ij}$.
 The parameters are 
 $m=1.0$, $\epsilon=0.00001$, $\delta=0.00001$,
$u_0=0.1$, $n_g=2$, and $ n_s=500$  \\
 1) for $g=0.2$, $k={1\over\kappa}tan(q\pi/2)$ with $ \kappa=10.0$ and 
$n1=40$, $n2=80$\\
 2) for $g= 0.6$, $k={q\Lambda m \over (1-q^2)\Lambda +m}$ with
$\Lambda=20.0$, and $n1=n2=58$. \label{srgtab5}}
\end{table}
Among the three parametrizations, the 
paramterization II is least sensitive to
$\sigma$. In order to check whether this behaviour is due to the presence of
$8 m^2$ in $\Sigma M^2_{ij}$ in the definition of $u_{\sigma ij}$ we present
the results in Table \ref{srgtab5} for parametrization II with $8 m^2$ subtracted
from $\Sigma M^2_{ij}$. It is clear that sensitivity to $\sigma$ is still
considerably less compared to the other two parametrizations. This may be due
to the fact that  $\Sigma M^2_{ij}$ is added to $\sigma^2$ in the definition
of $\sigma$.

\begin{table}[h]
\begin{center}
\begin{tabular}{||c|c|c|c|c|c|c||}
\hline \hline
  $g$ & $\sigma$   &  \multicolumn{5}{c||}  {$M^2$
(Parameterization II)} \\
\hline
\hline
      & 2.0 & 4.781 & 5.049 & 5.051 & 5.161 & 5.221 \\ 
  0.6  & 4.0 & 4.843 & 5.235 & 5.391 & 5.464 & 5.639 \\
    & 6.0 & 5.030 & 5.538 & 5.848 & 5.956 & 6.106 \\
\hline
\hline
\end{tabular}
\end{center}
\caption[Variation with $\sigma$ of the  full Hamiltonian(excluding the 
 imaginary term) after subtracting the $8m^2$ term from $\Sigma M^2_{ij}$ in 
the definition of $u_{\sigma ij}$ and  $n_g=1$.]
{ Variation with $\sigma$ of the  full Hamiltonian(excluding the 
 imaginary term) after subtracting the $8m^2$ term from $\Sigma M^2_{ij}$ in 
the definition of $u_{\sigma ij}$.
 The parameters
are 
 $m=1.0$, $\epsilon=0.00001$, $\delta=0.00001$,
$u_0=0.1$, $n_g=1$, and $ n_s=500$  
  $g= 0.6$, $k={q\Lambda m \over (1-q^2)\Lambda +m}$ with
$\Lambda=20.0$, and $n_1=n_2=58$. \label{srgtab6} }
\end{table}
 Note that in parametrization II sensitivity to $\sigma$ is
controlled also by additional parameters $u_0$ and $n_g$.  The sharpness
of the cutoff in this paramatrization depends on $n_g$. The cutoff
becomes sharper as one increases the value of $n_g$.  $\sigma$
dependence of the eigenvalues
for two different  $n_g$ are prestented in Tables \ref{srgtab5} and
\ref{srgtab6}. The other adjustable  parameter $u_0$ is a samll number
($u_0 <<1$) and
the sensitivity to
 $u_0$ of the lowest five eigenvalue for the coupling $g = 0.6$
is presented in Table \ref{srgtab7}.  
\begin{table}[h]
\begin{center}
\begin{tabular}{||c|c|c|c|c|c|c||}
\hline \hline
  $g$ & $u_0$   &  \multicolumn{5}{c||}  {$M^2$  (Parameterization II)} \\
\hline
     & 0.2 & 4.990 & 5.548 & 5.887 & 5.969 & 6.171\\
     & 0.1 & 4.802 & 5.222 & 5.350 & 5.469 & 5.626 \\
 0.6 & 0.05 & 4.746 & 5.076 & 5.091 & 5.253 & 5.313 \\
     & 0.01 & 4.813 & 5.003 & 5.029 & 5.069 & 5.130 \\
\hline
\hline
\end{tabular}
\end{center}
\caption[Variation with $u_0$ of the  full Hamiltonian after subtracting the 
$8m^2$ term from 
$\Sigma M^2_{ij}$ in the definition of $u_{\sigma ij}$.]
{ Variation with $u_0$ of the  full Hamiltonian(excluding the 
 imaginary term) after subtracting the $8m^2$ term from 
$\Sigma M^2_{ij}$ in the definition of $u_{\sigma ij}$.
The parameters are $m=1.0$, $\epsilon=0.00001$, $\delta=0.00001$,
$\sigma=4.0$, $n_g$=2, and $ n_s=500$  
  $g= 0.6$, $k={q\Lambda m \over (1-q^2)\Lambda +m}$ with
$\Lambda=20.0$, and $n1=n2=58$. \label{srgtab7}}
\end{table}

\section{Summary and Discussion }
From our results of Chapter \ref{chapbloch}, we know that
the attempt to solve (2+1) dimensional gauge theories using the Bloch 
effective Hamiltonian is unseccessful due to problem of uncancelled
infrared divergences. They arise out of vanishing energy denominators
and a more sophisticated tool is necessary to handle the bound state
problems in QCD.   
Similarity renormalization formalism attempts to solve the bound state
problem in a two step process. At the first step, coupling between low
and high energy degrees of
freedom are integrated out and ultraviolet renormalization carried out
perturbatively. In the second step, the effective Hamiltonian is
diagonalized non-perturbatively.

Here I  briefly summarize the main points addressed in this chapter. 
In order to have a better understanding of the numerical results, we have
performed analytical calculations with step function form for the
similarity factor. Many interesting results emerge from our analytical 
calculations. First of all, it is shown that due to the presence of 
instantaneous interactions in gauge theories on the 
light front, the logarithmic infrared divergence that
appeared in the Bloch formalism persists in two places, namely a part of the
self energy contribution and the Coloumb interaction that gives rise to the
logarithmically confining potential in the nonrelativistic limit.
However the terms that persist are precisely those
that produce a cancellation of resulting infrared divergences in the bound
state equation.
The rest of the infrared problem that appeared in the Bloch
formalism due to the vanishing energy denomnator problem is absent in 
the similarity formalism. 

Similarity transformation however prevents the cancellation of the most
severe ${1 \over (x-y)^2}$ singularity between instantaneous gluon exchange and
transverse gluon exchange interactions and produces ${ 1 \over
\sqrt{\delta}}$ divergences in the self energy and gluon exchange
contributions which cancel between the two in the bound state equation.
The resulting effective interaction between the quark and antiquark grows
linearly with large transverse separation but grows only with the square
root of the longitudinal separation. This produces severe violations of rotational
symmetry in the bound state spectrum.                
 We have also verified that no $ln
~ \delta$ divergence results from the ${ 1 \over x-y}$ singularity in the self
energy and gluon exchange contributions.     

In the G{\l}azek-Wilson formalism the exact form of the similarity factor
$f_\sigma$ is left unspecified. In the literature an exponential form has
been used in numerical calculations \cite{2Allen:2000kx}. For analytical 
calculations it
is convenient to choose a step function even though it is well known that it
is not suitable for quantitative calculations \cite{2wk}. 
There is also a
proposal due to Stan G{\l}azek which has two extra free parameters.
We have tested all three parametrizations. 
Our numerical results indeed show that step function choice always produces
non-smooth wavefunctions. Parameterization II costs us an extra integration to be
performed numerically but convergence is slightly better for small $g$ compared
to exponential form. All three parametrizations produce violations of
rotational symmetry even for small $g$. When an exponential form is used in
the G{\l}azek-Wilson formalism, the resulting effective Hamiltonian differs
from the Wegner form only by an overall factor that restricts large energy
diffrences between initial and final states. Numerically we have found this
factor to be insignificant.

We have studied the sensitivity of the low lying eigenvalues to the
similarity scale $ \sigma$. Since the effective Hamiltonian is calculated
only to order $g^2$ results do show sensitivity to $\sigma$. Among the three
parametrizations the form II is least sensitive to
$\sigma$ due to the functional form chosen. We have also studied the
sensitivity of eigenvalues to the parameters $u_0$ and $n_g$.  

The bound state equation has three parameters $m$,
$g^2$ and $\sigma$ with dimension of mass. The strength of 
the logarithmically confining interaction
is determined by $m$ and the strength of the rotational symmetry violating
effective interactions generated by similarity transformation is determined
by $\sigma$. For a given $g$ we expect the former to dominate over the 
latter for
$ m >> \sigma$. An examination of low lying eigenvalues and corresponding
wavefunctions show that this is borne out by our numerical calculations.

A major problem in the calculations is the slow convergence.
Compared to the Bloch formalism, in calculations with the
similarity formalism, 
various factors may contribute to this problem with the
Gauss quadrature points. One important factor 
is the presence of linear and square root confining interactions
generated by the similarity
transformation. It is well known that such interactions are highly singular
in momentum space. Another factor is the presence of ${ 1 \over
\sqrt{\delta}}$ divergences, the cancellation of which is achieved
numerically. It is of interest to carry out the same calculations with
numerical procedures other than the Gauss quadrature. However, one should
note that calculations in (3+1) dimensions employing basis functions and
splines have also yielded \cite{2Allen:2000kx} wavefunctions which show 
non-smooth structures.

An undesirable result of the similarity transformation carried out in
perturbation theory is the
violation of rotational symmetry. Our results show that this  violation
persists at all values of $g$ for $m=1$. Such a violation was also observed 
in (3+1)
dimensions. In that case the functional form of the logarithmic potential
generated by similarity transformation is the same in longitudinal and
transverse directions but the coefficients differ by a factor of two. Same
mechanism in (2+1) dimensions makes even the functional forms different. The
important questions are whether the confining interactions generated by the
similarity transformation are an artifact of the lowest order approximation
and if they are not, then, whether the violation of rotational symmetry will
diminish with higher order corrections to the effective Hamiltonian.  
Recall that matrix element between low and
 high energy degrees of freedom has been integrated out and the effective
low energy Hamiltonian determined only to order $g^2$. A clear answer will
emerge only after the determination of the effective Hamiltonian to fourth
order in the coupling.        

\chapter{Fermion Formulation on a Light-Front Transverse Lattice} \label{tlchap1} 
\section{Introduction}
In the previous two chapters we have discussed the meson bound state
problem in (2+1) dimensions in the light-front framework using Bloch and
similarity effective Hamiltonians  and have seen that light-front
framework provides the
opportunity for non-perturbative studies in Hamiltonian formalism.
In this chapter we introduce another
approach in light-front Hamiltonian framework. 
   
Till date, the most practiced non-perturbative technique is the
 lattice gauge theory \cite{3Wilson:1974sk}.
Using path integral formalism in Euclidean space, with no gauge fixing one
calculates the n-point Green
functions, but no direct informations about the bound state wavefunctions
are accessible in this approach.  Hamiltonian formalism provides  bound
state wavefunctions in a straight-forward way. Since
 boosts are kinemetical, one can have frame independent
description of the bound state wavefunctions in light-front framework.
One extra advantage of lattice gauge theory over
light-front formalism is full gauge invariance. 
Light front Hamiltonian formulation of transverse lattice QCD
\cite{3Bardeen:1976tm,3Bardeen:1980xx} is an optimum combination of
lattice gauge theory and light-front QCD. It uses the power of
light-front Hamiltonian formalism to produce the boost invariant
wavefunctions and the advantage of having gauge invariant ultraviolet cutoff
from lattice gauge theory.
With the gauge choice $A^{+}=A^0+A^3=0$ and the elimination of the
constrained variable $A^-=A^0-A^3$, it uses minimal gauge degrees of 
freedom in a manifestly gauge invariant formulation exploiting the 
residual gauge symmetry in this gauge. 

For Hamiltonian formalism light-front time ($x^+$) is kept continuous.
Due to the constraint equations in fermion and gauge degrees of freedom
in the light-front,
non-locality comes in the longitudinal direction.  Also there is no
ultraviolet divergences coming from small $x^-$ and hence $x^-$ is 
not latticized. Ultraviolet divergences come only from small transverse
separations which we want to regulate with gauge invariant cutoff. Thus, 
the  transverse plane ($x^{\perp} = (x^1,x^2)$) is discretized on a
square lattice. This defines the light-front transverse lattice (LFTL)
in (3+1) dimensions.
It is a promising
and  developing tool for non-perturbative investigations of QCD. 
So far encouraging results have been obtained
in the pure gauge sector \cite{3DalleyPRL} and in the meson sector with 
particle number truncation (for a recent review see,
Ref. \cite{3review}). 

It is well known that fermions on the lattice pose challenging
problems due to the doubling phenomenon. Light-front formulation of
field  theory has its own peculiarities concerning fermions because of
the presence of a constraint equation. As an example, the usual chiral
transformation on the four component fermion field is incompatible
with the constraint equation for nonzero fermion
mass \cite{3Wilson:1994fk}.  There have been previous studies of
fermions on the 
transverse lattice \cite{3stagger,3BK,3dalme,3buseal} in different
contexts. But properties and origin of  species doubling  of fermion on
light-front transverse lattice were not studied with proper care and
desired details. So, before embarking on any QCD calculation on
transverse lattice, it is better to understand  the
fermions on it in detail. 
 I devote this chapter to discuss different ways of
formulating fermion on a light-front transverse lattice and related
issues such as absence or origin of doubers, different ways of removing
doubers, relevant symmetry on light-front transverse lattice and so on 
\cite{3DH3}.   
 
As we shall see later in this chapter, 
the presence of the constraint equation in light front
field theory allows different methods to put fermions on a transverse
lattice. It is worthwhile to study all the different methods in order
to examine their strengths and weaknesses. 
Here I should also mention two important points. One, our ultimate aim
is to calculate QCD observables on LFTL where we deal with a Hamiltonian
acting on a Fock space. For a  reasonable size of Fock space, computing 
limitations will force us to be in a reasonably small lattice volume when we
deal with realistic problems. The second point is that the currently
practiced version of the transverse
lattice gauge theory uses {\it linear} link variables
\cite{3Bardeen:1976tm,3Bardeen:1980xx} justified on a
coarse lattice and recovering continuum physics is nontrivial.
Hence it is very important to carry out  detailed numerical investigation
of all the possible ways of formulating fermions on  LFTL with proper
attention to the finite volume effects.
  
In one of the approaches of treating fermions on the light front transverse
lattice,  we maintain as much transverse locality as possible on 
the lattice by using forward and backward lattice derivatives without
spoiling the  
hermiticity of the Hamiltonian. In this case
doublers are not present and the helicity flip term proportional
to the fermion mass in the full light front QCD becomes an irrelevant
term  in the 
free field limit. Thus in finite volume, 
depending on the boundary condition used, the two helicity states 
of the fermion may not be degenerate in the free field limit. However, 
we find that in the infinite volume limit the degeneracy is restored 
irrespective of the boundary condition. 

In the second approach \cite{3BK}, symmetric derivatives are used 
which results  in a Hamiltonian  with only
next to nearest neighbor interaction when we take the free field limit. 
As a consequence even and odd
lattice sites decouple and the fermions live independently of each other on the
two  sets of sites. As a result we get four species of
fermions  on a two dimensional lattice  as excitations around 
zero transverse momentum. Note that this is quite
different from what one gets in the conventional Euclidean lattice theory when
one uses symmetric derivatives. In that case, doublers have at least one
momentum component near the edge of the Brillouin zone. 
The doublers can be removed in more than one way.  We also study the 
staggered fermion formulation on the light front transverse lattice to eliminate 
two doublers and reinterpret the remaining two as  two
flavors.  In this  light front staggered fermion formulation, there is 
no flavor mixing in free field limit. But, in QCD, we get  irrelevant
flavor mixing terms.
An alternative which removes  doubling completely is to add the conventional Wilson 
term which generates many irrelevant interactions on the transverse
lattice. Among them, the  helicity flip interactions vanish but
the helicity non flip interactions survive in the free field limit.  
 
\section{Hamiltonian with forward and backward derivatives} \label{H-FB}
\subsection{Construction}
In this section we propose to use different lattice derivatives for
dynamical and constrained fermion field components  on the transverse lattice
in such a way that sacred Hermiticity of the Hamiltonian is
preserved.  
The starting point of our discussion is 
the fermionic part of the QCD Lagrangian density 
\be
{\cal L}_f = {\bar \psi}(i \gamma^\mu D_\mu -m ) \psi
\ee
with $ i D^\mu = i \partial^\mu - g A^\mu$. 
Since fermions on LFTL are our main concern in this chapter, we omit the pure
gauge part from QCD Lagrangian.
 
Moving to the light front coordinates we  impose the light-cone
gauge $A^{+}=0$ and introduce 
the transverse lattice by discretizing the transverse plane on a square 
lattice with lattice spacing $a$. Now, ${\cal L}_f$ on the LFTL can be
written as  
\be
{\cal L}_f &=& {\psi^+}^\dagger ( i \partial^- - g A^-) \psi^+ +
{\psi^-}^\dagger i \partial^+ \psi^- \nonumber \\
&~&- i {\psi^-}^\dagger \alpha_r D^f_r \psi^+ - i {\psi^+}^\dagger \alpha_r
D^b_r \psi^- \nonumber \\
&~& - m {\psi^-}^\dagger \gamma^0 \psi^+ - m {\psi^+}^\dagger \gamma^0
\psi^-. \label{lfb}
\ee
 Here $r=1,2$ and $D^{f/b}_r $  is the forward/backward covariant
lattice derivative defined as 
\be
D_r^f \eta(\bx)= { 1 \over  a} [ U_r(\bx) \eta(\bx + a \br) - \eta(\bx)]
\ee
and
\be
D_r^b \eta(\bx) = { 1 \over a} [ \eta(\bx) - U_r^\dagger(\bx -a \br)
\eta(\bx -a \br)],
\ee
where $a$ is the lattice constant and $\br$ is unit vector in the
direction $r=1,2$ and ${D^f_r}^{\dagger} =-D^b_r$. $U_r(\bx)$ is the
group valued lattice gauge field with the property $U_r^{\dagger}(\bx) 
=U_{-r}(\bx+a\br)$. In the  weak coupling limit 
\be
 U_r(\bx) \approx e^{iga A_r(\bx + a \br/2)}.
\ee
For notational convenience we suppress $x^-$ in the
arguments of the fields. 

Our goal here is to write the most local lattice
derivative.  That is why, instead of using the symmetric lattice
derivative,  in the above we have used the forward and backward
lattice derivatives. However,  
the Hermiticity of the Lagrangian (Hamiltonian)  requires that if one
of the  covariant lattice derivatives 
appearing in Eq. (\ref{lfb}) is  the forward derivative, the other has 
to be the backward derivative or vice versa.  

The constraint equation is
\be
i \partial^+ \psi^- = ( i \alpha_r D^f_r + \gamma^0 m) \psi^+.
\ee
 Eliminating the constrained field
component $\psi^-$ in terms of dynamical field $\psi^+$, we obtain 
\be
{\cal L}_f &=& {\psi^+}^\dagger ( i \partial^- - g A^-) \psi^+ 
     - m {\psi^+}^\dagger \gamma^0 \psi^-  
- i {\psi^+}^\dagger
\alpha_r D^b_r \psi^-             \nonumber \\
&=&  {\psi^+}^\dagger ( i \partial^- - g A^-) \psi^+ \nonumber \\
&~&~~~~ - {\psi^+}^\dagger [ i \alpha_r D^b_r + \gamma^0 m ]{ 1 \over i
\partial^+} [ i \alpha_s D^f_s + \gamma^0 m] \psi^+ .
\ee
The dynamical field $\psi^+$ can essentially be represented by two
components \cite{3hz} such  that
\be
\psi^+(x^-, x^{\perp}) = \left[ \begin{array}{l} \eta(x^-, x^{\perp})
\\ 0 \end{array} \right] ,
\ee
where $\eta$ is a two component field. Finally going over to the two 
component fields $\eta$, the Lagrangian density can be written as 
\be
{\cal L}_f&=&  {\eta}^\dagger ( i \partial^- - g A^-) \eta \nonumber \\
&~&~~~~ - \eta^\dagger [ i {\hat \sigma}_r D^b_r -i m ]
{ 1 \over i \partial^+} [ i {\hat \sigma}_s D^f_s + i m] \eta
~. \label{lfb2} 
\ee  
$ {\hat \sigma_1} = \sigma_2$ and $ {\hat \sigma_2}= - \sigma_1$ where
$\sigma_i$ are Pauli spin matrices.
Writing explicitly in terms of the link variables, the Lagrangian
density is 
\be {\cal L}_f &=& \eta^\dagger(\bx) ( i \partial^- - g A^-) \eta(\bx)
- m^2 \eta^\dagger(\bx) { 1 \over i \partial^+}\eta(\bx) \nonumber \\
& ~& ~~ -m \eta^\dagger(\bx) \sum_r {\hat
\sigma}_r {1 \over a}{1 \over  i \partial^+}\Big [
U_r(\bx) \eta(\bx + a \br) - \eta(\bx) \Big ] 
\nonumber \\
&~&~~- m \sum_r
\Big [
\eta^\dagger(\bx + a \br) U_r^\dagger(\bx) -
\eta^\dagger(\bx) \Big ] 
 {\hat \sigma}_r{ 1 \over a} { 1 \over 
i \partial^+} \eta(\bx) \nonumber \\
&~& ~- { 1 \over a^2}\sum_{r, s} [ \eta^\dagger(\bx + a \br) U_r^\dagger(\bx) -
\eta^\dagger(\bx)] {\hat \sigma}_r { 1 \over i \partial^+}{\hat
\sigma}_s [ U_s(\bx) \eta(\bx + a \bs) - \eta(\bx)]. \label{fbfull}
\ee

In the free limit the fermionic part of the 
 Hamiltonian becomes 
\be
P^-_{fb} &=& \int dx^- a^2 \sum_{\bx} {\cal H} \nonumber \\
& = & \int dx^- a^2 \sum_{\bx} \Bigg [ m^2 \eta^\dagger(\bx) 
{ 1 \over i \partial^+}\eta(\bx) \nonumber \\
&~&~~~~ - { 1 \over a^2}  \eta^\dagger(\bx) \sum_r
{ 1 \over i \partial^+} [ \eta(\bx + a \br) - 2 \eta(\bx)
+ \eta(\bx - a \br) \nonumber \\
&~&~~~~ + { 1 \over a^2} \eta^\dagger(\bx)\sum_r (am {\hat \sigma_r}) 
{ 1 \over i \partial^+} [ \eta(\bx + a \br) - 2 \eta(\bx)
+ \eta(\bx - a \br) ] \Bigg ].\label{fb}
\ee
 In order to get Eq. (\ref{fb}), we have assumed infinite transverse lattice
and accordingly  have used shifting of lattice points which is
equivalent of neglecting surface terms.  
The positive  sign in front of the last term would change if we had
switched  forward and backward  derivatives.
 One should note that
the cross term coming from the last line of Eq. (\ref{fbfull}) survives
in the free field limit. Explicitly,  the term is
\be
{ 1 \over a^2}\sum_{r \ne s}[ \eta^\dagger(\bx + a \br)-\eta(\bx)] {\hat
\sigma}_r { 1 \over i \partial^+}{\hat\sigma}_s [ \eta(\bx + a \bs) -
\eta(\bx)]. \label{cross}
\ee 
It produces extra helicity nonflip hoppings in the transverse plane. In
the continuum limit ($a \rightarrow 0$) this term does not survive and one
can recover the right continuum limit without this cross term.  If one  demands
 hypercubic (square) symmetry of
the  transverse lattice, then this cross term (Eq. (\ref{cross})) vanishes.
  More detailed
 discussion about this term is provided in Appendix \ref{APrsv}.
    
Because of the presence of ${\hat \sigma}_r$,  the last term of
Eq. (\ref{fb})  couples  
fermions of opposite helicities. Note that it is also linear in mass.
Such a helicity flip linear mass term is typical in continuum
light-front QCD.  
Here in free transverse lattice theory this 
 term arises from the interference of the first 
order derivative term and the mass term, due to the constraint equation. 
This is in contrast to the conventional lattice (see Appendix
\ref{APfblat}) where no  
helicity flip or chirality-mixing term arises in the free theory if we
use forward and backward lattice derivatives.   
 
\subsection{Absence of doubling}
Consider the Fourier transform in transverse space
\be
\eta(x^-, \bx) = \int {d^2 k \over (2 \pi)^2} e^{i {\bf k} \cdot \bx}
\phi_{\bf k}(x^-)
\ee
where
$ - {\pi \over a} \le k_1, k_2 \le + {\pi \over a} $.
Then  the helicity nonflip part of Eq. (\ref{fb}) becomes
\be
 P^-_{nf} &=& \int dx^- \int { d^2 k \over (2 \pi)^2} \int { d^2 p \over (2
\pi)^2}
\phi^\dagger_{\bf k} (x^-){ 1 \over i \partial^+} \phi_{\bf p}(x^-) a^2
\sum_\bx  e^{-i ({\bf k} - {\bf p}) \cdot \bx} \nonumber \\
&~&~~~~~~~~~~~~~~~~~~~~~~~~~~~~~~~~~~\Bigg [ m^2 -\sum_r{ 1 \over a^2} \Big 
[ e^{ i {\bf p} \cdot a \br} - 2 + 
e^{-i{\bf p} \cdot a \br} \Big ] \Bigg ].
\ee
Using
\be
a^2 \sum_\bx e^{i ({\bf k} - {\bf p}) \cdot \bx}
= ( 2 \pi)^2 ~ \delta^2({\bf k} - {\bf p})
\ee
we get, 
\be
P^-_{nf} = \int dx^- \int { d^2 k \over (2 \pi)^2} \phi^\dagger_{\bf k}(x^-) { 1 \over
i \partial^+}
\phi_{\bf k}(x^-) \Bigg [ m^2 + \sum_r k_r^2  \left ({\sin~ k_ra/2 \over k_r
a/2}\right )^2 \Bigg ]
\ee
where we have defined $ k_ra = {\bf k} \cdot  \br a$.
Note that the $ sine $ function vanishes at the origin $k_1, k_2=0$
but does not vanish at the edges of the Brillouin zone $ k_1, k_2 =
\pm {\pi\over a} $. 

Define $ {\tilde k}_r =  k_r  {\sin ~ k_ra/2 \over k_r a/2}$. In the
naive continuum limit $ {\tilde k}_r \rightarrow k_r$.
  
Now, let us consider the  full Hamiltonian (Eq. \ref{fb}) 
including the helicity flip term.
In the helicity space we have the following matrix structure for
$P^+P^-$ (since $P^-$ is inversely proportional to the total longitudinal
momentum $P^+$, we study the operator $P^+P^-$) 
\be
\begin{pmatrix} {m^2 + { 4 \over a^2} \sum_r \sin^2{k_ra \over 2} &
-{4m \over a} (i \sin^2{k_xa \over 2} + \sin^2 {k_ya \over 2}) \cr
 & \cr
{4 m \over a} (i \sin^2 {k_x a \over 2} - \sin^2 {k_y a \over 2}) &
 m^2 + { 4 \over a^2} \sum_r \sin^2{k_ra \over 2}} 
\end{pmatrix}
\ee
which leads to the eigenvalue equation
\be
{\cal M}^2 = m^2 + {4 \over a^2} \sum_r \sin^2{k_r a \over 2} \pm {4 m \over
a} \sqrt{\sum_r \sin^4{k_r a \over 2}}.\label{eval}
\ee

Third term in the above equation comes from the linear mass helicity
flip term. If the mass $m=0$, then it is obvious from Eq. (\ref{eval}) 
that ${\cal M}^2=0$ if and only if $k_1=k_2=0$.  For nonzero $m$, one
can also in general conclude that ${\cal M}^2=m^2$ only for the case
$k_1=k_2=0$.  Thus there are no fermion doublers in this case (for
physical masses $am<1$). 
 In the following for  specific choices of momenta we elaborate on this
further.

If one component of the momentum vanishes, then 
\be
{\cal M}^2 = m^2 + {4 \over
a}({ 1 \over a}\pm m)\sin^2{ka \over 2} \label{fbe1}
\ee
 where $k$ is the non-vanishing
momentum component. Thus for $am=1$, irrespective of the value of $k$
we get ${\cal M}^2=m^2$ which is unwanted. In general, for $am >1$, 
 ${\cal M}^2$ can become negative. It is important to recall
that physical particles have $ m < { 1 \over a}$ (the lattice cutoff)
and hence are free from the species doubling on the lattice.
With periodic boundary condition (discussed in the next subsection),
 allowed $k$ values are 
$ k_q a = \pm {2 \pi q \over 2n+1}$, with $q=1,2,3, ..., n$ 
for $2n+1$ lattice sites in each direction.  
Let $k_1=0$. For $ma=1.0$, Eq. (\ref{fbe1}) with 
the minus sign within the bracket gives ${\cal M}^2 =  m^2$ for all values
of $k_2$ and we get $2(2n+1)$-fold degenerate ground state with eigenvalue
$m^2.$\\ 
 The two spin states (spin up and down) are degenerate for
$k_1=k_2=0$. But if any one (or both) of the two 
transverse momenta is (are) nonzero then the
degeneracy is broken on the lattice by the spin flip term proportional to $m$.
So the total degeneracy of the lowest states for $ma=1.0 $ can 
be calculated in the
following  way: (a)
$~~k_1=k_2=0~~$: Number of states =2 (spin up and spin down),
(b)$~~k_1=0,~~k_2\ne 0~~$: Number of states =2$n$ and
(c)$~~k_1\ne 0,~~k_2=0~~$: Number of states =2$n$.
Note that $k_i$ can have $2n$ nonzero
values and there is no spin degeneracy  for any nonzero $k_i$. 
So, the total number of degenerate states = $2+2n+2n = 2(2n+1).$
But if $ma \ne 1$ we cannot have $m^2$ eigenvalue for nonzero $k_i$ and we
have only two (spin) degenerate states with eigenvalue  $m^2$. 
Again we see from 
Eq. (\ref{fbe1}) that if $ma > 1$, the kinetic energy term becomes negative
and the eigenvalues go below $m^2$. But $ma \ge 1$ means $m \ge {1\over a}$
(ultra violet lattice cutoff) and hence unphysical. 

\subsection{Numerical Investigation}
For numerical investigations we use  Discretized Light Cone 
Quantization (DLCQ) \cite{3BPP} for the
longitudinal direction ($ -L \le x^- \le +L$) and implement
antiperiodic boundary condition to avoid zero modes. Then,
\be
\eta(x^-, \bx) = { 1 \over \sqrt{2L}} \sum_\lambda \chi_\lambda
\sum_{l=1,3,5, \dots} [ b(l, \bx, \lambda)e^{-i { \pi l x^- / (2
L)}} + d^\dagger(l, \bx, -\lambda) e^{i { \pi l x^- / (2 L)}}] \label{fe}
\ee
with 
\be
\{ b (l, \bx, \lambda), b^\dagger(l', \bx', \lambda') \} =
 \{ d (l, \bx, \lambda), d^\dagger(l', \bx', \lambda') \} = 
\delta_{l l'} \delta_{\bx, \bx'} ~ \delta_{\lambda,
\lambda'}.
\ee
 In DLCQ with antiperiodic boundary condition, it is usual to
multiply the Hamiltonian $P^-$ by ${\pi \over L}$, 
so that $H={\pi \over L}P^-$ has the dimension of mass squared.

In DLCQ the Hamiltonian (Eq. \ref{fb}) can be written as,
\be
H_{fb}= H_0 + H_{hf} \label{fbdlcq}
\ee
where, the helicity nonflip part  
\be
H_0 &=& \sum_{\bz} \sum_\lambda \sum_l {a^2 m^2
\over l} \left [
b^\dagger(l,\bz,\lambda) b(l, \bz, \lambda)
+ d^\dagger(l,\bz,\lambda) d(l, \bz, \lambda) \right ] \nonumber \\
&~ &- \sum_{\bz} \sum_r \sum_\lambda \sum_{\lambda'}
\sum_l{ 1 \over l}
{\chi^\dagger}_{\lambda'}  \chi_\lambda 
 \Big [ b^\dagger(l, \bz, \lambda') b(l, \bz + a \br, \lambda) 
- 2 b^\dagger(l, \bz, \lambda') b(l, \bz, \lambda)\nonumber \\ 
&~&+ b^{\dagger}(l, \bz, \lambda') b(l, \bz - a \br, \lambda)
+d^\dagger(l, \bz, \lambda') d(l, \bz + a \br, \lambda) 
- 2 d^\dagger(l, \bz, \lambda') d(l, \bz, \lambda)\nonumber \\
&~&+ d^{\dagger}(l, \bz, \lambda') d(l, \bz - a \br, \sigma)
\Big ] \label{fbnf}
\ee
and helicity flip term
\be
H_{hf} &= & \sum_{\bz} \sum_r \sum_\lambda \sum_{\lambda'}
\sum_l{ 1 \over l}
{\chi^\dagger}_{\lambda'} [   am {\hat \sigma}^r ] \chi_\lambda 
~ \Big [ b^\dagger(l, \bz, \lambda') b(l, \bz + a \br, \lambda)\nonumber \\
&~& - 2 b^\dagger(l, \bz, \lambda') b(l, \bz, \lambda)
+ b^{\dagger}(l, \bz, \lambda') b(l, \bz - a \br, \lambda)
+ d^\dagger(l, \bz, \lambda') d(l, \bz + a \br, \lambda) \nonumber \\
&~& - 2 d^\dagger(l, \bz, \lambda') d(l, \bz, \lambda)
+ d^{\dagger}(l, \bz, \lambda') d(l, \bz - a \br, \sigma)
\Big ]. \label{fbf}
\ee

Let us now investigate the Hamiltonian (Eq. \ref{fbdlcq}) with
 two types of boundary conditions on the transverse lattice:
(1) fixed boundary condition and (2) periodic boundary condition.  
 For each transverse direction, 
we choose $2n+1$ lattice points ranging from $-n$ to $+n$ where
 fermions are allowed to hop.
For the study of the fermion spectra on the transverse lattice, the
longitudinal momentum plays a passive role and for the numerical
studies we choose the dimensionless longitudinal momentum ($l$) to be unity
which is kept fixed.  For a given set of lattice points in the
transverse space we diagonalize the Hamiltonian and compute both
eigenvalues and eigenfunctions.
\subsubsection{Boundary conditions}
{\bf Fixed boundary condition:}$~~$
 To implement fixed boundary condition  we add two more
points at the two ends and demand that the fermion remains fixed at
these lattice points.  Thus we consider $2n+3$ lattice points.
Let us denote the fermion wavefunction at the location $s$ by $u(s)$. We have 
$ u(s) \sim \sin (s-1)ka$ with $u(1)=u(2n+3)=0$.  Allowed values of
$k$ are $(2n+2)k_pa=p \pi$
with $ p=1,2,3, ...., 2n+1$ and $ k_p= {\pi \over (2n+2)a}p$. Thus the
minimum $k_p$ allowed is ${ \pi  \over a}{ 1 \over (2n+2)}$ and maximum $k_p$
allowed is ${\pi \over a} {(2n+1) \over (2n+2)}$. For example, for $n=1$
we have   $k_1={\pi \over 4a},~ k_2={2 \pi \over 4 a},~ k_3 = {3 \pi
\over 4 a}$, etc. 

{\bf Periodic boundary condition:}$~~$
Periodic boundary condition identifies the
$(2n+2)^{th}$ lattice point with the first lattice point. 
In this case we have the  fermion wavefunction $ u(s) \sim e^{iska}$ with the
condition $u(s) = u(s+L)$ where $L=2n+1$. Thus $ (2n+1)k_p a = \pm 2
\pi p$ so that $k_p = \pm { 2 \pi \over (2n+1)a}p $, $p=0,1,2, ...,
n$. Thus the minimum $k_p$ allowed is $0$ and the maximum $k_p$
allowed is $ { 2 n \pi  \over (2n+1) a}$.  
For $n=1$, we have, $ k_0=0, k_1=\pm {2 \pi \over 3 a}$, etc.
\subsubsection{Numerical results}
First we discuss the results for $H_0$ given in Eq. (\ref{fbnf}).
We  diagonalize the Hamiltonian using basis states defined at each
lattice point in a finite region in the transverse plane. Let us
denote a general lattice point in the transverse plane by
$(x_i,y_i)$.  For each choice of $n$ (measure of the linear lattice
size), we have  
$ -n \le x_i,y_i \le +n $. Thus for a given $n$, we have a $(2n+1) \times (2n+1)$
dimensional matrix for the Hamiltonian.
\begin{figure}[h]
\centering
\psfrag{n}{$n$}
\psfrag{M}{$M^2$}
\includegraphics[height=7cm,clip]{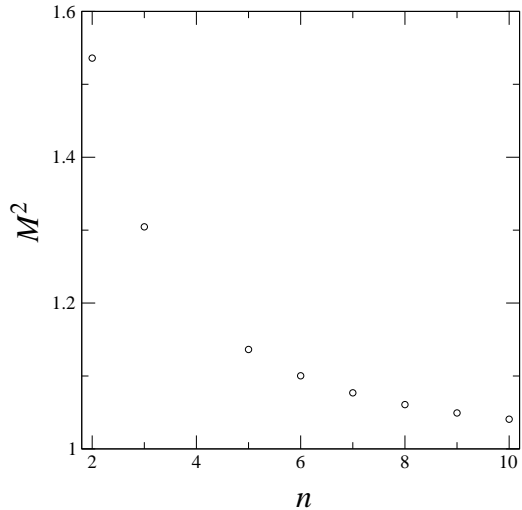}
\caption{Convergence of ground state eigenvalue (of
forward-backward Hamiltonian) versus $n$ ($m=1.0$). 
\label{tlfig1}}
\end{figure}
The boundary conditions do have significant effects at small
volumes. For example, a zero transverse momentum fermion at  finite
$n$ is not allowed with fixed boundary condition. But  with periodic
boundary condition,
 we can have zero transverse momentum fermion for any finite $n$. 
With fixed boundary condition, in
the infinite volume limit, we expect the lowest eigenstate to be the
zero transverse momentum fermion with the eigenvalue $m^2$. 
In Fig. \ref{tlfig1} we
show the convergence of the lowest eigenvalue as a function of $n$. 

\begin{figure}[h]
\centering
\fbox{\includegraphics[height=12cm]{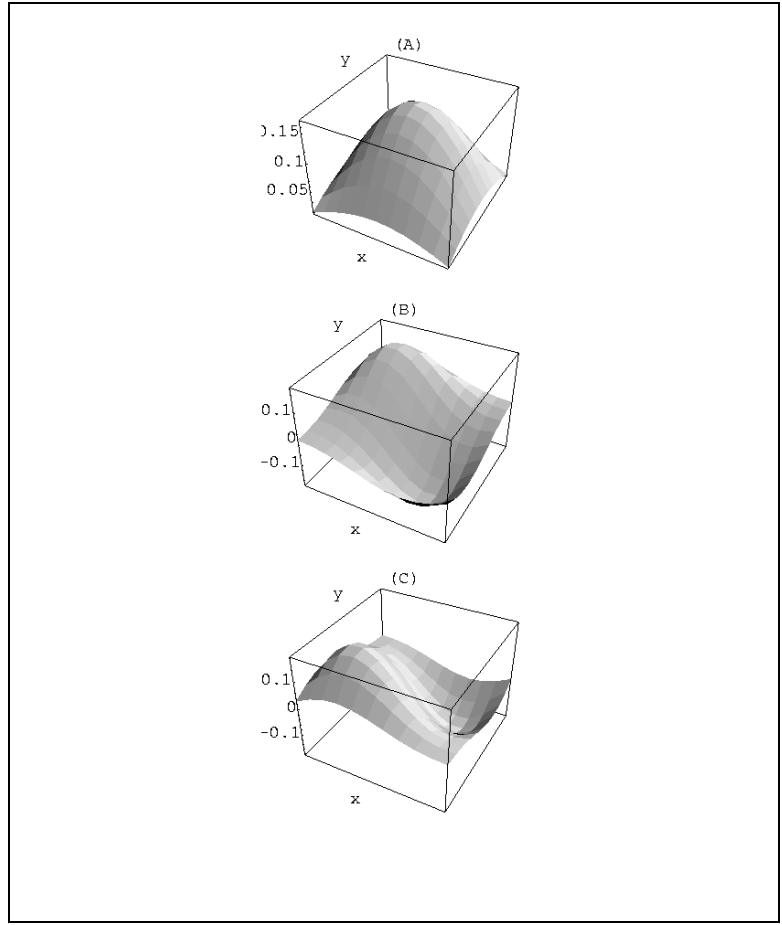}}
\caption[Eigenfunctions of first three states of the spin nonflip
Hamiltonian ($H_0$) with
forward and backward lattice derivatives] 
{Eigenfunctions of first three states of the Hamiltonian $H_0$
 with lattice points $n=5$. \label{tlfig2}}
\end{figure}
For a zero transverse momentum fermion, the probability amplitude to
be at any transverse location should be independent of the transverse
location. Thus we expect the eigenfunction for such a particle to be a
constant. At finite volume, with fixed boundary condition, we do get a 
nodeless wave function which
nevertheless is not a constant since it carries some non-zero
transverse momentum. All the excited states carry non-zero transverse
momentum in the infinite volume limit. All of them have nodes
characteristic of sine waves. The eigenfunctions corresponding to the
first three eigenvalues are shown in Fig. \ref{tlfig2} for the case of fixed
boundary condition.
With periodic boundary condition, for any $n$, we get a zero
transverse momentum fermion with a flat wave function.

\begin{figure}[h]
\centering
\psfrag{n}{$n$}
\psfrag{spin splitting}{spin splitting}
\includegraphics[height=10cm,clip]{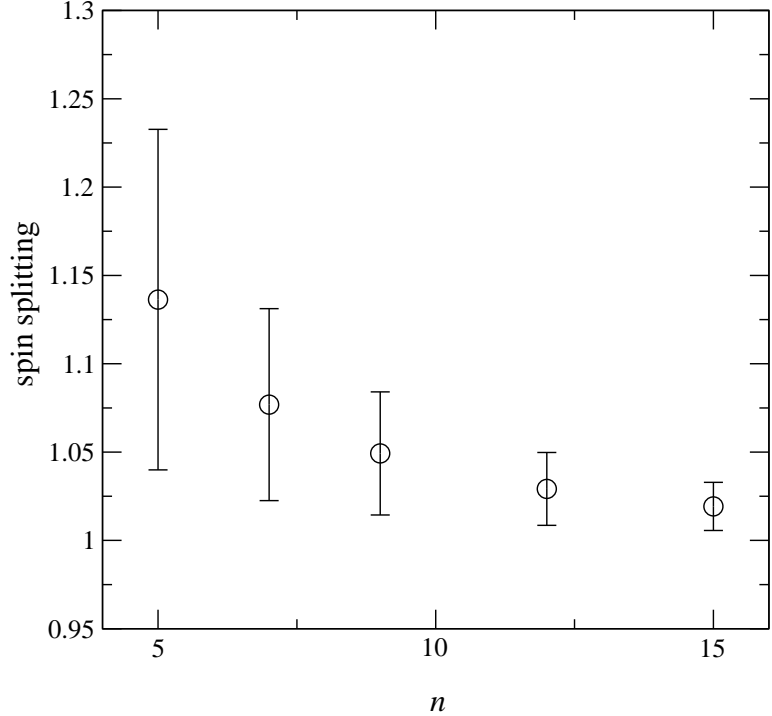}
\caption[Spin splitting of the ground state caused by the spin dependent
interaction as a function of $n$.]
{Spin splitting of the ground state caused by the spin dependent
interaction as a function of $n$. Circles represent the eigenvalues without
splitting and the vertical bars show the magnitude of splitting due to
spin flip interaction. \label{tlfig3}}
\end{figure}
Now, we consider the effect of helicity flip term. 
With fixed boundary condition the lowest eigenstate has non vanishing 
transverse momentum in finite volume. In the absence of helicity flip term
positive and negative helicity fermions are degenerate. The helicity flip 
term  lifts the degeneracy. The splitting is larger for larger transverse
momentum. 
In Fig. \ref{tlfig3} we present the level splitting  
for the helicity up and down 
fermions as a
function of $n$. As expected, the level splitting vanishes and we get
exact degeneracy in the infinite volume limit. 
For the periodic boundary condition, the lowest state has exactly zero
transverse momentum and we get two degenerate fermions for all $n$.

\section{Hamiltonian with symmetric derivative}
\subsection{Construction}
Now, let us discuss the other way of putting fermions on a transverse lattice.
Here we use the symmetric lattice derivative  defined by
\be
D_r \psi^\pm(\bx) = { 1 \over 2 a} [ U_r(\bx) \psi^\pm(\bx + a \br) - 
U_{-r}(\bx) \psi^\pm(\bx - a \br)].
\ee
In place of using forward and backward derivatives in Eq. (\ref{lfb}), 
we use the above symmetric derivative for all lattice
derivatives. Proceeding as in Sec. \ref{H-FB}, we arrive at the 
 fermionic part of the QCD Hamiltonian 
\be
P^-_{sd}&= &\int dx^- a^2 \sum_\bx
 m^2
\eta^\dagger(\bx) { 1 \over i \partial^+} \eta( \bx) \nonumber \\
 & ~ &  -\int dx^- a^2 \sum_\bx
\Bigg \{  m \frac{1}{2a}
\eta^\dagger(\bx) \sum_r {\hat \sigma}_r { 1 \over i \partial^+}  
\left [ U_{r}(\bx) \eta(\bx + a \br) - U_{-r}(\bx) \eta(\bx - a \br) \right ] 
\nonumber \\
& ~ & -  m  \frac{1}{2a}
 \sum_r \left [ \eta^\dagger(\bx - a \br) {\hat \sigma}_r U_{r}(\bx - a \br) - 
 \eta^\dagger(\bx + a \br) {\hat \sigma}_r U_{-r}(\bx + a \br)  \right ]
{ 1 \over i \partial^+} \eta(\bx) \Bigg \}  \nonumber \\
 & ~ & -\int dx^- a^2 \sum_\bx
 \frac{1}{4a^2} \sum_r 
\Big [ \eta^\dagger(\bx - a \br)  U_{r}(\bx - a \br) - 
\eta^\dagger(\bx + a \br)   U_{-r}(\bx + a \br) \Big ]
\nonumber \\
&~&~~~~~{ 1 \over i \partial^+} 
\Big [  U_{r}(\bx) \eta(\bx + a \br) - U_{-r}(\bx) \eta(\bx - a \br)\Big ] .
\label{qcdsd}
\ee 
In the free limit, the above Hamiltonian becomes
\be
P^-_{sd} &=&\int dx^- a^2 \sum_\bx \Bigg\{
 m^2 {\eta}^\dagger(\bx) { 1 \over i \partial^+}
\eta(\bx) \nonumber \\
&~& + { 1 \over 4 a^2} \sum_r [{\eta}^\dagger(\bx + a \br) - 
{\eta}^\dagger(\bx - a \br)]{ 1 \over i \partial^+}[\eta(\bx + a
\br) - \eta(\bx - a \br)]\Bigg\}.\label{sd}
\ee
In the free field limit the two linear mass terms cancel with each
other.

When we implement the constraint equation on the lattice and use
symmetric definition of the lattice derivative, it is important to
keep in mind that we have only {\it next to nearest neighbor}
interactions which can easily be seen from Eq. (\ref{sd}). 
Thus in each transverse direction even and odd lattice points are decoupled
and as a result we have four independent sub-lattices in two dimensional
transverse plane connecting ($x^1$ = even, $x^2$ = even), ($x^1$ = even, $x^2$
= odd), ($x^1$ = odd, $x^2$ = even) and   ($x^1$ = odd, $x^2$ = odd)
lattice points.

Let us now address the nature of the spectrum and the presence
and origin of doublers. 

\subsection{Fermion doubling}
For clarity, the Hamiltonian (Eq. \ref{sd}) can be  rewritten as
\be
 P^-_{sd}&=& \int dx^- a^2 \Big \{ \nonumber \\
&~& \sum_{x^1_e, x^2_e }\Big [
m^2 \eta^{\dagger}(\bx){1 \over i\partial^+}\eta(\bx)
-{1 \over 4 a^2} a^2  [ \eta^\dagger(\bx) { 1 \over i \partial^+}\sum_r [
\eta(\bx +2 a \br) ] + \eta(\bx -2 a \br)- 2\eta(\bx )]\Big ]
\nonumber \\
&+&\sum_{x^1_e, x^2_o}\Big [
m^2 \eta^{\dagger}(\bx){1 \over i\partial^+}\eta(\bx)
-{1 \over 4 a^2} a^2  [ \eta^\dagger(\bx) { 1 \over i \partial^+}\sum_r [
\eta(\bx +2 a \br) ] + \eta(\bx -2 a \br)- 2\eta(\bx )]\Big ]\nonumber\\
&+&\sum_{x^1_o, x^2_e}\Big [
m^2 \eta^{\dagger}(\bx){1 \over i\partial^+}\eta(\bx)
-{1 \over 4 a^2} a^2  [ \eta^\dagger(\bx) { 1 \over i \partial^+}\sum_r [
\eta(\bx +2 a \br) ] + \eta(\bx -2 a \br)- 2\eta(\bx )]\Big ]\nonumber \\
&+&\sum_{x^1_o, x^2_o}\Big [
m^2 \eta^{\dagger}(\bx){1 \over i\partial^+}\eta(\bx)
-{1 \over 4 a^2} a^2  [ \eta^\dagger(\bx) { 1 \over i \partial^+}\sum_r [
\eta(\bx +2 a \br) ] + \eta(\bx -2 a \br)- 2\eta(\bx )]\Big ]\Big
\},\nonumber \\
\ee
where $x^i_e$ stands for $x^i=$ even and $x^i_o$ stands for $x^i=$ odd.
Clearly the Hamiltonian is divided into four independent sub-lattices each
with lattice constant $2a$. As a result, a momentum component in each
sub-lattice is bounded by  ${\pi \over 2a}$ in magnitude. 
Again, going through the Fourier transform in each sub-lattice of the
transverse space, we arrive at
 the free particle dispersion relation for the light
front energy in each sector
\be
k^-_{\bf k} = { 1 \over k^+} [ m^2 + { 1 \over a^2}
\sum_r \sin^2~ k_ra ]. \label{potdoub}
\ee
For fixed $k_r$, in the limit $ a \rightarrow 0$ $, { 1 \over a^2}
\sin^2~k_ra \rightarrow k_r^2$ and we get the continuum dispersion
relation
\be
k^-_{\bf k} = { m^2 + {\bf k}^2 \over k^+}.
\ee
        
Because of the momentum bound of ${\pi \over 2a}$
 doublers cannot arise from
$ka=\pi$ in sharp contrast with Euclidean lattice gauge theory where
doubers come from $ka=\pi$ \cite{3montvay}. 
However, because of the decoupling 
of odd and even lattices,
one can get  zero transverse momentum fermions one each from the
four sub-lattices for a two dimensional transverse lattice.
Thus we expect a four fold degeneracy of zero transverse
momentum fermions.

\subsection{Numerical Investigation}
Using DLCQ for the longitudinal direction, Eq. (\ref{sd}) can be written as
\be
P_{sd}^- ={ L \over \pi} H_{sd} \equiv { L \over \pi} [ H_m + H_k]
\ee
where
\be
H_m  &= & a^2{m}^2 \sum_l \sum_\sigma \sum_\bz { 1 \over l} \nonumber \\
&~&~~~~ [ b^\dagger(l,\bz, \sigma) b(l,\bz, \sigma) + d^\dagger(l,\bz, \sigma)
d(l,\bz, \sigma) ]
\ee
and 
\be
H_k & = & \sum_l \sum_\sigma \sum_\bz \sum_r {1 \over l}\nonumber \\
&~& ~~~ \Bigg [ 
b^\dagger(l,\bz+ a \br, \sigma) b(l,\bz+a \br, \sigma) +
b^\dagger(l,\bz- a \br, \sigma) b(l,\bz- a \br, \sigma) \nonumber \\
&~&~~~- b^\dagger(l,\bz+ a \br, \sigma) b(l,\bz-a \br, \sigma)
-b^\dagger(l,\bz-a \br, \sigma) b(l,\bz+ a \br, \sigma) \nonumber \\
&~&~~~+d^\dagger(l,\bz+ a \br, \sigma) d(l,\bz+a \br, \sigma) +
d^\dagger(l,\bz- a \br, \sigma) d(l,\bz- a \br, \sigma) \nonumber \\
&~&~~~- d^\dagger(l,\bz+ a \br, \sigma) d(l,\bz-a \br, \sigma)
-d^\dagger(l,\bz-a \br, \sigma) d(l,\bz+ a \br, \sigma) \Bigg ].
\ee

For each transverse direction,  we have $2n+1$ lattice points where the 
fermions are
allowed to hop. Since even and odd lattice points are decoupled we need
to fix the boundary conditions separately for even and odd sub-lattices
in each transverse direction.
\subsubsection{Boundary conditions}
{\bf Fixed boundary condition:}$~~$
To implement the fixed boundary condition, we need to 
consider $2n+5$ lattice points. For one sub-lattice we have to fix
particles at $s=1$ and $s=2n+5$. We have, the wavefunction at location
$s$, $u_s \sim \sin~(s-1)ka$ which gives $u_s=0$ for $ s=1$. We also
need $u_s=0$ for $s=2n+5$. Thus $ (2n+4)k_pa =p \pi$, with $p=1,2,3,
..., n+1$. For $n=1$, allowed values of $k_p$ are $k_p={ \pi \over
6a}, { 2\pi \over 6a}$.

For the other sub-lattice, we fix the particles at $s=2$ and
$s=2n+4$. The wavefunction at location $s$, $u_s \sim
\sin~(s-2)ka$. $u_s=0$ for $s=2$ and $s=2n+4$. Thus $ (2n+2)k_pa = p
\pi$ 
with $ p = 1,2,3, ...., n$. For $n=1$, only allowed value
of $k$ is $k={\pi \over 4a}$.

Combining the two sub-lattices, for $n=1$, the allowed values of $k$ are
${\pi \over 6a}, {\pi \over 4a},$ and $ {2 \pi \over 6a}$. 

{\bf Periodic boundary condition:}$~~$
To implement the periodic boundary condition, we need to consider $2n+3$
lattice points when fermions can hop in $2n+1$ lattice points. 
For one sub-lattice $(2n+3)^{rd}$ lattice
point is identified with the lattice point 1. For the other sub-lattice
$(2n+2)^{nd}$ lattice point is identified with the lattice point
2. Wavefunction at point $s$, $u_s \sim e^{is ka}$. We require $
e^{ika} =e^{ i (2n+3)ka}$. Thus $k_pa = \pm { 2 \pi p \over
(2n+2)}$, $ p=0,1,2, ..., {n+1 \over 2}$.  For $n=1$, we have, $k_0=0, k_1=\pm
{\pi \over 2a}$.

For the other sub-lattice we require $e^{ 2ika} = e^{i(2n+2)ka}$. Thus
the allowed values of momentum $k_p$ are $ k_pa = \pm {\pi \over n}p$,
 $p = 0,1,2, ..., {n-1 \over 2}$. For $n=1$, allowed
value of $k=0$. Thus for $n=1$, taking the two
sub-lattices together, the allowed values of $k$ are $0,0, {\pi \over
2a}$.   

\subsubsection{Numerical results}

The results of matrix diagonalization in the case of the symmetric derivative 
with fixed boundary condition are presented in Figs. \ref{tlfig4},
\ref{tlfig5} and \ref{tlfig6}. 
\begin{figure}[hbt]
\centering
\psfrag{n}{$n$}
\psfrag{M}{ $M^2$}
\includegraphics[height=8cm,clip]{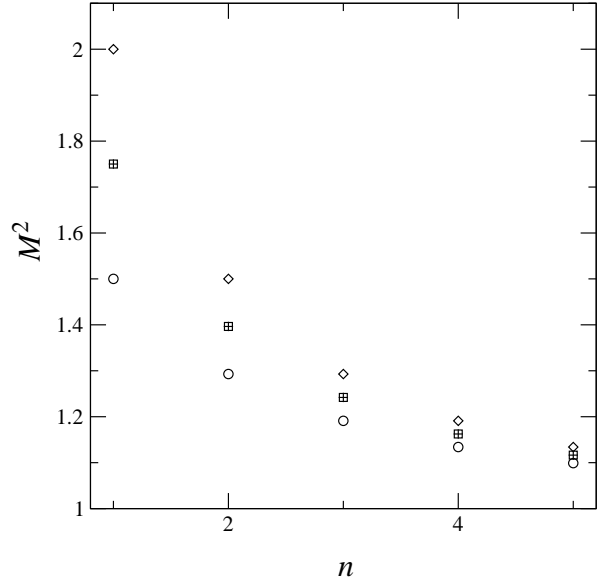}
\caption{First four eigenvalues of Hamiltonian with symmetric derivative
as a function of $n$. \label{tlfig4}}
\end{figure}
In Fig. \ref{tlfig4} we present the lowest four
eigenvalues as a function of $n$. 
At finite volume, the four states do
not appear exactly degenerate even though the even-odd and odd-even
states are always degenerate because of the hypercubic (square) symmetry 
in the transverse
plane. The four states become degenerate in the infinite volume
limit. 
\begin{figure}[h]
\centering
\fbox{\includegraphics[totalheight=12cm]{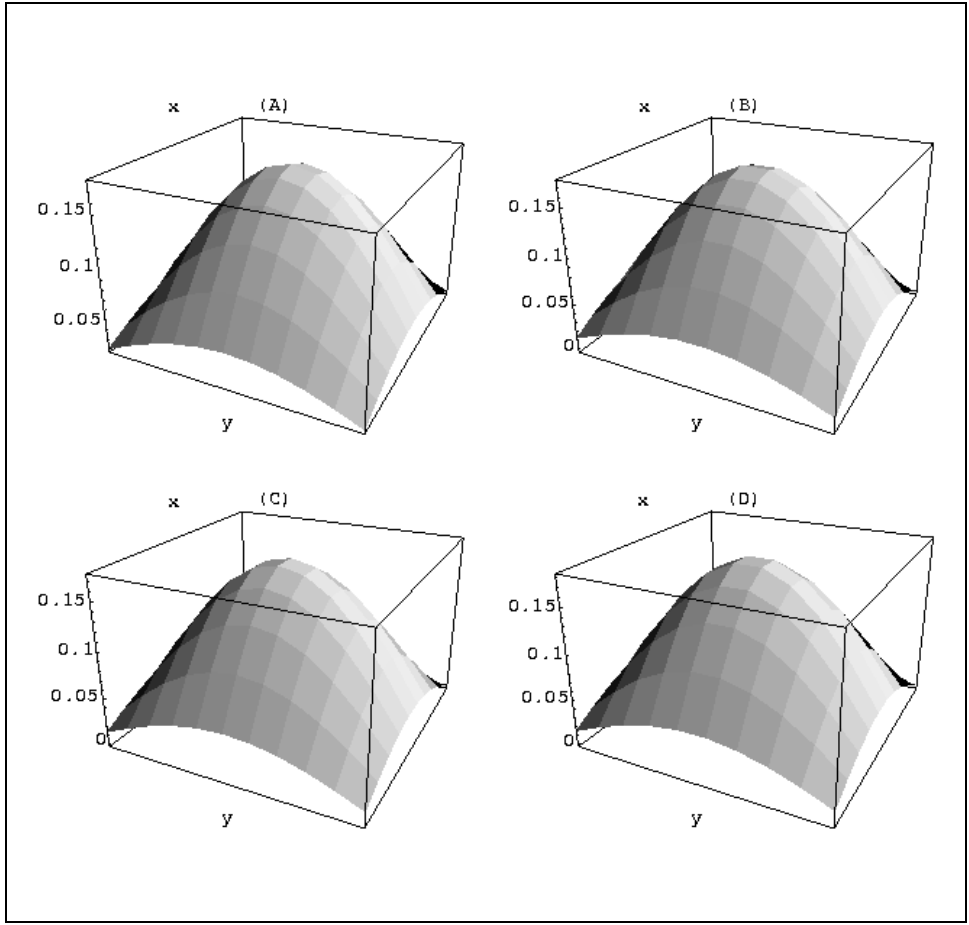}}
\caption{Eigenfunctions of first four (degenerate) states for the case
of fermion doubling with symmetric lattice derivative. \label{tlfig5}}
\end{figure}
The eigenfunctions of the lowest four states are presented in
Fig. \ref{tlfig5} for $n=5$ and fixed boundary condition. As they carry
 small nonzero transverse momenta they are not flat in the transverse
plane. But,
as they correspond to particle states, they are
nodeless and same in shape and magnitude. All other states in the
spectrum have one or more nodes. For 
example, in Fig. \ref{tlfig6} we show the eigenfunction corresponding to the
fifth eigenvalue for $n=5$ which clearly exhibits the node structure.
\begin{figure}[h]
\centering
\includegraphics[height=7cm]{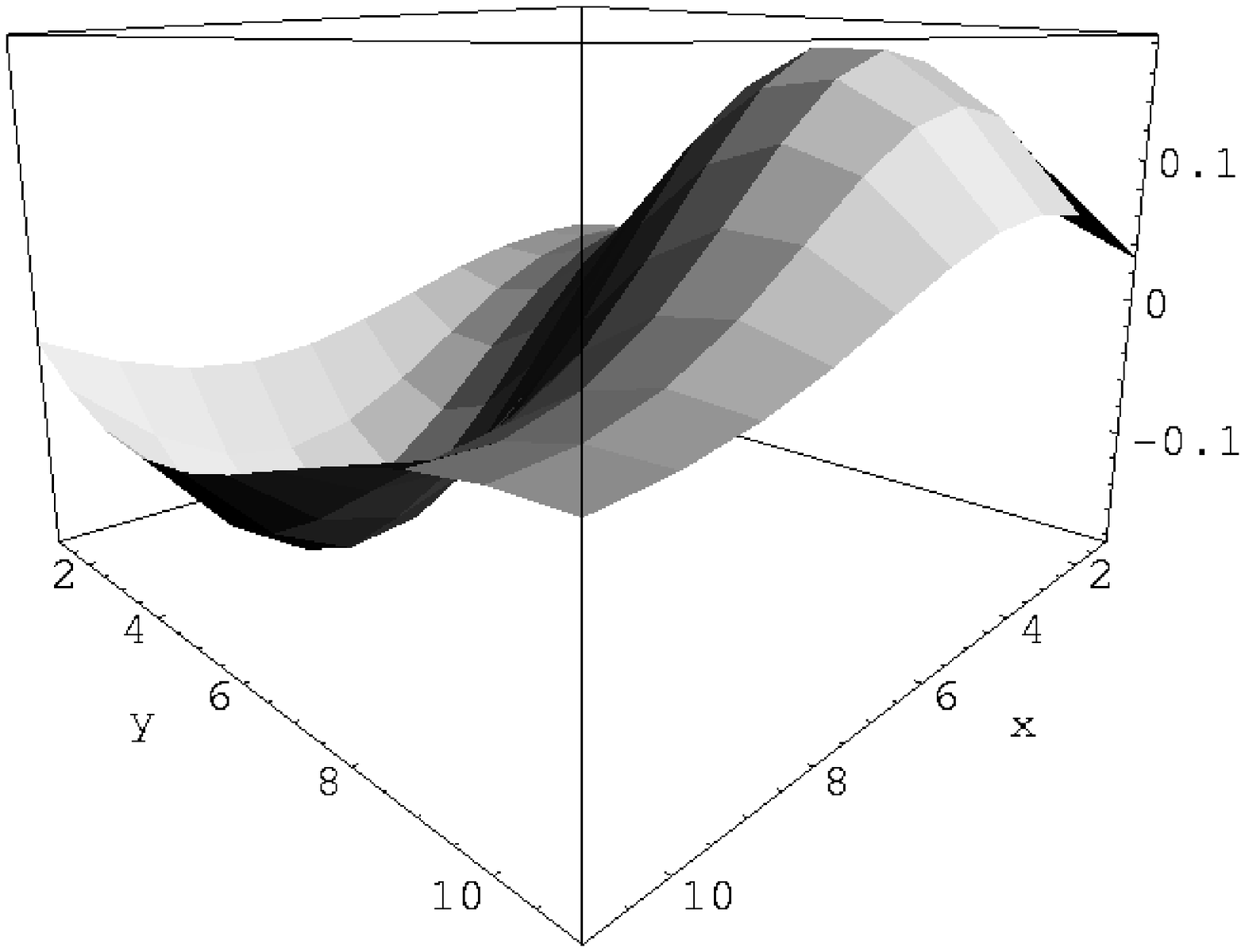}
    \caption{Eigenfunction corresponding to the fifth state. 
\label{tlfig6}}
\end{figure}

With periodic boundary condition we can achieve exactly zero transverse
momenta for any $n$ and hence  get four fold degenerate lowest
eigenvalue corresponding to four zero transverse momentum fermions. 
Corresponding wavefunctions are flat in transverse coordinate space and
excited states (nonzero $k$ states) show the expected node structures.

\section{Staggered fermion on the light-front transverse lattice}
\label{lfsf}
As we have seen in the previous section that the method of symmetric
derivatives results in fermion doublers, we now consider 
 two approaches to  remove the doublers. In this section we study an
approach similar to the staggered fermions in conventional lattice
gauge theory. In the next section we will take up the case of
 Wilson fermions.

In analogy with the Euclidean staggered formulation, define the spin 
diagonalization transformation
\be
\eta(x_1,x_2) = ({\hat\sigma}^1)^{x_1}({\hat\sigma}^2)^{x_2}
\chi(x_1,x_2). \label{sdt}
\ee
We see from the  QCD Hamiltonian given in Eq. (\ref{qcdsd}) with symmetric 
derivative that  in the interacting 
theory (except  for the linear mass term) and also  in the free fermion limit,
even and odd lattice sites are decoupled and the Hamiltonian 
 is already spin diagonal. So, it is very natural to try staggered fermion 
formulation on the light front transverse lattice.  In this section we
shall follow  
the Kogut-Susskind formulation \cite{3ks} and  present an elementary
configuration space analysis for two flavor interpretation.
After the spin transformation  the  linear mass term in the
Hamiltonian (Eq. \ref{qcdsd})  becomes:
\be
&~&\int dx^- a^2 \sum_\bx \Bigg \{  m \frac{1}{2a}
\chi^\dagger(\bx) \sum_r \phi(\bx,r) { 1 \over i \partial^+}  
\left [ U_{r}(\bx) \chi(\bx + a \br) - U_{-r}(\bx) \chi(\bx - a \br) \right ] 
\nonumber \\
& ~ & -  m  \frac{1}{2a}
 \sum_r \left [ \chi^\dagger(\bx - a \br) \phi(\bx,r)U_{r}(\bx - a \br) - 
 \chi^\dagger(\bx + a \br) \phi(\bx,r) U_{-r}(\bx + a \br)  \right ]
{ 1 \over i \partial^+} \chi(\bx) \Bigg \}  \nonumber\\
\ee
where, $\phi(\bx,r) = 1$ for $r=1$ and $\phi(\bx,r) = (-1)^{x_1}$ for $r=2$.
After spin diagonalization, the  full Hamiltonian in the free field limit becomes 
\be
  P^-_{sf}&=&\int dx^- a^2 \sum_\bx \Bigg\{ 
 m^2 {\chi}^\dagger(\bx) { 1 \over i \partial^+}
\chi(\bx) \nonumber \\
&~& + { 1 \over 4 a^2} \sum_r [{\chi}^\dagger(\bx + a \br) - 
{\chi}^\dagger(\bx - a \br)]{ 1 \over i \partial^+}[\chi(\bx + a
\br) - \chi(\bx - a \br)] \nonumber \\
&~& -{1\over 2a}m {\chi}^\dagger(\bx){ 1 \over i \partial^+} 
\sum_r \phi(\bx,r)
[\chi(\bx + a\br) - \chi(\bx - a \br)] \nonumber \\
&~& -{1\over 2a}m\sum_r [{\chi}^\dagger(\bx + a \br) - {\chi}^\dagger
(\bx - a \br)]
\phi(\bx,r){ 1 \over i \partial^+}\chi(\bx)
\Bigg\}.\label{stag}
\ee  
 The two linear mass terms cancel with each other in the free theory,
but since they are present
in the interacting theory we keep them to investigate the staggered
fermions. 

Since all the terms in Eq. (\ref{stag}) are spin diagonal,
 we can put only a single component field at each transverse site.
From now on, all the $\chi$'s and $\chi^{\dagger}$'s appearing in
Eq. (\ref{stag}) can be taken as single component fermion fields. 
Thus we have thinned the fermionic degrees of freedom by half.
Without loss of generality, we keep the helicity up component of $\chi$ at 
each lattice point.

\begin{figure}[hbtp]
\centering
\includegraphics[width=8cm, height=8cm]{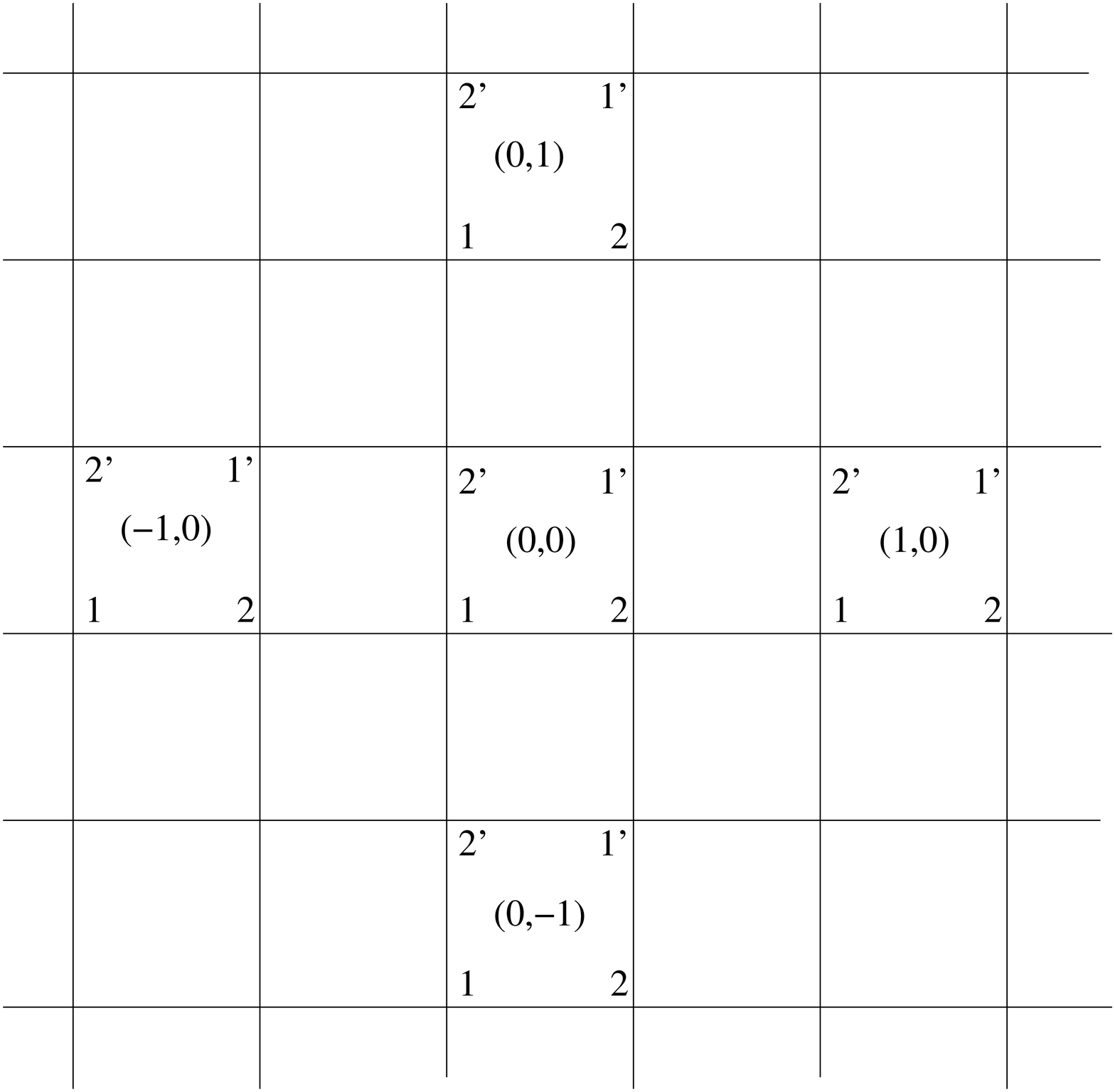}
    \caption{Staggered lattice. \label{tlfig7}}
\end{figure}
Apart from the linear mass term in Eq. (\ref{stag}), all the other
terms have the feature that fermion fields on the even and odd
lattices do not mix. Let us denote (see Fig. \ref{tlfig7}) the even-even lattice
 points by 1,
odd-odd lattice points by $1'$, odd-even lattice points by 2 and
even-odd lattice points by $2'$, and the corresponding fields by
$\chi_1$, $\chi_{1^{\prime}}$, $\chi_2$ and $\chi_{2^{\prime}}$. 
Then  the first of the linear mass terms    
\be
  \sum_{\bx}{\chi}^\dagger(\bx){ 1 \over i \partial^+} \sum_r \phi(\bx,r)
[\chi(\bx + a\br) - \chi(\bx - a \br)]
\ee
can be rewritten as (suppressing  factors of $a$ from now on),  
\be
 &~& {\chi_1}^\dagger{ 1 \over i \partial^+}( \nabla_1\chi_2+\nabla_2
\chi_{2^{\prime}})
+{\chi_2}^\dagger{ 1 \over i \partial^+}( \nabla_1\chi_1-\nabla_2\chi_
{1^{\prime} })\nonumber \\
&~&+{\chi_{1^{\prime }}}^\dagger{ 1 \over i \partial^+}( \nabla_1
\chi_{2^{\prime} } -
\nabla_2\chi_2)
+{\chi_{2^{\prime }}}^\dagger{ 1 \over i \partial^+}( \nabla_1\chi_{1^{\prime }}+
\nabla_2\chi_1) +B \label{sc}
\ee
where $\nabla_1$ and $\nabla_2$ are the symmetric derivatives in
the respective directions and  $B$ represents the
contribution from other blocks. Looking at Fig. \ref{tlfig7} 
it is apparent that these 
$\nabla_1$ and $\nabla_2$ can also be interpreted as a block
derivative, i.e., finite differences between block variables. For example,
$\nabla_1\chi_1 =\chi_1(1,0)-\chi_1(0,0)$ where $(1,0)$ and $(0,0)$ are
the block indices as shown in Fig. \ref{tlfig7}. 

Using Eq. (\ref{sdt}),  in terms of the nonvanishing components of $\eta$, we have
\be
\eta_1=\chi_1,~~\eta_2=i\chi_2,~~\eta_{ 1'}=i\chi_{ 1'},~~
\eta_{2' }=-\chi_{ 2'}.
\ee
An interesting feature of lattice points 1 and $1'$ is that fermion
fields $\eta_1$ and $\eta_{1'}$ have positive helicity. $\eta_2$ and
$\eta_{2'}$ have negative helicity.
In terms of $\eta$ fields the expression given in Eq. (\ref{sc})  can be written as
\be
&~& {\eta_1}^\dagger{ 1 \over i \partial^+}(-i \nabla_1\eta_2 -\nabla_2
\eta_{2^{\prime }})
+i{\eta_2}^\dagger{ 1 \over i \partial^+}( \nabla_1\eta_1+i\nabla_2\eta_{1^{
\prime }}) \nonumber \\
&~&+i{\eta_{1^{\prime }}}^\dagger{ 1 \over i \partial^+}(- \nabla_1
\eta_{2^{\prime} }+i\nabla_2\eta_2)
-{\eta_{2^{\prime}}}^\dagger{ 1 \over i \partial^+}(-i \nabla_1\eta_{1^{\prime }}+
\nabla_2\eta_1) + B. \label{stag2}
\ee
Now,  
\be
\eta(1)-\eta(0)&=& {1\over 2}(\eta(1)-\eta(-1)) +{1\over 2}(\eta(1)+\eta(-1) 
-2\eta(0)) \nonumber \\
&\equiv& {\hat\nabla}\eta(0) +{1\over 2} {\hat\nabla}^2\eta(0)~,
\ee
\be
\eta(0)-\eta(-1)&=& {1\over 2}(\eta(1)-\eta(-1)) -{1\over 2}(\eta(1)+\eta(-1) 
-2\eta(0)) \nonumber \\
&\equiv& {\hat\nabla}\eta(0) -{1\over 2} {\hat\nabla}^2\eta(0)
\ee
where $ {\hat\nabla} $ and  $ {\hat\nabla}^2 $ are respectively first order and 
second order block derivatives.
So, we can write the expression (\ref{stag2})  as
\be
&~& {\eta_1}^\dagger{ 1 \over i \partial^+}\Big \{-i({\hat \nabla}_1\eta_2 
-{1\over 2}{{\hat \nabla}_1}^2\eta_2)
-({\hat\nabla}_2\eta_{2^{\prime }} -{1\over 2}{{\hat\nabla}_2}^2
\eta_{2^{\prime }} \Big \} \nonumber \\
 & ~&+i{\eta_2}^\dagger{ 1 \over i \partial^+}\Big\{ ( {\hat\nabla}_1
\eta_1+{1\over 2} {{\hat\nabla}_1}^2\eta_1)
+i({\hat\nabla}_2\eta_{1^{\prime }} -{1\over 2}{{\hat\nabla}_2}^2
\eta_{1^{\prime }}) \Big \} \nonumber \\
 &~&+i{\eta_{1^{\prime }}}^\dagger{ 1 \over i \partial^+}\Big\{-( 
{\hat\nabla}_1\eta_{2^{\prime }}+{1\over 2}
{{\hat\nabla}_1}^2\eta_{2^{\prime }})
+i({\hat\nabla}_2\eta_2+{1\over 2}
{{\hat\nabla}_2}^2\eta_ 2 )\Big \}  \nonumber \\
&~&-{\eta_{2^{\prime }}}^\dagger{ 1 \over i \partial^+}\Big\{-i ({\hat\nabla}_1
\eta_{1^{\prime }}-{1\over 2}{{\hat\nabla}_1}^2\eta_{1^{\prime}})
+( {\hat\nabla}_2\eta_1 + {1\over 2} {{\hat\nabla}_2}^2\eta_1)\Big\}~.
\label{stag3} \ee
Let us introduce the fields
\be
u_1&=&{1\over {\sqrt 2}}(\eta_1+\eta_{1^{\prime}}) \nonumber \\
u_2&=&{1\over {\sqrt 2}}(\eta_2+\eta_{2^{\prime}}) \nonumber \\
\tilde{d}_1&=&{1\over {\sqrt 2}}(\eta_1-\eta_{1^{\prime}}) \nonumber \\
\tilde{d}_2&=&{1\over {\sqrt 2}}(\eta_2-\eta_{2^{\prime}}). \nonumber \\
\ee
Then, the first order derivative term in Eq. (\ref{stag3}) can be written as
\be
 u^{\dagger}{1\over i \partial^+}{\hat\sigma}^r{\hat\nabla}_r u +
d^{\dagger}{1\over i \partial^+}{\hat\sigma}^r{\hat\nabla}_r d 
= f^{\dagger}{1\over i \partial^+}{\hat\sigma}^r{\hat\nabla}_r f
\ee
where, $d={\hat\sigma}^1\tilde{d}$ and  
 the flavor isospin  doublet
\be
f = \left[ \begin{array}{l} u \\ d \end{array} \right] .
\ee
Similarly, we can write the second order block derivative term in
expression (\ref{stag3}) as
\be
{1\over 2}f^{\dagger}{1\over i \partial^+}\sigma^3{T}^r
{{\hat\nabla}_r}^2f
\ee
where, $T^r$s are the matrices in the flavor space defined as
\be
T^1 = -i\sigma^2,~~~
T^2 =  -i\sigma^1.
\ee
Similarly, the second term in Eq. (\ref{stag}) 
\be
 \sum_r [{\eta}^\dagger(\bx + a \br) - 
{\eta}^\dagger(\bx - a \br)]{ 1 \over i \partial^+}[\eta(\bx + a
\br) - \eta(\bx - a \br)]
\ee
reads as
\be
{\hat\nabla}_r f^{\dagger}{1\over i\partial^+}{\hat \nabla}_r f 
+{{\hat\nabla}_r}^2f^{\dagger}{1\over i \partial^+}{{\hat\nabla}_r}^2f 
+{i\over2}[{\hat\nabla}_rf^{\dagger}{1\over i \partial^+}\sigma^rT^r{{\hat
\nabla}_r}^2f +{{\hat\nabla}_r}^2f^{\dagger}{1\over i
\partial^+}\sigma^rT^r{{\hat  
\nabla}_r}f]~.
\ee
 The full Hamiltonian given in Eq.  (\ref{stag}) can 
now be written in two flavor notation (restoring factors of $a$) as
\be
P^-_{sf}=\int dx^- a^2\sum_x &\Big\{& 
 m^2f^{\dagger}{1\over i \partial^+}f
 +{1\over 4 }[
{\hat\nabla}_r f^{\dagger}{1\over i\partial^+}{\hat \nabla}_r f    
+a^2{{\hat\nabla}_r}^2f^{\dagger}{1\over i \partial^+}{{\hat\nabla}_r}^2f
\nonumber \\
&~& +{ia\over2}({\hat\nabla}_rf^{\dagger}{1\over i
\partial^+}\sigma^rT^r{{\hat
\nabla}_r}^2f +{{\hat\nabla}_r}^2f^{\dagger}{1\over i
\partial^+}\sigma^rT^r{{\hat
\nabla}_r}f)]\nonumber \\
&~&-{1\over 2}m (f^{\dagger}{1\over i
\partial^+}{\hat\sigma}^r{\hat\nabla}_r
 f + {a\over 2}f^{\dagger}{1\over i \partial^+}\sigma^3{T}^r
{{\hat\nabla}_r}^2f + h.c )\Big \}~.\label{flavor}
 \ee
The above simple exercise shows that applying the spin diagonalization 
on the symmetric derivative method, the number of doublers on the
transverse lattice can be reduced from four to two which can be
reinterpreted as two flavors.   Although in the free case given by
Eq. (\ref{flavor}) the second and third lines are separately zero
identically, we have kept these terms because in QCD similar terms
will survive.  These terms exhibit flavor mixing and also helicity
flipping. The flavor mixing terms are always irrelevant.

\section{Wilson fermion on the light-front transverse lattice}
Since doublers in the light front transverse lattice arise from the
decoupling of even and odd lattice sites, a term that
will couple these sites will remove the zero momentum doublers. However, 
 conventional doublers now may arise from the edges of the Brillouin zone.
 A second derivative term couples the even and odd lattice sites and also 
removes the conventional doublers. Thus,   
the term originally proposed by Wilson to
remove the doublers arising from $ka=\pi$ in the conventional lattice
theory will do the job \cite{3BK}. 

To remove doublers, add an irrelevant term to the Lagrangian density
\be
\delta {\cal L}(\bx) = {\kappa \over a} \sum_{r}{\bar \psi}(\bx) [U_r(\bx) \psi(\bx 
+ a\br) - 2 \psi(\bx)+ U_{-r}(\bx)\psi(\bx - a \br)]
\ee
where $\kappa$ is the Wilson parameter.
The constraint equation for $\psi^-$ in the presence of the Wilson term becomes
\be
i \partial^+ \psi^-(\bx) & = & m \gamma^o \psi^+(\bx) \nonumber \\
&~& + i {\alpha_r \over 2 a} [ U_r(\bx) \psi^+(\bx + a \br) -
U_{-r}(\bx) \psi^+(\bx - a \br)] \nonumber \\
&~& - { \kappa \over a} \gamma^0 [ U_r(\bx) \psi^+(\bx + a \br)- 2
\psi^+(\bx)
+ U_{-r}(\bx) \psi^+(\bx - a \br)].
\ee
The Wilson term  generates the following additional terms in the Hamiltonian 
 (\ref{qcdsd}):
\be
P^-_{w} & = &  -\int dx^- a^2 \sum_\bx
\Bigg \{  4 \frac{\kappa}{a} \frac{1}{2a}
\eta^\dagger(\bx) \sum_r {\hat \sigma}_r { 1 \over i \partial^+}  
\left [ U_{r}(\bx) \eta(\bx + a \br) - U_{-r}(\bx) \eta(\bx - a \br) \right ] 
\nonumber \\
& ~ & - 4 \frac{\kappa}{a}  \frac{1}{2a}
 \sum_r \left [ \eta^\dagger(\bx - a \br) {\hat \sigma}_r U_{r}(\bx - a \br) - 
 \eta^\dagger(\bx + a \br) {\hat \sigma}_r U_{-r}(\bx + a \br)  \right ]
{ 1 \over i \partial^+} \eta(\bx) \Bigg \}  \nonumber \\
 & ~ &+ \int dx^- a^2 \sum_\bx \Bigg \{
 \frac{\kappa}{a} \frac{1}{2a}\sum_r \sum_s 
\Big [ \eta^\dagger(\bx - a \br) U_{r}(\bx - a \br) + 
\eta^\dagger(\bx + a \br) U_{-r}(\bx + a \br) \Big ] \nonumber \\
&~&~~~~~{ 1 \over i \partial^+} {\hat \sigma}_s 
\Big [  U_{s}(\bx) \eta(\bx + a \bs) - U_{-s}(\bx) \eta(\bx - a \bs) \Big ] \nonumber \\
& ~ & - \frac{\kappa}{a} \frac{1}{2a}\sum_r \sum_s 
\Big [ \eta^\dagger(\bx - a \br) {\hat \sigma}_r U_{r}(\bx - a \br) - 
\eta^\dagger(\bx + a \br) {\hat \sigma}_r U_{-r}(\bx + a \br) \Big ]
\nonumber \\
&~&~~~~~{ 1 \over i \partial^+}  
\Big [  U_{s}(\bx) \eta(\bx + a \bs) + U_{-s}(\bx) \eta(\bx - a \bs)\Big ]
\Bigg \} \nonumber \\
& ~ & -\int dx^- a^2 \sum_\bx
\Bigg \{ \mu \frac{\kappa}{a} 
\eta^\dagger(\bx) { 1 \over i \partial^+}\sum_r    
\left [ U_{r}(\bx) \eta(\bx + a \br) 
+ U_{-r}(\bx) \eta(\bx - a \br) \right ] \nonumber \\
& ~ & +  \mu\frac{\kappa}{a}  
 \sum_r \left [ \eta^\dagger(\bx - a \br) U_{r}(\bx - a \br) + 
 \eta^\dagger(\bx + a \br) U_{-r}(\bx + a \br)  \right ]
{ 1 \over i \partial^+} \eta(\bx) \Bigg \} \nonumber \\
 & ~ & - \int dx^- a^2 \sum_\bx
  \frac{\kappa^2}{a^2} \sum_r \sum_s 
\Big [ \eta^\dagger(\bx - a \br)  U_{r}(\bx - a \br) + 
\eta^\dagger(\bx + a \br)  U_{-r}(\bx + a \br) \Big ] \nonumber \\
&~&~~~~~{ 1 \over i \partial^+}  
\Big [  U_{s}(\bx) \eta(\bx +a \bs)  + U_{-s}(\bx) \eta(\bx - a \bs)\Big ].
\ee
In addition, the factor $m^2$ in the free term in Eq. (\ref{qcdsd}) gets replaced by
$\mu^2=(m+4{\kappa \over a})^2$.

In the free limit the resulting Hamiltonian  goes over to
 
\be
 P^-_w &=& \int dx^- a^2 \sum_{\bx} \Bigg [ 
\mu^2 {\eta}^\dagger (\bx) { 1 \over i \partial^+} \eta(\bx)
\nonumber \\
&~& +
{ 1 \over 2 a} \sum_{r}[{\eta}^\dagger(\bx + a \br) - {\eta}^\dagger(\bx
-a \br)]{ 1 \over i \partial^+} {1 \over 2 a}[\eta(\bx + a \br) -
\eta(\bx - a \br)]
\nonumber \\
&~& + { \kappa^2 \over a^2}\sum_{r} [{\eta}^\dagger (\bx + a \br)
-2 {\eta}^\dagger(\bx) + {\eta}^\dagger(\bx -a \br)]
{ 1 \over i \partial^+}[\eta(\bx + a \br) - 2 \eta(\bx) +
\eta(\bx - a \br) ] \nonumber \\
&~& - 2 {\mu\kappa \over a}\sum_{r} {\eta}^\dagger(\bx) {1 \over i \partial^+}[
\eta(\bx + a \br) - 2 \eta(\bx) + \eta(\bx - a \br)] \Bigg ] . \label{w}
\ee
We  rewrite the free Hamiltonian (\ref{w}) as
\be
P_w^-=P_D^- + P_{OD1}^- +P_{OD2}^-. \label{wf_2}
\ee

The diagonal terms are
\be
P^-_D &=& \int dx^- a^2 \sum_{\bx} {\eta}^\dagger (\bx) { 1 \over i
\partial^+} \eta(\bx) \Big [ \mu^2 + { 1 \over a^2} + 8 \mu \kappa { 1 \over a} +
12 \kappa^2 { 1 \over a^2} \Big ].
\ee
The nearest neighbor interaction is
\be
P^-_{OD1} &=& -\int dx^- a^2 \sum_{\bx} \sum_{\br} \nonumber \\
&~& 
\Bigg [  (2 \mu\kappa { 1 \over a} + 4 {\kappa^2 \over a^2}) \Big [
{\eta}^\dagger (\bx) { 1 \over i \partial^+} \eta (\bx + a \br)
+ {\eta}^\dagger (\bx) { 1 \over i \partial^+} \eta (\bx + a \br) \Big ]
\Bigg ].
\ee
The next to nearest neighbor interaction is
\be P^-_{OD2} &=& \int dx^- a^2 \sum_{\bx} \sum_{\br} 
\Big \{ - { 1 \over 4 a^2} + { \kappa^2 \over a^2} \Big \}
\nonumber \\ 
&~& \Big [
{\eta}^\dagger(\bx + a \br) { 1 \over i \partial^+} \eta(\bx - a \br) +
{\eta}^\dagger(\bx - a \br) { 1 \over i \partial^+} \eta(\bx + a \br) 
\Big ]. \nonumber \\
\ee 

Using the Fourier transform in the transverse space, we get,
\be 
P^-_w&=&\int dx^- \int {d^2 k \over (2 \pi)^2)} \phi_{\bf k}^\dagger(x^-) {1 \over i
\partial^+} \phi_{\bf k}(x^-)  
\Bigg [ \mu^2 + \sum_r k_r^2 \Bigg ( {\sin k_ra \over k_r a} \Bigg
 )^2  \nonumber \\
&~&+ 2 a \mu \kappa \sum_r k_r^2 \Bigg ( {\sin k_ra/2 \over k_r a/2} \Bigg 
 )^2  
~+ a^2 \kappa^2  \sum_r k_r^4 \Bigg ( {\sin k_ra/2 \over k_r a/2} \Bigg 
 )^4 ~\Bigg ].\label{wil_ev}
\ee

Note that, as anticipated, Wilson term removes the doublers. Since the
Wilson term introduces nearest neighbor interactions, the sub-lattices
are now coupled to each other and  we have only one transverse
lattice.
The lowest eigenvalue in Eq. (\ref{wil_ev}) occurs only if all the $k_r$'s
are zero and there are no more doublers in the theory. 
\subsection{Numerical Investigation}
For our numerical investigation,
 we write the Hamiltonian (Eq. \ref{wf_2}) in DLCQ as
\be
H_{w} = H_D +H_{OD1} + H_{OD2}
\ee
with
\be
H_D &= & [ a^2{\mu}^2 + 1+ 8 a \mu \kappa + 12 \kappa^2] \sum_l 
\sum_\sigma \sum_\bz { 1 \over l} \nonumber \\
&~&~~~~ [ b^\dagger(l,\bz, \sigma) b(l,\bz, \sigma) + d^\dagger(l,\bz, \sigma)
d(l,\bz, \sigma) ],
\ee
\be
H_{OD1} & = & - [ 2 \kappa a \mu + 4 \kappa^2] 
\sum_l \sum_\sigma \sum_\bz \sum_r {1 \over l}\nonumber \\
&~& ~~~ \Bigg [ 
b^\dagger(l,\bz, \sigma) b(l,\bz+a \br, \sigma) +
b^\dagger(l,\bz, \sigma) b(l,\bz- a \br, \sigma) \nonumber \\
&~&~~~+d^\dagger(l,\bz, \sigma) d(l,\bz+a \br, \sigma) +
d^\dagger(l,\bz, \sigma) d(l,\bz- a \br, \sigma) 
 \Bigg ]
\ee 
and
\be
H_{OD2} & = & - [ { 1 \over 4} - \kappa^2] 
\sum_l \sum_\sigma \sum_\bz \sum_r {1 \over l}\nonumber \\
&~& ~~~ \Bigg [ 
b^\dagger(l,\bz+ a \br, \sigma) b(l,\bz-a \br, \sigma)
+b^\dagger(l,\bz-a \br, \sigma) b(l,\bz+ a \br, \sigma) \nonumber \\
&~&~~~+ d^\dagger(l,\bz+ a \br, \sigma) d(l,\bz-a \br, \sigma)
+d^\dagger(l,\bz-a \br, \sigma) d(l,\bz+ a \br, \sigma) \Bigg ]~.
\ee
 
\subsubsection{Boundary condition}
With the Wilson term added, we do not have decoupled sub-lattices. We have
both nearest neighbor and next-to-nearest neighbor interactions. 
Since with fixed boundary condition, the lowest four eigenvalues are not exactly
degenerate in finite volume, it is difficult to investigate the removal of degeneracy
by the addition of Wilson term.  
With periodic boundary condition, for a
lattice with $2n+1$ lattice points in each transverse direction, 
we identify the $(2n+2)^{th}$
lattice site with the first lattice site. Then for the Hamiltonian
matrix we get the following additional contributions.
\be
H = \begin{pmatrix} {. & . & . & . & ... & . & . & NN & N \cr
              . & . & . & . & ... & . & . & 0 & NN \cr
              . & . & . & . & ....& . & . & . & . \cr
                   ... \cr
                   ... \cr
               . & . & . & . & ....& . & . & . & . \cr
               NN & 0 & . & . & ... & . & . & . & . \cr
               N & NN & . & . & ... & . & . & . & .} 
\end{pmatrix}
\ee
The matrix elements $ NN = -{1 \over 4} + \kappa^2$ 
and $N=-2a\mu\kappa - 4 \kappa^2$.
For a given $n$, the allowed values of $k$ are $k_p a = \pm {2 \pi p
\over 2n+1}$, $p=0,1,2, ......$. Thus for $n=3$, we expect multiples
of ${2 \pi \over 7}$ apart from $0$. For $n=5$, apart from $0$,
allowed values of $k$ are multiples of ${2 \pi \over 11}$.  
  
\subsubsection{Numerical results}
Since the Wilson term connects even and odd lattices, the extra
fermions that appear at zero transverse momentum are removed once
Wilson term is added as we now have nearest and next to nearest
neighbor interactions. 
For large $n$, we get the expected spectra but,
numerical results suggest that the finite volume effect is larger for 
small $\kappa$ which is obvious because $\kappa$ is a mass-like
parameter.  For small $\kappa$ the wavefunctions become fat and
requires larger lattice to fit into.  
For example, with periodic boundary condition, 
for $n=3$, for $ \kappa=1.0, 0.5, 0.4$, we get the expected
harmonics but not for $\kappa=0.1$. The situation is similar for $n=5$. For
$n=10$, expected harmonics emerge even for $\kappa=0.1$ but not for
$\kappa=0.01$. Since all realistic calculations are done in a small
lattice, it is desirable to have the Wilson parameter $\kappa$ not too
small. 

\section{Doubling and symmetries on the light front transverse lattice}
Let us now try to understand the fermion doubling in terms of the  symmetries
of the transverse lattice Hamiltonians. We are aware that there are
rigorous theorems and anomaly arguments in the conventional lattice
gauge theories \cite{3NN} regarding presence of fermion doublers. In standard
lattice gauge theory, some chiral symmetry needs to be broken in the
kinetic part of the action to avoid the doublers. On the light-front, 
chirality means helicity. For example, a standard Wilson term
which is not invariant under chiral transformations in the
conventional lattice gauge theory, is chirally invariant on the light-front 
in the free field limit. The question is then why the Wilson term
removes the
doublers on the light-front transverse lattice. The argument that
there is nonlocality in the longitudinal direction cannot hold
because, in the first place, having nonlocality is not a guarantee for
removing doublers and secondly, there is no nonlocality on the
transverse lattice. One, therefore needs to find a reasoning that
involves the helicity in some way.

Because of  the constraint equation which is inconsistent
with the equal time chiral transformation in the presence of massive
fermions, we should distinguish between chiral symmetry in the equal
time formalism and in the light-front formalism.  For example, the free
massive  light-front Lagrangian involving only the dynamical degrees
of freedom  is  invariant under $\gamma_5$ transformation. On the
light-front, helicity takes over the notion of chirality even in
presence of fermion mass which can be understood in the following way.

In the two component representation \cite{3hz} in the light-front formalism,
let  us look at the objects
 $ \psi_L^+$
and $ \psi_R^+$.  We have
\begin{eqnarray}
\psi^+(x) = \pmatrix{ \eta(x) \cr
                    0 \cr}
\end{eqnarray}
with
\begin{eqnarray}
\eta(x) = \pmatrix{ \eta_1(x) \cr
                    \eta_2(x) \cr}
\end{eqnarray}
The projection operators are $ P_R = { 1 \over 2} (1 + \gamma^5)$ 
and $ P_L = { 1 \over 2} (1 -\gamma^5)$ with
\begin{eqnarray}
\gamma^5 = \pmatrix{ \sigma^3 & 0 \cr
                    0 & - \sigma^3 \cr}.
\end{eqnarray}
Then 
\begin{eqnarray}
\psi_R^+= P_R \psi^+ = \pmatrix{ \eta_1 \cr
                     0  \cr
                     0  \cr
                     0}
\end{eqnarray}
and 
\begin{eqnarray}
\psi_L^+= P_L \psi^+ = \pmatrix{ 0 \cr
                     \eta_2  \cr 
                     0  \cr
                     0}. 
\end{eqnarray}
Thus $\psi_R^+=P_R \psi^+$ represents a positive helicity fermion and 
$ \psi_L^+=P_L \psi^+$
represents a negative helicity fermion, even when the fermion is 
{\em massive}.
This makes sense since chirality {\em is} helicity even for a massive
fermion in front form. This is again to be contrasted with the instant
form. In that case the right handed and left handed fields defined by
$ \psi_R= P_R \psi= { 1 \over 2} (1 + \gamma^5) \psi $ and $ 
\psi_L = P_L \psi = { 1 \over 2} (1 -\gamma^5) \psi $ 
contain both positive helicity and negative helicity
states. Only  in the massless limit or in the infinite momentum limit,
 $\psi_R$ becomes the
positive helicity state and $\psi_L$ becomes the negative helicity state. 

As a passing remark, we would like to mention that  in continuum light 
front QCD there is a linear mass term  that allows  for  helicity flip 
interaction.

In  lattice gauge theory in the Euclidean or equal time formalism,
because of reasons connected to anomalies (the standard ABJ anomaly in 
vector-like gauge theories), there  has to be explicit chiral symmetry 
breaking  in the kinetic part of the action or Hamiltonian.
Translated to the light front transverse lattice formalism, 
this would then require
helicity flip in the kinetic part.   A careful observation of all the
above methods that get rid of fermion doublers on the light front
transverse lattice reveals that this is indeed true.
  
In particular, we draw attention to  the  even-odd helicity flip
transformation
\be
\eta(x_1,x_2) \rightarrow ({\hat\sigma}_1)^{x_1}({\hat\sigma}_2)^{x_2}
\eta(x_1,x_2) \label{eosf}
\ee
that was used in Sec. \ref{lfsf} for spin diagonalization.  
It should also
be clear that the form of the above transformation is not unique in
the sense that  one could exchange ${\hat\sigma}_1$ and 
${\hat\sigma}_2$ and their  exponents $x_1$ and $x_2$ could be 
changed by $\pm 1$.

Note that the Hamiltonians $P^-_{fb}$ given in Eq. (\ref{fb}) and   
$P^-_{w}$ given in Eq. (\ref{w}) that do not exhibit fermion doubling
are not invariant under the
transformation Eq. (\ref{eosf}). On the other hand  the Hamiltonian 
$P^-_{sd}$ given by Eq. (\ref{sd}) that exhibits fermion doubling is 
invariant under this transformation.
\section{Summary}
The presence of the constraint equation for fermions on the light front
gives rise to interesting possibilities of formulating fermions on
a transverse lattice. We have discussed in detail  the
transverse lattice Hamiltonians resulting from different approaches in
two different boundary conditions.
 
In the first approach, we have proposed to use  forward and backward 
lattice derivatives  respectively for $\psi^+$
and $\psi^-$ (or vice versa) so that the resulting Hamiltonian
is Hermitian. There is no fermion doubling. 
The helicity flip (chiral symmetry breaking) term proportional
to the fermion mass in the full light front QCD becomes an irrelevant
term  in the free field limit. With periodic boundary
condition one can get the helicity up and helicity down fermions to be
degenerate for any transverse lattice size $n$. With fixed boundary
condition, there is a 
splitting between the two states at any $n$ but the splitting vanishes
in the large volume limit.  

In the second approach,  symmetric
derivatives are used for both $\psi^+$ and $\psi^-$. This 
results in four fermion species. This is a consequence of the fact that the
resulting free Hamiltonian has only next to nearest neighbor interactions
and as a result even and odd lattice sites get decoupled. 
One way to remove doublers is to reinterpret them as flavors using
staggered fermion formulation on the light front. In QCD Hamiltonian,
it  generates  irrelevant flavor mixing interactions. However, in
the free field limit, there is no flavor mixing. Another way to remove 
the doublers is to add a Wilson term which generates many extra terms
in the Hamiltonian. In the free field limit, only the  helicity
nonflip terms survive. The  Wilson term 
couples  even and odd sites and  removes the
doublers. Numerically, we found that in small lattice volumes it is preferable to
have not too small values of the Wilson mass $\kappa /a$. 
   
Chiral symmetry in  light-front is different from the chiral symmetry
in equal time formalism. So, fermion doubling on LFTL should be related with the
{\it chiral symmetry  on  the light-front}. Light-front chirality  is equivalent
to helicity even for massive fermions. 
We have identified an even-odd helicity flip symmetry of the light front
transverse lattice Hamiltonian, absence of which
means removal of doublers in all the cases we have studied.



\chapter{Meson Bound States in Transverse Lattice 
QCD} \label{tlchap2}
\section{Introduction}
In the previous chapter we have discussed different ways of formulating
fermions  on a light-front transverse lattice.  We have also discussed
the advantages and disadvantages of all the methods in free field
limit. But, the true testing ground for the strengths and weaknesses of
different methods is the full QCD.  To complete this comparative study,
 in this chapter we consider the meson
bound  state problem in (3+1) dimensional light-front QCD with two
transverse directions ($x^1, x^2$) discretized on a square lattice.  

In this chapter we make a detailed comparison of 
two different light-front QCD 
Hamiltonians. One is the Hamiltonian with  fermions formulated using 
forward and backward lattice derivatives and the other Hamiltonian with fermions 
formulated with symmetric lattice derivative with Wilson term \cite{4BK} to 
get rid of doublers   on the 
transverse lattice. Light-front staggered lattice formalism \cite{4DH3, 4griffin}
to remove doublers 
in case of symmetric lattice derivative  that we 
have discussed in the previous 
Chapter is a different game altogether and will not be further investigated
 in this work.
 For our calculation, we adopt
the {\it one link approximation} in the meson sector 
which has been widely used in
the literature \cite{4dalme,4buseal,4review}. 
(Only very recently, the effect of additional links in the
meson sector has been investigated \cite{4vandaly}). 
One link approximation is too crude to reproduce physical observables.
 So, rather than fitting the parameters to reproduce any physical observable 
we concentrate here to 
investigate the effects of various coupling strengths on
the low-lying spectra and wave functions  and compare two 
different formulations.  

We use Discretized Light Cone Quantization (DLCQ) \cite{4dlcq} 
to address longitudinal
dynamics. Because of the presence of severe light-front infrared
divergences, a major concern here is the reliability of DLCQ results when
calculations are done at finite resolution $K$ and results are extrapolated
to the continuum  ($ K \rightarrow \infty$). In meson calculations so far, 
$K \leq 20$ have been chosen.  We perform a detailed study of
the continuum limit of DLCQ by performing calculations at larger values of $K$
\cite{4DH4}. 

In the meson sector, in the zero link approximation, at each transverse
location we have a two-dimensional field theory which in the large $N_c$ limit
(where $N_c$ is the number of colors) is
nothing but the 't Hooft model. In this well-studied model, excited states
are simply excitations of the $q {\bar q}$ pair, which contain nodes in
the wavefunctions. The picture changes when one link is included thereby
allowing fermions to hop. The admixture of $q {\bar q}$ link states with $q
{\bar q}$ states is controlled by the strengths of the particle number
changing 
interactions and the mass of the link field. One link approximation is a
priori justified for very massive links and/or weak particle changing
interaction since in this case low lying excited states are also $q{\bar q}$
excitations. Likewise, for large particle changing 
interaction strength and/or light
link mass, low lying excited states are $q{\bar q}$ link states. We explore the
spectra and wavefunctions resulting from the choice of various regions of
parameter space.

Details of the derivation of the fermionic part of the 
Hamiltonian are already discussed in the previous chapter.
 Here we give the details of the
gauge field part of the QCD Hamiltonian. Non-linear constraints on the unitary
link variables make it difficult to perform canonical quantization. We also
present the effective Hamiltonian when non-linear unitary variables are
replaced by linear variables.

\section{Gauge field part of the Lagrangian density}
 
The gauge field part of the Lagrangian density in the continuum is
\be
{ {\cal L}}_G = { 1 \over 2 g^2} Tr F_{\rho \sigma}F^{\rho \sigma}
\ee
where $ F^{\rho \sigma} = \partial^\rho A^\sigma - \partial^\sigma
A^\rho +  ~[ A^\rho, A^\sigma]$ with $ A^\rho = i g A^{\rho \alpha} T^\alpha$.
Here $ \rho, \sigma=0,1,2,3$ and $ \alpha=1,2, \ldots, 8.$ For ease of
notation we suppress the dependence of field variables on the longitudinal
coordinate in this section.
With the gauge choice $A^+=0$,  the Lagrangian density can be separated into
three parts,
\be
{\cal L}_G = {\cal L}_T + {\cal L}_L + {\cal L}_{LT}.
\ee
Here $ {\cal L}_T $ depends entirely on the lattice gauge field $U_r(\bx)$.
\be
{\cal L}_T = { 1 \over g^2 a^4 } \sum_{r \neq s} \Bigg \{
Tr \Big [
U_r(\bx) U_s(\bx + a \br) U_{-r}(\bx + a \br + a \bs)U_{-s}(\bx + a\bs) 
 -1 \Big ] \Bigg \},
\ee
$r,s =1,2$.
The  purely longitudinal part $ {\cal L}_{L}$ depends on the constrained
gauge field $A^-$, 
\be
{\cal L}_{L}= { 1 \over 8} (\partial^+ A^{- \alpha})^2
\ee
and the mixed part $ {\cal L}_{LT}$ depends both on lattice gauge field and the
constrained gauge field. 
\be
{\cal L}_{LT} &=&  { 1 \over 2 g^2} Tr \big [ F_{\mu r}F^{\mu r} \big ]
= { 1
\over  g^2 a^2 } Tr \big [ D_\mu U_r(\bx) (D^\mu U_r(\bx))^\dagger \big ]\nonumber \\
&=& {1 \over  g^2 a^2 }Tr \Big [\partial_\mu U_r (\bx)
{\partial^\mu} U_r^\dagger(\bx) + A_\mu(\bx)
[ U_r(\bx) \stackrel {\leftrightarrow}{\partial^\mu} U_r^\dagger (\bx)]
\nonumber \\
&~&~~~~+ A_\mu (\bx + a \br)
[ U_r^\dagger (\bx) \stackrel{\leftrightarrow}{\partial^\mu} U_r(\bx)] \Big ]
\ee
where $\mu = +, -$ only and the mixed covariant derivatives of the link variables 
are defined by (see Appendix \ref{APcov_der} for the derivation)
\be
D_\mu U_r(\bx) = \partial_\mu U_r(\bx) +  A_\mu U_r(\bx) - U_r(\bx)
A_\mu (\bx + a \br).
\ee
In the $A^+ =0$ gauge, we can write $ {\cal L}_{LT}$ as
\be
{\cal L}_{LT} = {1 \over g^2 a^2} Tr [ \partial_\mu U_r(\bx)
\partial^\mu U_r^\dagger(\bx)] + {1 \over 2 a^2}g A^{-\alpha}J^{+\alpha}_{LINK}.
\ee
Here the link current
\be J^{+\alpha}_{LINK} (\bx)= \sum_r {1 \over g^2}Tr \Bigg \{ T^\alpha 
[  U_r(\bx) i\stackrel{\leftrightarrow}{\partial^+} U_r^\dagger (\bx)+  
U_r^\dagger (\bx- a \br) 
i \stackrel{\leftrightarrow}{\partial^+} U_r (\bx - a \br)] \Bigg \}.
\ee

Substituting back the expression for $A^{-\alpha}$ from the constraint
equation
\be
(\partial^+)^2 A^{-\alpha} = \frac{2g}{a^2} \left (J^{+\alpha}_{LINK} - 
J^{+\alpha}_q) \right )
\ee
with 
\be J^{+\alpha}_q (\bx)= 2 \eta^\dagger(\bx) T^\alpha \eta(\bx) 
\ee
where $\eta$ is the dimensionless two-component lattice fermion field,
in the $A^{-\alpha}$ dependent terms in the Lagrangian density, namely,
\be
- { 1 \over 2} \frac{g}{a^2} A^{-\alpha} J^{+ \alpha}_q + 
{ 1 \over 8}  (\partial^+ A^{- \alpha})^2  + { 1 \over 2} \frac{g}{a^2}
A^{- \alpha}J^{+\alpha}_{LINK}
\ee
we generate the terms
\be
{g^2 \over 2 a^4} J^{+\alpha}_{LINK}\left ({ 1\over \partial^+}\right)^2
J^{+\alpha}_{LINK}+ {g^2 \over 2 a^4} \eta^\dagger T^\alpha  \eta   
\left ({ 1\over \partial^+}\right )^2  \eta^\dagger T^\alpha  \eta -
 { g^2 \over a^4} 
J^{+\alpha}_{LINK} \left ({ 1\over \partial^+}\right)^2 \eta^\dagger T^\alpha  \eta.
\ee
Collecting all the terms, the canonical Lagrangian density for
transverse lattice QCD is
\be
{\cal L}  &=& {\cal L}_f  +{ 1 \over a^4 g^2} Tr [ \partial_\mu U_r(\bx)
\partial^\mu U_r^\dagger(\bx)] \nonumber \\
&~& +  {1 \over a^4 g^2} \sum_{r \neq s}
\Bigg \{ Tr \Big [
U_r(\bx) U_s(\bx + a \br) U_{-r}(\bx + a \br + a \bs)U_{-s}(\bx + a
\bs) 
- 1\Big ] \Bigg \} \nonumber \\
&~& + {g^2 \over 2 a^4} J^{+\alpha}_{LINK}\left({ 1\over \partial^+}\right)^2
J^{+ \alpha}_{LINK}+ {1 \over 2 a^4} g^2 J^{+\alpha}_q   
\left ({ 1\over \partial^+}\right )^2  J^{+\alpha}_q \nonumber \\
&~&~~~~ - { g^2 \over a^4} 
J^{+\alpha}_{LINK} \left ({ 1\over \partial^+}\right )^2 
J^{+\alpha}_q, \label{fb_L}
\ee
where ${\cal L}_f$ is the fermionic part of the QCD Lagrangian density with 
forward and backward lattice derivatives  given by
\be
{\cal L}_f &=& {1\over a^2}\eta^\dagger (\bx) i \partial^- \eta (\bx) 
- {m^2\over a^2}
\eta^\dagger (\bx) { 1 \over i \partial^+} \eta(\bx) \nonumber \\ 
&~&  + im{1\over a^2} \eta^\dagger(\bx)  {\hat
\sigma}_s {1 \over a}{1 \over \partial^+}\Big [
U_s(\bx) \eta(\bx + a \bs) - \eta(\bx) \Big ] 
\nonumber \\
&~& +im {1\over a^2}
\Big [
\eta^\dagger(\bx + a \br) U_r^\dagger(\bx) -
\eta^\dagger(\bx) \Big ] 
 {\hat \sigma}_r{ 1 \over a} { 1 \over 
\partial^+} \eta(\bx) \nonumber \\
&~& - { 1 \over a^4} [ \eta^\dagger(\bx + a \br) U_r^\dagger(\bx) -
\eta^\dagger(\bx)] {\hat \sigma}_r { 1 \over i \partial^+}{\hat
\sigma}_s [ U_s(\bx) \eta(\bx + a \bs) - \eta(\bx)],
\ee
 or with symmetric lattice derivative with 
Wilson term given by
\be
{\cal L}_f & =& {1\over a^2}\eta^\dagger (\bx) i \partial^- \eta (\bx)
-{1\over a^2}\left(m+4 \frac{\kappa}{a}\right)^2
\eta^\dagger(\bx) { 1 \over i \partial^+} \eta(\bx)  \nonumber \\
&~& +{ 1 \over a^2} \left ( m+ 4 \frac{\kappa}{a} \right ) \frac{1}{2a}
 \Bigg \{ 
\eta^\dagger(\bx) \sum_r {\hat \sigma}_r { 1 \over i \partial^+}
\left [ U_{r}(\bx) \eta(\bx + a \br) - U_{-r}(\bx) \eta(\bx - a \br) \right ]
\nonumber \\
& ~ & -  \sum_r \left [ \eta^\dagger(\bx - a \br) {\hat \sigma}_rU_{r}(\bx - a \br)
- \eta^\dagger(\bx + a \br) {\hat \sigma}_r U_{-r}(\bx + a \br)  \right ]
{ 1 \over i \partial^+} \eta(\bx) \Bigg \} . \nonumber \\
&~& -{ 1 \over a^2}\Bigg \{
 \frac{\kappa}{a} \frac{1}{2a}\sum_r \sum_s
\Big [ \eta^\dagger(\bx - a \br) U_{r}(\bx - a \br) +
\eta^\dagger(\bx + a \br) U_{-r}(\bx + a \br) \Big ] \nonumber \\
&~&~~~~~{ 1 \over i \partial^+} {\hat \sigma}_s
\Big [  U_{s}(\bx) \eta(\bx + a \bs) - U_{-s}(\bx) \eta(\bx - a \bs) \Big ]
\nonumber \\
& ~ & - \frac{\kappa}{a} \frac{1}{2a}\sum_r \sum_s
\Big [ \eta^\dagger(\bx - a \br) {\hat \sigma}_r U_{r}(\bx - a \br) -
\eta^\dagger(\bx + a \br) {\hat \sigma}_r U_{-r}(\bx + a \br) \Big ]
\nonumber \\
&~&~~~~~{ 1 \over i \partial^+}
\Big [  U_{s}(\bx) \eta(\bx + a \bs) + U_{-s}(\bx) \eta(\bx - a \bs)\Big ]
\Bigg \}\nonumber \\
&~&+{ 1 \over a^2}\frac{1}{4a^2} \sum_r \sum_s
\Big [ \eta^\dagger(\bx - a \br) {\hat \sigma}_r U_{r}(\bx - a \br) -     
\eta^\dagger(\bx + a \br)  {\hat \sigma}_r U_{-r}(\bx + a \br) \Big ]     
\nonumber \\
&~&~~~~~{ 1 \over i \partial^+} {\hat \sigma}_s
\Big [  U_{s}(\bx) \eta(\bx + a \bs) - U_{-s}(\bx) \eta(\bx - a \bs)\Big ] 
\nonumber \\
&~&+ {1 \over a^2}\left ( m+ 4 \frac{\kappa}{a} \right ) \frac{\kappa}{a} \Bigg \{           
\eta^\dagger(\bx) { 1 \over i \partial^+}\sum_r
\left [ U_{r}(\bx) \eta(\bx + a \br)
+ U_{-r}(\bx) \eta(\bx - a \br) \right ] \nonumber \\
& ~ & + 
 \sum_r \left [ \eta^\dagger(\bx - a \br) U_{r}(\bx - a \br) +
 \eta^\dagger(\bx + a \br) U_{-r}(\bx + a \br)  \right ]
{ 1 \over i \partial^+} \eta(\bx) \Bigg \}  \nonumber \\
&~& +{1 \over a^2}\frac{\kappa^2}{a^2} \sum_r \sum_s
\Big [ \eta^\dagger(\bx - a \br)  U_{r}(\bx - a \br) +
\eta^\dagger(\bx + a \br)  U_{-r}(\bx + a \br) \Big ] \nonumber \\        
&~&~~~~~{ 1 \over i \partial^+}
\Big [  U_{s}(\bx) \eta(\bx +a \bs)  + U_{-s}(\bx) \eta(\bx - a \bs)\Big ].
\ee
 Here we use the
 two-component representation \cite{4hz} for the dynamical fermion field
\be
\psi^+(x^-, x^{\perp}) = \left[ \begin{array}{l} {1 \over a}\eta(x^-, x^{\perp})
\\ 0 \end{array} \right] 
\ee
where $\eta$ is the dimensionless two component lattice fermion field and 
${\hat \sigma}_1=\sigma_2, ~{\hat \sigma}_2 =-\sigma_1$.

\subsubsection{Linearization of the link fields}
Because of the nonlinear constraints $ U^\dagger U= 1$, $ det~U=1$, 
it is highly nontrivial to quantize the system. Hence 
Bardeen and Pearson \cite{4Bardeen:1976tm}
and Bardeen, Pearson, and Rabinovici \cite{4Bardeen:1980xx} proposed to 
replace the
nonlinear variables $U$ by {\it linear} variables $M$ where $M$ belongs to
$GL(N,{\cal C})$, i.e., we replace 
$ { 1 \over g} U_r(\bx) \rightarrow M_r(\bx)$.
  This  linearized approximation is somewhat 
meaningful only on a coarse lattice. Justification may come from the fact that
particle structure of a hadron is a long-distance property  which can be
described by  some effective variables without any explicit details of 
the microscopic variables \cite{4Bardeen:1980xx}. The $x^-$ independent residual 
gauge invariance on the transverse plane is still preserved with the linear gauge fields
(see Appendix \ref{APgauge} for details).
 Once we replace $U$ by $M$, many more terms are
allowed in the Hamiltonian.  
 Thus one needs to add an effective potential
$V_{eff}$ to the Lagrangian density
\be
V_{eff}=  - {\mu^2\over a^2} ~Tr (M^\dagger M) + \lambda_1 ~Tr[(M^\dagger M)^2] +
\lambda_2 ~[det~ M + H.c] + \ldots .
\ee

\section{Effective Hamiltonian}
\subsection{Hamiltonian with forward and backward derivatives}
Once we replace the nonlinear link fields by linear link field by including 
the effective potential, we  perform canonical quantization and construct the
effective Hamiltonian for transverse lattice QCD.

Thus, the effective Hamiltonian for QCD on the transverse
lattice, when fermions are put in with forward and backward lattice derivatives, becomes
\be
P^-_{fb} & = & P^-_{f~free} + P^-_V+ P^-_{fhf}+P^-_{hf} 
+ P^-_{chnf} \nonumber \\
&~& + P^-_{qqc} + P^-_{ggc} + P^-_{qgc} + P^-_{p}.
\ee
The free fermion part is
\be  P^-_{f~free} & = & \int dx^-  \sum_{\bx} (m^2 + {2 \over a^2})
\eta^\dagger (\bx) { 1 \over i \partial^+} \eta(\bx) .
\ee
The effective potential part is 
\be 
 P^-_V & = & 
\int dx^- a^2 \sum_\bx \Bigg ({ \mu^2\over a^2} ~Tr (M^\dagger M)
- \lambda_1 ~Tr[(M^\dagger M)^2] -
\lambda_2 ~[det~ M + H.c] + \ldots  \Bigg ). \nonumber \\
\ee
The free helicity-flip part is 
\be 
P^-_{fhf} & = & 2im\int dx^-\sum_{\bx}\sum_s ~\eta^\dagger(\bx)  {\hat
\sigma}_s {1 \over a}{1 \over \partial^+} \eta(\bx). 
\ee
Helicity flip associated with the fermion hop is
\be
P^-_{hf} & = & - img \int dx^-\sum_{\bx} \sum_s \eta^\dagger(\bx)  {\hat
\sigma}_s {1 \over a}{1 \over \partial^+}\Big [
M_s(\bx) \eta(\bx + a \bs) \Big ]
\nonumber \\
&~& -img \int dx^-\sum_{\bx}\sum_r
\Big [
\eta^\dagger(\bx + a \br)  ~M_r^\dagger(\bx) \Big ]
 {\hat \sigma}_r{ 1 \over a} { 1 \over
\partial^+} \eta(\bx).
\ee
Canonical helicity non-flip terms are
\be
P^-_{chnf} & = & - { g \over a^4}\int dx^- a^2\sum_{\bx} \sum_{rs} 
[ \eta^\dagger(\bx + a \br)  M_r^\dagger(\bx) ] 
{\hat \sigma}_r { 1 \over i \partial^+}{\hat
\sigma}_s [  \eta(\bx)] \nonumber \\
&~& - { g \over a^4}\int dx^- a^2\sum_{\bx} \sum_{rs} 
[\eta^\dagger(\bx)] {\hat \sigma}_r { 1 \over i \partial^+}{\hat
\sigma}_s [  M_s(\bx) \eta(\bx + a \bs)] \nonumber \\
&~& - { g^2 \over a^4} \int dx^- a^2\sum_{\bx}\sum_{rs} [ \eta^\dagger(\bx + a \br)
~M_r^\dagger(\bx)
] {\hat \sigma}_r { 1 \over i \partial^+}{\hat
\sigma}_s [  M_s(\bx) \eta(\bx + a \bs) ]. \label{canhnf}
\ee
The four-fermion instantaneous term is
\be
P^-_{qqc} & = & -2 {g^2\over a^2} \int dx^-  \sum_\bx
 \eta^\dagger(\bx) T^a \eta(\bx) { 1 \over (\partial^+)^2}
\eta^\dagger(\bx) T^a \eta(\bx) .
\ee
The four link instantaneous  term is
\be
P^-_{ggc} & = & - \frac{1}{2} {g^2 \over a^2}\int dx^-  \sum_\bx
  J^{+a}_{LINK}(\bx) { 1 \over
(\partial^+)^2}J^{+a}_{LINK}(\bx). 
\ee
The fermion - link instantaneous term is
\be
P^-_{qgc} & = &  2 {g^2\over a^2}\int dx^- \sum_\bx
 J^{+a}_{LINK}(\bx) { 1 \over (\partial^+)^2}
\eta^\dagger(\bx)
T^a \eta(\bx) .
\ee
The plaquette term is
\be 
P^-_{p} & = & -  {g^2 \over a^4} \int dx^- a^2\sum_\bx \sum_{r \neq s}
\Bigg \{ Tr \Big [
M_r(\bx) M_s(\bx + a \br) M_{-r}(\bx + a \br + a \bs)M_{-s}(\bx + a \bs)
- 1\Big ] \Bigg \}.\nonumber \\
~
\ee
Here
\be 
J^{+\alpha}_{LINK} (\bx)= \sum_r Tr \Bigg \{ T^\alpha
[  M_r(\bx) i \stackrel{\leftrightarrow}{ \partial^+} M_r^\dagger (\bx)+
M_r^\dagger (\bx- a \br)
i \stackrel{\leftrightarrow}{ \partial^+} M_r (\bx - a \br)] \Bigg \}.
\ee

\subsection{Hamiltonian with the Wilson term}
When one uses symmetric derivatives for the fermion fields,  
doublers arise as a result of the decoupling of even and odd lattice sites. 
Here we use the Wilson term to remove the doublers.
In this
subsection, the details of the structure of the Hamiltonian resulting with the
modification of the Wilson term are presented.

 The effective  Hamiltonian for this case can be written as
\be
 P^- & = & P^-_{f~free} + P^-_V+ P^-_{hf} + P^-_{whf}  \nonumber \\
&~& +P^-_{chnf} + P^-_{wnf1} + P^-_{wnf2} \nonumber \\
&~& + P^-_{qqc} + P^-_{ggc} + P^-_{qgc} + P^-_{p}.
\ee
The free fermion part is
\be
P^-_{f~free}= \int dx^- a^2 \sum_\bx
{1\over a^2}\left(m+4 \frac{\kappa}{a}\right)^2
\eta^\dagger(\bx) { 1 \over i \partial^+} \eta(\bx) .
\ee
The helicity flip part is
\be
P^-_{hf} & = &  -g \int dx^-  \sum_\bx
\Bigg \{ \left ( m+ 4 \frac{\kappa}{a} \right ) \frac{1}{2a}
\eta^\dagger(\bx) \sum_r {\hat \sigma}_r { 1 \over i \partial^+}  
\left [ M_{r}(\bx) \eta(\bx + a \br) - M_{-r}(\bx) \eta(\bx - a \br) \right
] 
\nonumber \\
& ~ & - \left ( m+ 4 \frac{\kappa}{a} \right ) \frac{1}{2a}
 \sum_r \left [ \eta^\dagger(\bx - a \br) {\hat \sigma}_rM_{r}(\bx - a \br)
- 
 \eta^\dagger(\bx + a \br) {\hat \sigma}_r M_{-r}(\bx + a \br)  \right ]
{ 1 \over i \partial^+} \eta(\bx) \Bigg \} . \nonumber \\
\ee
The Wilson term induced helicity flip part
\be
P^-_{whf} & = & g^2\int dx^-  \sum_\bx \Bigg \{
 \frac{\kappa}{a} \frac{1}{2a}\sum_r \sum_s 
\Big [ \eta^\dagger(\bx - a \br) M_{r}(\bx - a \br) + 
\eta^\dagger(\bx + a \br) M_{-r}(\bx + a \br) \Big ] \nonumber \\
&~&~~~~~{ 1 \over i \partial^+} {\hat \sigma}_s 
\Big [  M_{s}(\bx) \eta(\bx + a \bs) - M_{-s}(\bx) \eta(\bx - a \bs) \Big ]
\nonumber \\
& ~ & - \frac{\kappa}{a} \frac{1}{2a}\sum_r \sum_s 
\Big [ \eta^\dagger(\bx - a \br) {\hat \sigma}_r M_{r}(\bx - a \br) - 
\eta^\dagger(\bx + a \br) {\hat \sigma}_r M_{-r}(\bx + a \br) \Big ]
\nonumber \\
&~&~~~~~{ 1 \over i \partial^+}  
\Big [  M_{s}(\bx) \eta(\bx + a \bs) + M_{-s}(\bx) \eta(\bx - a \bs)\Big ]
\Bigg \}.
\ee
The canonical helicity non-flip term arising from fermion constraint is 
\be
P^-_{chnf} & = & -g^2 \int dx^-  \sum_\bx
 \frac{1}{4a^2} \sum_r \sum_s 
\Big [ \eta^\dagger(\bx - a \br) {\hat \sigma}_r M_{r}(\bx - a \br) - 
\eta^\dagger(\bx + a \br)  {\hat \sigma}_r M_{-r}(\bx + a \br) \Big ]
\nonumber \\
&~&~~~~~{ 1 \over i \partial^+} {\hat \sigma}_s 
\Big [  M_{s}(\bx) \eta(\bx + a \bs) - M_{-s}(\bx) \eta(\bx - a \bs)\Big ] .
\ee
The Wilson term induced helicity non flip terms are  
\be
P^-_{wnf1} & = & -g \int dx^-  \sum_\bx
\Bigg \{\left ( m+ 4 \frac{\kappa}{a} \right ) \frac{\kappa}{a}
\eta^\dagger(\bx) { 1 \over i \partial^+}\sum_r    
\left [ M_{r}(\bx) \eta(\bx + a \br) 
+ M_{-r}(\bx) \eta(\bx - a \br) \right ] \nonumber \\
& ~ & + \left ( m+ 4 \frac{\kappa}{a} \right ) \frac{\kappa}{a}
 \sum_r \left [ \eta^\dagger(\bx - a \br) M_{r}(\bx - a \br) + 
 \eta^\dagger(\bx + a \br) M_{-r}(\bx + a \br)  \right ]
{ 1 \over i \partial^+} \eta(\bx) \Bigg \}. \nonumber \\
\ee
and 
\be
P^-_{wnf2} & = & - g^2 \int dx^-  \sum_\bx
  \frac{\kappa^2}{a^2} \sum_r \sum_s 
\Big [ \eta^\dagger(\bx - a \br)  M_{r}(\bx - a \br) + 
\eta^\dagger(\bx + a \br)  M_{-r}(\bx + a \br) \Big ] \nonumber \\
&~&~~~~~{ 1 \over i \partial^+}  
\Big [  M_{s}(\bx) \eta(\bx +a \bs)  + M_{-s}(\bx) \eta(\bx - a \bs)\Big ].
\ee

Comparing the Hamiltonians with a) forward-backward derivative and b)
symmetric derivative with the Wilson term we notice that the only
differences are in the particle number changing interactions, namely, helicity flip
 and helicity non-flip terms. 
\section{Meson bound state in one link approximation}
\subsection{Relevant interactions}
In one link approximation, for either Hamiltonian, the four link
instantaneous  term and the plaquette term do not contribute and only
the link mass term of the effective potential contributes.  Further, in the
case of the forward-backward Hamiltonian, the helicity non-flip part
proportional to $g^2$ does not contribute. For the Wilson term modified
Hamiltonian, the Wilson term induced helicity flip part $P^-_{whf}$, the
canonical helicity non-flip term $ P^-_{cnhf}$  and the term proportional to
$\kappa^2$ in the Wilson term induced helicity non-flip part do not
contribute. Thus in the case of the Wilson term 
modified Hamiltonian the entire
fermion hopping with no helicity flip arises from the Wilson term.
In the case of forward and backward derivatives, terms are also present 
in one link approximation which violate hypercubic
symmetry on the transverse lattice (see Appendix \ref{APrsv}).
 They become irrelevant in the continuum
limit when  the linear variables $M$ are replaced by non-linear variables $U$. 
 We have removed them
entirely from the Hamiltonian in the present investigation.   The Hamiltonian matrix 
elements in DLCQ for both forward-backward and symmetric derivative with Wilson 
term are explicitly given in Appendix \ref{APme}.
\subsection{Comparison with one gluon exchange in the continuum}
It is interesting to compare the one link approximation on the transverse
lattice with the one gluon exchange approximation in the continuum. In the
latter, a major source of singularity  is the $\frac{k^\perp}{k^+}$ term in
the quark - gluon vertex where $k^\perp$ ($k^+$) is the gluon transverse
(longitudinal) momentum. This originates from the $A^-J^+_q$ interaction term in
the Hamiltonian via $\frac{1}{\partial^+}\partial^\perp \cdot A^\perp$
contribution to the constrained field $A^-$. This term gives rise to
quadratic ultraviolet divergence in the transverse plane
accompanied by linear divergence in the longitudinal direction in fermion
self energy. 
On the transverse lattice, $ \partial^+ A^- \propto \frac{1}{\partial^+}
J^+_{LINK}$ so that $ A^- J^+_q \rightarrow J^+_{LINK}
\frac{1}{(\partial^+)^2} J^+_q$. Thus a term which gives rise to severe
divergence structure in the continuum gets buried in the fermion-link
instantaneous interaction term which gives rise to a term in the gauge boson
fermion vertex in the continuum in Abelian theory. In the non-Abelian gauge
theory this gives rise to a term in the quark-gluon vertex and also to the
instantaneous quark-gluon interaction in the continuum. 

The transfer of the troublesome term from quark-gluon vertex in the continuum
theory to quark - link instantaneous interaction term in the lattice theory
has an interesting consequence. In the continuum theory, the
addition of a gluon
mass term by hand spoils the cancellation of the light front singularity between one
gluon exchange and the instantaneous  four - fermion interaction. On the
transverse lattice, this cancellation is absent anyway with or without a link
mass term.
\subsection{Longitudinal dynamics and effects of transverse hopping}

We first consider the dynamics in the absence of any link. In this case,
fermions cannot hop, and at each transverse location we have (1+1)
dimensional light front QCD which reduces to the 't Hooft model in the large
$N_c$ limit. In this case quark and antiquark at the same transverse
position interact via the spin independent instantaneous interaction 
which, in the non-relativistic limit reduces to the linear potential in the 
longitudinal direction. The only parameters in the theory are the dimensionless fermion
mass $m_f = am$ and the gauge coupling $g$.   The spectrum consists of
a ground state and a tower of excited states corresponding to the
excitations of the $q {\bar q}$ pair.

Next consider the inclusion of  the $q {\bar q}$ link states. There are four
independent amplitudes corresponding to whether the quark is on the left or
right of the antiquark or, above or below the antiquark. 
With non-zero mass of the link, these states lie above the ground
state of pure quark - antiquark system. Further the  
$q$,  ${\bar q}$  and link (which are 
frozen at their transverse positions)
undergo fermion - link instantaneous interactions in the longitudinal
direction which further increases the mass of $q {\bar q}$ link states.
Now the quark or antiquark can hop via helicity flip or helicity non-flip.
Here we find a major difference between the Hamiltonians resulting from 
forward-backward derivative  and symmetric derivative. Let us first consider 
the helicity flip hopping term in the forward-backward case 
\be
 P^-_{hf} & = &  - img \int dx^- \sum_{\bx} \sum_r \Big [ \eta^\dagger(\bx)  {\hat \sigma}_r
{1 \over a}{1 \over \partial^+}\eta(\bx + a \br) + \eta^\dagger(\bx + a \br) 
 ~M_r^\dagger(\bx)
 {\hat \sigma}_r{ 1 \over a} { 1 \over
\partial^+} \eta(\bx) \Big ]. \nonumber \\
~\label{fbhf}
\ee
If we consider transition from two particle to three particle state by a quark hop, then the 
first term in Eq. (\ref{fbhf}) corresponds to $\mid 2\rangle \rightarrow \mid 3 a\rangle$ and 
the second term corresponds to  $\mid 2\rangle \rightarrow \mid 3 b\rangle$. 
The helicity flip term in symmetric derivative case,
 after making some shifts in 
lattice points, can be written as 
\be
P^-_{hf} & = &  -g \left ( m+ 4 \frac{\kappa}{a} \right )\frac{1}{2a}
\int dx^- \sum_{\bx}\sum_r 
\nonumber \\
&~& \Bigg [ \Bigg \{
\eta^\dagger(\bx) {\hat \sigma}_r { 1 \over i \partial^+}  
 M_{r}(\bx) \eta(\bx + a \br) 
-\eta^\dagger(\bx ) {\hat \sigma}_rM_{r}(\bx){ 1 \over i \partial^+} \eta(\bx +a\br) 
\Bigg \} \nonumber \\
&~& -\Bigg \{ \eta^\dagger(\bx) {\hat \sigma}_r { 1 \over i \partial^+}
 M_{-r}(\bx) \eta(\bx - a \br) 
 -\eta^\dagger(\bx) {\hat \sigma}_r M_{-r}(\bx) 
{ 1 \over i \partial^+} \eta(\bx-a\br) \Bigg \} \Bigg ].  \label{symhf}
\ee

 For the Hamiltonian with symmetric derivative, a
quark or antiquark hopping accompanied by helicity flip has opposite signs
for forward and backward hops. On the other hand, hopping accompanied by
helicity non-flip have the same signs. As a result, there is no
interference between helicity flip and helicity non-flip interactions \cite{4BK}. 
In the case of the Hamiltonian with forward-backward derivative,  quark or
antiquark hopping accompanied by helicity flip has the same sign for forward
and backward hops.
 As a consequence  the helicity non-flip hop can 
interfere with the helicity flip hop.
 This has immediate 
consequences for the
spectrum. In the case with symmetric derivative, in lowest order
perturbation theory, the helicity zero states mix with 
each other which causes a
splitting in their eigenvalues resulting in the singlet state lower than the
triplet state. On the other hand, helicity plus or minus one states do not mix with each
other or with helicity zero states resulting in a two fold degeneracy. 
In the case with forward and backward
derivatives all helicity states mix with each other resulting in the complete
absence of degeneracy. Obviously, one has the freedom to tune the free parameters 
to minimize the splitting. 

\subsection{Singularities, divergence and counterterms}

Since the transverse lattice serves as an  ultraviolet regulator, we need to
worry about only light front longitudinal momentum singularities.    
\subsubsection{Tree level}

We take all the terms in the Hamiltonian to be normal 
ordered. At tree level this leaves us with singular factors of the form $\frac{1}{(k)^2}$
in the normal ordered four fermion and fermion link instantaneous 
interactions. The singularities are removed by adding
the counterterms used in the
previous work \cite{4dalme} on transverse lattice.
The explicit forms of the
counterterms are given in Appendix \ref{appc} in the appropriate places.   
\subsubsection{Self energy corrections}
In the one link approximation, a quark  can make a  forward (backward) hop
followed by a backward (forward) hop resulting in self energy corrections.
In a single hop, helicity flip or non-flip can occur. In the case of
symmetric derivatives, helicity flip cannot interfere 
with helicity non-flip,
and as a consequence, self energy corrections are diagonal in helicity space.
In the case of forward and backward derivatives, the
 interference is nonzero
resulting in self energy corrections, both diagonal and 
off-diagonal in the
helicity space. Similar self energy corrections are generated for an
antiquark
also. These self energy corrections contain a 
logarithmic light
front infrared divergence which must be removed by  counterterms. In 
Appendix \ref{APcounter} we present the explicit form of counterterms in 
the two cases
separately.  In previous works on one link approximation \cite{4BK,4dalme,4buseal},
these counterterms were not implemented. For low $K$ values one may not feel the
divergence, but as one increases $K$ the need of self energy counterterms are 
readily felt. 

\section{Numerical Results}
 We diagonalize the dimensionless matrix $a^2 P^-$. We further
divide the matrix elements by $g^2 C_f$ which is the 
strength of the matrix elements
for four fermion and fermion - link instantaneous interactions.
Now,  define the
constant $G$ with dimension of mass  by  $ G^2 =\frac{g^2}{a^2} C_f$.
 DLCQ yields the eigenvalue ${\cal M}^2= {M^2 \over G^2}$.

 The dimensionless couplings are introduced \cite{4dalme} as
follows. Fermion mass $ m_f=  m/G $, link  mass $ \mu_b =  \mu/G $, 
particle number conserving helicity flip coupling 
$ m_f/(aG)=m_f C_1 $, particle number non-conserving helicity flip 
$ \sqrt{N} g m_f/(aG) = m_f C_2 $, and particle number non-conserving helicity non-flip
$\sqrt{N}g /(a^2 G^2) =C_3 $. In the case of the
Wilson term modified Hamiltonian,
we have fermion mass term $ m_f = (m+ 4 \kappa/a)/G $, helicity-flip
coupling $  \sqrt{N}g m_f /(2aG) = m_f{\tilde C}_2$, and
helicity non-flip
coupling $ \sqrt{N}g m_f \kappa /(aG) =m_f {\tilde C}_3 $.  
\begin{figure}[h]
\centering
\fbox{\includegraphics[height=10cm]{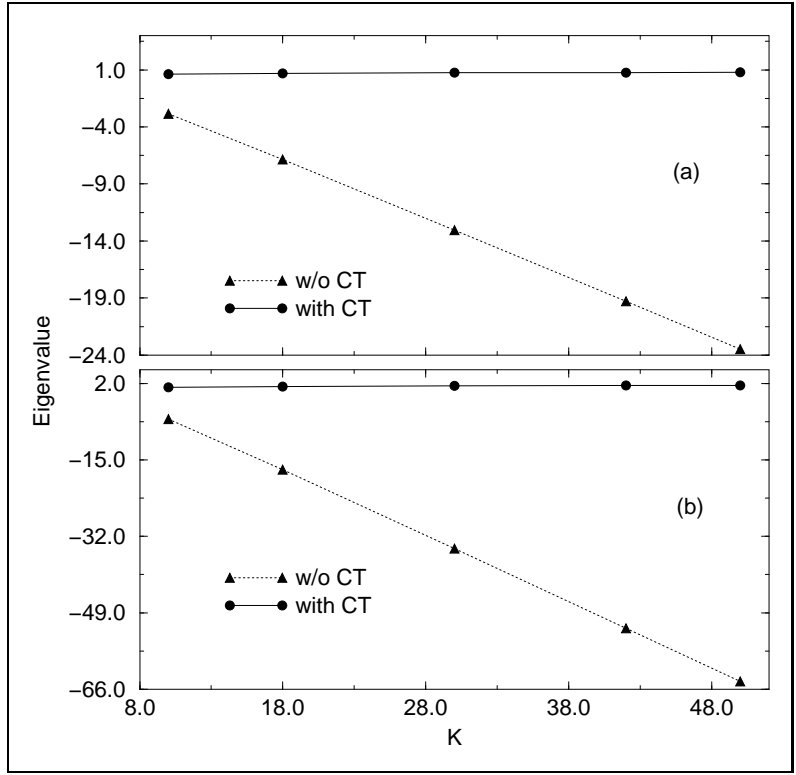}}
\caption{ Effect of counterterm on the ground state eigenvalue.   
(a) With and without the counterterm in the $ q {\bar q}$ sector for $m_f=0.3$. 
(b) With and without the counterterm in the $ q {\bar q}$ link sector for $m_f=0.3$ 
 and $\mu_b=0.2$.}
\label{CT}
\end{figure}

All the results presented here were obtained on a small cluster of computers using the
 Many Fermion Dynamics (MFD) code \cite{4mfd} implementing the Lanczos diagonalization 
method in parallel 
environment. For low K values, the  results were checked against an independent code 
running on a single processor.

\subsubsection{Cancellation of divergences}
As we already mentioned, we encounter $\frac{1}{(k^+)^2}$ singularities 
with instantaneous four fermion and instantaneous fermion - link interactions
which give rise to linear divergences.
We remove the divergences by adding appropriately chosen counterterms.
We have numerically checked the removal of linear divergence by counterterms 
in DLCQ.
First we consider only $q {\bar q}$ states with instantaneous interaction. 
We study the ground state eigenvalue as a function of $K$
with and without the counterterm. Results are presented in 
Fig. \ref{CT} (a). Next we consider only 
$q {\bar q}$ link states with fermion-link
instantaneous interaction with and without the counterterms. The behavior of 
ground state eigenvalue as a
function of $K$ is presented in Fig. \ref{CT} (b). In both cases, it is
evident that the counterterms are efficient in removing the divergence. 

\subsection{$q {\bar q}$ at the same transverse location}
Next  we study the spectrum of the Hamiltonian in the absence of any links.
Since, in this case, the Hamiltonian depends
only on the dimensionless ratio $\frac{m_f}{g}$ we fix $g=1$ and vary $m_f$
to study the spectra. 
The Hamiltonian matrix is diagonalized for various
values of $K$. 
The convergence of the ground state eigenvalue as a
function of $K$ is presented
in Table \ref{table1qq}.
 \begin{table}[h]	
\begin{center}
\begin{tabular}{||c|c|c|c||}
\hline \hline
  K & \multicolumn{3}{c||}  {Eigenvalue (${\cal M}^2$)} \\
\cline{2-4}
  & $m_f=0.3$ & $m_f = 0.9$ & $m_f = 3.0$\\
\hline
   10 & 0.620 & 4.547 & 39.233\\
  18 &   0.693 & 4.664 & 39.861\\
  30  &   0.745 & 4.724 & 40.053 \\
  50 &   0.788  & 4.762 &  40.163\\
  78 &   0.819 &  4.783  &  40.220\\
  98 &   0.832 & 4.791 &  40.241\\
$K\rightarrow\infty$ & 0.869 & 4.820 & 40.285 \\
\hline \hline
\end{tabular}
\end{center}
\caption{Ground state eigenvalue (in units of $G^2$) for $q {\bar q}$ sitting 
at the same transverse location.
\label{table1qq}}
\end{table}

\begin{figure}[h]	
\centering
\fbox{\includegraphics[height=10cm]{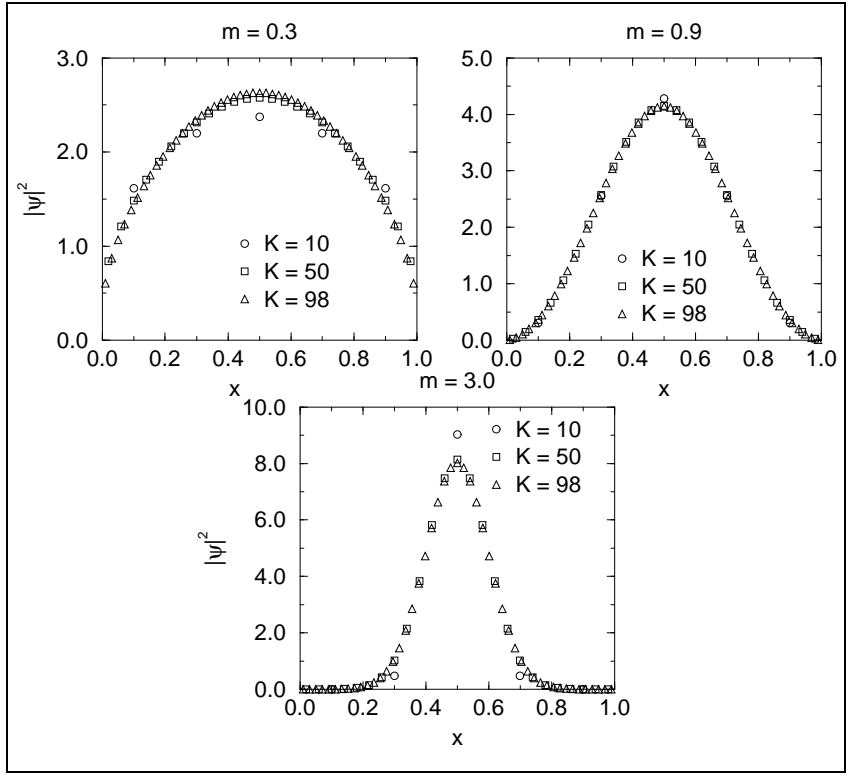}}
\caption{Quark distribution function $ \mid \psi(x) \mid^2$ of
the ground state  in the $q {\bar q}$
approximation for three choices of quark masses with coupling constant $g=1.0$.}
\label{qqbwave}
\end{figure}

The ground state wavefunction squared as a function of the
longitudinal momentum fraction $x$ is plotted in Fig.  \ref{qqbwave}. 
The convergence of
the wavefunction has a very different behavior as a function of fermion 
mass $m_f$. As can be seen from
this figure, the convergence in $K$ is from above for heavy $m_f$   and from
below for light $m_f$. As a consequence the wavefunction  is almost 
independent of $K$
when $m_f$ is of order $g$.
     
\subsection{Results of the one link approximation}
We encountered logarithmic infrared divergences due to self energy
corrections and, in Appendix \ref{APcounter}, we discuss
 the associated
counterterms. In Fig. \ref{fullself} we show the effect of self 
energy counterterms 
on the ground state energy in the two Hamiltonian cases we studied.   
\begin{figure}[h]
\centering
\fbox{\includegraphics[height=10cm]{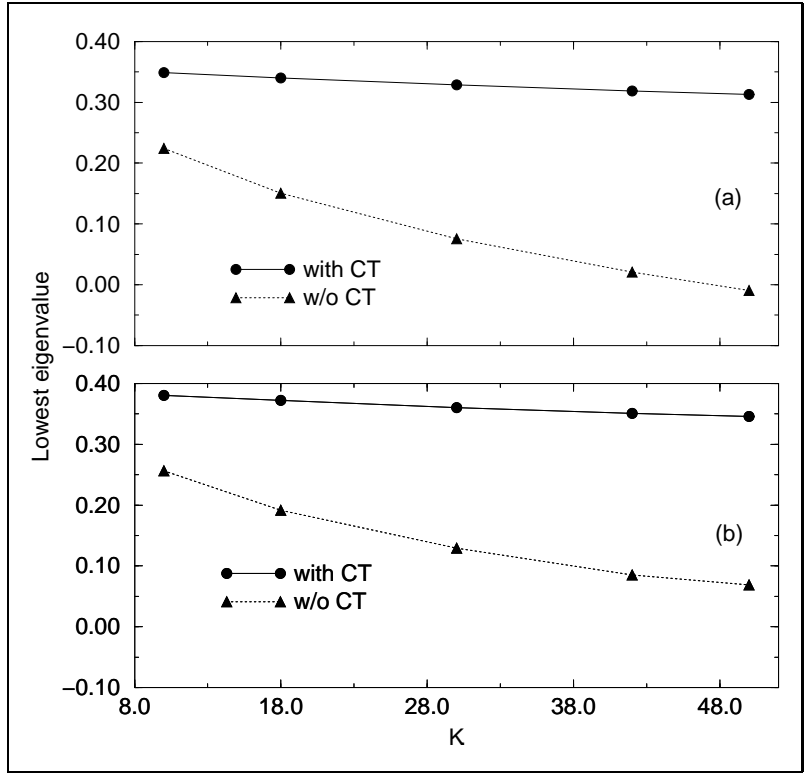}}
\caption{Effect of self energy counterterms on the ground state eigenvalue
in the case of (a) symmetric derivative  with ${\tilde C}_2=0.4,~ {\tilde C}_3 = 0.1$
and (b) forward-backward derivative with $C_2=0.4,~  C_3 = 0.01$.
 $m_f =0.3, ~\mu_b=0.2$ for both cases.}
\label{fullself}	
\end{figure}

 The convergence of lowest four 
eigenvalues with $K$ for the Hamiltonian with forward-backward and 
symmetric lattice derivatives is shown in Table \ref{table2qqg} for $m_f = 0.3$,
 $\mu_b = 0.2$.  We also show the results extrapolated to $K\rightarrow\infty$.
The convergence of the eigenvalues in $K$ is very slow and one really needs to
go for large $K$.
\begin{table}[h]
\begin{center}
\begin{tabular}{||c|c|c|c|c||c|c|c|c||}
\hline \hline
  &\multicolumn{4}{c||} {Forward-backward } &
\multicolumn{4}{c||} {Symmetric }\\
 & \multicolumn{4}{c||} {($ C_2=0.01,~ C_3=0.4$)} &
\multicolumn{4}{c||} { (${\tilde C}_2=0.1,~{\tilde C}_3=0.4$)}\\
\cline{2-9}
 K & ${\cal M}_1^2$ & ${\cal M}_2^2$ &  ${\cal M}_3^2$ & ${\cal M}_4^2$
& ${\cal M}_1^2$ & ${\cal M}_2^2$ &  ${\cal M}_3^2$ & ${\cal M}_4^2$\\
\hline
10 & 0.38041 & 0.4800 & 0.4899 & 0.5996 & 0.3486 & 0.4507 & 0.4507 & 0.5980 \\
 18 & 0.3722 & 0.4968 &  0.5110&  0.6447 & 0.3402 & 0.4673 & 0.4673 & 0.6409  \\
 30 & 0.3606 & 0.5027 & 0.5210 &  0.6680 & 0.3288 & 0.4702 & 0.4702 & 0.6620 \\
42  & 0.3511  & 0.5029 & 0.5240 & 0.6765 & 0.3189 & 0.4677 & 0.4677 & 0.6682 \\
50  &  0.3457 & 0.5019 & 0.5246 & 0.6790 & 0.3130  & 0.4651 & 0.4651 & 0.6693\\
$K\rightarrow \infty$ & 0.3243 & 0.5022 & 0.5313 & 0.6979 & 0.2913 & 0.4589 & 0.4589 & 0.6837 \\
\hline \hline
\end{tabular}
\end{center}
\caption{Lowest four eigenvalues (in units of $G^2$) in one link approximation.\label{table2qqg}}
\end{table}

The quark distribution function for the ground state and the fifth state
for the set of parameters  $m_f = 0.3$, $\mu_b = 0.2,~  C_2 = 0.4,
~ C_3 = 0.01$ and $K = 30$ is presented in Fig. \ref{qfull}. 
\begin{figure}[h]
\centering
\fbox{\includegraphics[height=12cm]{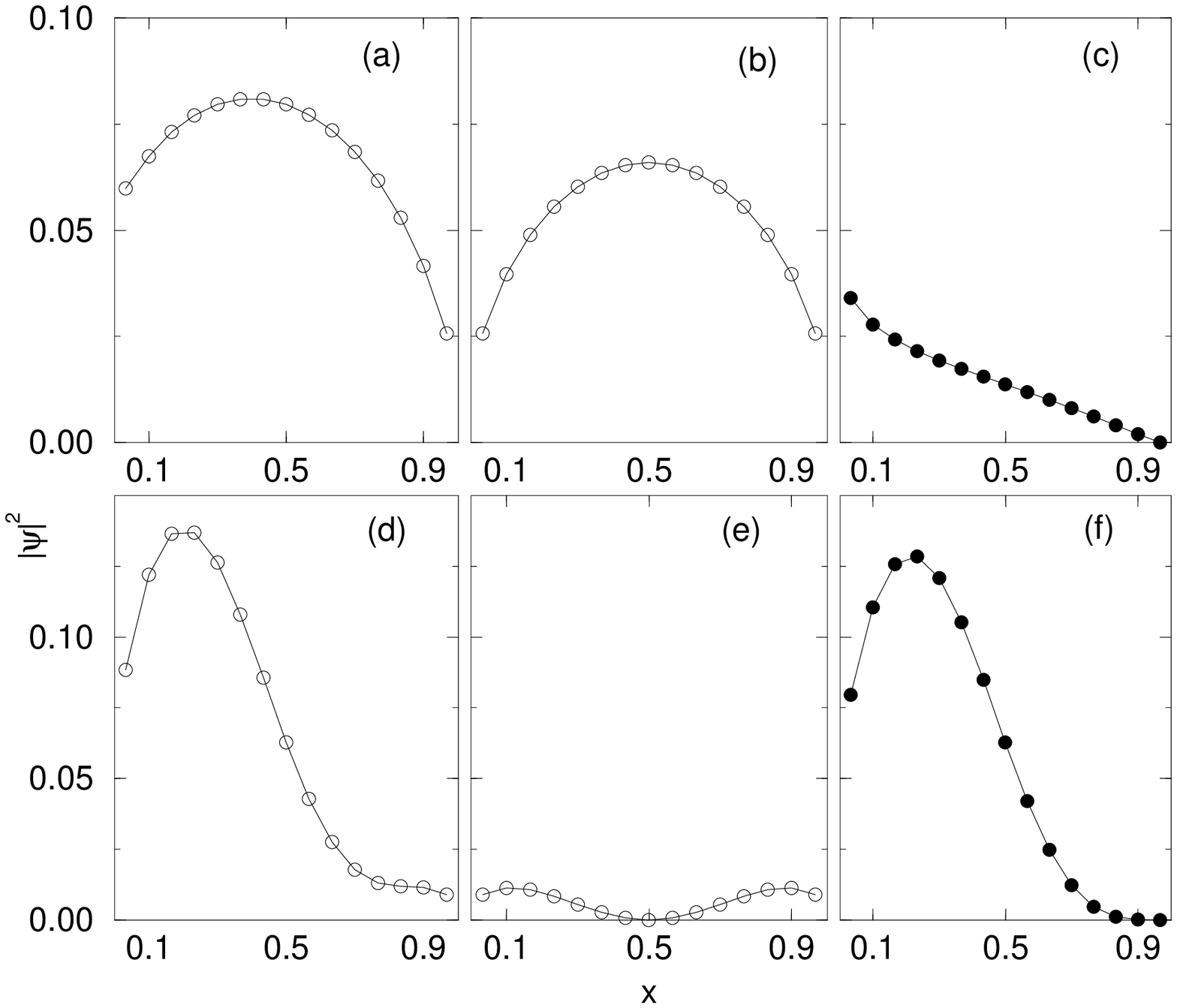}}
\caption{ (a) Quark distribution function $ \mid \psi(x) \mid^2$ of the
ground state in the one link
approximation, (b) $q {\bar q}$ contribution to the ground state,
(c) $q {\bar q}$ link contribution to the ground state. (d)  Quark
distribution function $ \mid \psi(x) \mid^2$ of the
fifth eigenstate in the one link
approximation, (e) $q {\bar q}$ contribution to the fifth eigenstate,
(f) $q {\bar q}$ link contribution to the fifth eigenstate. The parameters are
$m_f =0.3, ~\mu_b=0.2, ~ C_2=0.4,~ C_3 = 0.01$ and $K = 30$.}
\label{qfull}
\end{figure}
In this figure we also present separately the contribution from two particle
and three particle states. As expected, the contribution from 
the three particle
state peaks at smaller $x$ compared to the two particle state. The exact
location of this peak depends on the link mass. The ground state is dominated by
two particle sector while the fifth state is dominated by three particle sector.
We have shown only one of the first four states since they are similar 
looking states
with different spin contents. Fig. \ref{qfull} warns us  that
one link approximation is inadequate to study the excited states.  

\begin{figure}[h]
\centering
\fbox{\includegraphics[height=10cm]{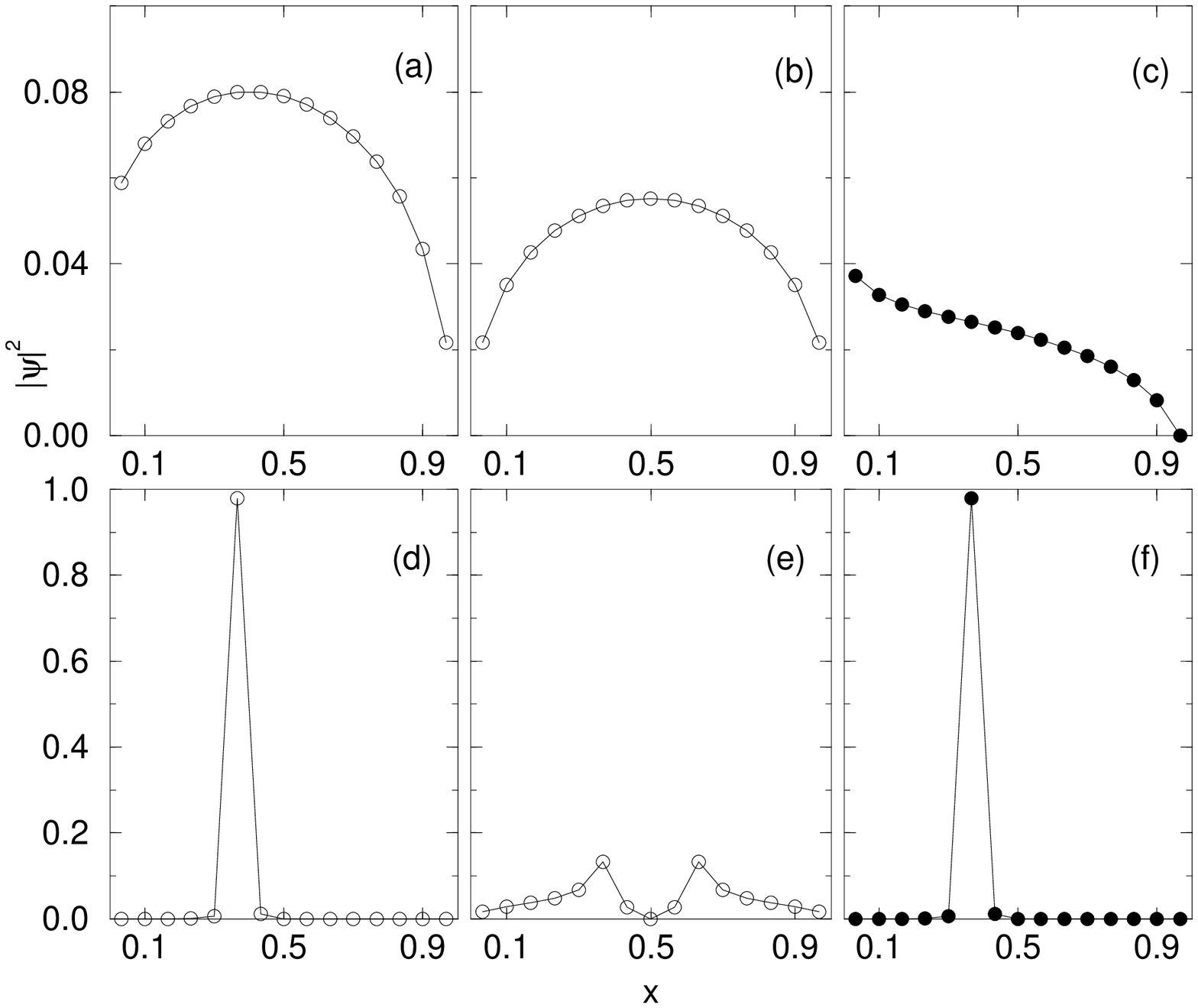}}
\caption{ Without the fermion - link instantaneous interaction:
(a) Quark distribution function $ \mid \psi(x) \mid^2$ of the
ground state in the one link
approximation, (b) $q {\bar q}$ contribution to the ground state,
(c) $q {\bar q}$ link contribution to the ground state. (d)  Quark
distribution function $ \mid \psi(x) \mid^2$ of the
fifth eigenstate in the one link
approximation, (e) $q {\bar q}$ contribution to the fifth eigenstate
multiplied by $10^{4}$,
(f) $q {\bar q}$ link contribution to the fifth eigenstate. Parameters are the 
same as in Fig. \ref{qfull}. }
\label{q3pfree}
\end{figure}

It is interesting to see the effect of fermion - link instantaneous
interaction on the low lying eigenvalues. In its absence, there is no
confining interaction in the longitudinal direction in the $q {\bar q}$
link sector. Furthermore, the mass of the lowest state  in this sector
corresponds to the threshold mass in this sector. Since its mass is lowered,
it mixes more strongly with the $q {\bar q}$ sector in the ground state and
from Fig. \ref{q3pfree} we see that the ground state gets comparable 
contribution from both sectors. The
fifth state now corresponds to an almost free $q {\bar q}$ link state with
infinitesimal $q {\bar q}$ component as shown in Fig. \ref{q3pfree}.

\section{Summary and Discussion}
 We have performed an  investigation  of $q {\bar q}$
states using two different light front Hamiltonians in the one link
approximation. The Hamiltonians correspond to two different ways of
formulating fermions on the transverse lattice, namely, (a) 
forward and backward
derivatives for $\psi^{+}$ and $ \psi^{-}$ respectively or vice versa and (b)
symmetric derivatives for both $\psi^{+}$ and $ \psi^{-}$. In the latter,
fermion doubling is present which is removed by an addition of the Wilson
term. In this case there is no interference between helicity flip hop and
helicity non-flip hop and, as a result, the $q {\bar q}$ component of the
ground state wavefunction which has helicity plus or minus one are
degenerate. In the former case, interference between helicity flip and
helicity non-flip leads to the absence of degeneracy in the low lying spectra. 
One can recover approximate degeneracy of helicity plus or minus one
components only by keeping the strength of the helicity non-flip hopping
very small.

Since the one link approximation is very crude and our motivation was to study 
and compare different fermions on the transverse lattice other than the assessment of the
transverse lattice approach itself,
we have not attempted a detailed fit to low lying  states in the meson
sector. Instead, we have explored the effects of various coupling strengths
on the low lying spectra and associated wavefunctions. In this work,
longitudinal dynamics is handled by DLCQ. We have performed a detailed study
of various convergence issues in DLCQ  using a wide range of $K$ values.

We summarize our results as follows. We have shown the effectiveness of
appropriate counterterms in the $q {\bar q}$ and $ q {\bar q}$ link sector
to regulate the instantaneous fermion and fermion - link interactions
respectively. We have also checked the cancellation of logarithmic
divergences due to self energy effects. In the limit where fermions are 
frozen on the transverse
lattice but undergo instantaneous longitudinal interaction, we have studied
the convergence of ground state wavefunction with respect to $K$ for three
typical values of the fermion mass. We have studied how the presence or
absence of fermion - link instantaneous interaction in the $q {\bar q}$ link
sector affects the wavefunction of low lying states. We have also studied  
the consequences of the interference of helicity flip and helicity non-flip
hopping in the Hamiltonian with forward-backward derivatives.
This interference is  absent
in the symmetric derivative case.





\chapter{Summary, Conclusions and Future Outlook}
Hadronic bound state problem is
one of the most challenging tasks in nonperturbative QCD. 
 Since it is very important to have direct access to the bound state wavefunctions
which is essential to calculate QCD observables, Hamiltonian approach is the most 
suitable candidate to address the problem of  bound states.
The simple structure of the vacuum in light-front QCD makes it possible to carry out 
a Hamiltonian analysis of the bound states in the Fock space language. 
In this thesis, we have investigated two nonpeturbative techniques in light-front 
QCD, namely, similarity renormalization group (SRG) approach and light-front transverse
lattice (LFTL) approach in the context of meson bound state problem.
Unless and until we have complete control over the intricacies of the
techniques in use, the dream to explore the nonperturbative QCD with full
confidence will remain unrealized.  In this work, we have performed a critical evaluation
of the two nonperturbative approaches mentioned above and  assessed 
them in terms of their strengths and weaknesses. 

 We have first investigated Bloch 
effective Hamiltonian in the context of meson bound states in (2+1) dimensional QCD.
There we encounter infrared divergences due to vanishing energy denominators 
in the bound state equation and SRG becomes 
mandatory to get rid of the divergences.  The investigation of the $q{\bar q}$ bound 
states with  Bloch effective Hamiltonian serves as a benchmark  for comparative study 
of the same problem with SRG generated effective Hamiltonian which is a modification over
the  Bloch Hamiltonian.  
Bound state equation in SRG scheme is free from the problem of uncanceled divergences
coming from vanishing energy denominators. We have also compared three different 
choices for similarity factor. To have better understanding, we have 
performed analytic calculation in the lowest order with step function similarity 
factor which in (2+1) dimensions  generates 
{\it linear confinement along the transverse direction } (for large $x^1$, $V(x^-,x^1) 
\sim x^1$) while only {\it square root confinement along the longitudinal direction} 
(for large $x^-$, $V(x^-,x^1)  \sim \sqrt{x^-}$) and thus breaks the rotational
 symmetry. Here one should recall that in (3+1) dimensions, in the lowest order, 
SRG generates logarithmic confining potential which also violates rotational symmetry. 
If the confinement generated by similarity transformation of the Hamiltonian
is not an artifact of the lowest order approximation, one might hope that the violation
of the rotational symmetry will diminish with higher order corrections to the 
effective Hamiltonian.  Higher order calculations are thus very important and 
illuminative in this context. It is also expected that higher order calculations in (2+1)
dimensions will be much easier than in (3+1) dimensions  due to simpler divergence 
structures and less demand of computing resources. 

We have studied another nonperturbative approach, the light-front transverse lattice 
formulation in this thesis. It is still a developing subject and is  a very potential tool
for nonperturbative investigations.  Only very recently efforts are being
 given  to formulate fermions  and to study meson bound states on the LFTL.  
 Lattice formulation of fermions is complicated due to generation of extra species
and needs special treatments to have a meaningful description of fermions in the discrete
world. Thus it is very much important to know their origin and way(s) to overcome them 
before attempting any realistic QCD calculation on a lattice. 
We have shown  that the origin of doublers on LFTL is completely different
from usual lattice gauge theory. In the usual lattice gauge theory doublers come
from the end of the Brillouin zone. But,  when one 
uses symmetric lattice derivative to formulate fermions on a LFTL, doublers  prop
up due to decoupling of odd and even lattice points. We have also studied two
different ways of removing doublers, namely, Wilson fermion and staggered fermion
on LFTL. We have {\it proposed} another way of  formulating
fermions on LFTL  by using forward and backward lattice derivatives in such a way
that the Hermiticity of the Hamiltonian is preserved. In our method,
{\it there is no generation of  extra fermion species}. In this case, the helicity flip
term proportional to fermion mass in full QCD becomes an irrelevant term in the free 
field  limit.

To assess which one of these two fermion formulations is better than the other
we have  compared them  in the context of  the meson 
bound state problem in (3+1) dimensional QCD.
In the zero link approximation i.e., when $q$ and ${\bar q}$ can sit only at the same 
transverse location, in the limit of large number of colors, it reduces to the 't Hooft 
model with linearly confining 
instantaneous interaction along the longitudinal direction.
In the one link approximation, $q$ and ${\bar q}$ can be separated at most by one lattice
point.  With this approximation,
the major difference between these two methods (in the case of symmetric
derivative we add a Wilson term to remove the doublers) is the interference
between helicity flip and nonflip hopping in the case of Hamiltonian with
forward and backward derivatives. As a  consequence of this interference 
the degeneracy structure of the bound state spectrum of the Hamiltonian with forward 
and backward derivatives is different from that with symmetric lattice derivative.

{\it Since this subject is still under development, there are many  unresolved questions 
one need to answer}.  
The transverse dynamics and the structure of the mesons are too constrained by the one 
link approximation  and  investigations are essential with more than one links.
 By construction, quarks in one link 
approximation are confined in the transverse directions. 
The true nature of confining potential in the transverse
directions can only be realized when sufficient number of links will be included.
These issues can be investigated as an extension of our work. A systemic 
study of the effects of sea quarks also need to be undertaken.  In this
thesis, all the studies are 
done with light quarks, it will also be  highly 
interesting to address the  problem of mesons containing one light and one heavy quark 
in the context of heavy quark effective theory on the transverse lattice.
 
A major unsettled issue in the transverse lattice formulation is the
continuum limit of the theory when nonlinear link variables are replaced by linear
link variables. The lack of well defined transformation rule between linear and nonlinear link
variables  makes it impossible to come back to nonlinear theory once the links are replaced
by linear link variables and  there is no straightforward way to take
the continuum ($a\rightarrow 0$) limit. Thus, as an alternative on a coarse lattice, one  
needs to search for a
trajectory in the parameter space with minimal violation of Lorentz
invariance.  Another disadvantage of linearization of the links is that
one needs to include more and  more terms in the effective potential
when  more and more links are include and calculations are viable only in the limit of 
large number of colors which suppress the higher order terms in the effective potential.
 It will be interesting to investigate the light-front
quantization problem with non-linear constraints. 
The kinetic energy term of the nonlinear link variables is similar to the
nonlinear $\sigma$ model. In this respect,
the study of nonlinear $\sigma$ model on the light-front appears worthwhile.

%
%
\appendix
\chapter{Notations and Conventions}\label{APnotcon}
In this appendix we provide the notations and conventions used in different
chapters of this thesis.

For completeness let us start from the definition of light-front
coordinates.  The light-front coordinates in (3+1) dimensions are defined by
\be
x^\pm =x^0 \pm x^3,~~~~~~x^\perp = \{ x^1, x^2 \}
\ee
and any four vector in light-front is denoted as
\be
V^{\mu} = (V^{+},V^{-},V^{\perp}) . 
\ee
The metric tensors are as follows
\be g^{\mu \nu} = \left(\begin{array}{lrrr}0 & 2 & 0 & 0 \\
                          2 &\,\, 0 & 0 & 0 \\
                          0 & 0 & -1 & 0 \\
                          0 & 0 & 0 & -1 \end{array}\right) \, , 
\,\,\,\quad\quad
g_{\mu \nu} = \left(\begin{array}{lrrr}0 & {1 \over 2} & 0 & 0 \\
                          {1 \over 2} & \,\,0 & 0 & 0 \\
                          0 & 0 & -1 & 0 \\
                          0 & 0 & 0 & -1 \end{array}\right) \, , 
\ee
so that
\be
 x_{-}= {1 \over 2} x^{+}  , \; \; x_{+} = {1 \over 2} x^{-} 
\ee
and 
the scalar product 
\be
 x\cdot y = {1 \over 2} x^{+}y^{-}+{1 \over 2}x^{-}y^{+} - x^{\perp}\cdot y^{\perp}  .
\ee
The light-front partial derivatives are
\be \partial^{+}= 2 \partial_{-}= 2 {\partial \over \partial x^{-}}  .\ee
\be \partial^{-}= 2 \partial_{+}= 2 {\partial \over \partial x^{+}}  . \ee
We define the integral operators
\be {1 \over \partial^{+}} f(x^{-})= {1 \over 4}\, \int \, dy^{-}
\epsilon(x^{-}-y^{-}) \, f(y^{-})  , 
\ee
\be  
\Big ( {1 \over \partial^{+}} \Big)^{2} f(x^{-}) =  { 1 \over 8} \, \int \,
dy^{-} \mid x^{-} - y^{-} \mid \, f(y^{-})  . 
\ee
The $\gamma$ matrices are define as
\be
\gamma^\pm =\gamma^0 \pm \gamma^3 .
\ee
We choose the representation for gamma matrices such that
\be \gamma^{+} =  \left(\begin{array}{cr} 0 &  0\\
                                    2 i I & 0 \end{array}\right) ,\,\,\,
\quad\quad{\gamma^-} = \left(\begin{array}{cr} 0 & -2 i I  \\
                               0 &  0  \end{array}\right), \nonumber
\ee
\be
\gamma^r = \left(\begin{array}{cr} -i {\hat\sigma}^r & 0 \\
                               0 &  i{\hat\sigma}^r  \end{array}\right) ,\,\,\,
\quad\quad{\gamma^5} = \left(\begin{array}{cr} \sigma^3 &  0\\
                                    0 & -\sigma^3 \end{array}\right) ,
\ee
 ${\hat\sigma}^1 =\sigma^2$ and ${\hat\sigma}^2 =-\sigma^1$ where $\sigma^r$
are Pauli matrices
and $I$ is a two component identity matrix.
 The projection operators
\be
  \Lambda^{\pm}  \; =  \; {1 \over 4} \gamma^{\mp} \gamma^{\pm}
 = {1 \over 2} \gamma^{0} \gamma^{\pm} \  
\ee
 in explicit form are
\be
\Lambda^+ = \left(\begin{array}{cr} I &  0\\
                                    0 & 0 \end{array}\right) ,\,\,\,
\quad\quad{\Lambda^-}  = \left(\begin{array}{cr} 0 &  0\\
                                    0  & I \end{array}\right).
\ee
So, the fermionic fields
\be
\psi^\pm =\Lambda^\pm \psi
\ee
can be written as
\be
\psi^+ =\pmatrix{\eta \cr 0}~~~~~\psi^- =\pmatrix{0 \cr \xi}
\ee
where $\eta$ and $\xi$ are two component fields.

{\it In (2+1) dimensions} the light-front coordinates are defined as
\be
x^\mu = (x^\pm =x^0 \pm x^2, ~x^1)
\ee
and accordingly the $\gamma$ matrices
\be
\gamma^\pm =\gamma^0 \pm \gamma^2 .
\ee
We use the two component representation for $\gamma$ matrices
\be
\gamma^0 =\sigma_2 = \pmatrix{ 0 & -i \cr
                               i & 0 }, ~~\gamma^1 = i \sigma_3 =
\pmatrix{i & 0 \cr
         0 & -i},~~  \gamma^2 = i \sigma_1 = \pmatrix{0 & i \cr
                                                    i & 0 }
\ee
so that
\be
\gamma^{\pm} = \gamma^0 \pm \gamma^2,~~ \gamma^+=\pmatrix{ 0 & 0 \cr
                                                 	  2i & 0}, ~~
\gamma^- = \pmatrix{0 & -2i \cr
                    0 & 0},
\ee
and
\be
\Lambda^\pm = { 1 \over 4} \gamma^\mp \gamma^\pm, ~~ \Lambda^+ = \pmatrix{1
& 0 \cr
0 & 0}, ~~ \Lambda^- = \pmatrix{0 & 0 \cr
                                0 & 1}.
\ee
Fermion field operator $ \psi^\pm = \Lambda^\pm \psi$.  We have
\be
\psi^+ = \pmatrix{ \xi \cr
                    0}, ~~ \psi^- = \pmatrix{0 \cr
                                             \eta}
\ee
where $\xi$ and $\eta$ are single component fields.

\chapter{Bloch Perturbation Theory for Effective Hamiltonian}\label{APbloch}
Here, we  present the detailed formalism of the Bloch perturbation 
theory\footnote{ C. Bloch, Nucl.\ Phys.\ {\bf 6}, 329 (1958).
Here we follow the treatment of R. J. Perry,
Ann.\ Phys.\ {\bf 232} (1994) 116
[hep-th/9402015] (reader can find here many examples of perturbative
 calculations); B. D. Jones and R. J. Perry, 
Phys. Rev. D {\bf 55}, 7715 (1997).}
 to calculate the effective Hamiltonian.

Consider a Hamiltonian $H$ defined at a cutoff $ \Lambda$.
Let us try to lower the cutoff to $ \lambda$. In general, the 
cutoff could be in energy and/or particle number.
Let us denote by
$Q$ the operator that projects on to all of the states removed when the
cutoff is lowered. Let $ P = I - Q$.  We have
\be
Q^2 =Q,~~ P^2=P, ~~ PQ=QP=0.
\ee
Our purpose is to find an effective Hamiltonian $ H_{eff}$ that produces the
same eigenvalues in the sub space $P$ as the original Hamiltonian $H$.

Introduce an
operator $R$ that satisfies
\be
Q \mid \psi \rangle = RP \mid \psi \rangle
\ee
for all eigenstates of the Hamiltonian that have support in the subspace
$P$. $R$ gives the part of $\mid \psi \rangle$ outside the space projected
by
$P$ in terms of the part of $ \mid \psi \rangle$  inside the space. Require
that $R$ gives zero acting on states outside the subspace. This means
$R=RP$, $R=QR$, $R^2=0$. From $R=QR$, we have, $PR=0$.  Note also that
$R^\dagger \neq R$.  

Start from the set of equations (projections\footnote{In Bloch-Horowitz formalism [C. Bloch,  J. Horowitz,  Nucl.\ Phys.\
 {\bf 8}, 91 (1958)] also, one has the same projection operators $P$ and $Q$, but
not $R$.  Substitution of $Q \mid \psi \rangle$ from Eq. (\ref{bqp}) into 
Eq. (\ref{bpp}) gives the Bloch-Horowitz effective Hamiltonian.} 
of Schr\"odinger equation  by $P$ and $Q$)
\be
PHP \mid \psi \rangle + PHQ \mid \psi \rangle &=& EP \mid \psi \rangle,
\label{bpp} \\
QHP \mid \psi \rangle + QHQ \mid \psi \rangle &=& EQ \mid \psi \rangle.
\label{bqp}
\ee

From Eq. (\ref{bpp}), 
\be
RPHP \mid \psi \rangle + RPHQRP \mid \psi \rangle = ERP \mid \psi \rangle.
\ee
From Eq. (\ref{bqp}),
\be
QHP \mid \psi \rangle + QHQRP \mid \psi \rangle = ERP \mid \psi \rangle.
\ee
Subtracting,
\be
RH_{PP} - H_{QQ}R + R H_{PQ}R - H_{QP} =0.
\ee
We have introduced the notations, $PHP = H_{PP}$ and so on. 
Put $ H = h + v$ with $[h,Q]=0$. Then
\be
Rh_{PP} - h_{QQ}R - v_{QP} + R v_{PP} - v_{QQ}R + R v_{PQ}R=0
\ee
which shows that $R$ starts first order in $v$.

We start from the eigenvalue equation,
\be
H(P+Q) \mid \psi \rangle = E (P+Q) \mid \psi \rangle.
\ee
i.e., 
\be
H(P+R) P \mid \psi \rangle = E (P+R) P \mid \psi \rangle.
\ee
Multiplying from the left by $(P+R^\dagger)$ we have,
\be
(P+R^\dagger) H(P+R) P \mid \psi \rangle = E (P+R^\dagger)(P+R)P \mid \psi
\rangle.
\ee
Using $ PR=0$, $ R^\dagger P=0$, $
(P+R^\dagger) (P+R) = P + R^\dagger R.$
Thus we can rewrite the eigenvalue equation as
\be
&&\!\!\!\!\!\!\!\!\!\!\!\!\!\!\!\!\!\!\!\!
\Big [ { 1 \over 1+ R^\dagger R} \Big ]^{1 \over 2} (P+R^\dagger) H(P+R) 
\Big [ { 1 \over 1+ R^\dagger R} \Big ]^{1 \over 2} [1 + R^\dagger R]^{1
\over
2} P \mid \psi \rangle  \nonumber \\
&&~~~~~~~~~~~~~~~~~~~~~~~~~~~~~~~~~~~~~~~~~~~~~~~
= E [ 1 + R^\dagger R]^{1 \over 2}P \mid \psi
\rangle.
\ee
i.e.,
\be
H_{eff} \mid \phi \rangle = E \mid \phi \rangle
\ee
where
\be
\mid \phi \rangle = [1+R^\dagger R]^{1 \over 2} P \mid \psi \rangle \label{bwf}
\ee
and
\be
H_{eff} = \Big [ { 1 \over 1+ R^\dagger R} \Big ]^{1 \over 2}
(P+R^\dagger)H(P+R) \Big [ { 1 \over 1+ R^\dagger R} \Big ]^{1 \over 2}.
\label{H_eff}
\ee

Our next task is to generate a perturbative expansion. Denote free
eigenstates in $P$ by $ \mid a \rangle$, $ \mid b \rangle$, etc.  Denote
free
eigenstates in $Q$ by $ \mid i \rangle$, $ \mid j \rangle$, etc.
Then
\be
h_{PP} \mid a \rangle &&= \epsilon_a \mid a \rangle, \nonumber \\
h_{QQ} \mid i \rangle && = \epsilon_i \mid i \rangle.
\ee
Let us compute $R$ to lowest orders in the perturbation theory.
Let us write $ R=R_1+R_2 + \ldots$ where the subscript denotes orders in
$v$. A straightforward calculation leads to
\be
\langle i \mid R_1 \mid a \rangle &&= { \langle i \mid v_{QP} \mid a \rangle
\over \epsilon_a - \epsilon_i}, \\
\langle i \mid R_2 \mid a \rangle &&= - \sum_b { \langle b \mid v \mid a
\rangle  \langle i \mid v \mid b \rangle \over (\epsilon_a - \epsilon_i)
(\epsilon_b - \epsilon_i)} +
\sum_j { \langle i \mid v \mid j
\rangle  \langle j \mid v \mid a \rangle \over (\epsilon_a - \epsilon_i)
(\epsilon_a - \epsilon_j)} .
\ee 
Note that the energy denominators in the matrix elements of $R$ involve only
the difference between  free energies of states in $P$ and $Q$ subspaces. 
This difference  may approach to zero and give rise the ``vanishing energy
denominator problem'' discussed in chapter \ref{chapbloch}.

Our next task is to develop a perturbation theory expansion for the
effective Hamiltonian to a given order.

We start from the expression for the effective Hamiltonian (\ref{H_eff}).
 Remember that $R_1 \sim O(v)$, $R_2 \sim O(v^2)$.

To order $v$, $H_{eff} = PHP$ and hence
\be
\langle a \mid H_{eff} \mid b \rangle = \langle a \mid (h+v) \mid b \rangle.
\ee
To second order in $v$, we have
\be
H_{eff} = [ 1 - { 1 \over 2} R^\dagger R] [ PHP + PHR + R^\dagger HP
+ R^\dagger H R] [ 1 - { 1 \over 2} R^\dagger R].
\ee
From $R^\dagger H R$ we get,
\be
\langle a \mid R^\dagger H R \mid b \rangle = \sum_i \epsilon_i
{ \langle a \mid v \mid i \rangle \langle i \mid v \mid b \rangle \over
(\epsilon_a - \epsilon_i)(\epsilon_b - \epsilon_i)}.
\ee
From $PHR$ and $R^\dagger HP$ terms we get
\be
\sum_i \langle a \mid H \mid i \rangle \langle i \mid R_1 \mid b \rangle+
\sum_i \langle a \mid R_1^\dagger \mid i \rangle \langle i \mid H \mid b
\rangle \\
= \sum_i \Big [ { \langle a \mid v \mid i \rangle \langle i \mid v \mid b
\rangle \over \epsilon_a - \epsilon_i} +
{ \langle a \mid v \mid i \rangle \langle i \mid v \mid b
\rangle \over \epsilon_b - \epsilon_i}.
\ee
Due to the normalization of the states, the effective Hamiltonian also
gets contribution from the  normalization factors.
From the {\it normalization factors} we get
\be
 - { 1 \over 2} R^\dagger R PHP - { 1 \over 2} PHP R^\dagger R =
- { 1 \over 2} (\epsilon_a + \epsilon_b) \sum_i 
{ \langle a \mid v \mid i \rangle \langle i \mid v \mid b
\rangle \over (\epsilon_a - \epsilon_i)(\epsilon_b - \epsilon_i)} 
\ee
Adding everything, to second order, we have,
\be
\langle a \mid H_{eff} \mid b \rangle = { 1 \over 2} 
\sum_i \langle a \mid v \mid i \rangle \langle i \mid v \mid b \rangle \Big
[{ 1 \over \epsilon_a - \epsilon_i} + { 1 \over \epsilon_b - \epsilon_i}
\Big ].
\ee
If $a=b$, this expression reduces to the familiar second order energy shift.

Why Bloch formalism is preferred over Bloch-Horowitz formalism?

In the former, eigenstates of the effective Hamiltonian are orthonormalized
projections of the original eigenstates. In the latter, they are not.
The Bloch  wavefunctions are defined by Eq. (\ref{bwf}) and the
 Bloch-Horowitz wavefunctions  by   
$\mid \phi \rangle = P \mid \psi \rangle$.
Consider two ortho normalized eigenstates of the original Hamiltonian $ \mid
\psi_1 \rangle$ and $ \mid \psi_2 \rangle$ with $ \langle \psi_1 \mid \psi_2
\rangle =0$. However, $P \mid \psi_1 \rangle$ and $ P \mid \psi_2 \rangle$
need not be orthogonal, i.e., $ \langle \psi_1 \mid PP \mid \psi_2 \rangle = 
\langle \psi_1 \mid P \mid \psi_2 \rangle \neq 0$.
Construct $ \mid {\tilde \psi_1} \rangle = [1+ R^\dagger R]^{1 \over 2} P
\mid \psi_1 \rangle$, $ \mid {\tilde \psi_2} \rangle = [1+ R^\dagger R]^{1
\over 2}P
\mid \psi_2 \rangle$.
 Then
\be
\!\!\!\!\!\!\!\!
\langle {\tilde \psi_1} \mid {\tilde \psi_2} \rangle = \langle \psi_1 \mid P
\mid \psi_2 \rangle + \langle \psi_1 \mid PR^\dagger R P \mid \psi_2 \rangle
=\langle \psi_1 \mid(P+Q)\mid \psi_2 \rangle
= \langle \psi_1 \mid \psi_2 \rangle.
\ee  

\chapter{Details of Numerical Procedure to Diagonalize the Effective
Hamiltonian}\label{APnumpro}
We convert the  bound state integral equation to a matrix eigenvalue 
equation  by discretizing the integrations using Gauss quadrature points for 
both  Bloch effective and similarity renormalized Hamiltonians. Here, we
elaborate  the parametrizations and diagonalization procedure we have used.

{\it Parametrization}: The light-front variables are parametrized in the
following ways in our numerical calculations. The full $k$-interval is
divided into $n_1$ quadrature points. 
 $k$ is defined by two
different ways. One definition is
\be 
k={u \Lambda m\over (1-u^2)\Lambda +m},
\label{k1}
\ee
 where $\Lambda$ is the 
ultraviolet cutoff and $u$'s are the quadrature points lying between $-1$
and $+1$, so that $k$ goes from $-\Lambda$ to $+\Lambda$. The other
definition is
\be 
k={1\over \kappa}tan({u \pi\over 2}),
\label{k2}
\ee
 here $\kappa$ is a parameter that can be tuned
to adjust the ultraviolet cutoff. 
The second definition (\ref{k2}) of $k$
is very suitable 
for weak coupling calculations where we need maximum points to be
concentrated near  $k=0$ and get better convergence  than 
 the first definition (\ref{k1}).

 The longitudinal momentum fraction $x$ ranges 
from $0$ to $1$. We divide all $x$- integrations in our calculations  
into two parts, $x$ ranging from 0 to 0.5 and $x$ ranging from 0.5 to 1
and  discretize each $x$-interval into $n_2$ quadrature points
 with the
parametrization
\be
x={1+v+2\epsilon(1-v) \over 4},~~~~~ \epsilon \le x\le 0.5,\\  
x={3+v-2\epsilon (1+v) \over 4},~~~~~0.5\le x \le 1-\epsilon,
\ee
where $v$'s are the Gauss-quadrature points lying between $-1$ and $+1$ 
and $\epsilon(\rightarrow 0)$ is introduced to handle
end-point singularities in $x$ as mentioned in the main text in Chapters
\ref{chapbloch} and \ref{chapsrg}.

To handle the infrared diverging terms we put the
 cutoff $|x-y|\ge \delta$
 and at 
the end we take the limit $\delta\rightarrow 0$. Numerically, it means
that the result should converge as one decreases $\delta$ if there is no net
infrared divergence in the theory.

{\it Diagonalization}: After discretization, solving the integral equation 
becomes 
a matrix diagonalization problem. The diagonalization has been performed 
by using the packed storage {\it LAPACK}\footnote{ E. Anderson {\it et al.},
{\it LAPACK Users' Guide},  
     third edition  
     (Society for Industrial and Applied Mathematics, Philadelphia, 1999).
Available on the internet at the URL:
http://www.netlib.org/lapack/lug/index.html.}
 routines 
{\it DSPEVX} for the reduced model (real symmetric matrix) and {\it ZHPEVX}
for the full Hamiltonian (Hermitian matrix).    

\chapter{Nonrelativistic Bound State Equation}\label{APnrbe}
In order to elucidate the implications of rotational symmetry in the (2+1)
dimensional world,
we review the nonrelativistic bound state equation in
this appendix.

For clarity, in this appendix we restore the superscript to the transverse
component, namely, $ k=k^1$, $q=q^1$, etc. We also use the notation
$ {\bf k} = (k^1, k^2)$, etc.
To discuss the nonrelativistic limit of the reduced model defined in 
Sec. \ref{reduce},
make the variable change,
$ x={ 1 \over 2} \Big ( 1 + {k^2 \over  E(k)} \Big )$,
$ y={ 1 \over 2} \Big ( 1 + {q^2 \over E(q)} \Big )$, where $E(p) =
\sqrt{m^2 + (p^1)^2 + (p^2)^2}$. So far, no approximations have been made.
We have,
\be
{ 1 \over x(1-x)} ~& =&~ 4 \Big [ 1 - \Big ({k^2 \over E(k)}\Big )^2 \Big
]^{-1}  \approx
4 \Big [ 1 + \Big ({k^2 \over m} \Big )^2 \Big ] , \nonumber \\
m(x-y) ~& =&~{m \over 2} \Big ({k^2 \over E(k)} - {q^2 \over E(q)} \Big )
\approx { 1 \over 2} (k^2 - q^2),  \nonumber \\
ky - qx ~&=&~ { 1 \over 2} (k^1 - q^1) + { 1 \over 2} \Big
({k^1 q^2 \over E(q)} - {q^1 k^2 \over E(k)} \Big ) \approx  { 1 \over 2}
(k^1 -q^1), \nonumber \\
k(1-y) - q(1-x) ~ &=&~ { 1 \over 2} (k^1 - q^1) - { 1 \over 2} \Big
({k^1 q^2 \over E(q)} - {q^1 k^2 \over E(k)} \Big ) \approx  { 1 \over 2}
(k^1 -q^1), \nonumber \\
{\partial y \over \partial q^2} ~ & =& ~ {1 \over 2}{(q^1)^2 + m^2 \over
[E(q)]^{3 \over 2}} \approx { 1 \over 2 E(q)}.
\ee
The $\approx$ equality holds in the nonrelativistic limit $ \mid {\bf k}
\mid, \mid {\bf q} \mid << m$.
Introducing the binding  energy ${\bar B}$ by $ M^2 = 4m^2 (1- {\bar B})$,
the bound state equation in momentum space in
the non-relativistic limit is given by
\be
\Big [ {\bar B} + {{\bf 
k}^2 \over m^2} \Big ] \psi({\bf k}) = {g^2 \over 4
\pi^2 m} C_f \int d{\bf q} { \psi({\bf q}) - \psi({\bf k}) \over ({\bf k} -
{\bf q})^2}.
\ee
Fourier transforming to coordinate space, with the momentum in the self
energy integral cutoff by fermion mass $m$, one arrives at the coordinate
space bound state equation
\be
\Big \{ -{1\over m} {\partial^2 \over \partial r^2} ~+ ~{ 4 l^2 - 1 \over 4 m r^2}~ + ~
{g^2 \over 2 \pi}~C_f~ ~ [\gamma_E~ + ln~mr]\Big \} \psi(r) ~=~ E ~ \psi(r)
\ee
where $E = -m{\bar B}$ and  $ \psi({\bf r}) = r^{- {1 \over 2}} \psi(r)
e^{\pm i l\phi}$ and
 $\gamma_E$ is the Euler constant.
We note that rotational symmetry implies two-fold degeneracy for $ l \neq 0$
states.
\chapter{Similarity Renormalization Theory for the Effective Hamiltonian}
\label{APsrg}
As it was promised in chapter \ref{chapsrg}, we present the detailed derivation
of the effective Hamiltonian using similarity renormalization group
approach in this appendix.

Since the  renormalization group transformation based on  integrating
out the high energy states encounters nearly degenerate states,  an
alternative way of calculating effective Hamiltonian was in demand.
    The solutions were  proposed by
G{\l}azek and Wilson and Wegner independently\footnote{S. ~D.~G{\l}azek and
K.~G.~Wilson, Phys.\ Rev.\ D {\bf 48}, 5863 (1993); {\bf 49}, 4214 (1994);
 F.~Wegner, Ann.\ Phys.\ (Leipzig) {\bf 3}, 77 (1994).}. 

Starting from a cutoff Hamiltonian $H_B$ which includes canonical terms and
counterterms we wish to arrive at an effective Hamiltonian $H_{\sigma}$
defined at the scale $\sigma$ via a similarity transformation 
\begin{eqnarray}
H_\sigma = S_\sigma ~ H_B ~ S_\sigma^\dagger
\end{eqnarray}
where $S_\sigma$ is chosen to be
unitary.

The boundary condition is $ Limit_{\sigma \rightarrow \infty} ~H_\sigma
= H_B$.

Introduce anti-Hermitian generator of infinitesimal changes of scale
$T_\sigma$ through
\begin{eqnarray}
S_\sigma = {\cal T}~ e^{\int_{\sigma}^{\infty}d\sigma' ~ T_{\sigma'}}
\end{eqnarray}
where $ {\cal T}$ puts operators in order of increasing scale. 

For infinitesimal change (lowering) of scale, $  S_\sigma 
= 1 - T_\sigma ~ d\sigma$
and $S_\sigma^\dagger = 1+ T_\sigma ~ d\sigma$. Then we arrive at the
infinitesimal form of the transformation
\begin{eqnarray}
{d H_\sigma \over d\sigma} = [ H_\sigma, T_\sigma]. \label{diffform}
\end{eqnarray}
This equation which has been called the flow equation of the Hamiltonian is
the starting point of the investigations. 

The basic goal of the transformation $S_\sigma$ is that $H_\sigma$ should be
band diagonal relative to the scale $ \sigma$. Qualitatively this means that
matrix elements of $H_\sigma$ involving energy jumps much larger
than $ \sigma$ should be zero. $T_\sigma$ still remains arbitrary to a great
extent. It is instructive to go through the steps of the derivation which
leads to the G{\l}azek-Wilson choice.

We write $ H_B = H_{B0}+H_{BI}$ where $H_{B0}$ is the free part and $H_{BI}$
is the interaction part of the bare cutoff Hamiltonian.
A brute force way of achieving our goal is to {\it define} the matrix
elements $ H_{I \sigma ij} = f_{\sigma ij} H_{BI  ij}$ 
where we have introduced the function $ f_{\sigma ij} =f(x_{\sigma ij})$
with $x$ a function of  ${\sigma^2}$ and  $ \Delta M^2_{ij}$.     
The function $f(x)$ should be chosen
as follows:
\begin{eqnarray}
\nonumber
{\rm when} \ {\sigma^2 } >> \Delta M^2_{ij} ,
&&\quad  f(x) = 1\qquad \qquad \qquad
{\it (near\ diagonal\ region)}; \nonumber \\
{\rm when} \  {\sigma^2} << \Delta M^2_{ij}, 
&& \quad  f(x)
 =0  \qquad \qquad \qquad {\it (far \ off\  diagonal\  region)};
 \nonumber \\
{\rm in \ between}  \qquad \qquad  f(x) ~~~~&& ~~~~~
{\rm drops \ from \ 1 \ to \ 0}
\qquad   
{\it (transition \ region)}  .  
\end{eqnarray}
Here $\Delta M^2_{ij} (=M^2_i - M^2_j)$ denotes the difference of invariant
masses of states
$i$ and $j$. Because of the properties of $f$, $H_{I \sigma ij}$ is band 
diagonal. What is wrong with such a choice of inserting form factors by hand
at the interaction vertices? First of all, we  simply discard degrees
of freedom above $\sigma$. Secondly,
$H_\sigma$ will have very strong dependence on
$\sigma$. Thirdly, to ensure that $H_\sigma$ has no ultraviolet cutoff
dependence, $H_B$ should contain canonical and counterterms. But, in
light front Hamiltonian field theory, because of the complexities due to
renormalization, a priori we do not know the structure of counterterms.

Note that in the definition of $H_{\sigma}$ given in Eq. (\ref{diffform}) 
the form
of $T_{\sigma}$ is still unspecified. 
In fact, a wide variety of choices are
possible. In the following, we consider the choices made by G{\l}azek 
and Wilson and Wegner. The price we have to pay
for the use of flow equations is that it will generate complicated
interactions even if the starting Hamiltonian has only simple interactions.
For example, starting with a Hamiltonian which has only 2 particle
interaction, the transformation will generate  3 particle interactions, 4
particle interactions, etc.

\section{G{\l}azek-Wilson Formalism}

Writing $ H_\sigma = H_0 + H_{I\sigma}$, noting that the free Hamiltonian
$H_0$ does not depend on $\sigma$ and taking matrix elements in free
particle states, we have,
\begin{eqnarray}
[H_\sigma, T_\sigma]_{ij} = (P_i^- - P_j^-) T_{\sigma ij} +
[H_{I\sigma},T_\sigma]_{ij}
\end{eqnarray}
where $ H_0 \mid i \rangle = P_i^- \mid i \rangle$, etc. .
i.e.,
\be
{1 \over f_{\sigma ij}}{d H_{I \sigma ij} \over d\sigma} = 
{ 1 \over f_{\sigma ij}} [H_{I\sigma},T_\sigma]_{ij} + {1 \over f_{\sigma
ij}} (P_i^- - P_j^-) T_{\sigma ij}.
\ee
Since we want $H_{I \sigma ij}$ to be band diagonal, it is advantageous to
trade 
${1 \over f_{\sigma ij}}{d H_{I \sigma ij} \over d\sigma}$ for
${ d \over d \sigma} \left [ { 1 \over f_{\sigma ij}}H_{I\sigma ij}\right ]$
which on integration has the chance to ensure that $H_{I\sigma ij}$
is band diagonal, we use
\begin{eqnarray}
{ d \over d \sigma} \left [ { 1 \over f_{\sigma ij}}H_{I\sigma ij}\right ]
+ {1 \over f^2_{\sigma ij}}{df_{\sigma ij} \over d\sigma} H_{I\sigma ij} =
{ 1 \over f_{\sigma ij}}{dH_{I\sigma ij} \over d \sigma} 
\end{eqnarray}
and arrive at 
\begin{eqnarray}
{d \over d \sigma} \left [ { 1 \over f_{\sigma ij}} H_{I\sigma ij} \right ]
&&=
 { 1 \over f_{\sigma ij}}(P_i^- -
P_j^-)T_{\sigma ij} \nonumber \\
&&~~ + { 1 \over f_{\sigma ij}} [H_{I\sigma},
T_{\sigma}]_{ij} - { 1 \over f^2_{\sigma ij}}{d f_{\sigma ij} \over d\sigma}
H_{I\sigma ij}.\label{me1}
\end{eqnarray}
Still $T_{\sigma ij}$ is not defined. We next convert this equation into two
equations, one defining the flow of $H_{I \sigma ij}$ 
and other defining $T_{\sigma ij}$.
Recalling the starting equation Eq. (\ref{diffform}) we add and subtract
$[H_{I\sigma},T_\sigma]_{ij}$ to the r.h.s. and arrive at  
\begin{eqnarray}
{d \over d \sigma} \left [ { 1 \over f_{\sigma ij}} H_{I\sigma ij} \right ]
&&=
[H_{I\sigma},T_\sigma]_{ij} + { 1 \over f_{\sigma ij}}(P_i^- -
P_j^-)T_{\sigma~ij} \nonumber \\
&&~~ + { 1 \over f_{\sigma ij}}(1-f_{\sigma ij}) [H_{I\sigma},
T_{\sigma}]_{ij} - { 1 \over f^2_{\sigma ij}}{d f_{\sigma ij} \over d\sigma}
H_{I\sigma ij}.\label{me}
\end{eqnarray}
G{\l}azek and Wilson choose $T_\sigma$ to be
\begin{eqnarray}
T_{\sigma ij} = { 1 \over P_j^- - P_i^-}\left [ (1 - f_{\sigma ij})
[H_{I\sigma},T_\sigma]_{ij}- { d \over d\sigma} ({\rm ln}~f_{\sigma ij})
H_{I\sigma ij} \right ]. \label{gwt}
\end{eqnarray}
Then from Eq. (\ref{me}), we have,
\begin{eqnarray}
{d \over d \sigma}\left [ {1\over f_{\sigma ij}} H_{I\sigma ij} \right ] = 
[H_{I\sigma},T_{\sigma}]_{ij}. \label{gwh}
\end{eqnarray}
 
Integrating Eq. (\ref{gwh}) from $\sigma$ to $\infty$, we arrive at,
\begin{eqnarray}
H_{I\sigma ij} = f_{\sigma ij} \left [ H_{IB ij} - \int_\sigma^\infty
d\sigma' [H_{I\sigma'},T_{\sigma'}]_{ij} \right ]. \label{gwhs}
\end{eqnarray}

Note that $H_{I\sigma ij}$ is zero in the far off-diagonal
region. This is clear from the solution given in Eq. (\ref{gwhs}) since
$f(x)$ vanishes when $x \geq 2/3$. 

$T_{\sigma ij}$ vanishes in the near diagonal region. When $i$ is close to
$j$, $f_{\sigma ij}=1$ and both $(1 - f_{\sigma ij})$ and ${d \over
d\sigma}f_{\sigma ij}$ vanishes. It follows, then, from Eq. (\ref{gwt}) that
$T_{\sigma ij}$ vanishes in the near-diagonal region. This guarantees that a
perturbative solution to $H_{I\sigma ij}$ in terms of $H_{BI ij}$ will never
involve vanishing energy denominators.

The effective Hamiltonian can be calculated up to any order of
perturbation theory from Eq. (\ref{gwhs}) by iterative method. Here, we  
 derive the effective Hamiltonian to second order in
perturbation theory. 
Using
\begin{eqnarray}
H_{I\sigma ik}^{(1)}  \simeq f_{\sigma ik} H_{BI ik} 
\ee
and
\be
T_{\sigma kj}~ && \simeq {1 \over P_j^- - P_k^-} \left \{ - { d \over d
\sigma} 
({\rm ln} ~f_{\sigma kj}) f_{\sigma kj} H_{BI kj} \right \}
\end{eqnarray}
in Eq. (\ref{gwhs}), a straightforward calculation leads to
\begin{eqnarray}
H_{I\sigma ij}^{(2)} = - \sum_k H_{BI ik} H_{BI kj} \left [
{g_{\sigma ijk} \over P_k^- - P_j^-} + { g_{\sigma jik} \over P_k^- - P_i^-}
 \right ], \label{gw2eh}
\end{eqnarray}
where
\begin{eqnarray}
g_{\sigma ijk} &&= f_{\sigma ij} ~ \int_{\sigma}^\infty d\sigma' ~
f_{\sigma' ik}~
{ d \over d \sigma'} f_{\sigma' jk}, \nonumber \\
g_{\sigma jik} &&= f_{\sigma ij}~ \int_{\sigma}^\infty d\sigma'~
f_{\sigma'jk}
{ d \over d \sigma'} f_{\sigma'ik}.\label{APgfact}
\end{eqnarray}
We find that the effective Hamiltonian in similarity perturbation theory is
a modification of the effective 
Hamiltonian in Bloch perturbation theory\footnote{Detail discussion of
Bloch effective perturbation theory is given in Appendix \ref{APbloch}.}. 

\section{Wegner Formalism}
In the Wegner formalism\footnote{For applications of Wegner formalism
 in condensed matter physics and quantum field theory 
see F.~J.~Wegner,
Nucl.\ Phys.\ Proc.\ Suppl.\  {\bf 90}, 141 (2000);
E.~L.~Gubankova and F.~Wegner,
hep-th/9708054;
E.~L.~Gubankova and F.~Wegner,
Phys.\ Rev.\ D {\bf 58}, 025012 (1998)
[hep-th/9710233].}, 
the flow equation is given by
\be
{d H(l) \over d l} = [ \tau (l),H(l)].
\ee 
Wegner chooses 
\be
\tau(l) = [H_d, H] = [H_d, H_r]
\ee
where $H_d$ is the diagonal part of the Hamiltonian and $H_r$ is the rest,
i.e., $H=H_d+ H_r$. Here the word diagonal is used in the particle number
conserving sense. It is important to note that $H_d$ is not the free part
of the Hamiltonian and both $H_d$ and $H_r$ depend on the length scale $l$.

The light front Hamiltonian has dimension of $(mass)^2$ and hence 
$\tau$ has the dimension of $(mass)^4$, $l$ has dimension of 
${ 1 \over (mass)^4}$.

Expanding in powers of the coupling constant, 
\be
H = H_d^{(0)}+ H_r^{(1)}+ H_d^{(2)} + H_r^{(2)} + \ldots
\ee
where the superscript denotes the order in the coupling constant,
\be
\tau(l) = [H_d^{(0)}, H_r^{(1)}] + [H_d^{(0)}, H_r^{(2)}] + \ldots ~ .
\ee  
Then, to second order,
\be
{d H \over d l} = [[H_d^{(0)}, H_r^{(1)}], H_d^{(0)}] + [[H_d^{(0)},
H_r^{(1)}], H_r^{(1)}] + [[H_d^{(0)}, H_r^{(2)}], H_d^{(0)}] + \ldots ~ .
\ee
Introduce
the eigenstates of $H_d^{(0)}$,
\be
H_d^{(0)} \mid i \rangle = P^{-}_{i} \mid i \rangle. 
\ee
Then, to second order,
\be
{d H_{lij} \over dl} = - (P^-_i - P^-_j)^2 H_{rij}^{(1)} + [ \tau_l^{(1)},
H_r^{(1)}]_{ij} - (P^-_i - P^-_j)^2 H_{rij}^{(2)} + \ldots~ .
\ee
To first order in the coupling,
\be
{ d H_{rij} \over dl} = - (P^-_i - P^-_j)^2  H_{rij}^{(1)}
\ee
which on integration yields
\be
H_{rij}^{(1)} (\sigma) = e^{ - {(P^-_i - P^-_j)^2 \over \sigma^4}}
H_{rij}^1(\Lambda)
\ee
where we have introduced the energy scale $\sigma$ via $ l = { 1 \over
\sigma^4}$ and
used the fact that $l=0$ corresponds to the original bare cutoff. We notice
the emergence of the similarity factor $ f_{\sigma ij} = e^{ - {(P^-_i -
P^-_j)^2 \over \sigma^4}} $. 

If we are interested only in particle number conserving (diagonal) part of
the effective interaction, to second order we have,
\be
{d H_{lij} \over dl} = [ \tau_l^{(1)},
H_r^{(1)}]_{ij}
\ee
Using
\be
\tau_{lij}^{(1)} = (P^-_j - P^-_i) H_{rij}^{(1)},
\ee
the effective interaction generated to second order in the diagonal sector
is
\be   
H_{lij} = \sum_k H^B_{ik} H^B_{kj} { (P^-_i - P^-_k) + (P^-_j - P^-_k) \over
(P^-_i - P^-_k)^2 + (P^-_j - P^-_k)^2 } \Bigg [ 1 -
e^{ - \Big \{(P^-_i - P^-_k)^2 + (P^-_j - P^-_k)^2 \Big \}/\sigma^4 } \Bigg
].
\ee
Even though the second order formula is very similar to the one in
G{\l}azek-Wilson formalism when an exponential form is chosen for the
similarity factor (see Sec. IV),
we note a slight difference. In the G{\l}azek-Wilson formalism, since the
purpose is to bring the Hamiltonian into a band diagonal form, even in the
particle number conserving sectors the large jumps in energies do not appear
by construction. In the version of the Wegner formalism presented here the
purpose is to bring the Hamiltonian in the block diagonal form in particle
number sector so that large jumps in energies are allowed by the effective
Hamiltonian. Note that small energy denominators do not appear in both
formalisms. 
\chapter{Violations of Hypercubic Symmetry on Transverse
Lattice}\label{APrsv}
The canonical helicity non-flip interactions given in Eq.
(\ref{canhnf}) for $ r \neq s$  break the  hypercubic symmetry on the
transverse lattice. 
For interacting theory this is also true for the  Hamiltonian with 
symmetric derivative.  In the free field limit they do not survive for
 Hamiltonian with symmetric derivative but for forward-backward derivative 
they survive. In that case, 
in the free field limit they reduce to 
\be
\frac {1}{a^2}~\int dx^- \sum_{\bx}\sum_{r \neq s}\Bigg [ 
\eta^\dagger(\bx + a \br) {\hat \sigma}_r {\hat \sigma}_s \frac{1}{
\partial^+} \eta(\bx) \nonumber \\
+ \eta^\dagger(\bx) {\hat \sigma}_r {\hat \sigma}_s \frac{1}{
\partial^+} \eta(\bx + a \bs) \nonumber \\
- \eta^\dagger(\bx + a \br) {\hat \sigma}_r {\hat \sigma}_s 
\frac{1}{
\partial^+} \eta(\bx + a \bs) \Bigg ].
\ee
Going to the transverse momentum space via
\be
\eta(x^-, x^\perp) = \int d^2 k^\perp e^{i k^\perp \cdot x^\perp} ~
\phi_{k^{\perp}}(x^-)
\ee
we get
\be
- \frac{2}{a^2} \int dx^- \int d^2 k^\perp  \phi^\dagger _{k^{\perp}}(x^-)
\sigma_3 \frac{1}{ i \partial^+} \phi_{k^{\perp}}(x^-) \nonumber \\
\Big [ \sin (k_ya)  - \sin (k_xa) + \sin (k_x a - k_y a) \Big ]. 
\ee
Thus the violations of hypercubic symmetry are of the order of the lattice
spacing $a$. Sign in front of this term changes if we switch forward and  
backward derivatives.

In case of interacting theory with symmetric lattice derivative,
 these terms do not come in one link approximation but will come in 
if one considers more than one link. But in case
of forward and backward derivative, they also appear in one link
approximation.
 In our numerical studies presented 
 in this work, 
we have set the coefficients of hypercubic symmetry violating terms to zero. 

\chapter{Fermions With Forward-Backward Derivatives in Conventional Lattice 
Theory}\label{APfblat}
In chapter \ref{tlchap1} we have discussed the light-front transverse 
lattice formulation of fermions with forward and backward lattice derivatives.
Let us discuss the situation in conventional lattice gauge theory in this
appendix.

In discretizing the Dirac action in conventional lattice theory the
use of forward or backward derivative for $\partial_{\mu}$ leads to 
non-hermitian action. The hermiticity can be preserved in the following 
way \footnote{In this appendix we follow the treatment of
 H. Banerjee and Asit K. De, ~Nucl.\ Phys.\ B (Proc. Suppl.) {\bf 53}, 
641 (1997).}.

In the chiral representation
\be
\gamma^0=\left[ \begin{array}{ll} 0 & -I
\\ -I & 0 \end{array} \right],~~\gamma^i=\left[ \begin{array}{ll} 0 & 
\sigma^i
\\ -\sigma^i & 0 \end{array} \right],~~\gamma^5=\left[ \begin{array}{ll} 
I & 0\\ 0 & -I \end{array} \right].
\ee
The Dirac operator in Minkowski space
\be 
i\gamma^{\mu}\partial_{\mu} \equiv \left[ \begin{array}{ll} 0 & 
-i\sigma^{\mu}\partial_{\mu}
\\ -i{\bar \sigma}^{\mu}\partial_{\mu} & 0 \end{array} \right],
\ee
where, $\sigma^{\mu} =(I,{\bf \sigma}),~~{\bar \sigma}^{\mu} 
=(I,-{\bf \sigma})$. 
For massive Dirac fermions, this leads to the structure
\be
-i\sigma^{\mu}\partial_{\mu}\psi_R -m\psi_L \label{one} \\
-i\sigma^{\mu}\partial_{\mu}\psi_L -m\psi_R. \label{two} 
\ee
For discretization we  replace $\partial_{\mu}$ in Eq. (\ref{one}) by
forward derivative
\be \Delta^f_{\mu}=(\delta_{y,x+\mu}-\delta_{y,x})/a
\ee
and in Eq. (\ref{two}) by backward derivative
\be 
\Delta^b_{\mu}=(\delta_{y,x}-\delta_{y,x-\mu})/a.
\ee
This leads to the structure  
\be
i\gamma^{\mu}\partial_{\mu} - m = i\gamma_{\mu}\Delta^s_{\mu}
-i\gamma_{\mu}\gamma_5\Delta^a_{\mu} - m
\ee
which results in hermitian action.
Here,
\be
\Delta^s_{\mu}&=&(\delta_{y,x+\mu}-\delta_{y,x-\mu})/2a \nonumber\\
\Delta^a_{\mu}&=&(\delta_{y,x+\mu}+\delta_{y,x-\mu}-
2\delta_{y,x})/2a.
\ee
Note that irrelevant helicity nonflip second order derivative term is 
produced in this method of discretization. In contrast, the corresponding 
term in the transverse lattice depends linearly on $m$ and flips helicity.
One can trace this difference to the presence of the constraint equation 
in the light front theory. 

Writing lattice derivatives in this fashion in
the conventional lattice theory
eliminates the doublers from the edges of the Brillouin zone 
but some non-covariant doublers prop up from other places.   

\chapter{Mixed Covariant Derivative}\label{APcov_der}
Here we derive the form of mixed covariant derivative used in the
transverse lattice formulation.

In a {\it local} field theory,
we cannot simply compare objects at distances. As Feynman reminds us,
\textquotedblleft {\em We
must take into account of the rotation of the frame by transporting $U(x^\mu
+ \Delta x^\mu)$ back to $x^\mu$ before making a
comparison}\textquotedblright \footnote{R.~P.~Feynman,
in {\it Weak and Electromagnetic
Interactions at High Energies}, (eds.) Roger Balian, Christopher H.
Llewellyn
Smith, (Elsevier North-Holland Pub. Co., Amsterdam 1977).}. 
Hence the total change is
given by
\be
D^\mu U_{r}(x^\mu, \bx) \Delta x_\mu &= &
R(\bx)^{-1} U_{r}(x^\mu + \Delta x^\mu, \bx) R(\bx + a \br) - U_{r}(x^\mu,
\bx)
\nonumber \\
&= & [ 1 +  A^\mu(\bx) \Delta x_\mu] U_{r}(x^\mu + \Delta x^\mu, \bx)
[1 -  A^\mu(\bx + a \br)\Delta x_\mu] -
U_{r} (x^\mu, \bx) \nonumber \\
&= & U_{r}(x^\mu, \bx) + \partial^\mu U_{r}(x^\mu, \bx) \Delta x_\mu -
U_{r} (x^\mu, \bx) \nonumber \\
&~~~~& +  A^\mu(\bx) U_{r}(x^\mu, \bx) \Delta x_\mu - 
U_{r}(x^\mu, \bx) A^\mu(\bx + a \br) \Delta x_\mu \nonumber \\
&= & \Big \{ [ \partial^\mu +  A^\mu(\bx)]U_{r}(x^\mu, \bx) - 
U_{r}(x^\mu, \bx) A^\mu(\bx + a \br)\Big \} \Delta x_\mu.
\ee

\chapter{Transverse Gauge Invariance}\label{APgauge}
Let us explicitly verify the transverse gauge invariance of the lattice
theory. The theory is invariant under the gauge transformations 
\be
\eta(\bx) \rightarrow \eta'(\bx) = G^\dagger(\bx) \eta(\bx)
\ee
and
\be
M_r(\bx) \rightarrow  M_r'(\bx) = G^\dagger(\bx) M_r(\bx) G(\bx + a \br)
\ee
where 
\be G(\bx) = e^{-i T^a \theta^a(\bx)}.
\ee
For infinitesimal transformation,
\be
G(\bx) \approx 1 - i T^a \theta^a(\bx)
\ee
and
\be
\eta(\bx) \rightarrow \eta'(\bx) = \eta(\bx) + i T^a\theta^a (\bx) \eta(\bx)
\label{ifet}
\ee
and
\be
M_r(\bx)_{pq}  \rightarrow  M_r'(\bx)_{pq} = M_r(\bx)_{pq} + i T^a_{pl}
M_r(\bx)_{lq} \theta^a(\bx) - i M_r(\bx)_{pl} T^a_{lq} \theta^a(\bx + a
\br).\label{ifmt}
\ee
 
In quantum theory the gauge transformations are generated by the operator
\be
{\cal G} = e^{{i \over 2} \sum_{\by} Q^a(\by) \theta^a(\by)}
\ee
with
\be
Q^a(\by) &=& \int dy^- \Bigg [ 
Tr \Big \{ T^a \sum_{r'}
\Big ( 
M_{r'}(\by) i \stackrel{\leftrightarrow}{\partial^+} M^\dagger_{r'}(\by) +
M^\dagger_{r'}(\by - a \br') i
\stackrel{\leftrightarrow} {\partial^+} M_{r'}(\by - a \br') \Big )
 \Big \} \nonumber \\
&~& \qquad \qquad \qquad 
- 2 \eta^\dagger(\by) T^a \eta(\by) \Bigg ]
\ee
so that
\be
\eta(\bx) \rightarrow \eta'(\bx) = {\cal G} \eta(\bx){\cal G}^\dagger
\label{fet}
\ee
and
\be
M_r(\bx) \rightarrow  M_r'(\bx) = {\cal G} M_r(\bx) {\cal G}^\dagger.
\label{fmt}
\ee

For infinitesimal $\theta^a$, using $ [A,BC] =A\{B,C\} - \{A,C\}B $ and the
canonical commutation relations for the fermion field operator, we readily
verify Eq. (\ref{ifet}). Using $[A,BC] = A[B,C]+[A,C]B$ and canonical
commutation relation for the link field, we also verify Eq. (\ref{ifmt}).

Next we look at the behavior of fermion and link creation and annihilation
operators under transverse gauge transformations in order to construct gauge
invariant multiparticle states. 

The gauge transformation on the link variable is
\be
M_{r}(\bx) \rightarrow M'_{r}(\bx) = G^\dagger(\bx) M_{r}(\bx) G(\bx + a
\br)
\ee
where $G$ belongs to $SU(N)$. 
Thus
\be
A_{r}(\bx) \rightarrow A'_{r}(\bx) = G^\dagger(\bx) A_{r}(\bx) G(\bx + a
\br),~~~
B^\dagger_{r}(\bx) \rightarrow {B^{\dagger}}'_{r}(\bx) = 
G^\dagger(\bx) B^\dagger_{r}(\bx) G(\bx + a \br). \nonumber \\
\ee
Thus $ Tr(A^\dagger B^\dagger)$, $ Tr(AB)$, $Tr(A^\dagger A)$, $Tr(B^\dagger
B)$ are locally gauge invariant operators. A locally gauge invariant two
link state is $ Tr(A^\dagger B^\dagger) \mid 0 \rangle$. 

Recall that the current has the structure
\be
J^{+a} \sim Tr (T^a M M^\dagger) \qquad {\rm with} \qquad M \sim A +
B^\dagger.
\ee
The interaction term $ J^{+a} ({ 1 \over \partial^+})^2 J^{+a} $ has many
terms. Consider one term 
\be
 Tr (T^a A A^\dagger) \, Tr(T^a B^\dagger B) & = & Tr((A^\dagger T^a A)
\,Tr(B T^a
B^\dagger) = T^{a}_{mp} T^a_{st}  A^\dagger_{nm} A_{pn} B_{rs}
B^\dagger_{tr} \nonumber \\
& \Rightarrow & T^a_{mp} T^a_{st} \delta_{nr} \delta_{mt} \delta_{ps}
\delta_{nr} = (T^a T^a)_{mm}.
\ee 

On the other hand, consider the ``pair creation term'' 
\be
 Tr(T^a AB) Tr(T^a B^\dagger A^\dagger) & = & Tr(BT^a A) Tr(A^\dagger T^a
B^\dagger) \nonumber \\
&=& T^a_{mp} T^a_{st} B_{nm} A_{pn} A^\dagger_{rs} B^\dagger_{tr}
\nonumber \\
&~& \Rightarrow
T^a_{mp} T^a_{st} \delta_{mp} \delta_{ts} = 0.
\ee
Thus pair creation or pair destruction terms  (even if they conserve
particle number) do not contribute if we restrict ourselves to a two link
gauge invariant sector.

Next let us look at the two component fermion field $ \eta(\bx)$. The 
transformation of $ \eta(\bx)$ is
\be
\eta(\bx) \rightarrow \eta(\bx)'  = G^\dagger(\bx) \eta(\bx).
\ee
Since $ \eta(\bx) \approx b(\bx) + d^\dagger(\bx) $, we have
\be
b(\bx) \rightarrow b(\bx)' = G^\dagger(\bx) b(\bx) \, \, {\rm and} \, \,  
 d^\dagger(\bx) \rightarrow d^{\dagger}(\bx) ' = G^\dagger(\bx)
d^\dagger(\bx) .
\ee 
Next consider how to form gauge invariant two-particle 
($q {\bar q}$) and three-particle ($ q {\bar q} $ 
link) states. A gauge invariant $ q {\bar q}$ state is
 $ b^\dagger(\bx) d^\dagger(\bx) \mid 0 \rangle$. Gauge invariant three
particle 
states are\\ $ b^\dagger(\bx) B^\dagger_{r}(\bx) d^\dagger(\bx + a \br) \mid 0
\rangle$ 
and $ b^\dagger(\bx +a \br) A^\dagger_{r}(\bx) d^\dagger(\bx) \mid 0
\rangle$ ($=b^\dagger(\bx +a \br) B^\dagger_{-r}(\bx + a \br)
d^\dagger(\bx) \mid 0 \rangle $).

\chapter{Hamiltonian Matrix Elements in One Link Approximation}\label{APme}
\section{Structure of terms in DLCQ}
\label{appa}
We use  DLCQ for the
longitudinal dimension ($ -L \le x^- \le +L$) and  implement
anti periodic boundary condition for  the two component fermion field, 
\be
\eta_c(x^-, \bx) = { 1 \over \sqrt{2L}} \sum_\lambda \chi_\lambda
\sum_{m=1,3,5, \dots} [ b_c(m, \bx, \lambda)e^{-i  \pi m x^- /(2
L)} + d_c^\dagger(m, \bx, -\lambda) e^{i  \pi m x^- /(2 L)}]
\ee
with
\be
\{ b_c (m, \bx, \lambda), b_c^\dagger(m', \bx', \lambda') \} =
 \{ d_c (m, \bx, \lambda), d_c^\dagger(m', \bx', \lambda') \} =
\delta_{m m'} \delta_{\bx, \bx'} \delta_{c,c'} \delta_{\lambda,
\lambda'}.
\ee
The link field has periodic boundary condition (with the omission of the 
zero
momentum mode),
\be
M_{r~pq}(x^-, \bx) = { 1 \over \sqrt{4 \pi}} 
\sum_{m=1,2,3, \dots} \frac{1}{\sqrt{m}}[ B_{-r~pq}(m, \bx + a \br)e^{-i
\pi m x^- / 
L} + B_{r~pq}^\dagger(m, \bx) e^{i  \pi m x^- /L)}]
\ee
with
\be
[ B_{r~pq} (m, \bx), B_{r'~ts}^\dagger(m', \bx') ] =
\delta_{m m'} \delta_{\bx, \bx'} \delta_{r,r'} \delta_{ps} \delta_{qt}.
\ee

The Hamiltonian $P^- = { L \over \pi}H$. 

In the following subsection we give the explicit structure of terms in the
Hamiltonian in the forward-backward case in DLCQ restricting to those 
relevant for the one link approximation. 


\subsubsection{Mass terms}
Mass terms:

\be
H_{f ~ free} & = &  m^2  \sum_\bx \sum_c\sum_\lambda \sum_n { 1
\over n} \Big [  b_c^\dagger(n, \bx, \lambda) b_c(n, \bx ,\lambda)
+ d_c^\dagger(n, \bx,
\lambda) d_c(n, \bx, \lambda) \Big ].
\ee
\be
H_{LINK ~ free} = \frac{\mu^2}{2} \sum_{\bx} \sum_{\br} \sum_n
\frac{1}{n}
\left [ B_r^\dagger(m,\bx) B_r(m,\bx) + B_{-r}^\dagger(m,\bx + a \br)
B_{-r}(m, \bx + a \br) \right ].
\ee   
%
\subsubsection{Four fermion  instantaneous term}
The four fermion instantaneous term which gives rise to a linear potential
in the
color singlet state
\be
2 { g^2 \over \pi a^2} \sum_{c c'c'' c'''}
\sum_{\lambda \lambda'\lambda'' \lambda'''}  \sum_\bx
\delta_{\lambda \lambda'} \delta_{\lambda'' \lambda'''}
\sum_{m_1 m_2 m_3 m_4} ~~~~~~~~~~~~~~~~~~~~~~~~~~
\nonumber \\
\times ~b^\dagger_c(m_1, \bx, \lambda)
d^\dagger_{c'''}(m_4, \bx, -\lambda''') b_{c'}(m_2, \bx, \lambda')
d_{c''}(m_3, \bx, \lambda''') ~~~~~~~~\nonumber \\
~~~~~~~\times ~{ 1 \over (m_3 -m_4)^2} \delta_{m_1+m_4,
m_2+m_3}~. 
\ee
\subsubsection{Helicity flip terms}
Particle number conserving terms:

\be
{ m g \over a} \sum_r  \sum_\bx \sum_{\lambda_{1}, \lambda_{2}}
~\chi^\dagger_{\lambda_{1}} ~{\hat \sigma}_r ~\chi_{\lambda_2}~
\sum_{m
_{1}} {1 \over m_{1}} ~~~~~~~ \nonumber \\
\left [ b^\dagger_c(m_{1}, \bx, \lambda_{1}) b_c(m_{1}, \bx, \lambda_{2}) +
d^\dagger_c(m_{1}, \bx, - \lambda_{2}) d_c(m_{1}, \bx, - \lambda_{2})
\right].
\ee

Particle number non conserving terms:
a typical term is
\be
{ m g \over a} { 1 \over \sqrt{4 \pi}}
\sum_r  \sum_\bx \sum_{\lambda_1, \lambda_2}~
~\chi^\dagger_{\lambda_{1}} ~{\hat \sigma}_r ~\chi_{\lambda_2}~
\sum_{m_{1}m_{2}m_{3}} {1 \over \sqrt{m_3}} { 1 \over 2m_3+m_2}
\delta_{m_{1} - m_{2}, 2 m_{3}} \nonumber \\
b_c^\dagger (m_1, \bx, \lambda_1) B_{-r c c '}(m_3, \bx + a \br)
b_{c'}(m_2, \bx + a \br, \lambda_2).
\ee
\subsubsection{Helicity non flip terms}

Two operators:

\be
{ 2 \over a^2}  \sum_\bx \sum_\lambda \sum_n { 1 \over n}
\left [ b^\dagger_{c}(n,\bx,\lambda) b_{c}(n,\bx, \lambda) +
d^\dagger_{c}(n,\bx,\lambda) d_{c}(n,\bx, \lambda \right ] .
\ee
Three operators:

A typical term is
\be
- g{ 1 \over a^2} { 1 \over \sqrt{4 \pi}} \sum_r  \sum_\bx
\sum_{\lambda}  \sum_{m_1 m_2
m_3} {1 \over \sqrt{m_3}} { 1 \over 2m_3 + m_2} \delta_{m_1 - m_2, 2m_3}
\nonumber \\
b^\dagger_c (m_1, \bx, \lambda) B_{-rcc'}(m_3, \bx + a \br)
b_{c'}(m_2, \bx + a \br, \lambda).
\ee
%
\subsubsection{Fermion - link instantaneous term}
A typical term is
\be
2 { g^2 \over 4 \pi} { 1 \over a^2}  \sum_\bx \sum_{r} \sum_{cc'c''}
\sum_{dd'} T^\alpha_{cc'} T^\alpha_{dd'} \sum_{m_{1}m_{2}m_{3}m_{4}}
\frac{1}{\sqrt{m_{3}}} \frac{1}{\sqrt{m_{4}}}\nonumber \\
b_d^\dagger(m_1, \bx, \lambda_1) b_{d'}(m_2, \bx, \lambda_2) B_{-r
c'c''}(m_3, \bx + a \br) B^\dagger_{-r c''c}(m_4, \bx + a \br) \nonumber
\\
(-)(m_3+ m_4)/(m_1 - m_2)^2 ~~\delta_{m_{1} - m_{2}, 2m_{3} - 2
m_{4}}~.~~~~~~~~
\ee
\section{States in DLCQ}
\label{appb}
We will consider states of zero transverse momentum.
In the one - link approximation, the gauge invariant states are
 $q {\bar q}$ state
\be  \mid 2 \rangle &= &{ 1 \over \sqrt{N}}~{1 \over \sqrt{V}} ~
\sum_d ~ \sum_{\by(q)}~\sum_{\by({\bar q})} ~ \delta_{\by(q), \by({\bar q})}
\nonumber \\
&~&~~~~b^\dagger_d(n_1, \by(q), \sigma_1)~ 
d^\dagger_d(n_2, \by({\bar q}), \sigma_2)~\mid 0 \rangle 
\ee
and the $q {\bar q} ~{\rm link}$ states 
\be
\mid 3a \rangle &=& { 1 \over N}~{1 \over \sqrt{V}} ~ \frac{1}{\sqrt{2}}
~\sum_{dd'}~ \sum_s~\sum_{\by(q)}~\sum_{\by({\bar q})}~ \sum_{\by(l)}~
\delta_{\by(l),\by(q)}~\delta_{\by(q),\by({\bar q})- a \bs} \nonumber \\
&~&~~~~b^\dagger_{d}(n_1, \by({q}), \sigma_1)~ B^\dagger_{s
dd'}(n_3, \by({l}))~ d^\dagger_{d'}(n_2, \by({\bar q}), \sigma_2)~ 
~\mid 0
\rangle \nonumber 
\ee
and
\be
\mid 3b \rangle &=& { 1 \over N}~{1 \over \sqrt{V}} ~ \frac{1}{\sqrt{2}}
~\sum_{dd'}~ \sum_s~\sum_{\by(q)}~\sum_{\by({\bar q})}~ \sum_{\by(l)}~
\delta_{\by(l),\by(q)}~\delta_{\by(q),\by({\bar q})+ a \bs} \nonumber \\
&~&~~~~b^\dagger_{d}(n_1, \by(q), \sigma_1)~ B^\dagger_{-s
dd'}(n_3, \by(l))~ d^\dagger_{d'}(n_2, \by({\bar q}), \sigma_2)~ 
\mid 0
\rangle . 
\ee
We shall consider transition from these initial states to the following
final states:
The $q {\bar q}$ state
\be
\langle 2' \mid &=& { 1 \over \sqrt{N}}~
~{1 \over \sqrt{V}} ~
\sum_{e}
\sum_{\bz(q)}~\sum_{\bz({\bar q})} ~ \delta_{\bz(q), \bz({\bar q})} 
\nonumber \\
&~&~~~~\langle 0 \mid  
d_e(n_2', \bz({\bar q}), \sigma_2')~
 b_e(n_1', \bz({ q}), \sigma_1')
\ee
and the $q {\bar q} ~{\rm link}$ states 
\be
\langle 3a' \mid &=& { 1 \over N}~ 
{1 \over \sqrt{V}} ~ \frac{1}{\sqrt{2}} \sum_{ee'}~ \sum_t
~ \sum_{\bz(q)}~\sum_{\bz({\bar q})}~ \sum_{\bz(l)}~
\delta_{\bz(l),\bz(q)}~\delta_{\bz(q),\bz({\bar q})- a \bt} \nonumber \\
&~&~~~~\langle 0 \mid  
d_{e}(n_2', \bz({\bar q}), \sigma_2')~ 
B_{te e'}(n_3', \bz(l))~ b_{e'}(n_1', \bz(q), \sigma_1')~
\ee
and
\be
\langle 3b' \mid &=& { 1 \over N} ~{1 \over \sqrt{V}} ~ \frac{1}{\sqrt{2}}
~\sum_{ee'}~
\sum_t
~ \sum_{\bz(q)}~\sum_{\bz({\bar q})}~ \sum_{\bz(l)}~
\delta_{\bz(l),\bz(q)}~\delta_{\bz(q),\bz({\bar q})+ a \bt} \nonumber \\
&~&~~~~\langle 0 \mid  d_{e}(n_2', \bz({\bar q}), \sigma_2')~ 
B_{-te e'}(n_3', \bz(l)) ~b_{e'}(n_1', \bz({q}), \sigma_1')~
 \nonumber \\
\ee 
\section{Forward-backward derivatives: Matrix Elements in DLCQ} 
\label{appc}

\subsection{Transitions from two particle state}
\subsubsection{To two particle state}
Let us consider transitions to the two particle state:
We have, from the free particle term,
\be
\langle 2' \mid H_{f~free} \mid 2 \rangle = m^2  \left ( { 1 \over
n_1} + { 1 \over n_2} \right )~{\cal N}_2
\ee
where 
\be
{\cal N}_2 = \delta_{n_1,n_1'}~
 \delta_{\sigma_1, \sigma_1'}~ 
 \delta_{n_2,n_2'} ~ \delta_{\sigma_2, \sigma_2'}~ .
\ee
From the four fermion instantaneous  term we get
\be
\langle 2' \mid H_{qqc} \mid 2 \rangle &=& - 2 {g^2 \over \pi a^2} 
~C_f   ~ \delta_{n_1 + n_2,n_1' + n_2'}~ 
{ 1 \over (n_1 - n_1')^2 } \nonumber \\
&~&~\delta_{\sigma_1, \sigma_1'} ~ 
\delta_{\sigma_2, \sigma_2'} 
\ee
where $ C_f = {N^2 - 1 \over 2 N}$~.

To implement the regulator prescription for $\frac{1}{(k^+)^2}$, we add the
counterterm matrix elements
\be
\langle 2' \mid H_{CT} \mid 2 \rangle &=&
2 {g^2 \over \pi a^2}~C_f   ~ \delta_{n_1 + n_2,n_1' + n_2'}~
\sum_{n_{loop}=1}^{K} \frac{1}{(n_1 - n_{loop})^2}~
~\delta_{\sigma_1, \sigma_1'} ~\delta_{\sigma_2, \sigma_2'}. 
\ee
Here the term $ n_{loop}=n_1$ is dropped from the sum.

From the helicity flip term we get
\be
\langle 2' \mid H_{hf1} \mid 2 \rangle &=& - 2 { 1 \over a}  \sum_s
\left [ {m  \over n_1} ~\chi^{\dagger}_{\sigma_{1}'}~ 
{\hat \sigma}_s ~\chi_{\sigma_{1}} ~\delta_{\sigma_2, \sigma_2'} ~+
~{m  \over n_2} ~ \chi^\dagger_{-\sigma_{2}} ~
{\hat \sigma}_s ~\chi_{-\sigma_{2}'} ~\delta_{\sigma_1, \sigma_1'} \right ]
~
{\cal N}_{hf} \nonumber \\
\ee
with 
\be
{\cal N}_{hf} = \delta_{n_1,n_1'} 
~\delta_{n_2,n_2'} ~.
\ee
From the helicity non-flip term we get
\be
\langle 2' \mid H_{hnf}(1) \mid 2 \rangle &=&  2 { 1 \over a^2}  
 \left  ( { 1 \over n_1} + { 1 \over n_2} \right ) ~ {\cal N}_2~.
\ee
\subsubsection{To three particle state}
\subsubsection{To the state $\mid 3 a \rangle$}
From the helicity flip term we get
\be
\langle 3 a' \mid H_{hf2} \mid 2 \rangle & = & { mg \over a} 
 ~\sqrt{N}~\frac{1}{V}\frac{1}{\sqrt{2}}~{ 1 \over \sqrt{4 \pi}}~ \sum_t 
\chi^\dagger_{\sigma_1'}~{\hat \sigma}_t ~\chi_{\sigma_1}~
\delta_{\sigma_2, \sigma_2'} \nonumber \\
&~&  \delta_{n_2,n_2'}~{ \delta_{n_1' + 2n_3', 
n_1} \over n_1'}~ {1 \over \sqrt{n_3'}} ~ 
 ~ \sum_{\bz(q)}~ 
\sum_{\by(q)}~ \delta_{\bz(q),\by(q)- a \bt}\nonumber \\
&~&  +~ { mg \over a}  ~\sqrt{N}~\frac{1}{V}\frac{1}{\sqrt{2}}~
{ 1 \over \sqrt{4 \pi}} 
\sum_t~\chi^\dagger_{-\sigma_2}~{\hat \sigma}_t ~\chi_{-\sigma_2'}~
\delta_{\sigma_1, \sigma_1'} \nonumber \\
&~&  \delta_{n_1,n_1'}~{ \delta_{n_2' + 2n_3', 
n_2} \over n_2}~ { 1 \over \sqrt{n_3'}}~
 ~ \sum_{\bz({\bar q})}~ 
\sum_{\by({\bar q})}~ \delta_{\bz({\bar q}),\by({\bar q})+ a \bt}.
\ee 

From the helicity non-flip term we get
\be
\langle 3 a' \mid H_{hnf}(2) 
\mid 2 \rangle & = & -g{ 1 \over a^2}
~\sqrt{N}~\frac{1}{V}~\frac{1}{\sqrt{2}}~ { 1 \over \sqrt{4 \pi}}
~\delta_{\sigma_1, \sigma_1'}
\delta_{\sigma_2, \sigma_2'} \nonumber \\
&~&  \delta_{n_2,n_2'}~{ \delta_{n_1'+ 2n_3',
n_1} \over n_1'}~ {1 \over \sqrt{n_3'}} ~ 
~ \sum_t~\sum_{\bz(q)}~
\sum_{\by(q)}~ \delta_{\bz(q),\by(q)- a \bt}
 \nonumber \\
&~&  -g~ { 1 \over a ^2}  ~\sqrt{N} 
~\frac{1}{V}~\frac{1}{\sqrt{2}}~ { 1 \over \sqrt{4 \pi}}~
\delta_{\sigma_2, \sigma_2'}~
\delta_{\sigma_1, \sigma_1'} \nonumber \\
&~&  \delta_{n_1,n_1'}~{ \delta_{n_2' + 2n_3',n_2} \over n_2}~ 
{ 1 \over \sqrt{n_3'}}
~\sum_t~ \sum_{\bz({\bar q})}~
\sum_{\by({\bar q})}~ \delta_{\bz({\bar q}),\by({\bar q})+ a \bt}~
.
\ee 
\subsubsection{To the state $ \mid 3 b \rangle$}
 
From the helicity flip term we get
\be
\langle 3 b' \mid H_{hf2} \mid 2 \rangle & = & { mg \over a} 
 ~\sqrt{N}~\frac{1}{V}~\frac{1}{\sqrt{2}}~ { 1 \over \sqrt{4 \pi}} ~
~\sum_t~\chi^\dagger_{\sigma_1'}~{\hat \sigma}_t ~\chi_{\sigma_1}~
\delta_{\sigma_2, \sigma_2'} \nonumber \\
&~&  \delta_{n_2, n_2'}~
{ \delta_{n_1' + 2 n_3',
n_1} \over n_1}~ { 1 \over \sqrt{n_3'}} 
~\sum_{\bz({q})}~
\sum_{\by({ q})}~ \delta_{\bz({ q}),\by({q})+ a \bt}~
 \nonumber \\
&~&  +~ { mg \over a}  ~\sqrt{N}
~\frac{1}{V}~\frac{1}{\sqrt{2}}~  {1 \over \sqrt{4 \pi}}~
~\sum_t
\chi^\dagger_{-\sigma_2}~{\hat \sigma}_t ~\chi_{-\sigma_2'}~
\delta_{\sigma_1, \sigma_1'} \nonumber \\
&~&  \delta_{n_1, n_1'}~{ \delta_{n_2' + 2 n_3',
n_2} \over  n_2'}~ {1 \over \sqrt{n_3'}} 
~\sum_{\bz({\bar q})}~
\sum_{\by({\bar q})}~ \delta_{\bz({ \bar q}),\by({\bar q})- a \bt}~
.
\ee 
From helicity non-flip term we get
\be
\langle 3 b' \mid H_{hnf}(3) \mid 2 \rangle 
& = & - g{ 1 \over a^2} ~\sqrt{N}
~\frac{1}{V}~\frac{1}{\sqrt{2}}~  {1 \over \sqrt{4 \pi}}~
\delta_{\sigma_1, \sigma_1'}~
\delta_{\sigma_2, \sigma_2'} \nonumber \\
&~&  \delta_{n_2,n_2'}~
{ \delta_{n_1' + 2 n_3',
n_1} \over n_1}~ { 1 \over \sqrt{n_3'}} ~
\sum_t ~~\sum_{\bz({q})}~
\sum_{\by({ q})}~ \delta_{\bz({ q}),\by({q})+ a \bt}~
 \nonumber \\
&~&  -g~ { 1 \over a^2} ~\sqrt{N}
~\frac{1}{V}~\frac{1}{\sqrt{2}}~  { 1 \over \sqrt{4 \pi}}
\delta_{\sigma_2, \sigma_2'}~
\delta_{\sigma_1, \sigma_1'} \nonumber \\
&~&  \delta_{n_1, n_1'}~
{ \delta_{n_2' + 2 n_3',
n_2} \over  n_2'}~ { 1 \over \sqrt{n_3'}} ~ 
\sum_t~~\sum_{\bz({\bar q})}~
\sum_{\by({\bar q})}~ \delta_{\bz({ \bar q}),\by({\bar q})- a \bt}~
.
\ee 
\subsection{Transitions from three particle ($q~ {\bar q} 
~{\rm link}$) state $ \mid 3 a \rangle $}
\subsubsection{To three particle state}
From the free particle term, we get
\be
\langle 3a' \mid H_{free} \mid 3 a \rangle = \Bigg ( m^2  \Big ( { 1
\over n_1} + { 1 \over n_2} \Big ) + { 1 \over 2} \mu^2  { 1 \over n_3}
\Bigg ) {\cal N}_3
\ee
with 
\be
{\cal N}_3 & = &  \delta_{n_1,n_1'}~
\delta_{n_2, n_2'}~ \delta_{n_3,n_3'} 
 ~\delta_{\sigma_1, \sigma_1'}~ 
\delta_{\sigma_2, \sigma_2'}~.
\ee 

Diagonal contribution from the four  
fermion instantaneous term to the three 
particle state vanishes due to the vanishing trace of 
the generators 
of $SU(N)$.

Contribution from the fermion - link instantaneous term  
\be
\langle 3 a' \mid H_{qgc}(1) \mid 3a \rangle & = & -  {g^2  \over 
\pi}
~{ 1 \over
a^2}  C_f ~ \delta_{n_1+ 2 n_3, n_1'+2 n_3'}~
 \delta_{n_2,n_2'} \nonumber \\
&~&~~~ { 1 \over \sqrt{n_3}\sqrt{n_1 - n_1' + 2 n_3}} ~ 
{ (n_1 - n_1'+4n_3 ) \over (n_1 -n_1')^2}~ \frac{1}{\sqrt{2}}
 ~ \delta_{\sigma_1, \sigma_1'}~ \delta_{\sigma_2,
\sigma_2'} \nonumber \\
&~& -  { g^2 \over  \pi} ~{ 1 \over
a^2}  C_f ~ \delta_{n_2 + 2 n_3, n_2'+2 n_3'}~
 \delta_{n_1,n_1'} \nonumber \\
&~&~~~ { 1 \over \sqrt{n_3}\sqrt{n_2- n_2'+ 2 n_3}} ~ 
{ (n_2 - n_2'+4n_3)
 \over (n_2 -n_2')^2}~ \frac{1}{\sqrt{2}}
 ~ \delta_{\sigma_1, \sigma_1'}~ \delta_{\sigma_2,
\sigma_2'}~. \nonumber \\
\ee
Counterterm matrix elements in DLCQ to implement the regulated prescription
for $\frac{1}{(k^+)^2}$
\be
\langle 3 a' \mid H_{CT}(1) \mid 3a \rangle & = &   {g^2  \over
\pi}
~{ 1 \over
a^2}  C_f ~ \delta_{n_1+ 2 n_3, n_1'+2 n_3'}~
 \delta_{n_2,n_2'}~ \delta_{\sigma_1, \sigma_1'}~ \delta_{\sigma_2,
\sigma_2'} \nonumber \\
&~&~~ \Bigg [ \sum_{n_{loop}=1}^{{n_1}_{max}}{ 1 \over \sqrt{n_3}\sqrt{n_1-
n_{loop}+ 2 n_3}} 
\frac{(n_1 - n_{loop}+4n_3)}{(n_1 - n_{loop})^2}\frac{1}{\sqrt{2}}
\nonumber \\
&~&~~ + \sum_{n_{loop}=1}^{{n_2}_{max}} 
{ 1 \over \sqrt{n_3}\sqrt{n_2-
n_{loop}+ 2 n_3}}
\frac{(n_2 - n_{loop}+4n_3)}{(n_2 - n_{loop})^2} \frac{1}{\sqrt{2}}\Bigg ]~
, \label{flct} \nonumber \\
\ee
where ${n_1}_{max}< n_1+ 2 n_3$ and  ${n_2}_{max}< n_2+ 2 n_3$.
 
The contribution from the helicity flip term that conserves particle number
is
\be
\langle 3 a' \mid H_{hf}(1) \mid 3a \rangle & = &
-2 ~{ m \over a}  ~  \delta_{n_1,n_1'}~
 \delta_{n_2,n_2'}~\delta_{n_3, n_3'}  \nonumber \\
&~& ~~~~ \Bigg [ { 1 \over n_1} ~ 
\sum_r ~\chi^\dagger_{\sigma_1'} ~{\hat \sigma_r}~ 
~\chi_{\sigma_{1}} ~ \delta_{\sigma_2, \sigma_2'} ~ +~
{ 1 \over n_2} ~ \sum_r~ \chi^\dagger_{-\sigma_2} {\hat \sigma_r} 
\chi_{-\sigma_{2}'} ~ \delta_{\sigma_1, \sigma_1'} \Bigg ] . \nonumber \\
\ee
The contribution from the helicity non-flip term that conserves particle
number
is
\be
\langle 3 a' \mid H_{hnf}(1) \mid 3a \rangle & = & { 2 \over a^2}  
\left ( { 1 \over n_1} + { 1 \over n_2} \right )
{\cal N}_3 ~.
\ee

\subsubsection{To two particle state}
From the helicity flip term we get
\be
\langle 2' \mid H_{hf2} \mid 3a \rangle & = & 
{ mg \over a}  ~\sqrt{N}~ \frac{1}{V}~ \frac{1}{\sqrt{2}}~
{ 1 \over \sqrt{ 4 \pi}} 
\sum_s~\chi^\dagger_{\sigma_1'}~{\hat \sigma}_s ~\chi_{\sigma_1}~
\delta_{\sigma_2, \sigma_2'} \nonumber \\
&~&  \delta_{n_2, n_2'} ~{ \delta_{n_1', 
n_1 +2
n_3} \over n_1}~ { 1 \over \sqrt{n_3}} ~ 
\sum_{\bz({q})}~
\sum_{\by({ q})}~ \delta_{\bz({ q}),\by({q})+ a \bs}~
 \nonumber \\
&~& ~ +~ { mg \over a} ~\sqrt{N}~ 
~ \frac{1}{V}~ \frac{1}{\sqrt{2}}~
{ 1 \over \sqrt{4 \pi}}~
\sum_s~\chi^\dagger_{-\sigma_2}~{\hat \sigma}_s ~\chi_{-\sigma_2'}~
\delta_{\sigma_1, \sigma_1'} \nonumber \\
&~&  \delta_{n_1, n_1'}~{ \delta_{n_2', n_2+ 2n_3 } \over n_2'}~ { 1 \over
\sqrt{n_3}} ~ 
\sum_{\bz({\bar q})}~
\sum_{\by({ \bar q})}~ \delta_{\bz({ \bar q}),\by({\bar q}) - a \bs}~
.
\ee 
From the helicity non-flip term we get
\be
\langle 2' \mid H_{hnf} \mid 3a \rangle & = & 
-g{ 1 \over a^2}  ~\sqrt{N}~ \frac{1}{V}~ \frac{1}{\sqrt{2}}~
{ 1 \over \sqrt{4 \pi}}~
\delta_{\sigma_1,\sigma_1'}~
\delta_{\sigma_2, \sigma_2'} \nonumber \\
&~&  \delta_{n_2, n_2'}~{ \delta_{n_1', 
n_1 +
2 n_3} \over n_1}~ { 1 \over \sqrt{n_3}} ~ 
\sum_s \sum_{\bz({q})}~
\sum_{\by({ q})}~ \delta_{\bz({ q}),\by({q})+ a \bs}~
 \nonumber \\
&~&  -g~ { 1 \over a^2}  ~\sqrt{N}~ \frac{1}{V}~ \frac{1}{\sqrt{2}}~
{ 1 \over \sqrt{4 \pi}}~
\delta_{\sigma_2,\sigma_2'}~
\delta_{\sigma_1, \sigma_1'} \nonumber \\
&~&  \delta_{n_1, n_1'}~{ \delta_{n_2', n_2+2n_3 
} \over n_2'}~ { 1 \over \sqrt{n_3}} ~
\sum_{\bz({\bar q})}~
\sum_{\by({ \bar q})}~ \delta_{\bz({ \bar q}),\by({\bar q}) - a \bs}~
.
\ee 

\subsection{Transitions from three particle ($q ~{\bar q}~ 
{\rm link}$) state $ \mid 3 b \rangle $}
\subsubsection{To three particle state}
From the free particle term, we get
\be
\langle 3b' \mid H_{free} \mid 3 b \rangle = \Bigg ( m^2  \Big ( { 1
\over n_1} + { 1 \over n_2} \Big ) + { 1 \over 2} \mu^2  { 1 \over n_3}
\Bigg ) {\cal N}_3
\ee
with 
\be
{\cal N}_3 & = & \delta_{n_1, n_1'}~
\delta_{n_2, n_2'}~ \delta_{n_3, n_3'} 
  ~\delta_{\sigma_1, \sigma_1'}~ 
\delta_{\sigma_2, \sigma_2'}~.
\ee 

The diagonal contribution from the four fermion instantaneous term to the
three 
particle state vanishes due to the vanishing trace 
of the generators 
of $SU(N)$.

The contribution from the fermion - link instantaneous term  is
\be
\langle 3 b' \mid H_{qgc}(1) \mid 3b \rangle & = & -  ~{g^2 \over 
\pi}
~{ 1 \over
a^2}  C_f ~ \delta_{n_1 + 2 n_3, n_1' + 2 n_3'}~
 \delta_{n_2, n_2'} \nonumber \\
&~&~~~ 
{ 1 \over \sqrt{n_3}\sqrt{n_1 - n_1' + 2 n_3}} ~
{ (n_1 - n_1'+4n_3 ) \over (n_1 -n_1')^2}~ \frac{1}{\sqrt{2}}
 ~ \delta_{\sigma_1, \sigma_1'}~ \delta_{\sigma_2,
\sigma_2'} \nonumber \\
&~& -  ~{g^2 \over  \pi}~~{ 1 \over
a^2}  C_f ~ \delta_{n_2 +2 n_3, n_2' + 2 n_3'}~
 \delta_{n_1,n_1'} \nonumber \\
&~&~~~ { 1 \over \sqrt{n_3}\sqrt{n_2 - n_2' + 2 n_3}} ~
{ (n_2 - n_2'+4n_3 ) \over (n_2 -n_2')^2}~ \frac{1}{\sqrt{2}}
 ~ \delta_{\sigma_1, \sigma_1'}~ \delta_{\sigma_2,
\sigma_2'}
\ee
Here also we have the counterterm matrix 
elements given in Eq. (\ref{flct}).

The contribution from the helicity flip term that conserves particle number
is
\be
\langle 3 b' \mid H_{hf}(1) \mid 3b \rangle & = &
-2 ~{ m \over a}  ~ \delta_{n_1,n_1'}~
 \delta_{n_2, n_2'}~\delta_{n_3, n_3'}  \nonumber \\
&~& ~~~~ \Bigg [ { 1 \over n_1} ~ 
\sum_r ~\chi^\dagger_{\sigma_1'} {\hat \sigma_r} 
~\chi_{\sigma_{1}} ~ \delta_{\sigma_2 \sigma_2'} ~ +~
{ 1 \over n_2} ~ \sum_r~ \chi^\dagger_{-\sigma_2} {\hat \sigma_r} 
\chi_{-\sigma_{2}'} ~ \delta_{\sigma_1 \sigma_1'} \Bigg ] .
\ee
The contribution from the helicity non-flip term that conserves particle
number
is
\be
\langle 3 b' \mid H_{hnf}(1) \mid 3b \rangle & = & { 2 \over a^2}  
\left ( { 1 \over n_1} + { 1 \over n_2}  \right )
{\cal N}_3 ~.
\ee
\subsubsection{To the two particle state}
From the helicity flip term we get
\be
\langle 2' \mid H_{hf2} \mid 3b \rangle & = & 
{ mg \over a} ~\sqrt{N}~ \frac{1}{V}~ \frac{1}{\sqrt{2}}~
{ 1 \over \sqrt{4 \pi}}~
\sum_s~ \chi^\dagger_{\sigma_1'}~{\hat \sigma}_s ~\chi_{\sigma_1}~
\delta_{\sigma_2, \sigma_2'} \nonumber \\
&~&  \delta_{n_2, n_2'}~{ \delta_{n_1', 
n_1+ 2 
n_3} \over n_1'}~ { 1 \over \sqrt{n_3}} ~ 
\sum_{\bz({q})}~
\sum_{\by({ q})}~ \delta_{\bz({ q}),\by({q})- a \bs}~
 \nonumber \\
&~&  +~ { mg \over a} ~\sqrt{N}~ 
\frac{1}{V}~ \frac{1}{\sqrt{2}}~
{ 1 \over \sqrt{4 \pi}}~
\sum_s~ \chi^\dagger_{-\sigma_2}~{\hat \sigma}_s ~\chi_{-\sigma_2'}~
\delta_{\sigma_1, \sigma_1'} \nonumber \\
&~&  \delta_{n_1, n_1'}~{ \delta_{n_2', 2 n_3 +
n_2} \over n_2}~ { 1 \over \sqrt{n_3}} ~ 
\sum_{\bz({\bar q})}~
\sum_{\by({ \bar q})}~ \delta_{\bz({ \bar q}),\by({\bar q}) + a \bs}~
~.
\ee 
From the helicity non-flip term we get
\be
\langle 2' \mid H_{hnf} \mid 3b \rangle & = & 
-g{ 1 \over a^2} ~\sqrt{N}~ \frac{1}{V}~ \frac{1}{\sqrt{2}}
{ 1 \over \sqrt{4 \pi}}~
\delta_{\sigma_1,\sigma_1'}~
\delta_{\sigma_2, \sigma_2'} \nonumber \\
&~&  \delta_{n_2, n_2'}~{ \delta_{n_1', 
n_1 + 2
n_3} \over n_1'}~ { 1 \over \sqrt{ n_3}} ~ 
\sum_s~\sum_{\bz({q})}~
 \sum_{\by({ q})}~ \delta_{\bz({ q}),\by({q})- a \bs}~
 \nonumber \\
&~&  -g~ { 1 \over a^2} ~\sqrt{N}~ \frac{1}{V}~ \frac{1}{\sqrt{2}}
{ 1 \over \sqrt{4 \pi}}~
\delta_{\sigma_2,\sigma_2'}~
\delta_{\sigma_1, \sigma_1'} \nonumber \\
&~&  \delta_{n_1, n_1'}~{ \delta_{n_2',  n_2 + 2 n_3 
} \over n_2}~ { 1 \over \sqrt{n_3}} ~ 
\sum_s~\sum_{\bz({\bar q})}~
 ~ \sum_{\by({ \bar q})}~ \delta_{\bz({ \bar q}),\by({q})+ a \bs}~ 
~.
\ee 
\section{Symmetric derivatives and Wilson term: Matrix elements in DLCQ}
\label{appd}
In this section, we list only those matrix elements that differ from the
forward-backward case.
\subsection{Transitions from the two particle state}
\subsubsection{To the state $ \mid 3 a \rangle$}
{ Helicity flip}:
\be
\langle  3a' \mid  P^-_{whf} \mid 2  \rangle & = &
\left (m + 4 \frac{\kappa}{a} \right) \frac{1}{2a} ~ \sqrt{N} 
~\frac{1}{V}~\frac{1}{{\sqrt{2}}}~{ 1 \over
\sqrt{4 \pi}} 
\sum_t ~ \chi^\dagger_{\sigma_{1}'}~
{\hat \sigma}_t ~\chi_{\sigma_{1}} ~\delta_{\sigma_{2}, \sigma_{2}^{'}}
 \nonumber \\
&~& \sum_{\by(q)}~
\sum_{\bz(q)} ~
 \delta_{\bz(q), \by(q)-a \bt}  \nonumber \\
&~&~~~ { 1 \over \sqrt{n_3^{'}}}\left 
( { 1 \over n_1} - {1 \over n_1^{'}} \right )~
 \delta_{n_2,n_2^{'}} ~ \delta_{n_1^{'} + 2 n_3^{'},
n_1} \nonumber \\
&+& \left(m + 4 \frac{\kappa}{a}\right)\frac{1}{2a} ~ \sqrt{N} 
~\frac{1}{V}~\frac{1}{{\sqrt{2}}}
{ 1 \over
\sqrt{4
\pi}}~
\sum_t~\chi^\dagger_{-\sigma_{2}}~
{\hat \sigma}_t ~\chi_{-\sigma_{2}'}~ \delta_{\sigma_{1}, 
\sigma_{1}^{'}}
 \nonumber \\
&~& \sum_{\by({\bar q})} 
~\sum_{\bz({\bar q})} 
~ \delta_{\bz({\bar q}), \by({\bar q})+ a \bt} 
\nonumber \\
&~&~~~ { 1 \over \sqrt{n_3^{'}}}
\left ( { 1 \over n_2^{'}} - {1 \over n_2^{}} \right )~
 \delta_{n_1,n_1^{'}}~  \delta_{n_2^{'} + 2 n_3^{'},
n_2}~. 
\ee
{ Helicity non-flip}:

\be
\langle 3a' \mid  P^-_{wnf1} \mid 2 \rangle & = &
 -\left(m + 4 \frac{\kappa}{a} \right) \frac{\kappa}{a} ~ \sqrt{N} 
~\frac{1}{V}~\frac{1}{{\sqrt{2}}}
{ 1 \over \sqrt{4 \pi}}
 \delta_{\sigma_{2}, \sigma_{2}^{'}}~\delta_{\sigma_{1}, 
\sigma_{1}^{'}}
 \nonumber \\
&~& \sum_t~ \sum_{\by(q)}~
\sum_{\bz(q)} 
\delta_{\bz(q), \by(q)-a \bt}  \nonumber \\
&~&~~~ { 1 \over \sqrt{n_3^{'}}}\left ( { 1 \over n_1} + {1 \over
n_1^{'}} \right )~
 \delta_{n_2,n_2^{'}} ~ \delta_{n_1^{'} + 2 n_3^{'},
n_1} \nonumber \\
&-& \left(m + 4 \frac{\kappa}{a} \right)\frac{\kappa}{a} ~ \sqrt{N} 
~\frac{1}{V}~\frac{1}{{\sqrt{2}}}
{ 1 \over \sqrt{4 \pi}}
 \delta_{\sigma_{2}, \sigma_{2}^{'}~}\delta_{\sigma_{1}, 
\sigma_{1}^{'}}
 \nonumber \\
&~& \sum_t~ \sum_{\by({\bar q})}~
~ \sum_{\bz({\bar q})} 
~~~\delta_{\bz({\bar q}), \by({\bar q})+a \bt} 
\nonumber \\
&~&~~~ { 1 \over \sqrt{n_3^{'}}}\left ( { 1 \over n_2^{'}} + {1 \over
n_2^{}} \right )~
 \delta_{n_1,n_1^{'}}~  \delta_{n_2^{'} + 2n_3^{'},
n_2}~. 
\ee
\subsubsection{To the state $ \mid 3 b \rangle $}
{ Helicity flip}:
 \be
 \langle 3b' \mid  P^-_{whf} \mid 2  \rangle
&=& \left(m + 4 \frac{\kappa}{a}\right) \frac{1}{2a} ~ \sqrt{N} 
\frac{1}{V} \frac{1}{\sqrt{2}}
{1 \over \sqrt{4 \pi}}~
\sum_t ~\chi^\dagger_{\sigma_{1}'}~
{\hat \sigma}_t ~\chi_{\sigma_{1}} \delta_{\sigma_{2}, \sigma_{2}^{'}}
 \nonumber \\
&~& \sum_{\by(q)}~
\sum_{\bz(q)}~ 
~~~  \delta_{\bz(q), \by(q)+a \bt} 
\nonumber \\
&~&~~~ { 1 \over \sqrt{n_3^{'}}}\left (- { 1 \over n_1} + {1 \over
n_1^{'}} \right )~
 \delta_{n_2,n_2^{'}}  ~\delta_{n_1^{'} + 2 n_3^{'},
n_1}~ \nonumber \\
&+& \left(m + 4 \frac{\kappa}{a}\right)\frac{1}{2a} ~ \sqrt{N} 
\frac{1}{V} \frac{1}{\sqrt{2}}
{ 1 \over \sqrt{4 \pi}} 
\sum_t~ \chi^\dagger_{-\sigma_{2}}~
{\hat \sigma}_t ~\chi_{-\sigma_{2}'} ~\delta_{\sigma_{1}, 
\sigma_{1}^{'}}
 \nonumber \\
&~& \sum_{\by({\bar q}}~ \sum_{\bz({\bar q})}
~~~  \delta_{\bz({\bar q}), \by({\bar q}) -a \bt} 
\nonumber \\
&~&~~~ { 1 \over \sqrt{n_3^{'}}}
  \left ( -{ 1 \over n_2^{'}} + {1 \over n_2} \right )~
 ~\delta_{n_1, n_1^{'}} ~ \delta_{n_2^{} + 2 n_3^{},
n_2^{'}}~. 
\ee
{ Helicity non-flip}:
\be
 \langle 3a' \mid P^-_{wnf1} \mid 2  \rangle 
&=&-\left(m + 4 \frac{\kappa}{a}\right) \frac{\kappa}{a} ~ \sqrt{N} 
\frac{1}{V} \frac{1}{\sqrt{2}}
{ 1 \over \sqrt{4 \pi}}
 \delta_{\sigma_{2}, \sigma_{2}^{'}} ~\delta_{\sigma_{1}, \sigma_{1}^{'}}
 \nonumber \\
&~& \sum_t~\sum_{\by(q)}~
~\sum_{\bz(q)}~ 
~~~  \delta_{\bz(q), \by(q)+a \bt} 
\nonumber \\
&~&~~~ { 1 \over \sqrt{n_3^{'}}}
 \left ( { 1 \over n_1} + {1 \over n_1^{'}} 
\right )~
 \delta_{n_2, n_2^{'}} ~\delta_{n_1^{'} + 2 n_3^{'},
n_1}~
 \nonumber \\
&-& \left(m + 4 \frac{\kappa}{a}\right)\frac{\kappa}{a} ~ \sqrt{N} 
\frac{1}{V} \frac{1}{\sqrt{2}}~
{ 1 \over \sqrt{4 \pi}}
 \delta_{\sigma_{1}, \sigma_{1}^{'}} ~\delta_{\sigma_{2}, \sigma_{2}^{'}}
 \nonumber \\
&~& \sum_t ~ \sum_{\by({\bar q})}
~ \sum_{\bz({\bar q})}  
~~~ \delta_{\bz({\bar q}), \by({\bar q}) - a \bt} 
\nonumber \\
&~&~~~ { 1 \over \sqrt{n_3^{'}}}     \left ( { 1 \over n_2^{'}} + {1 \over
n_2} \right )~
~ \delta_{n_1,n_1^{'}} ~ \delta_{n_2^{'} + 2 n_3^{'},
n_2}~. 
\ee

\subsection{Transitions from three particle state $\mid 3 a \rangle $ to two
particle state}
{ Helicity flip}:
\be
\langle 2' \mid P^-_{whf} \mid 3a  \rangle
&=&\left(m + 4 \frac{\kappa}{a} \right) \frac{1}{2a}  \sqrt{N} 
\frac{1}{V} \frac{1}{\sqrt{2}}~
\frac{1}{\sqrt{4 \pi}}
\sum_s~ \chi^\dagger_{\sigma_{1}'}~
{\hat \sigma}_s ~\chi_{\sigma_{1}}~ \delta_{\sigma_{2}, \sigma_{2}^{'}}
 \nonumber \\
&~& \sum_{\bz(q)}
\sum_{\bz(q)} 
~~~ \delta_{\bz(q),\by(q) + a \bs)} 
\nonumber \\
&~&~~~ { 1 \over \sqrt{n_3^{}}} \left ( { 1 \over n_1^{'}} - {1 \over
n_1^{}} \right )~
 \delta_{n_2,n_2^{'}}~  \delta_{n_1 + 2 n_3,
n_1^{'}}~ \nonumber \\
& + & \left(m + 4 \frac{\kappa}{a}\right)\frac{1}{2a} \sqrt{N} 
\frac{1}{V} \frac{1}{\sqrt{2}}~
{1 \over \sqrt{4 \pi}}
\sum_s ~ \chi^\dagger_{-\sigma_{2}}~
{\hat \sigma}_s ~\chi_{-\sigma_{2}'}~ \delta_{\sigma_{1}, \sigma_{1}^{'}}
 \nonumber \\
&~& \sum_{\bz({\bar q})}
\sum_{\by({\bar q})} 
~~ \delta_{\bz({\bar q}),
\by({\bar q}) - a \bs)} 
\nonumber \\
&~&~~~ { 1 \over \sqrt{n_3}}\left ( { 1 \over n_2} - {1 \over
n_2^{'}} \right )~
 \delta_{n_1^{'}, n_1} ~ \delta_{n_2^{'} + 2 n_3^{'},
n_2}~. 
\ee

{ Helicity non-flip}:

\be
\langle 2' \mid  P^-_{wnf1} \mid 3a \rangle
&=& -\left(m + 4 \frac{\kappa}{a}\right) \frac{\kappa}{a}  \sqrt{N} 
\frac{1}{V} \frac{1}{\sqrt{2}}~
{ 1 \over \sqrt{4 \pi}}
 \delta_{\sigma_2, \sigma_{2}^{'}} ~\delta_{\sigma_1, \sigma_{1}^{'}}
 \nonumber \\
&~& \sum_s~ \sum_{\bz(q)}
\sum_{\by(q)}
~~~  \delta_{\bz(q),\by(q) + a \bs)} 
\nonumber \\
&~&~~~ { 1 \over \sqrt{n_3}}\left ( { 1 \over n_1} + {1 \over
n_1^{'}} \right )~
 \delta_{n_2,n_2^{'}} ~ \delta_{n_1 + 2 n_3,
n_1^{'}}~ \nonumber \\
&-& \left(m + 4 \frac{\kappa}{a}\right)\frac{\kappa}{a}  \sqrt{N} 
\frac{1}{V} \frac{1}{\sqrt{2}}~
{ 1 \over \sqrt{4 \pi}} 
 \delta_{\sigma_2, \sigma_{2}^{'}}\delta_{\sigma_1, \sigma_{1}^{'}}
\nonumber \\
&~& \sum_{\bz({\bar q})} ~
\sum_{\by({\bar q})} ~
~\delta_{\bz({\bar q}),\by({\bar q}) + a \bs)} 
\nonumber \\
&~&~~~ { 1 \over \sqrt{n_3}}\left ( { 1 \over n_2^{'}} + {1 \over
n_2} \right )~
 ~\delta_{n_1,n_1^{'}} ~ \delta_{n_2^{} + 2 n_3^{},
n_2^{'}}~.
\ee

\subsection{Transitions from three particle state $\mid 3 b \rangle $ to two 
particle state}

{ Helicity flip}:
\be
\langle 2' \mid P^-_{whf} \mid 3b \rangle
 &=&\left(m + 4 \frac{\kappa}{a}\right) \frac{1}{2a}  \sqrt{N} 
\frac{1}{V} \frac{1}{\sqrt{2}}~
{ 1 \over \sqrt {4 \pi}}
\sum_s ~ \chi^\dagger_{\sigma_{1}'}~
{\hat \sigma}_s ~\chi_{\sigma_{1}}~ \delta_{\sigma_2, \sigma_{2}^{'}}
 \nonumber \\
&~& \sum_{\bz(q)}~
\sum_{\by(q)}~ 
~ \delta_{\bz(q),\by(q) - a \bs)}
\nonumber \\
&~&~~~ { 1 \over \sqrt{n_3}}\left ( { 1 \over n_1} - {1 \over
n_1^{'}} \right )~
 \delta_{n_2,n_2^{'}}~  \delta_{n_1 + 2 n_3,
n_1^{'}}~ \nonumber \\
&+& \left(m + 4 \frac{\kappa}{a}\right)\frac{1}{2a}  \sqrt{N} 
\frac{1}{V} \frac{1}{\sqrt{2}}~
{ 1 \over \sqrt{4 \pi}}
\sum_s ~ \chi^\dagger_{-\sigma_{2}}~
{\hat \sigma}_s ~\chi_{-\sigma_{2}'}~ \delta_{\sigma_1, \sigma_{1}^{'}}
 \nonumber \\
&~& \sum_{\bz({\bar q})} 
\sum_{\by({\bar q})}  
~\delta_{\bz({\bar q}),\by({\bar q}) + a \bs)}
\nonumber \\
&~&~~~ { 1 \over \sqrt{n_3}}\left ( { 1 \over n_2^{'}} - {1 \over
n_2} \right )~
 \delta_{n_1,n_1^{'}}~  \delta_{n_2^{} + 2 n_3^{},
n_2^{'}}~. 
\ee

{ Helicity non-flip}:
\be
\langle 2' \mid  P^-_{whf} \mid 3b \rangle
&=&-\left(m + 4 \frac{\kappa}{a}\right) \frac{\kappa}{a}  \sqrt{N} 
\frac{1}{V} \frac{1}{\sqrt{2}}~
{ 1 \over \sqrt{4 \pi}} 
 \delta_{\sigma_2, \sigma_{2}^{'}}~\delta_{\sigma_1, \sigma_{1}^{'}}
 \nonumber \\
&~& \sum_s~ \sum_{\bz(q)} 
~\sum_{\by(q)}~
 \delta_{\bz(q),\by(q) - a \bs)} 
\nonumber \\
&~&~~~ { 1 \over \sqrt{n_3}}\left ( { 1 \over n_1} + {1 \over
n_1^{'}} \right )~
 \delta_{n_2,n_2^{'}} ~ \delta_{n_1 + 2 n_3,
n_1^{'}}~ \nonumber \\
&-& \left(m + 4 \frac{\kappa}{a}\right)\frac{\kappa}{a}  \sqrt{N} 
\frac{1}{V} \frac{1}{\sqrt{2}}~
{ 1 \over \sqrt{4 \pi}}
 \delta_{\sigma_2, \sigma_{2}^{'}}~\delta_{\sigma_1, \sigma_{1}^{'}}
 \nonumber \\
&~& \sum_s~ \sum_{\bz({\bar q})}
~\sum_{\by({\bar q})}~~ 
~\delta_{\bz({\bar q}),\by({\bar q}) + a \bs)} 
\nonumber \\
&~&~~~ { 1 \over \sqrt{n_3}}\left ( { 1 \over n_2^{'}} + {1 \over
n_2} \right )~
 \delta_{n_1,n_1^{'}} ~ \delta_{n_2^{} + 2 n_3^{},
n_2^{'}}~. 
\ee 
\section{Self energy counterterms} \label{APcounter}
In this section we list the self energy counterterms. 
\subsubsection{Symmetric derivatives case}
The counterterm for self energy for a quark or an antiquark with
longitudinal
momentum $n_1$ due to double helicity flip hops
\be
CT_1 = \frac{2}{n_1}\sum_{n_{1}'=1}^{n_{1}} \frac{1}{n_{1}'}
\frac{(n_1 - n_1')^2}{\mu^2 n_1 n_1' + m^2 (n_1 - n_1')^2}.
\ee
The counterterm for self energy for a quark or an antiquark with
longitudinal 
momentum $n_1$ due to double helicity non-flip hops
\be
CT_2 = \frac{2}{n_1}\sum_{n_{1}'=1}^{n_{1}} \frac{1}{n_{1}'}
\frac{(n_1 + n_1')^2}{\mu^2 n_1 n_1' + m^2 (n_1 - n_1')^2}.
\ee

\subsubsection{Forward and backward derivative case}
 In this case we have three types of contributions: (1) helicity flip acting
twice, (2) helicity non-flip acting twice and (3) interference of helicity
flip and helicity non-flip hops. The first two are diagonal in helicity
space but the last one is off-diagonal in helicity space.

The transition from state $ \mid 2  \rangle $ to state $ \mid 3 a \rangle $
and
back due to a quark hop gives rise to longitudinal infrared divergence. In
this case the
counterterm due to double helicity flip is
\be 
CT_3 = 2 \sum_{n_{1}'=1}^{n_{1}} \frac{1}{n_{1}'}
\frac{n_1}{\mu^2 n_1 n_1' + m^2 (n_1 - n_1')^2}. \label{ctq}
\ee
The counterterm due to double helicity non-flip is the same without the
factor
of 2. The transition from state $ \mid 2  \rangle $ to state $ \mid 3 b
\rangle $ and 
back due to a quark hop does not give rise to longitudinal infrared
divergence.
Similarly the transition from state $ \mid 2  \rangle $ to state $ \mid 3 a
\rangle $ and 
back due to an antiquark hop does not give rise to longitudinal infrared
divergence.
The transition from state $ \mid 2  \rangle $ to state $ \mid 3 b \rangle $
and 
back due to an antiquark hop gives rise to longitudinal infrared divergence
which requires  counterterms the explicit forms of which are the same as in
the quark case for the transition from  $ \mid 2  \rangle $ to state $ \mid
3
a \rangle $.
Lastly we consider counterterms for self energy contributions arising from
the interference of helicity flip and helicity non-flip hopping. The
counterterms have the same structure as in the case of helicity non-flip
transitions accompanied by the following extra factors.
Since we have two possibilities namely helicity flip followed by helicity
non-flip and vice versa and these two contributions are the same, we get a
factor of two. We also get a  factor 
$ \chi^\dagger_{s'} {\hat \sigma^\perp} \chi_{s}$
where $s (s')$ is the initial (final) helicity and ${\hat \sigma^1} =
\sigma^2$, ${\hat \sigma^2} =
- \sigma^1$. 


 \centerline{\large{\bf  List of Publications}}
\begin{dinglist}{10}
\item 1. {\bf Ab initio results for the broken phase of scalar light front field theory}\\
Dipankar Chakrabarti, A. Harindranath, L. Martinovi\v c, G. B.
  Pivovarov and J. P. Vary\\
{\it hep-th/0310290, submitted for publication}.
\item 2. {\bf Kinks in discrete light cone quantization}\\
Dipankar Chakrabarti, A. Harindranath, L. Martinovi\v c and J. P. Vary\\
{\it hep-th/0309263, submitted for publication}.
\end{dinglist}
\begin{dinglist}{43}
\item 3. {\bf A study of $q{\bar q}$ states in transverse lattice QCD using
alternative fermion formulations }\\
Dipankar Chakrabarti, A. Harindranath and J. P. Vary\\
{\it hep-ph/0309317, to be published in Phys. Rev. D}.
\item 4. {\bf Fermions on the light front transverse lattice}\\
Dipankar Chakrabarti, Asit K. De and A. Harindranath\\
{\it Phys. Rev. D67(2003), 076004;
hep-th/0211145}.
\item 5. {\bf Mesons in (2+1) dimensional light front QCD. II. Similarity
renormalization approach}\\
Dipankar Chakrabarti and A. Harindranath\\
{\it Phys. Rev. D65(2002), 045001;
hep-th/0110156}.
\item 6. {\bf Mesons in (2+1)-dimensional
light-front QCD: Investigation of a Bloch effective Hamiltonian}\\
 Dipankar Chakrabarti and A. Harindranath\\
{\it Phys. Rev. D64(2001),105002;
 hep-th/0107188}.
\end{dinglist}
\begin{dinglist}{10}
\item 7. {\bf Quark transversity
distribution in perturbative QCD: light-front Hamiltonian approach}\\
Asmita Mukherjee and Dipankar Chakrabarti\\
{\it Phys. Lett. B506(2001), 283;
hep-ph/0102003}.
\item 8. {\bf A numerical experiment
in DLCQ: microcausality, continuum limit and all that}\\
Dipankar Chakrabarti, Asmita Mukherjee, Rajen Kundu, A. Harindranath\\
{\it Phys. Lett. B480(2000), 409;
hep-th/9910108}.
\end{dinglist}
\vskip0.2cm
\begin{dinglist}{43}
\item included in the thesis\footnote{Though the {\it official} spelling of my name is
{\bf Dipankar Chakravorty}, the papers are published with the spelling 
{\bf Dipankar Chakrabarti}.}.
\end{dinglist}

\end{document}